\documentclass[11pt,a4paper]{article}

\usepackage{amssymb}

\usepackage[dvips]{graphicx}

\begin{document}

\title{\bf The NSVZ $\beta$-function and the Schwinger--Dyson
equations for ${\cal N}=1$ SQED with $N_f$ flavors, regularized by
higher derivatives}

\author{K.V.Stepanyantz}

\maketitle

\begin{center}

{\em Moscow State University, Faculty of Physics,\\
Department of Theoretical Physics.\\
$119991$, Moscow, Russia}

\end{center}

\begin{abstract}
The effective diagram technique based on the Schwinger--Dyson
equations is constructed for ${\cal N}=1$ SQED with $N_f$ flavors,
regularized by higher derivatives. Using these effective diagrams,
it is possible to derive the exact NSVZ relation between the
$\beta$-function and the anomalous dimension of the matter
superfields exactly in all loops, if the renormalization group
functions are defined in terms of the bare coupling constant. In
particular, we verify that all integrals which give the
$\beta$-function defined in terms of the bare coupling constant
are integrals of double total derivatives and prove some
identities relating Green functions.
\end{abstract}

\unitlength=1cm



\section{Introduction}
\hspace{\parindent}

The existence of ultraviolet divergences is a long standing
problem of quantum field theory. An important step towards solving
this problem is a discovery of supersymmetry
\cite{Golfand:1971iw,Volkov:1973ix}. It is well known that the
behavior of supersymmetric theories in the ultraviolet region is
better due to non-renormalization theorems. In particular, the
${\cal N}=4$ supersymmetric Yang--Mills theory is finite
\cite{Grisaru:1982zh,Mandelstam:1982cb,Brink:1982pd,Howe:1983sr}.
Divergences in ${\cal N}=2$ supersymmetric Yang--Mills theories
exist only in the one-loop approximation \cite{Howe:1983wj}. Even
in ${\cal N}=1$ supersymmetric theories the superpotential is not
renormalized \cite{Grisaru:1979wc}. However, the $\beta$-function
in ${\cal N}=1$ supersymmetric Yang--Mills theories receives
quantum corrections in all orders. Nevertheless, this
$\beta$-function is related with the anomalous dimension of the
matter superfields. This relation is called the exact Novikov,
Shifman, Vainshtein, and Zakharov (NSVZ) $\beta$-function
\cite{Novikov:1983uc,Jones:1983ip,Novikov:1985rd}. In the original
papers this $\beta$-function was obtained using arguments based on
the structure of instanton contributions (for review, see
\cite{Shifman:1999mv}) or on the supermultiplet structure of
anomalies. In particular, in the lowest orders the relation
between the $\beta$-function and anomalies was investigated in
\cite{Jones:1983ip,Jones:1983vk,Jones:1984cx,Vainshtein:1985ed}
and exactly in all orders in
\cite{Jones:1983ip,Novikov:1985mf,Shifman:1986zi}. This was done
using the Adler--Bardeen theorem \cite{Adler:1969er} for the axial
anomaly, a relation between the anomaly of energy-momentum tensor
trace and a $\beta$-function, and a supermultiplet structure of
anomalies. Another derivation of the exact NSVZ $\beta$-function
based on anomalies was made in \cite{Grisaru:1983hc}. This
$\beta$-function was also obtained in \cite{Kraus:2002nu} using
the non-renormalization theorem for the topological term. In
\cite{ArkaniHamed:1997mj} the rescaling anomaly is used for
explanation of the higher order corrections to the NSVZ
$\beta$-function.

In this paper we consider the ${\cal N}=1$ supersymmetric
electrodynamics (SQED) with $N_f$ flavors, for which the NSVZ
$\beta$-function is written as
\cite{Vainshtein:1986ja,Shifman:1985fi}

\begin{equation}
\beta(\alpha) = \frac{\alpha^2
N_f}{\pi}\Big(1-\gamma(\alpha)\Big).
\end{equation}

\noindent The NSVZ $\beta$-function can be compared with the
results of explicit calculations in the lowest orders of the
perturbation theory. In the one- and two-loop approximations a
$\beta$-function for supersymmetric theories was first calculated
in \cite{Ferrara:1974pu} and \cite{Jones:1977zr}, respectively,
using the dimensional regularization
\cite{'tHooft:1972fi,Bollini:1972ui,Ashmore:1972uj,Cicuta:1972jf}.
However, because the dimensional regularization breaks
supersymmetry \cite{Delbourgo:1974az}, most calculations in
supersymmetric theories were made with the dimensional reduction
proposed in \cite{Siegel:1979wq} (see \cite{Mihaila:2013wma} for a
recent review). Using this regularization and the
$\overline{\mbox{DR}}$-scheme, which is a modification of the
$\overline{\mbox{MS}}$-scheme \cite{Bardeen:1978yd}, the
$\beta$-function of the ${\cal N}=1$ supersymmetric Yang--Mills
theory with matter was calculated in the three-
\cite{Avdeev:1981ew,Jack:1996cn} and four-loop
\cite{Harlander:2006xq} approximations. The results coincide with
the NSVZ $\beta$-function in one- and two-loop approximations
(where the $\beta$-function is scheme-independent). In the higher
loops the NSVZ $\beta$-function can be obtained after a special
redefinition of the coupling constant
\cite{Jack:1996cn,Jack:1996vg}. (Using such a redefinition the
result for the four-loop $\beta$-function was correctly predicted
in \cite{Jack:1998uj} before the explicit calculation made in
\cite{Harlander:2006xq}.)

However, the regularization by the dimensional reduction is not
self-consistent \cite{Siegel:1980qs}. The inconsistencies can be
removed only if one breaks the manifest supersymmetry
\cite{Avdeev:1982np,Avdeev:1982xy}. Therefore, supersymmetry can
be broken by higher orders quantum corrections
\cite{Avdeev:1982np}. This was verified explicitly: in the ${\cal
N}=2$ supersymmetric Yang--Mills theory without matter superfields
obtaining the three-loop $\beta$-function by different methods
(using various Green functions) gives different results
\cite{Avdeev:1982np,Velizhanin:2008rw}. (The calculation made in
\cite{Velizhanin:2008rw} showed that this does not take place for
the ${\cal N}=1$ supersymmetric Yang--Mills theory at three-loop
level, as it was argued in \cite{Avdeev:1982np}.) In the ${\cal
N}=4$ supersymmetric Yang--Mills theory the dimensional reduction
does not break supersymmetry even in the four-loop approximation
\cite{Velizhanin:2010vw}. Nevertheless, with the dimensional
reduction one can expect breaking of supersymmetry by quantum
corrections in higher loops (see table 1 in \cite{Avdeev:1982np}).

Although the dimensional reduction is the most popular
regularization for calculations in supersymmetric theories, other
methods are also used. For example, using a method based on the
operator product expansion two-loop $\beta$-functions of scalar,
spinor, and ${\cal N}=1$ supersymmetric electrodynamics were
calculated in \cite{Shifman:1985tj}. With the differential
renormalization \cite{Freedman:1991tk} a two-loop $\beta$-function
of the ${\cal N}=1$ supersymmetric Yang-Mills theory was found in
\cite{Mas:2002xh}. Another regularization used for calculations in
supersymmetric theories is the higher covariant derivative
regularization proposed in \cite{Slavnov:1971aw,Slavnov:1972sq}.
This regularization was subsequently generalized to the
supersymmetric case in \cite{Krivoshchekov:1978xg,West:1985jx}. It
can be also applied in ${\cal N}=2$ supersymmetric theories
\cite{Krivoshchekov:1985pq,Buchbinder:2014wra}. The higher
covariant derivative regularization leads to loop integrals which
have complicated structure. That is why this regularization is not
frequently used for explicit calculations. However, it is quite
possible. For example, a one-loop $\beta$-function of the
(non-supersymmetric) Yang--Mills theory was calculated in
\cite{Martin:1994cg}. After essential corrections introduced in
the subsequent papers \cite{Asorey:1995tq,Bakeyev:1996is} the
well-known one-loop result \cite{Gross:1973id,Politzer:1973fx} was
reobtained (although the original calculation made in
\cite{Martin:1994cg} gave a different result). One can prove that
at the one-loop level the higher covariant derivative
regularization always produces the same result for a
$\beta$-function as the dimensional regularization
\cite{Pronin:1997eb}.

Quantum corrections obtained with the higher covariant derivative
regularization in supersymmetric theories appear to have an
interesting feature: the $\beta$-function defined in terms of the
bare coupling constant is given by integrals of total derivatives
with respect to a loop momentum
\cite{Soloshenko:2003nc,Pimenov:2009hv,Stepanyantz:2011wq,Stepanyantz:2011zz}
and even by integrals of double total derivatives
\cite{Smilga:2004zr,Stepanyantz:2011bz,Stepanyantz:2012zz,Stepanyantz:2012us}.
Thus, it is possible to calculate one of the loop integrals
analytically and reduce a number of the integrations over loop
momentums. At least, in the Abelian case this allows to prove that
the $\beta$-function and the anomalous dimension of the matter
superfields defined in terms of the bare coupling constant satisfy
the NSVZ relation
\cite{Stepanyantz:2011jy,Kataev:2013eta,Kataev:2013csa}. (For a
fixed regularization) these renormalization group functions are
scheme independent (see, e.g., \cite{Kataev:2013csa}), so that the
NSVZ $\beta$-function is obtained for an arbitrary renormalization
prescription. However, if the renormalization group functions are
defined by the standard way in terms of the renormalized coupling
constant, they depend on the subtraction scheme
\cite{Vladimirov:1975mx}. In this case the NSVZ $\beta$-function
is obtained in a special subtraction scheme. If the theory is
regularized by higher derivatives, such a scheme can be obtained
in all orders by imposing simple boundary conditions on the
renormalization constants \cite{Kataev:2013eta,Kataev:2013csa}. So
far there is no similar prescription in the case of using the
dimensional reduction, and the NSVZ scheme should be constructed
in each order of the perturbation theory after calculating the
renormalization group functions.

Thus, using the higher covariant derivative regularization one can
naturally construct the scheme in which the $\beta$-function
coincides with the exact NSVZ $\beta$-function at least in the
Abelian case. Certainly, it is desirable to generalize the results
to the non-Abelian case. However, in the non-Abelian case the
calculations with the higher covariant derivative regularization
were performed only in the one- and two-loop approximations, where
the $\beta$-function is scheme-independent. Nevertheless, in both
cases the $\beta$-function appears to be given by integrals of
double total derivatives and coincide with the NSVZ expression.
This allows to suggest that the structure of quantum corrections
in the non-Abelian case is similar to the case of ${\cal N}=1$
SQED. However, the method used in \cite{Stepanyantz:2011jy} (which
was proposed in \cite{Stepanyantz:2005wk}) is not convenient for
generalizing the results to the non-Abelian case. Possibly, using
this method one can prove the factorization of integrands into the
double total derivatives, but obtaining the exact $\beta$-function
by this method seems to be a very complicated problem. Even in
${\cal N}=1$ SQED for this purpose it is necessary to compare
coefficients of different Feynman diagrams
\cite{Stepanyantz:2011jy}. From the other side, the NSVZ
expression naturally appears in case of using another method
proposed in \cite{Stepanyantz:2004sg}. It is based on substituting
solutions of the Ward (or Slavnov--Taylor) identities into the
Schwinger--Dyson equations. The Schwinger--Dyson equations can be
used for making calculations in a certain approximation as in
\cite{Broadhurst:1999he}, where the four-loop anomalous dimension
of quenched QED was obtained by this method. However, they can
also allow to find results which are exact in all orders. In
particular, in Abelian supersymmetric theories by using the
Schwinger--Dyson equation it is possible to present the two-point
Green function of the gauge superfield as a sum of two effective
diagrams. One of them is related with the two-point Green
functions of the matter superfields and gives the exact NSVZ
$\beta$-function. The second effective diagram cannot be expressed
in terms of these two-point Green functions. However, calculations
made in \cite{Pimenov:2007md,Pimenov:2006cu} show that this
effective diagram (in the limit of the vanishing external
momentum) is always given by integrals of total derivatives and
vanishes, as it was suggested in \cite{Stepanyantz:2004sg}. This
feature was not so far explained within a method based on using
the Schwinger--Dyson equations that is the main obstacle for
deriving the exact NSVZ $\beta$-function by this method. Thus, it
is desirable to understand, why the second effective diagram
vanishes, especially because it seems that the considered
technique can be generalized to the the non-Abelian case.
(Vanishing of this diagram can be interpreted as a special
identity relating some Green functions.)

In this paper we complete derivation of the NSVZ $\beta$-function
started in \cite{Stepanyantz:2004sg}. In particular, we directly
prove that the second effective diagram vanishes, and the
$\beta$-function is given by integrals of double total
derivatives. The method used in this paper seems to be applicable
in the non-Abelian case. That is why throughout the paper we try
to use the notation, which can be also used for non-Abelian
supersymmetric Yang--Mills theories.

The paper is organized as follows: In Section \ref{Section_SQED}
we describe ${\cal N}=1$ SQED with $N_f$ flavors, regularized by
higher derivatives, and introduce the notation. In Section
\ref{Section_Effective_Diagrams} we write the Schwinger--Dyson
equations for the considered theory and present the
$\beta$-function (and its derivative with respect to a specially
introduced parameter $g$) as a sum of effective diagrams. In
Section \ref{Section_Commutators} we prove that the
$\beta$-function is given by integrals of total derivatives and is
equal to the exact NSVZ $\beta$-function. Also in this section we
present a direct proof of a special identity for Green functions,
which was proposed in \cite{Stepanyantz:2004sg}. In Section
\ref{Section_Double_Commutators} we prove that the
$\beta$-function is given by integrals of double total derivatives
and present the derivation of the exact NSVZ $\beta$-function from
this fact. A large number of technical details are collected in
appendixes.

\section{${\cal N}=1$ SQED with $N_f$ flavors, regularized by higher derivatives}
\hspace{\parindent}\label{Section_SQED}

In this paper we derive the NSVZ $\beta$-function for ${\cal N}=1$
SQED with $N_f$ flavors in all orders using the technique based on
the Schwinger--Dyson equations. It is convenient to write the
action for this theory in terms of superfields, because in this
case supersymmetry is a manifest symmetry
\cite{West:1990tg,Buchbinder:1998qv}. In this notation in the
massless limit the action is given by

\begin{equation}
S = \frac{1}{4e_0^2}\mbox{Re}\int d^4x\,d^2\theta\,W^a W_a +
\frac{1}{4} \sum\limits_{\alpha=1}^{N_f} \int
d^4x\,d^4\theta\,\Big(\phi_\alpha^* e^{2V}\phi_\alpha +
\widetilde\phi_\alpha^* e^{-2V} \widetilde\phi_\alpha\Big).
\end{equation}

\noindent In order to regularize this theory, we modify its action
by adding a term with higher derivatives
\cite{Slavnov:1971aw,Slavnov:1972sq}:

\begin{equation}
S_{\mbox{\scriptsize reg}} = \frac{1}{4e_0^2}\mbox{Re}\int
d^4x\,d^2\theta\,W^a R(\partial^2/\Lambda^2) W_a + \frac{1}{4}
\sum\limits_{\alpha=1}^{N_f} \int
d^4x\,d^4\theta\,\Big(\phi_\alpha^* e^{2V}\phi_\alpha +
\widetilde\phi_\alpha^* e^{-2V} \widetilde\phi_\alpha\Big).
\end{equation}

\noindent The higher derivatives are included into the function
$R$, which satisfies the conditions $R(0)=1$ and $R(\infty)=
\infty$. For example, it is possible to choose $R = 1+
\partial^{2n}/\Lambda^{2n}$. The term with higher derivatives
increases a degree of the momentum in the propagator of the gauge
superfield. As a consequence, most loop integrals become
convergent in the ultraviolet region. An accurate analysis shows
that after introducing the higher derivative term divergences
remain only in the one-loop approximation \cite{Faddeev:1980be}.
In order to cancel these remaining one-loop divergences, one
should insert the Pauli--Villars determinants $\det(V,M_I)$ into
the generating functional \cite{Slavnov:1977zf}:

\begin{equation}\label{Preliminary_Generating_Functional}
Z = \int DV\, D\phi\, D\widetilde\phi\,\prod\limits_{I=1}^n
(\det(V,M_I))^{c_I N_f}\exp\Big(i S_{\mbox{\scriptsize
reg}}[V,\phi,\widetilde\phi] + i S_{\mbox{\scriptsize gf}}[V] + i
S_{\mbox{\scriptsize source}} \Big).
\end{equation}

\noindent For fixing a gauge

\begin{equation}\label{Gauge_Fixing}
S_{\mbox{\scriptsize gf}} = - \frac{1}{64 e_0^2}\int
d^4x\,d^4\theta\, \Big(V R(\partial^{2}/\Lambda^{2}) D^2 \bar D^2
V + V R(\partial^{2}/\Lambda^{2}) \bar D^2 D^2 V\Big)
\end{equation}

\noindent is added to the classical action, while ghosts can be
omitted in the Abelian case. The masses of the Pauli--Villars
fields should be proportional to the parameter $\Lambda$:

\begin{equation}\label{Scheme}
M_I = a_I\Lambda,
\end{equation}

\noindent where $a_I$ are some real constants which do not depend
on the bare coupling constant. The coefficients $c_I$ should
satisfy the conditions

\begin{equation}\label{C_Conditions}
\sum\limits_{I=1}^n c_I =1;\qquad \sum\limits_{I=1}^n c_I M_I^2
=0,
\end{equation}

\noindent which ensure cancelation of the remaining one-loop
divergences. For simplicity, in this paper we use the following
choice of this coefficients:

\begin{equation}
c_I = (-1)^{P_I+1},
\end{equation}

\noindent where $P_I$ is an integer. In this case for even $P_I$
we can present the Pauli--Villars determinants as functional
integrals over the commuting (chiral) Pauli--Villars superfields.
For odd $P_I$ the Pauli--Villars superfields are anticommuting.
Therefore, $P_I$ is a Grassmannian parity of the Pauli--Villars
superfields, and

\begin{equation}
\prod\limits_{I=1}^{n} \Big(\det(V,M_I)\Big)^{c_I N_f} = \int
\prod\limits_{\alpha=1}^{N_f} \prod\limits_{I=1}^{n} D\phi_{\alpha
I} D\widetilde\phi_{\alpha I}\,\exp(iS_{\mbox{\scriptsize PV}}),
\end{equation}

\noindent where the action for the Pauli--Villars superfields is

\begin{eqnarray}\label{PV_Action}
&& S_{\mbox{\scriptsize PV}} =\sum\limits_{I=1}^n
\sum\limits_{\alpha=1}^{N_f} \Bigg\{\frac{1}{4} \int
d^8x\,\Big(\phi_{\alpha I}^* e^{2V}\phi_{\alpha I} +
\widetilde\phi_{\alpha I}^* e^{-2V}\widetilde\phi_{\alpha I} \Big)
\nonumber\\
&&\qquad\qquad\qquad\qquad\qquad + \Big(\frac{1}{2} \int
d^4x\,d^2\theta\,M_I \phi_{\alpha I} \widetilde\phi_{\alpha I} +
\frac{1}{2}\int d^4x\,d^2\bar\theta\,M_I \phi_{\alpha I}^*
\widetilde \phi_{\alpha I}^* \Big)\Bigg\}\qquad
\end{eqnarray}

\noindent with

\begin{equation}
\int d^8x \equiv \int d^4x\,d^4\theta.
\end{equation}

In order to simplify subsequent equations and make the
calculations similar to a non-Abelian case, we also introduce the
notation

\begin{equation}
\phi_i \equiv (\phi_{\alpha I},\,\widetilde\phi_{\alpha I});\qquad
\phi^{*i} \equiv (\phi_{\alpha I}^*,\,\widetilde \phi_{\alpha
I}^*),\quad i=1,\ldots 2(n+1)N_f,
\end{equation}

\noindent where the usual fields $\phi_\alpha$ and
$\widetilde\phi_\alpha$ by definition correspond to $I=0$. The sum
of mass terms can be written as

\begin{eqnarray}
&& S_m \equiv \frac{1}{2} \sum\limits_{I=0}^n
\sum\limits_{\alpha=1}^{N_f} \Big(\int d^4x\,d^2\theta\,M_I
\phi_{\alpha I} \widetilde \phi_{\alpha I} + \mbox{c.c.} \Big)
\nonumber\\
&&\qquad\qquad\qquad\qquad\qquad \equiv \frac{1}{4}\int
d^4x\,d^2\theta\,M^{ij} \phi_{i} \phi_{j} + \frac{1}{4}\int
d^4x\,d^2\bar\theta\,M^*_{ij} \phi^{*i} \phi^{*j},\qquad
\end{eqnarray}

\noindent where $M_0=0$, because the usual fields, which
corresponds to $I=0$, are considered in the massless limit. Due to
the gauge invariance the mass matrix satisfies the equation

\begin{equation}\label{MT_Symmetry}
(T)_m{}^i M^{mj} + (T)_{m}{}^j M^{im} = (-1)^{P_j} (M T)^{j\,i} +
(M T)^{ij} = 0,
\end{equation}

\noindent where

\begin{equation} (T)_i{}^j \equiv \left(
\begin{array}{cc}
1 & 0 \\
0 & -1
\end{array}
\right)\cdot \delta_{\alpha\beta}\cdot \delta_{IJ} \equiv
(-1)^{P_i} (T)^j{}_i
\end{equation}

\noindent is a generator of the $U(1)$ group in the considered
representation and $P_i$ is a Grassmanian parity of the superfield
$\phi_i$.

It is convenient to introduce sources both for the usual
superfields $V$, $\phi$, and $\widetilde\phi$ and for the
Pauli--Villars superfields:

\begin{equation}
S_{\mbox{\scriptsize source}} \equiv \int d^8x\,V J + \Big(\int
d^4x\,d^2\theta\, \phi_{i} j^i + \int d^4x\,d^2\bar\theta\,
\phi^{*i} j_i^* \Big).
\end{equation}

Eq. (\ref{Preliminary_Generating_Functional}) is a standard
definition of the generating functional for the considered theory,
regularized by higher derivatives. However, it is convenient to
use the background field method and introduce some auxiliary
sources. In the Abelian case for this purpose we make the
substitution $V \to V + \mbox{\boldmath$V$}$, where
$\mbox{\boldmath$V$}$ is the background field. Also in the kinetic
terms of the matter superfields we introduce the auxiliary real
parameter $g$ according to the prescription

\begin{equation}\label{Introducing_Of_G}
e^{2V} \to 1 + g (e^{2V}-1);\qquad\quad e^{-2V} \to 1 + g
(e^{-2V}-1).
\end{equation}

\noindent Then the usual kinetic terms are obtained for $g=1$. It
is important that this substitution is made only for the quantum
gauge field, which is an integration variable in the generating
functional. Moreover, we introduce the auxiliary sources
$\phi_{0i}$ and $\widetilde\phi_{0i}$\footnote{It is important
that we do not impose the chirality condition on the fields
$\phi_{0i}$ and $\widetilde\phi_{0i}$.} for each pair of the
matter superfields (including the Pauli--Villars fields) according
to the prescription

\begin{eqnarray}\label{Phi0}
&& S_{\scriptsize \mbox{matter}}\to \frac{1}{4}
\sum\limits_{I=0}^n \sum\limits_{\alpha=1}^{N_f} \int d^8x\,\Big[
(\phi_{\alpha I}^*+\phi_{0\alpha I}^*) e^{2\mbox{\scriptsize
\boldmath$V$}}\Big(1 + g
(e^{2V}-1)\Big) (\phi_{\alpha I} +\phi_{0\alpha I})\nonumber\\
&&\qquad\qquad\qquad\qquad\qquad\quad +(\widetilde\phi_{\alpha
I}^*+\widetilde\phi_{0\alpha I}^*) e^{-2\mbox{\scriptsize
\boldmath$V$}}\Big(1 + g (e^{-2V}-1)\Big) (\widetilde\phi_{\alpha
I} +\widetilde\phi_{0\alpha I}) \Big] + S_m.\qquad\quad
\end{eqnarray}

\noindent From Eq. (\ref{Phi0}) we see that, by definition, the
parameter $g$ is present only in vertices containing internal
lines of the gauge superfield. It is important that introducing
the parameter $g$ we break the quantum gauge invariance. As a
consequence, it is impossible to use Ward identities for Green
functions containing external lines of the quantum gauge field.
However, the background gauge invariance

\begin{equation}\label{Background_Gauge_Invariance}
\mbox{\boldmath$V$} \to \mbox{\boldmath$V$} - \frac{1}{2}(A +
A^*);\qquad V \to V;\qquad \phi \to e^{A} \phi;\qquad
\widetilde\phi \to e^{-A} \widetilde\phi,
\end{equation}

\noindent where $A$ is an arbitrary chiral superfield, is
unbroken.

Thus, the generating functional is given by the following
expression:

\begin{equation}\label{Generating_Functional}
Z \equiv e^{iW} = \int D\mu \exp\Big(i S_{\mbox{\scriptsize
total}} + i S_{\mbox{\scriptsize gf}} + i S_{\mbox{\scriptsize
source}} \Big),
\end{equation}

\noindent where $D\mu$ denotes the integration measure and

\begin{eqnarray}
&& S_{\mbox{\scriptsize total}} = \frac{1}{4e_0^2}\mbox{Re}\int
d^4x\,d^2\theta\,W^a R(\partial^2/\Lambda^2) W_a +
\frac{1}{4e_0^2}\mbox{Re}\int
d^4x\,d^2\theta\,\mbox{\boldmath$W$}^a
\mbox{\boldmath$W$}_a\nonumber\\
&& + \frac{1}{4} \sum\limits_{I=0}^n \sum\limits_{\alpha=1}^{N_f}
\int d^8x\,\Big[ (\phi_{\alpha}^*+\phi_{0\alpha}^*)
e^{2\mbox{\scriptsize \boldmath$V$}} \Big(1 + g (e^{2V}-1)\Big)
(\phi_{\alpha} +\phi_{0\alpha})
+(\widetilde\phi_{\alpha}^*+\widetilde\phi_{0\alpha}^*)
e^{-2\mbox{\scriptsize \boldmath$V$}}\qquad\nonumber\\
&& \times \Big(1 + g (e^{-2V}-1)\Big) (\widetilde\phi_{\alpha}
+\widetilde\phi_{0\alpha}) \Big]_I + \sum\limits_{I=0}^n
\sum\limits_{\alpha=1}^{N_f}\Big(\frac{1}{2}\int
d^4x\,d^2\theta\,M \phi_{\alpha} \widetilde\phi_{\alpha} +
\mbox{c.c.}\Big)_I,
\end{eqnarray}

\noindent where $\mbox{\boldmath$W$}_a = \bar D^2 D_a
\mbox{\boldmath$V$}/4$ is the field strength for the background
gauge superfield $\mbox{\boldmath$V$}$. (It is easy to see that
terms linear in the quantum field $V$ can be omitted. Also, it is
not necessary to introduce the regulator in the part of the action
which depends only on the background field.) The effective action
is defined by the standard way as

\begin{equation}\label{Gamma}
\Gamma[V,\mbox{\boldmath$V$},\phi_{i}] = W - S_{\mbox{\scriptsize
source}},
\end{equation}

\noindent where the sources should be expressed in terms of fields
through solving the equations

\begin{equation}\label{Sourcers->Fields}
\phi_{i} = (-1)^{P_i} \frac{\delta W}{\delta j^i};\qquad \phi^{*i}
= (-1)^{P_i} \frac{\delta W}{\delta j^*_i};  \qquad V =
\frac{\delta W}{\delta J}.
\end{equation}

\noindent Differentiating the effective action we obtain

\begin{equation}\label{Fields->Sources}
j^{i} = - \frac{\delta \Gamma}{\delta \phi_i};\qquad j^*_{i} = -
\frac{\delta \Gamma}{\delta \phi^{*i}};\qquad J = - \frac{\delta
\Gamma}{\delta V}.
\end{equation}

\noindent Below we will see that it is not convenient to consider
$(\phi_{i},V)$ as independent variables. A more convenient choice
is $(\phi_{i},J)$, where $J$ is a source for the quantum gauge
superfield $V$. That is why below instead of the effective action
we will mostly use the Routhian

\begin{equation}
\gamma[J,\mbox{\boldmath$V$},\phi_{i}] = W - \Big( \int
d^4x\,d^2\theta\,\phi_i j^i + \mbox{c.c.}\Big),
\end{equation}

\noindent where it is necessary to express only the sources $j^i$
in terms of $J$ and $\phi_{i}$.

Due to the background gauge invariance
(\ref{Background_Gauge_Invariance}) the two-point function of the
background gauge superfield is transversal:

\begin{equation}\label{D_Definition}
\Gamma^{(2)}_{\scriptsize \mbox{\boldmath$V$}}  = -
\frac{1}{16\pi} \int
\frac{d^4p}{(2\pi)^4}\,d^4\theta\,\mbox{\boldmath$V$}(\theta,-p)\,\partial^2\Pi_{1/2}
\mbox{\boldmath$V$}(\theta,p)\, d^{-1}(\alpha_0,\Lambda/p),
\end{equation}

\noindent where $\alpha_0 = e_0^2/4\pi$ is a bare coupling
constant and the supersymmetric transversal projector is given by

\begin{equation}
\partial^2\Pi_{1/2}  = -\frac{1}{8} D^a \bar D^2 D_a.
\end{equation}

\noindent In this paper we will calculate the $\beta$-function
defined in terms of the bare coupling constant

\begin{equation}\label{Beta_Definition}
\beta\Big(\alpha_0(\alpha,\Lambda/\mu)\Big) =
\frac{d\alpha_0(\alpha,\Lambda/p)}{d\ln\Lambda}
\Big|_{\alpha=\mbox{\scriptsize const}},
\end{equation}

\noindent where $\alpha = \alpha(\alpha_0,\Lambda/p)$ is a
renormalized coupling constant. It is determined by the
requirement that the function
$d(\alpha_0(\alpha,\Lambda/\mu),\Lambda/p)$ is finite in the limit
$\Lambda\to \infty$. The anomalous dimension can be defined
similarly:

\begin{equation}\label{Gamma_Definition}
\gamma\Big(\alpha_0(\alpha,\Lambda/\mu)\Big) \equiv - \frac{d \ln
Z(\alpha,\Lambda/\mu)}{d
\ln\Lambda}\Big|_{\alpha=\mbox{\scriptsize const}},
\end{equation}

\noindent where $Z$ is a renormalization constant for the matter
superfield, which is constructed by requiring finiteness of the
function $Z G$ in the limit $\Lambda\to \infty$. It is easy to see
that the $\beta$-function (\ref{Beta_Definition}) and the
anomalous dimension (\ref{Gamma_Definition}) do not depend on a
choice of the renormalized coupling constant $\alpha$ and the
renormalization constant $Z$ (see, e.g., \cite{Kataev:2013eta}).
The renormalization group functions (\ref{Beta_Definition}) and
(\ref{Gamma_Definition}) differ from the standard ones defined in
terms of the renormalized coupling constant

\begin{eqnarray}\label{RG_Renormalized}
&& \widetilde\beta\Big(\alpha(\alpha_0,\Lambda/\mu)\Big) \equiv
\frac{d\alpha(\alpha_0,\Lambda/\mu)}{d\ln\mu}\Big|_{\alpha_0=\mbox{\scriptsize
const}};\nonumber\\
&& \widetilde\gamma\Big(\alpha(\alpha_0,\Lambda/\mu)\Big) \equiv
\frac{d\ln Z(\alpha(\alpha_0,\Lambda/\mu),
\Lambda/\mu)}{d\ln\mu}\Big|_{\alpha_0=\mbox{\scriptsize const}},
\end{eqnarray}

\noindent which are scheme-dependent. However
\cite{Kataev:2013eta,Kataev:2013csa}, the functions
(\ref{Beta_Definition}) and (\ref{Gamma_Definition}) can be
obtained from the renormalization group functions
(\ref{RG_Renormalized}) by imposing the boundary conditions

\begin{equation}\label{NSVZ_Scheme}
Z_3(\alpha,x_0) = 1;\qquad Z(\alpha,x_0)=1
\end{equation}

\noindent on the renormalization constants, where $x_0$ is an
arbitrary fixed value of $\ln\Lambda/\mu$.\footnote{These boundary
conditions are imposed only in a single point. They should not be
confused with the condition $Z_3 =1$ following from the conformal
symmetry (see, e.g., \cite{Kataev:2013vua}), which is valid for
arbitrary values of $\ln\Lambda/\mu$.}

In order to find the $\beta$-function (\ref{Beta_Definition}) we
calculate the expression

\begin{equation}\label{We_Calculate}
\frac{d}{d\ln \Lambda}\,
\Big(d^{-1}(\alpha_0,\Lambda/p)-\alpha_0^{-1}\Big)\Big|_{p=0} = -
\frac{d\alpha_0^{-1}}{d\ln\Lambda} =
\frac{\beta(\alpha_0)}{\alpha_0^2},
\end{equation}

\noindent where $\Lambda$ and $\alpha$ are considered as
independent variables. This expression is well defined if the
right hand side is expressed in terms of the bare coupling
constant $\alpha_0$. The left hand side of the expression
(\ref{We_Calculate}) can be obtained from the two-point Green
function of the background gauge superfield after the substitution

\begin{equation}\label{V_Substitution}
\mbox{\boldmath$V$}(x,\theta) \to \bar \theta^{\dot a}
\bar\theta_{\dot a} \theta^b \theta_b \equiv \theta^4.
\end{equation}

\noindent Strictly speaking, the part of the effective action
corresponding to the two-point function of the gauge superfield is
infinite after this substitution, because it is proportional to

\begin{equation}
\int d^4x \to \infty.
\end{equation}

\noindent However, this procedure can be rigorously formulated by
inserting a regulator $I(x)$

\begin{equation}
\mbox{\boldmath$V$}(x,\theta) \to \bar \theta^{\dot a}
\bar\theta_{\dot a} \theta^b \theta_b \cdot I(x)\equiv \theta^4
\cdot I(x) \approx \theta^4,
\end{equation}

\noindent which is approximately equal to 1 at finite $x^\mu$ and
tends to 0 at the large scale $R \to \infty$. Then in the leading
order in $R$ the considered part of the effective action is
proportional to

\begin{equation}
{\cal V}_4 \equiv \int d^4x\,I^2 \sim R^4 \to \infty.
\end{equation}

\noindent All terms containing the derivatives of the regulator
$I$ are suppressed as $1/R\Lambda \to 0$ and can be omitted. That
is why below we do not explicitly write the regulator $I$, but
assume that ${\cal V}_4$ is finite and tends to infinity. Actually
this corresponds to taking the limit of the vanishing external
momentum $p \sim R^{-1}\sim ({\cal V}_4)^{-1/4} \to 0$:

\begin{equation}\label{We_Calculate2}
\frac{1}{2\pi} {\cal V}_4\cdot \frac{d}{d\ln \Lambda}\,
\Big(d^{-1}(\alpha_0,\Lambda/p)-\alpha_0^{-1}\Big)\Big|_{p=0} =
\frac{1}{2\pi} {\cal V}_4\cdot
\frac{\beta(\alpha_0)}{\alpha_0^2}\Big|_{p=0} =
\frac{d(\Delta\Gamma^{(2)}_{\scriptsize
\mbox{\boldmath$V$}})}{d\ln\Lambda}
\Big|_{\mbox{\boldmath$V$}(x,\theta)=\theta^4},
\end{equation}

\noindent where

\begin{equation}
\Delta\Gamma \equiv \Gamma - \frac{1}{4e_0^2}\mbox{Re}\int
d^4x\,d^2\theta\,\mbox{\boldmath$W$}^a \mbox{\boldmath$W$}_a.
\end{equation}

The expressions for the two-point Green functions of the matter
superfields can be found using arguments based on the chirality.
Taking into account that the two-point functions of the matter
superfields constructed from $\Gamma$ and $\gamma$ evidently
coincide, they can be written as

\begin{equation}\label{A_Definition}
A_{xy} \equiv \left(
\begin{array}{cc}
{\displaystyle \frac{\delta^2\gamma}{\delta(\phi_i)_x \delta
(\phi^{*j})_y}\vphantom{\Bigg(}} & {\displaystyle
\frac{\delta^2\gamma}{\delta(\phi_i)_x \delta
(\phi_j)_y}} \\
{\displaystyle \frac{\delta^2\gamma}{\delta(\phi^{*i})_x \delta
(\phi^{*j})_y}\vphantom{\Bigg(}} & {\displaystyle
\frac{\delta^2\gamma}{\delta(\phi^{*i})_x \delta (\phi_j)_y}}
\end{array}
\right) = \left(
\begin{array}{cc}
{\displaystyle  G_j{}^i \frac{\bar D_x^2 D_x^2}{16} \delta^8_{xy}
\vphantom{\Bigg(}} & {\displaystyle - \frac{1}{4}
(MJ)^{j\,i} \bar D_x^2 \delta^8_{xy}} \\
{\displaystyle - \frac{1}{4} (MJ)_{j\,i}^* D_x^2 \delta^8_{xy}
\vphantom{\Bigg(}} & {\displaystyle G^j{}_i \frac{D_x^2 \bar
D_x^2}{16} \delta^8_{xy}}
\end{array}
\right),
\end{equation}

\noindent where the fields are set to 0, $G_i{}^j$ and $(M
J)^{ij}$ are functions of $\partial^2$, and in our notation

\begin{eqnarray}\label{G_And_J}
&& G_i{}^j \equiv (-1)^{P_i} G^j{}_i = \delta_{\alpha\beta} \cdot
\delta_{IJ}\cdot \left(\begin{array}{cc} 1 & 0\\
0 & 1
\end{array}\right) G_I(\partial^2);\nonumber\\
&& (MJ)^{ij} = (MJ)^*_{ij} = \delta_{\alpha\beta} \cdot
\delta_{IJ}\cdot  \left(\begin{array}{cc} 0 & M_I\\
(-1)^{P_I} M_I & 0
\end{array}\right) J_I(\partial^2).
\end{eqnarray}

\noindent ($2\times 2$ matrixes correspond to the fields $\phi$
and $\widetilde\phi$. The function $J$ is real as a consequence of
the $CP$-invariance.) By definition the matrix $A^{-1}$
constructed from the inverse Green functions satisfies the
condition

\begin{equation}\label{A-1_Definition}
\int d^8y\, (A^{-1})_{xy} \left(
\begin{array}{cc}
0 & \bar D^2/8\partial^2\\
D^2/8\partial^2 & 0
\end{array}
\right)_y A_{yz} = \left(
\begin{array}{cc}
0 & - \bar D^2/2\\
- D^2/2 & 0
\end{array}
\right) \delta^8_{xz}.
\end{equation}

\noindent An explicit expression for the matrix $A^{-1}$ can be
easily found:

\begin{eqnarray}\label{A-1_Explicit}
&& (A^{-1})_{xy} \equiv \left(
\begin{array}{cc}
{\displaystyle \Big(\frac{\delta^2\gamma}{\delta(\phi_i)_x \delta
(\phi^{*j})_y}\Big)^{-1}\vphantom{\Bigg(}} & {\displaystyle
\Big(\frac{\delta^2\gamma}{\delta(\phi_i)_x \delta
(\phi_j)_y}\Big)^{-1}} \\
{\displaystyle \Big(\frac{\delta^2\gamma}{\delta(\phi^{*i})_x
\delta (\phi^{*j})_y}\Big)^{-1}\vphantom{\Bigg(}} & {\displaystyle
\Big(\frac{\delta^2\gamma}{\delta(\phi^{*i})_x \delta
(\phi_j)_y}\Big)^{-1}}
\end{array}
\right)\nonumber\\
&&\quad\qquad\qquad\qquad\qquad\quad = - \frac{1}{\partial^2 G^2 +
|M J|^2} \left(
\begin{array}{cc}
{\displaystyle  G^j{}_i \frac{\bar D_x^2 D_x^2}{4} \delta^8_{xy}
\vphantom{\Bigg(}} & {\displaystyle (MJ)^*_{ij} \bar
D_x^2 \delta^8_{xy}} \\
{\displaystyle (MJ)^{ij} D_x^2 \delta^8_{xy} \vphantom{\Bigg(}} &
{\displaystyle G_j{}^i \frac{D_x^2 \bar D_x^2}{4} \delta^8_{xy}}
\end{array}
\right),\qquad
\end{eqnarray}

\noindent where the operator $\partial^2 G^2 + |M J|^2$ is defined
by the following prescription:

\begin{equation}\label{Propagator}
(\partial^2 G^2 + |M J|^2)_i{}^k \equiv \partial^2 G_i{}^j G_j{}^k
+ (M J)_{j\,i}^* (MJ)^{j\,k} = \delta_{\alpha\beta}\cdot
\delta_{IJ}\cdot \left(\begin{array}{cc} 1 & 0\\
0 & 1
\end{array}\right) (\partial^2 G_I^2 + M_I^2 J_I^2).
\end{equation}

\section{Schwinger--Dyson equations}
\hspace{\parindent}\label{Section_Effective_Diagrams}

Making the change of variables $\phi_{i} \to \phi_{i} + A_i$,
where $A_i$ are arbitrary chiral superfields, in the generating
functional (\ref{Generating_Functional}), we obtain the equation

\begin{equation}\label{Phi0_SD_Equation}
-\frac{1}{2} \bar D^2 \frac{\delta\Gamma}{\delta\phi_{0i}} -
\frac{\delta\Gamma}{\delta\phi_{i}} + \frac{1}{2} M^{ij} \phi_{j}
= 0.
\end{equation}

\noindent This equation can be considered as a Schwinger--Dyson
equation for the matter superfields.

Similarly, the Schwinger--Dyson equation for the two-point Green
function of the gauge superfield can be written as
\cite{Stepanyantz:2004sg}

\begin{eqnarray}\label{SD_Correlator}
&& \frac{\delta(\Delta\Gamma)}{\delta \mbox{\boldmath$V$}_x} =
\frac{1}{2}\sum\limits_{I=0}^n \sum\limits_{\alpha=1}^{N_f}
\Big\langle (\phi_\alpha^* +
\phi_{0\alpha}^*)e^{2\mbox{\scriptsize \boldmath$V$}}\Big(1 + g
(e^{2V}-1)\Big)(\phi_\alpha + \phi_{0\alpha})\nonumber\\
&&\qquad\qquad\qquad\qquad\qquad\qquad\qquad -
(\widetilde\phi_\alpha^* + \widetilde\phi_{0\alpha}^*)
e^{-2\mbox{\scriptsize \boldmath$V$}}\Big(1 + g (e^{-2V}-1)\Big)
(\widetilde\phi_\alpha + \widetilde\phi_{0\alpha})
\Big\rangle_I,\qquad
\end{eqnarray}

\noindent where

\begin{equation}
\langle A \rangle \equiv \frac{1}{Z} \int D\mu\, A[V,\phi_{i}]
\exp\Big(i S_{\mbox{\scriptsize total}} + i S_{\mbox{\scriptsize
gf}} +iS_{\mbox{\scriptsize Source}}\Big)
\end{equation}

\noindent and the sources should be expressed in terms of the
fields using Eq. (\ref{Sourcers->Fields}). Because in this paper
we use the background field method, Eq. (\ref{SD_Correlator}) can
be simply obtained by differentiation of the effective action with
respect to the background field $\mbox{\boldmath$V$}$. It is easy
to see that the Schwinger--Dyson equation (\ref{SD_Correlator})
can be equivalently rewritten in terms of derivatives with respect
to the sources

\begin{eqnarray}\label{SD_Equation}
&& \frac{\delta(\Delta\Gamma)}{\delta \mbox{\boldmath$V$}_x} = 2
(T)^j{}_i \Big(\frac{1}{i} \frac{\delta}{\delta (j^j)_x}
\frac{\delta\Gamma}{\delta (\phi_{0i})_x} +
\frac{\delta\Gamma}{\delta (\phi_{0i})_x} (\phi_j+\phi_{0j})_x
\Big)\nonumber\\
&&\qquad\qquad\qquad\qquad\qquad = 2 (T)^j{}_i \Big( \frac{1}{i}
\frac{\delta}{\delta (j^*_i)_x}\frac{\delta\Gamma}{\delta
(\phi^{*j}_{0})_x} + \frac{\delta\Gamma}{\delta (\phi_{0}^{*j})_x}
(\phi^{*i}+\phi_{0}^{i*})_x\Big),\qquad
\end{eqnarray}

\noindent where the derivatives with respect to the sources are
constructed according to the prescription

\begin{eqnarray}\label{J_Derivative}
&& \frac{\delta}{\delta (j^i)_x} = - \int d^8y\,\Bigg[
\Big(\frac{\delta^2\Gamma}{\delta(\phi_{i})_x \delta
(\phi^{*j})_y}\Big)^{-1} \frac{\bar
D_y^2}{8(\partial^2)_y}\frac{\delta}{\delta
(\phi^{*j})_y}\nonumber\\
&&\qquad\qquad\qquad +
\Big(\frac{\delta^2\Gamma}{\delta(\phi_{i})_x \delta
(\phi_{j})_y}\Big)^{-1}
\frac{D_y^2}{8(\partial^2)_y}\frac{\delta}{\delta (\phi_{j})_y} +
\Big(\frac{\delta^2\Gamma}{\delta(\phi_{i})_x \delta
V_y}\Big)^{-1} \frac{\delta}{\delta V_y}\Bigg].\qquad
\end{eqnarray}

\begin{figure}[h]
\begin{center}
\begin{picture}(0,1.8)
\put(-7.7,0.6){${\displaystyle\Delta\Gamma^{(2)}_{\mbox{\scriptsize
\boldmath$V$}} =\frac{1}{2} \int d^8x\,d^8y\,\mbox{\boldmath$V$}_x
\mbox{\boldmath$V$}_y \frac{\delta^2(\Delta\Gamma)}{\delta
\mbox{\boldmath$V$}_x \delta \mbox{\boldmath$V$}_y} =}$}
\put(3.1,0.6){+} \put(-1.7,-1.4){\vector(1,1){1.2}}
\put(0.9,-1.2){\vector(0,1){0.9}}
\put(3.3,-1.3){\vector(-1,1){1.2}}
\put(5.8,-1.3){\vector(-1,2){0.6}} \put(-2.8,-2){${\displaystyle
\frac{\delta^2\Gamma}{\delta \phi_{0x}^* \delta\phi_z}}$}
\put(2.7,-2){${\displaystyle \frac{\delta^3\Gamma}{\delta
\mbox{\boldmath$V$}_y\delta\phi^*_{w}\delta\phi_v}}$}
\put(5,-2){${\displaystyle \frac{\delta^3\Gamma}{\delta
\mbox{\boldmath$V$}_y\delta\phi^*_{0x}\delta\phi_z}}$}
\put(0,-2){${\displaystyle \Big(\frac{\delta^2\Gamma}{\delta\phi_z
\delta\phi_w^*}\Big)^{-1}}$}
\put(-0.9,0.0){\includegraphics[scale=0.88]{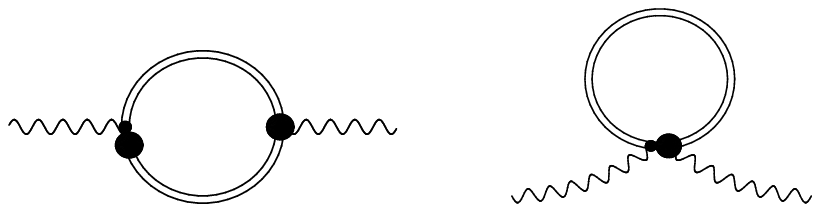}}
\end{picture}
\end{center}
\vspace*{18mm} \caption{The Schwinger--Dyson equation for the
two-point function of the gauge superfield. Below we present
Feynman rules (for simplicity, in the massless case). In the
massive case the effective diagrams are the same.}
\label{Figure_Graphical_SD_Equation}
\end{figure}

\noindent In order to verify this equation it is necessary to
apply it to $(j^k)_z$ taking into account Eqs.
(\ref{Fields->Sources}) and (\ref{A-1_Definition}).
Differentiating Eq. (\ref{SD_Equation}) with respect to
$\mbox{\boldmath$V$}_y$ and setting all fields (including
$\phi_0$) to 0, we obtain the Schwinger--Dyson equation for the
two-point Green function of the gauge superfield. Details of this
calculation are presented in Appendix \ref{Appendix_SD_Equation}.
Following \cite{Stepanyantz:2004sg} it is convenient to formulate
the result in the graphical form. It is presented in Fig.
\ref{Figure_Graphical_SD_Equation}. In this figure external lines
correspond to the gauge superfield $\mbox{\boldmath$V$}$, and, for
simplicity, expressions for vertices and propagators are written
for the massless case. In the analytical form the Schwinger--Dyson
equation presented in Fig. \ref{Figure_Graphical_SD_Equation} is
written as

\begin{eqnarray}\label{SD_Equation_For_2V_Function}
&& \frac{1}{2} \int d^8x\,d^8y\,\mbox{\boldmath$V$}_x
\mbox{\boldmath$V$}_y \frac{\delta^2(\Delta\Gamma)}{\delta
\mbox{\boldmath$V$}_y \delta \mbox{\boldmath$V$}_x} = -i (T)^j{}_i
\int d^8x\,d^8y\,d^8z\,\mbox{\boldmath$V$}_x \mbox{\boldmath$V$}_y
\Bigg( \frac{\mbox{\boldmath$\delta$}}{ \mbox{\boldmath$\delta$}
(j^j)_x} \frac{\delta^2\Gamma}{\delta \mbox{\boldmath$V$}_y
\delta(\phi_{0k})_z}\qquad\nonumber\\
&& \times \Big[\delta^8_{xz} \delta^i_k +
\frac{\mbox{\boldmath$\delta$}}{\mbox{\boldmath$\delta$} (j^k)_z}
\frac{\delta\gamma}{\delta (\phi_{0i})_x}  \Big] +
\frac{\mbox{\boldmath$\delta$}}{\mbox{\boldmath$\delta$} (j^j)_x}
\frac{\delta^2\Gamma}{\delta \mbox{\boldmath$V$}_y
\delta(\phi^{*k}_{0})_z}\cdot
\frac{\mbox{\boldmath$\delta$}}{\mbox{\boldmath$\delta$}
(j_{k}^*)_z} \frac{\delta\gamma}{\delta (\phi_{0i})_x}
\Bigg),\qquad
\end{eqnarray}

\noindent where all fields are set to 0, and we use the notation

\begin{equation}\label{Bold_Derivative}
\frac{\mbox{\boldmath$\delta$}}{ \mbox{\boldmath$\delta$} (j^j)_x}
= - \int d^8w\,\Bigg(\Big(\frac{\delta^2\gamma}{\delta (\phi_j)_x
\delta(\phi_m)_w}\Big)^{-1}_{\phi,\phi_0,V=0} \frac{\delta}{\delta
(\phi_{0m})_w} + \Big(\frac{\delta^2\gamma}{\delta (\phi_j)_x
\delta(\phi^{*m})_w}\Big)^{-1}_{\phi,\phi_0,V=0}
\frac{\delta}{\delta (\phi_{0}^{*m})_w}\Bigg).
\end{equation}

\noindent Note that all fields here are set to 0 in contrast to
Eq. (\ref{J_Derivative}). Due to this condition the derivatives
$\mbox{\boldmath$\delta$}/\mbox{\boldmath$\delta$} j$
(anti)commute. However, below we will not usually write explicitly
the condition $\phi,\phi_0,V=0$ as in Eq. (\ref{Bold_Derivative}).

The Schwinger--Dyson equation (\ref{SD_Equation_For_2V_Function})
can be simplified after the substitution (\ref{V_Substitution}).
(As we already mentioned above, this substitution automatically
gives $p=0$.) For this purpose it is convenient to use the
identity

\begin{eqnarray}\label{GreenLine_Identity}
&&\hspace*{-8mm} (T)^j{}_i \int d^8x\,(\theta^4)_x
\Bigg\{\frac{\mbox{\boldmath$\delta$}}{\mbox{\boldmath$\delta$}
(j^j)_x} \frac{\delta}{\delta (\phi_{0i})_x} + \int
d^8y\,\Bigg[\Big(\frac{\mbox{\boldmath$\delta$}}{\mbox{\boldmath$\delta$}
(j^k)_y} \frac{\delta\gamma}{\delta (\phi_{0i})_x}\Big)
\frac{\mbox{\boldmath$\delta$}}{\mbox{\boldmath$\delta$} (j^j)_x}
\frac{\delta}{\delta (\phi_{0k})_y} +
\Big(\frac{\mbox{\boldmath$\delta$}}{ \mbox{\boldmath$\delta$}
(j^*_k)_y}\frac{\delta\gamma}{\delta (\phi_{0i})_x}\Big)\nonumber\\
&&\hspace*{-8mm} \times
\frac{\mbox{\boldmath$\delta$}}{\mbox{\boldmath$\delta$} (j^j)_x}
\frac{\delta}{\delta (\phi_{0}^{*k})_y}\Bigg] \Bigg\} = \int
d^8x\,\Bigg\{ (T)^j{}_i \Big(\Big[\theta^a\theta_a
\bar\theta^{\dot b} \frac{\bar D_{\dot b} D^2}{4\partial^2} +
2i\bar\theta^{\dot a} (\gamma^\mu)_{\dot a}{}^{b} \theta_b
\frac{\partial_\mu}{\partial^2} - \frac{D^2}{4\partial^2} \theta^a
\theta_a \Big]
\frac{\mbox{\boldmath$\delta$}}{\mbox{\boldmath$\delta$} j^j}\Big)
\frac{\delta}{\delta \phi_{0i}}\nonumber\\
&&\hspace*{-8mm} - i (M T)^{ij}\, \bar\theta^{\dot a}
(\gamma^\mu)_{\dot a}{}^b \theta_b
\Big(\frac{D^2\partial_\mu}{16\partial^4}
\frac{\mbox{\boldmath$\delta$}}{\mbox{\boldmath$\delta$} j^i}\Big)
\frac{\mbox{\boldmath$\delta$}}{ \mbox{\boldmath$\delta$} j^j}
\Bigg\},
\end{eqnarray}

\begin{figure}[h]
\begin{picture}(0,2.3)
\put(1,0){\includegraphics[scale=0.7]{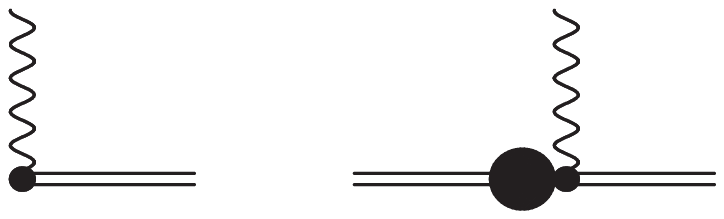}}
\put(8.5,1.3){\includegraphics[angle=180]{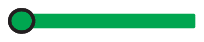}}
\put(3.4,0.6){$+$} \put(7.3,0.65){$\vector(1,0){1}$}
\put(1.95,1.7){$\theta^4$} \put(5.8,1.7){$\theta^4$}
\put(11.5,0.6){+ \mbox{terms without $\bar\theta$}}
\put(9.1,1.05){$1$} \put(10.9,1.05){$2$}
\end{picture}
\caption{The sum of two effective lines.}
\label{Figure_Sum_Of_Lines}
\end{figure}

\noindent which is proved in Appendix \ref{Appendix_Sum_Of_Lines}.
It is convenient to define the operator which contains all terms
of the first degree in $\bar\theta$ in Eq.
(\ref{GreenLine_Identity}):

\begin{eqnarray}\label{GreenLine_Definition}
&& \mbox{GreenLine}[1,2] \equiv \int d^8x\,\Big((T)^j{}_i
\Big(\theta^a\theta_a \bar\theta^{\dot b} \frac{\bar D_{\dot b}
D^2}{4\partial^2} + 2i\bar\theta^{\dot a} (\gamma^\mu)_{\dot
a}{}^{b} \theta_b \frac{\partial_\mu}{\partial^2} \Big)
\frac{\mbox{\boldmath$\delta$}}{\mbox{\boldmath$\delta$}_2 j^j}
\cdot \frac{\delta}{\delta_1 \phi_{0i}}\qquad\nonumber\\
&& - i (M T)^{ij}\, \bar\theta^{\dot a} (\gamma^\mu)_{\dot a}{}^b
\theta_b \Big(\frac{D^2\partial_\mu}{16\partial^4}
\frac{\mbox{\boldmath$\delta$}}{\mbox{\boldmath$\delta$}_2
j^i}\Big)\cdot \frac{\mbox{\boldmath$\delta$}}{
\mbox{\boldmath$\delta$}_1 j^j}\Big).
\end{eqnarray}

\begin{figure}[p]
\vspace*{1cm}
\begin{picture}(0,0)
\put(0.25,0.25){\includegraphics[angle=180]{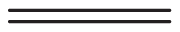}}
\put(2.8,0){${\displaystyle \mbox{UsualLine}[1,2]= \int d^8x\,
\Big(\frac{\delta}{\delta_2 \phi_{0i}} \cdot
\frac{\mbox{\boldmath$\delta$}}{\mbox{\boldmath$\delta$}_1 j^i} +
\frac{\delta}{\delta_2 \phi^{*i}_{0}} \cdot
\frac{\mbox{\boldmath$\delta$}}{\mbox{\boldmath$\delta$}_1 j^*_i}
\Big)}$}
\end{picture}\\
\begin{picture}(0,1.2)
\put(-0.4,0.75){\includegraphics[angle=180]{line_green.eps}}
\put(2.8,0){${\displaystyle \mbox{GreenLine}[1,2] = \int
d^8x\,\Big((T)^j{}_i \Big(\theta^a\theta_a \bar\theta^{\dot b}
\frac{\bar D_{\dot b} D^2}{4\partial^2} + 2i\bar\theta^{\dot a}
(\gamma^\mu)_{\dot a}{}^{b} \theta_b
\frac{\partial_\mu}{\partial^2} \Big) }$}
\put(6.5,-1.1){${\displaystyle \times
\frac{\mbox{\boldmath$\delta$}}{\mbox{\boldmath$\delta$}_2 j^j}
\cdot \frac{\delta}{\delta_1 \phi_{0i}} - i (M T)^{ij}\,
\bar\theta^{\dot a} (\gamma^\mu)_{\dot a}{}^b \theta_b
\Big(\frac{D^2\partial_\mu}{16\partial^4}
\frac{\mbox{\boldmath$\delta$}}{\mbox{\boldmath$\delta$}_2
j^i}\Big)\cdot \frac{\mbox{\boldmath$\delta$}}{
\mbox{\boldmath$\delta$}_1 j^j}\Big) }$}
\end{picture}\\
\begin{picture}(0,2.2)
\put(0.25,0.5){\includegraphics[angle=180]{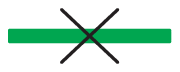}}
\put(2.8,0){${\displaystyle \mbox{GreenWithCross}[1,2] = - \int
d^8x\,\bar\theta^{\dot a} (\gamma^\mu)_{\dot a}{}^b \theta_b
\Big(i(M T)^{ij} \Big(\frac{D^2\partial_\mu}{16\partial^4}
\frac{\mbox{\boldmath$\delta$}}{\mbox{\boldmath$\delta$}_2
j^i}\Big)\cdot \frac{\mbox{\boldmath$\delta$}}{
\mbox{\boldmath$\delta$}_1 j^j}\Big)}$}
\end{picture}\\
\begin{picture}(0,1.1)
\put(-0.4,0.75){\includegraphics[angle=180]{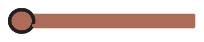}}
\put(2.8,0){${\displaystyle \mbox{BrownLine}_\mu[1,2] = \int
d^8x\,\Big((T)^j{}_i \Big(-\frac{2i \partial_\mu}{\partial^2} -
(\gamma_\mu)^{a{\dot b}} \theta_a \frac{\bar D_{\dot b}
D^2}{4\partial^2}\Big) }$} \put(8.2,-1.1){${\displaystyle\times
\frac{\mbox{\boldmath$\delta$}}{\mbox{\boldmath$\delta$}_2 j^j}
\cdot \frac{\delta}{\delta_1 \phi_{0i}} + i (M T)^{ij}
\Big(\frac{D^2\partial_\mu}{16\partial^4}
\frac{\mbox{\boldmath$\delta$}}{\mbox{\boldmath$\delta$}_2
j^i}\Big)\cdot \frac{\mbox{\boldmath$\delta$}}{
\mbox{\boldmath$\delta$}_1 j^j}\Big) }$}
\end{picture}\\
\begin{picture}(0,2.2)
\put(0.25,0.5){\includegraphics[angle=180]{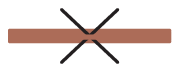}}
\put(2.8,0){${\displaystyle \mbox{BrownWithCross}_\mu[1,2] = i(M
T)^{ij} \int d^8x\, \Big(\frac{D^2\partial_\mu}{16\partial^4}
\frac{\mbox{\boldmath$\delta$}}{\mbox{\boldmath$\delta$}_2
j^i}\Big)\cdot \frac{\mbox{\boldmath$\delta$}}{
\mbox{\boldmath$\delta$}_1 j^j} }$}
\end{picture}\\
\begin{picture}(0,1.1)
\put(-0.4,0.75){\includegraphics[angle=180]{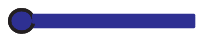}}
\put(2.8,0){${\displaystyle \mbox{BlueLine}^{\dot b}[\alpha;1,2] =
\int d^8x\, \alpha\,(T)^j{}_i \Big(\frac{\bar D^{\dot b}
D^2}{8\partial^2}
\frac{\mbox{\boldmath$\delta$}}{\mbox{\boldmath$\delta$}_2
j^j}\Big) \cdot \frac{\delta}{\delta_1 \phi_{0i}}}$}
\end{picture}\\
\begin{picture}(0,1.1)
\put(-0.4,0.75){\includegraphics[angle=180]{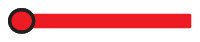}}
\put(2.8,0){${\displaystyle \mbox{RedLine}^{\dot b}[1,2] = \int
d^8x\, \Big((T)^j{}_i \Big(\theta^a \theta_a \frac{\bar D^{\dot b}
D^2}{\partial^2} - i (\gamma^\mu)^{a{\dot b}} \theta_a \frac{\bar
D^2 D^2\partial_\mu}{2\partial^4}\Big) }$}
\put(6.8,-1.1){${\displaystyle \times
\frac{\mbox{\boldmath$\delta$}}{\mbox{\boldmath$\delta$}_2 j^j}
\cdot \frac{\delta}{\delta_1 \phi_{0i}} - i (M T)^{ij}
(\gamma^\mu)^{a{\dot b}} \theta_a
\Big(\frac{D^2\partial_\mu}{4\partial^4}
\frac{\mbox{\boldmath$\delta$}}{\mbox{\boldmath$\delta$}_2
j^i}\Big)\cdot \frac{\mbox{\boldmath$\delta$}}{
\mbox{\boldmath$\delta$}_1 j^j}\Big) }$}
\end{picture}\\
\begin{picture}(0,2.2)
\put(0.25,0.5){\includegraphics[angle=180]{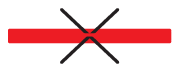}}
\put(2.8,0){${\displaystyle \mbox{RedWithCross}^{\dot b}[1,2] = -
i (M T)^{ij} \int d^8x\,(\gamma^\mu)^{a{\dot b}} \theta_a
\Big(\frac{D^2\partial_\mu}{4\partial^4}
\frac{\mbox{\boldmath$\delta$}}{\mbox{\boldmath$\delta$}_2
j^i}\Big)\cdot \frac{\mbox{\boldmath$\delta$}}{
\mbox{\boldmath$\delta$}_1 j^j}}$}
\end{picture}\\
\begin{picture}(0,1.1)
\put(0.18,-0.1){\includegraphics{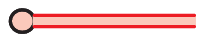}}
\put(2.8,0){${\displaystyle \mbox{PinkLine}^{\dot b}[1,2] = \int
d^8x\, \Big((T)^j{}_i \Big(\theta^a \theta_a \frac{\bar D^{\dot b}
D^2}{2\partial^2} - i (\gamma^\mu)^{a{\dot b}} \theta_a \frac{\bar
D^2 D^2\partial_\mu}{8\partial^4}\Big) }$}
\put(6.8,-1.1){${\displaystyle \times
\frac{\mbox{\boldmath$\delta$}}{\mbox{\boldmath$\delta$}_2 j^j}
\cdot \frac{\delta}{\delta_1 \phi_{0i}} - i (M T)^{ij}
(\gamma^\mu)^{a{\dot b}} \theta_a
\Big(\frac{D^2\partial_\mu}{16\partial^4}
\frac{\mbox{\boldmath$\delta$}}{\mbox{\boldmath$\delta$}_2
j^i}\Big)\cdot \frac{\mbox{\boldmath$\delta$}}{
\mbox{\boldmath$\delta$}_1 j^j}\Big) }$}
\end{picture}\\
\begin{picture}(0,2.2)
\put(0.25,-0.2){\includegraphics{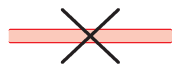}}
\put(2.8,0){${\displaystyle \mbox{PinkWithCross}^{\dot b}[1,2] = -
i (M T)^{ij} \int d^8x\,(\gamma^\mu)^{a{\dot b}} \theta_a
\Big(\frac{D^2\partial_\mu}{8\partial^4}
\frac{\mbox{\boldmath$\delta$}}{\mbox{\boldmath$\delta$}_2
j^i}\Big)\cdot \frac{\mbox{\boldmath$\delta$}}{
\mbox{\boldmath$\delta$}_1 j^j} }$}
\end{picture}\\
\begin{picture}(0,1.1)
\put(0.16,0.3){\includegraphics[angle=180]{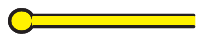}}
\put(2.8,0){${\displaystyle \mbox{YellowLine}_\mu[\alpha;1,2] =
\int d^8x\, \alpha\, \Big( 2i (T)^j{}_i
\Big(\frac{\partial_\mu}{\partial^2}
\frac{\mbox{\boldmath$\delta$}}{\mbox{\boldmath$\delta$}_2
j^j}\Big) \cdot \frac{\delta}{\delta_1 \phi_{0i}} }$}
\put(10.6,-1){${\displaystyle - i (M T)^{ij}
\Big(\frac{D^2\partial_\mu}{16\partial^4}
\frac{\mbox{\boldmath$\delta$}}{\mbox{\boldmath$\delta$}_2
j^i}\Big) \cdot \frac{\mbox{\boldmath$\delta$}}{
\mbox{\boldmath$\delta$}_1 j^j}\Big) }$}
\end{picture}\\
\begin{picture}(0,2.0)
\put(0.25,0.3){\includegraphics[angle=180]{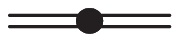}}
\put(2.8,0){${\displaystyle \mbox{LineWithDot}[\alpha;1,2] \equiv
\frac{i}{4} \int d^8x\, \alpha\, \Big(
\frac{\mbox{\boldmath$\delta$}}{\mbox{\boldmath$\delta$}_2 j^*_i}
\frac{\mbox{\boldmath$\delta$}}{\mbox{\boldmath$\delta$}_1 j^i} +
\frac{\mbox{\boldmath$\delta$}}{\mbox{\boldmath$\delta$}_1 j^*_i}
\frac{\mbox{\boldmath$\delta$}}{\mbox{\boldmath$\delta$}_2 j^i} +
M^{ij} \Big(\frac{D^2}{8\partial^2}
\frac{\mbox{\boldmath$\delta$}}{\mbox{\boldmath$\delta$}_2
j^i}\Big) }$} \put(2.9,-1){${\displaystyle \times
\frac{\mbox{\boldmath$\delta$}}{\mbox{\boldmath$\delta$}_1 j^j} +
M^*_{ij} \Big(\frac{\bar D^2}{8\partial^2}
\frac{\mbox{\boldmath$\delta$}}{\mbox{\boldmath$\delta$}_2
j^*_i}\Big)
\frac{\mbox{\boldmath$\delta$}}{\mbox{\boldmath$\delta$}_1 j^*_j}
+ M^{ij} \Big(\frac{D^2}{8\partial^2}
\frac{\mbox{\boldmath$\delta$}}{\mbox{\boldmath$\delta$}_1
j^i}\Big)
\frac{\mbox{\boldmath$\delta$}}{\mbox{\boldmath$\delta$}_2 j^j} +
M^*_{ij} \Big(\frac{\bar D^2}{8\partial^2}
\frac{\mbox{\boldmath$\delta$}}{\mbox{\boldmath$\delta$}_1
j^*_i}\Big)
\frac{\mbox{\boldmath$\delta$}}{\mbox{\boldmath$\delta$}_2 j^*_j}
\Big) }$}
\end{picture}
\vspace*{1.2cm} \caption{Definitions of some effective lines which
are used in this paper. Expressions for the lines with a cross are
obtained after extracting terms proportional to masses of the
Pauli--Villars fields.}\label{Figure_Feynman_Rules}
\end{figure}

\noindent We will graphically denote this operator by a green
effective line with the ends $1$ and $2$. In Fig.
\ref{Figure_Sum_Of_Lines} the identity (\ref{GreenLine_Identity})
is presented in the graphical form. The indexes $1$ and $2$ in the
right hand side of Eq. (\ref{GreenLine_Definition}) point the
vertices to which the corresponding derivatives act. (Sometimes we
will omit these indexes if they coincide.) Actually this
expression can be considered as a modification of the effective
propagator (multiplied by two derivatives with respect to the
$\phi_0$). Also we will also use other effective lines. Our
notation is presented in Fig. \ref{Figure_Feynman_Rules}. Note
that in the case of using color lines we do not sometimes
explicitly draw the external lines. Instead of them we draw a
small circle, to which we attach corresponding $\theta$-s.

\begin{figure}[h]
\begin{center}
\begin{picture}(0,1.8)
\put(-6,0.6){${\displaystyle \frac{1}{2\pi}{\cal V}_4\cdot
\frac{\beta(\alpha_0)}{\alpha_0^2} =
\frac{d}{d\ln\Lambda}}\Bigg($} \put(2.4,0.6){+} \hspace*{-1.8cm}
\includegraphics[scale=0.88]{figure1.eps}
\put(0.3,0.6){$\Bigg)\Bigg|_{\mbox{\scriptsize \boldmath$V$}\to
\theta^4}$} \put(-11.5,-1){${\displaystyle =
\frac{d}{d\ln\Lambda}}$}
\put(-10.6,-2){\includegraphics[scale=0.4]{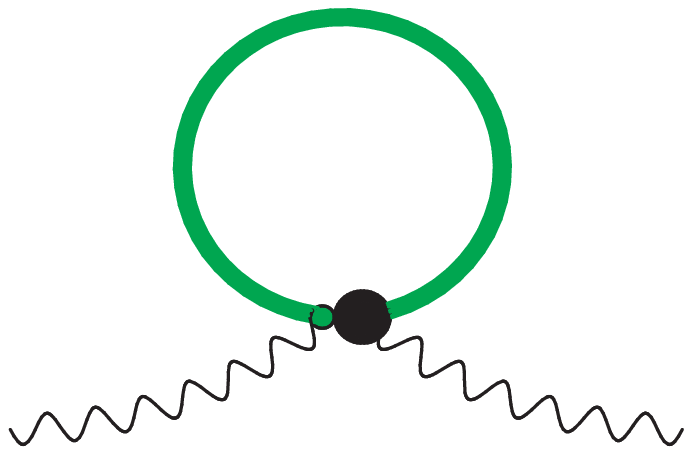}}
\put(-7.6,-1){${\displaystyle +\ \delta }$}
\end{picture}
\end{center}
\vspace*{12mm} \caption{Obtaining the $\beta$-function (defined in
terms of the bare coupling constant) from the Schwinger--Dyson
equation. The additional term $\delta$ is given by Eq.
(\ref{Small_Delta}).} \label{Figure_SD_Equation}
\end{figure}

Using the identity (\ref{GreenLine_Identity}) it is possible to
rewrite the Schwinger--Dyson equation
(\ref{SD_Equation_For_2V_Function}) in a different form. The
result is presented in Fig. \ref{Figure_SD_Equation}. In the
analytical form it can be written as

\begin{equation}\label{SD_Equation_With_Green_Line}
\frac{1}{2} \int d^8x\, d^8y\,(\theta^4)_x (\theta^4)_y
\frac{d}{d\ln\Lambda} \frac{\delta^2(\Delta\Gamma)}{\delta
\mbox{\boldmath$V$}_x \delta \mbox{\boldmath$V$}_y} = -i
\frac{d}{d\ln\Lambda}\int d^8y\,(\theta^4)_y\,
\mbox{GreenLine}\cdot \frac{\delta\Gamma}{\delta
\mbox{\boldmath$V$}_y} + \delta,
\end{equation}

\noindent where

\begin{equation}\label{Small_Delta}
\delta \equiv i \int d^8x\,d^8y\,(\theta^4)_y (T)^j{}_i
\Big(\frac{D^2}{4\partial^2} \theta^a \theta_a
\frac{\mbox{\boldmath$\delta$}}{\mbox{\boldmath$\delta$}
j^j}\Big)_x \frac{\delta^2\Gamma}{\delta (\phi_{0i})_x \delta
\mbox{\boldmath$V$}_y}.
\end{equation}

\begin{figure}[h]
\begin{center}
\begin{picture}(0,3)
\put(-5.5,0.4){\includegraphics[angle=180]{line_green.eps}}
\put(-0.5,1.5){\includegraphics[angle=180]{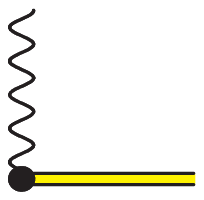}}
\put(3,1.5){\includegraphics[angle=180]{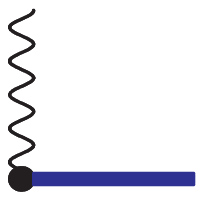}}
\put(-1.8,-0.4){$=$} \put(2.2,-0.4){$+$}
\put(-1.0,1.7){$\bar\theta^{\dot a} (\gamma^\mu)_{\dot a}{}^b
\theta_b$} \put(2.7,1.7){$2\theta^a\theta_a \bar\theta^{\dot b}$}
\put(0.5,0.0){$\mu$} \put(4.1,-0.05){${\dot b}$}
\end{picture}
\end{center}
\caption{The green effective line can be presented as a sum of the
blue and yellow effective lines.}\label{Figure_Green=Blue+Yellow}
\end{figure}

The green effective line can be presented as a sum of the blue and
yellow lines, see Fig. \ref{Figure_Green=Blue+Yellow}:

\begin{equation}\label{Splitting_Of_Line}
\mbox{GreenLine}[1,2] = 2\cdot \mbox{BlueLine}_{\dot b}[\theta^a
\theta_a \bar\theta^{\dot b};1,2]+
\mbox{YellowLine}_\mu[\bar\theta^{\dot a} (\gamma^\mu)_{\dot
a}{}^b \theta_b;1,2],
\end{equation}

\noindent where

\begin{eqnarray}\label{BlueLine_Definition}
&& \mbox{BlueLine}^{\dot b}[\alpha;1,2] \equiv \int d^8x\,
\alpha\,(T)^j{}_i \Big(\frac{\bar D^{\dot b} D^2}{8\partial^2}
\frac{\mbox{\boldmath$\delta$}}{\mbox{\boldmath$\delta$}_2
j^j}\Big) \cdot \frac{\delta}{\delta_1
\phi_{0i}};\\
&& \mbox{YellowLine}_\mu[\alpha;1,2] \equiv \int d^8x\, \alpha\,
\Big( 2i (T)^j{}_i \Big(\frac{\partial_\mu}{\partial^2}
\frac{\mbox{\boldmath$\delta$}}{\mbox{\boldmath$\delta$}_2
j^j}\Big) \cdot \frac{\delta}{\delta_1 \phi_{0i}} - i (M T)^{ij}
\Big(\frac{D^2\partial_\mu}{16\partial^4}
\frac{\mbox{\boldmath$\delta$}}{\mbox{\boldmath$\delta$}_2
j^i}\Big) \cdot \frac{\mbox{\boldmath$\delta$}}{
\mbox{\boldmath$\delta$}_1 j^j}\Big).\nonumber
\end{eqnarray}

\noindent Using Eq. (\ref{Splitting_Of_Line}) and the identity

\begin{equation}\label{Splitting_Of_Additional_Term}
\frac{D^2}{4\partial^2} \theta^a\theta_a = \theta^a\theta_a
\frac{D^2}{4\partial^2} + \Big(\theta^a \frac{D_a}{\partial^2} -
\frac{1}{\partial^2}\Big)
\end{equation}

\begin{figure}[h]
\begin{center}
\begin{picture}(0,1.8)
\put(-7.5,0.05){${\displaystyle \frac{d}{d\ln\Lambda}}$}
\put(-7.0,-1){\includegraphics[scale=0.4]{figure4.eps}}
\put(-4.2,0){${\displaystyle +\, \delta \ =}$}
\put(-2.7,0){${\displaystyle \Big(\frac{d}{d\ln\Lambda}}$}
\put(0.9,0){${\displaystyle +\delta_1\Big)}$}
\put(-1.9,-1){\includegraphics[scale=0.4]{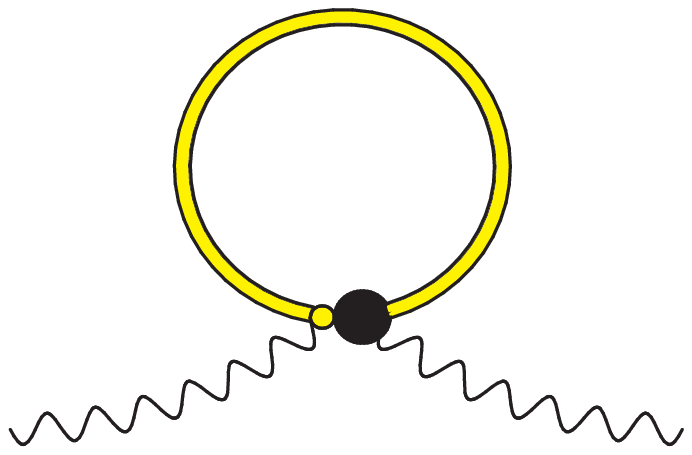}}
\put(2.2,0){$+$} \put(2.7,0){${\displaystyle
\Big(\frac{d}{d\ln\Lambda}}$} \put(6.3,0){${\displaystyle
+\delta_2\Big)}$}
\put(3.6,-1){\includegraphics[scale=0.4]{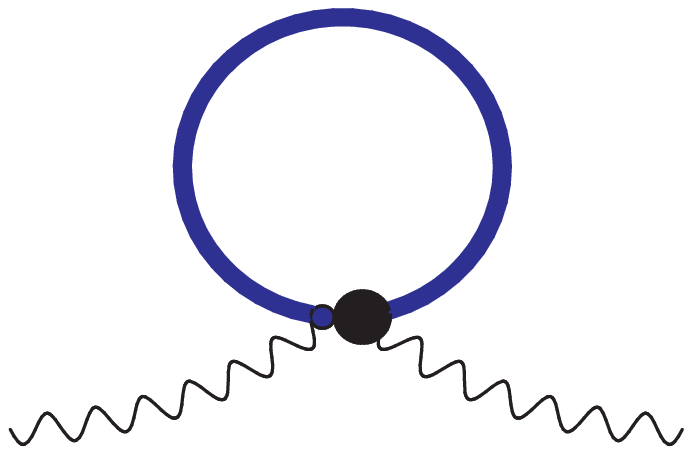}}
\put(-4.1,-1.2){$\theta^4$} \put(1.1,-1.2){$\theta^4$}
\put(6.5,-1.2){$\theta^4$} \put(-2.4,-1.2){$\bar\theta^{\dot a}
(\gamma^\mu)_{\dot a}{}^b \theta_b$}
\put(3.5,-1.2){$2\theta^a\theta_a \bar\theta^{\dot b}$}
\put(-0.1,-1.0){$\vector(0,-1){2.3}$}
\put(0.1,-2.2){\cite{Stepanyantz:2004sg}}
\put(5.25,-0.5){$\vector(0,-1){1.6}$} \put(5.2,-2.5){$0$}
\put(5.5,-1.8){(this paper)} \put(-7.0,-4.0){${\displaystyle N_f
{\cal V}_4 \frac{d}{d\ln\Lambda} \int \frac{d^4q}{(2\pi)^4}
\frac{4}{q^2} \frac{d}{dq^2}\Bigg(2\ln G -\sum\limits_{I=1}^n
c_I\Big(\ln(q^2 G^2 + M^2 J^2) + \frac{M^2 J}{q^2 G^2 + M^2
J^2}\Big)_I\Bigg)}$}
\end{picture}
\vspace*{3.8cm}
\end{center} \caption{The result of substituting the solution of
Ward identities into the Schwinger--Dyson equation (for the
diagram with the yellow effective line).} \label{Figure_SD_Result}
\end{figure}

\noindent it is possible to split the effective diagram presented
in Fig. \ref{Figure_SD_Equation} into two parts. This is shown in
Fig. \ref{Figure_SD_Result}. In this figure we use the notation

\begin{eqnarray}\label{Small_Deltas}
&& \delta_1 \equiv i \int d^8x\,d^8y\,(\theta^4)_y (T)^j{}_i \Big(
\frac{\theta^a D_a - 1}{\partial^2}
\frac{\mbox{\boldmath$\delta$}}{\mbox{\boldmath$\delta$}
j^j}\Big)_x \frac{\delta^2\Gamma}{\delta (\phi_{0i})_x \delta
\mbox{\boldmath$V$}_y};\nonumber\\
&& \delta_2 \equiv i \int d^8x\,d^8y\,(\theta^4)_y (T)^j{}_i
\Big(\theta^a \theta_a \frac{D^2}{4\partial^2}
\frac{\mbox{\boldmath$\delta$}}{\mbox{\boldmath$\delta$}
j^j}\Big)_x \frac{\delta^2\Gamma}{\delta (\phi_{0i})_x \delta
\mbox{\boldmath$V$}_y} = \delta-\delta_1.\quad
\end{eqnarray}

\noindent The result for the effective diagram with the yellow
effective line (including $\delta_1$) can be expressed in terms of
the anomalous dimension of the matter superfield
\cite{Stepanyantz:2004sg}.\footnote{In \cite{Stepanyantz:2004sg}
only the case $N_f=1$ is considered.} For this purpose it is
necessary to substitute the solution of the Ward identity for the
effective vertex. The result is also presented in Fig.
\ref{Figure_SD_Result}. It is given by an integral of a total
derivative with respect to a loop momentum. Such a structure
allows to reduce a number of momentum integrations and relate the
$\beta$-function with the anomalous dimension of the matter
superfield. Really, calculating the integral in the
four-dimensional spherical coordinates we obtain

\begin{eqnarray}
&&\hspace*{-4mm} N_f {\cal V}_4 \frac{d}{d\ln\Lambda} \int
\frac{d^4q}{(2\pi)^4} \frac{4}{q^2} \frac{d}{dq^2}\Bigg(2\ln G
-\sum\limits_{I=1}^n c_I \Big(\ln(q^2 G^2 + M^2 J^2) + \frac{M^2
J}{q^2 G^2 + M^2 J^2}\Big)_I\Bigg)\nonumber\\
&&\hspace*{-4mm} = \frac{1}{2\pi} {\cal V}_4 \cdot
\frac{N_f}{\pi} \Big(1-\gamma(\alpha_0)\Big),
\end{eqnarray}

\noindent where we take into account that for a function $f(q^2)$
which rapidly decreases at the infinity

\begin{equation}
\int \frac{d^4q}{(2\pi)^4} \frac{1}{q^2} \frac{df}{dq^2} =
\frac{1}{16\pi^2} \int\limits_0^\infty dq^2\frac{d f}{dq^2} =
\frac{1}{16\pi^2}\Big(f(\infty)- f(0)\Big) = - \frac{1}{16\pi^2}
f(0).
\end{equation}

\noindent (The functions considered here rapidly decrease at the
infinity due to the higher derivative regularization.) As a
consequence, the contribution of the diagram with the yellow
effective line gives the exact NSVZ $\beta$-function

\begin{equation}
\beta(\alpha_0) = \frac{\alpha_0^2
N_f}{\pi}\Big(1-\gamma(\alpha_0)\Big).
\end{equation}

\begin{figure}[h]
\bigskip
\begin{center}
\vspace*{1cm}
\begin{picture}(0,0)
\put(-3.3,0.05){${\displaystyle 2\cdot \frac{d}{d\ln\Lambda}}$}
\put(-2,-0.8){\includegraphics[scale=0.4]{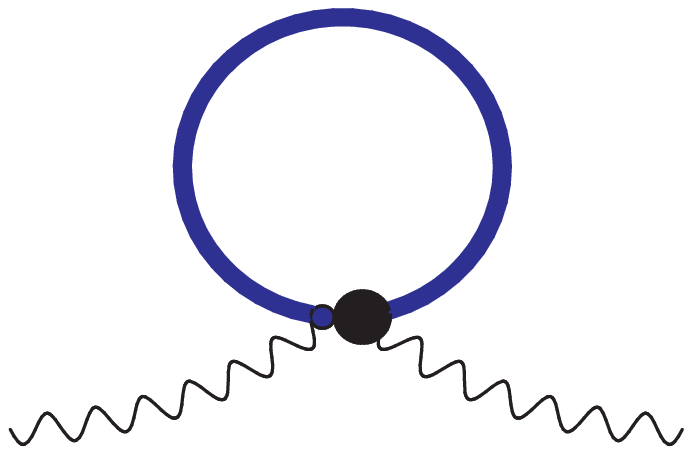}}
\put(-2.2,-1){$\theta^a \theta_a \bar\theta^{\dot b}$}
\put(0.8,-1){$\theta^4$} \put(1,0){${\displaystyle +\ \delta_2\ =\
0.}$}
\end{picture}
\vspace*{7mm}
\end{center}
\caption{This identity was suggested in \cite{Stepanyantz:2004sg}.
In this paper this equality is proved.}
\label{Figure_New_Identity}
\end{figure}

However, using this method it is impossible to calculate the
diagram with the blue effective line in Fig.
\ref{Figure_SD_Result}. Some explicit calculations in the lowest
(three- and four-) loops \cite{Pimenov:2006cu} show that this
diagram plus $\delta_2$ is also given by an integral of a total
derivative and vanishes. In the graphical form this is presented
in Figs. \ref{Figure_SD_Result} and \ref{Figure_New_Identity}.
This equality can be considered as a nontrivial relation between
Green functions \cite{Stepanyantz:2004sg}. It was proved
indirectly in \cite{Stepanyantz:2011jy} using a method proposed in
\cite{Stepanyantz:2005wk}. In particular, it is possible to prove
that the integrand corresponding to this diagram is a total
derivative. In the analytical form the equality presented in Fig.
\ref{Figure_New_Identity} can be written as

\begin{equation}\label{New_Identity_Analytical_Form}
-i \frac{d}{d\ln\Lambda} \int d^8x\,d^8y\,(\theta^4)_y (T)^j{}_i
\Big(\theta^a\theta_a \bar\theta^{\dot b} \frac{\bar D_{\dot b}
D^2}{4\partial^2} - \theta^a \theta_a \frac{D^2}{4\partial^2}
\Big)_x \frac{\mbox{\boldmath$\delta$} }{\mbox{\boldmath$\delta$}
(j^j)_x} \frac{\delta}{\delta (\phi_{0i})_x}
\frac{\delta\Gamma}{\delta \mbox{\boldmath$V$}_y} = 0.
\end{equation}

In this paper we prove this identity directly. Moreover, we prove
that the $\beta$-function is given by integrals of double total
derivatives. In order to do this, it is necessary to use two
ideas.

1. First, it is necessary to rewrite the effective vertices in the
diagrams presented in Fig. \ref{Figure_SD_Equation} or Fig.
\ref{Figure_New_Identity} using the Schwinger--Dyson equation one
more time. This procedure was first proposed in \cite{Shevtsova}.
Let us, for example, start with Eq.
(\ref{SD_Equation_With_Green_Line}) and substitute the expression
for $\delta\Gamma/\delta \mbox{\boldmath$V$}_y$ from the
Schwinger--Dyson equation (\ref{SD_Equation}). It is convenient to
write the result in terms of the Routhian $\gamma$, because in
this case the number of effective diagrams is less. Details of
this calculation are presented in Appendix
\ref{Appendix_Second_SD}. The result can be written in the
following form:

\begin{equation}\label{SD_Equation_With_Two_Green_Lines}
\frac{1}{2} \int d^8x\, d^8y\,(\theta^4)_x (\theta^4)_y
\frac{d}{d\ln\Lambda} \frac{\delta^2 \Delta\Gamma}{\delta
\mbox{\boldmath$V$}_x \delta \mbox{\boldmath$V$}_y} = -2
\frac{d}{d\ln\Lambda} (\mbox{GreenLine})^2 \cdot \gamma + \Delta,
\end{equation}

\noindent where

\begin{eqnarray}\label{Delta_Definition}
&& \Delta \equiv -i \frac{d}{d\ln\Lambda} \int
d^8x\,d^8y\,(\theta^4)_x \Big(\frac{D^2}{4\partial^4}\Big)_y
\Big(\frac{\delta^2\gamma}{\delta (\phi_i)_x \delta
(\phi_j)_y}\Big)^{-1}\times\nonumber\\
&&\qquad\qquad\quad \times \Big\{C(R)_k{}^i M^{j\,k} \delta^8_{xy}
- (M T)^{im} (M T)^{lj}  \Big(\frac{D^2}{32\partial^2}\Big)_x
\Big(\frac{\delta^2\gamma}{\delta (\phi_m)_x \delta
(\phi_l)_y}\Big)^{-1}\Big\}\qquad
\end{eqnarray}

\noindent with

\begin{equation}\label{C(R)_Definition}
C(R)_i{}^m = (T)_i{}^k (T)_k{}^m = \left(
\begin{array}{cc}
1 & 0 \\
0 & 1
\end{array} \right) \cdot \delta_{\alpha\beta}\cdot \delta_{IJ}.
\end{equation}

\noindent ($\Delta$ can be graphically interpreted as a one-loop
effective diagram. However, in order to avoid too large number of
effective diagrams we write this term explicitly.) Note that this
expression does not contain infrared divergences due to the
differentiation with respect to $\ln\Lambda$, which should be made
before the momentum integration.

\begin{figure}[h]
\begin{center}
\includegraphics[scale=0.7]{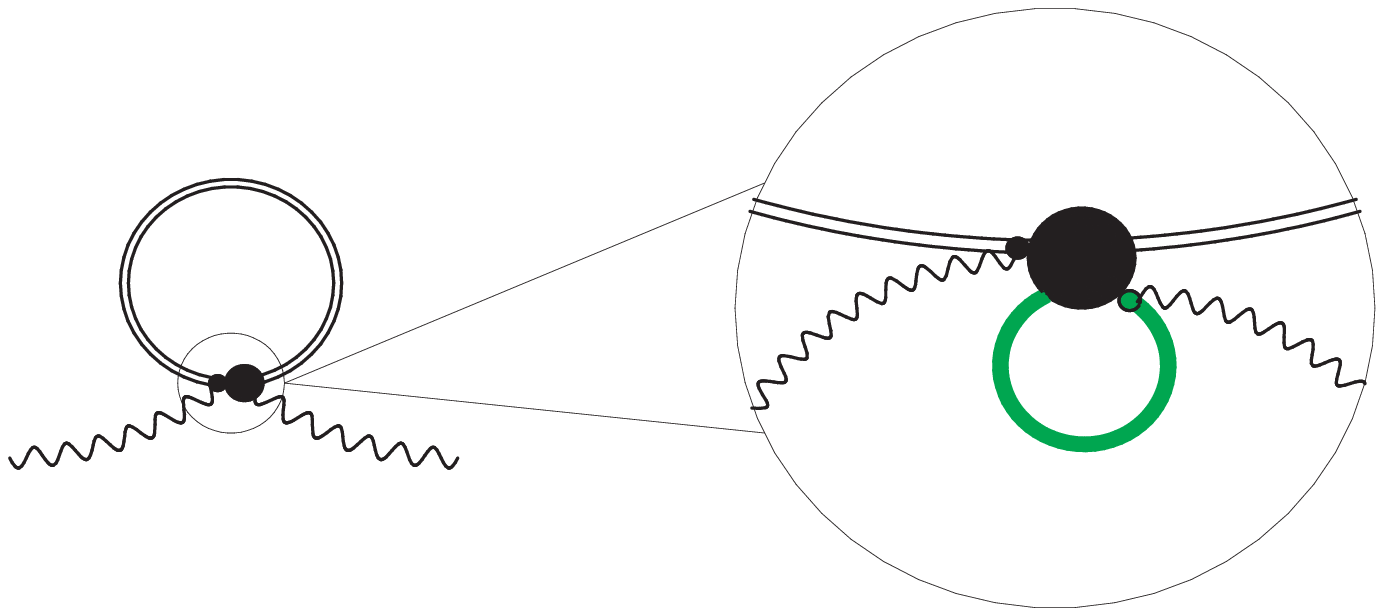}
\end{center}
\vspace*{-7mm} \caption{Applying the Schwinger--Dyson equation to
the effective vertex we can see "the inner structure" of the
effective diagram. (The large circles can correspond to any
effective line.)}\label{Figure_Zoom}
\end{figure}

\begin{figure}[h]
\begin{center}
\begin{picture}(0,1.8) \put(-4.05,0.6){${\displaystyle
\frac{1}{2\pi} {\cal V}_4 \cdot \frac{\beta(\alpha_0)}{\alpha_0^2}
= -2 \frac{d}{d\ln\Lambda}}$} \put(4.6,0.6){$+\ \Delta$}
\put(0.4,0.1){\includegraphics[scale=0.88]{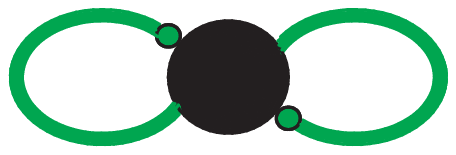}}

\put(2.0,0){\vector(-3,-2){3}} \put(2.5,0){\vector(1,-3){0.7}}

\put(-7.1,-3){${\displaystyle \Delta_1\ -\
2\,\frac{d}{d\ln\Lambda} }$} \put(0.5,-3){${\displaystyle  -\
4\,\frac{d}{d\ln\Lambda} }$}
\put(-4.5,-3.5){\includegraphics[scale=0.88]{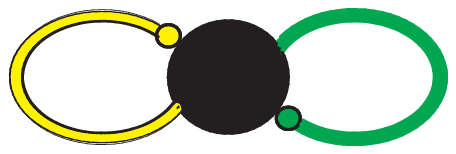}}
\put(2.3,-3.5){\includegraphics[scale=0.88]{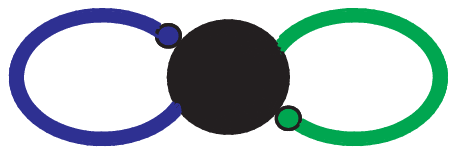}}
\put(6.54,-3){$+ \ \Delta_2$}

\put(3.8,-3.8){\vector(1,-3){0.4}}
\put(-3.4,-3.8){\vector(-3,-2){1.5}}
\put(-7.0,-5.5){${\displaystyle \beta(\alpha_0)=\frac{\alpha_0^2
N_f}{\pi}\Big(1-\gamma(\alpha_0)\Big)}$}
\put(4.2,-5.5){${\displaystyle 0}$}

\end{picture}

\vspace*{5.3cm}
\end{center}
\caption{Applying the Schwinger--Dyson equation one more time, it
is possible to obtain that a $\beta$-function is determined by the
two-loop effective diagrams.}\label{Figure_Double_Diagram}
\end{figure}

A simple (qualitative) graphical interpretation of results
obtained in Appendix \ref{Appendix_Second_SD} is presented in Fig.
\ref{Figure_Zoom}. (A white line can be substituted by any other
effective line.) In particular, if this rule is applied to the
diagram in the left hand side of the equation presented in Fig.
\ref{Figure_SD_Result}, then a $\beta$-function will be determined
by the two-loop effective diagram presented in Fig.
\ref{Figure_Double_Diagram}. As earlier, it is convenient to split
this effective diagram into two parts using Eq.
(\ref{Splitting_Of_Line}). The result is graphically presented in
Fig. \ref{Figure_Double_Diagram}, where (see Appendix
\ref{Appendix_Subsection_New_Identity})

\begin{equation}
\Delta_2 = \delta_2;\qquad \Delta_1 = \Delta-\Delta_2.
\end{equation}

\noindent In Appendix \ref{Appendix_Subsection_New_Identity} the
expression $\Delta_2$ is also written in terms of the functions
$G$ and $ MJ$. $\Delta_1$ can be easily found using Eq.
(\ref{Delta_Definition}). After simple transformations we obtain

\begin{eqnarray}\label{Delta1}
&& \Delta_1 = -i C(R)_k{}^i \frac{d}{d\ln\Lambda} \int d^8x\,
d^8y\, (\theta^4)_y \Big(\frac{D^2}{4\partial^4}\Big)_x
\Big(\frac{\delta^2\gamma}{\delta (\phi_i)_x \delta
(\phi_j)_y}\Big)^{-1} \nonumber\\
&&\ \ \times \Big\{M^{jk} \delta^8_{xy} - M^{jn} M^{mk}
\Big(\frac{D^2}{32\partial^4}\Big)_x
\Big(\frac{\delta^2\gamma}{\delta (\phi_m)_x \delta
(\phi_n)_y}\Big)^{-1} - \frac{\delta^2\gamma}{\delta (\phi_{k})_x
\delta (\phi_{0j})_y} \Big\}; \\
\label{Delta2} && \Delta_2 = - 2i C(R)_i{}^j \frac{d}{d\ln\Lambda}
\int d^8x\, d^8y\,(\theta^4)_y \frac{1}{\partial^2}
\Big(\frac{\delta^2\gamma}{\delta (\phi_{j})_x
(\phi_{k})_y}\Big)^{-1} \frac{\delta^2\gamma}{\delta (\phi_{0i})_x
\delta (\phi_{0k})_y}.\qquad\quad
\end{eqnarray}

2. Now let us proceed to the second idea. An attempt to present
all two-loop effective diagrams in Fig.
\ref{Figure_Double_Diagram} as integrals of total derivatives
encounters considerable problems. The reason can be understood
from the results of \cite{Stepanyantz:2011jy}. The matter is that
the total derivative in this case nontrivially depends on the
number of vertices in a diagram. Therefore, it seems impossible to
write the total derivatives in the form of effective diagrams.
However, the solution can be found. For this purpose we introduce
the parameter $g$ according to the prescription
(\ref{Introducing_Of_G}).

Let us differentiate the upper diagram in Fig.
\ref{Figure_Double_Diagram} with respect to the parameter $g$
using the identity

\begin{eqnarray}\label{Gamma_Derivative}
&&\hspace*{-6mm} \frac{\partial\gamma}{\partial\ln g} =
\frac{1}{2} \int d^8x\, \Big\{ -\frac{1}{2} (\phi^{*i} +
\phi_0^{*i}) (\phi_i + \phi_{0i}) + (\phi_i + \phi_{0i})_x
\frac{\delta\gamma}{\delta (\phi_{0i})_x} + (\phi^{*i} +
\phi_{0}^{*i})_x \frac{\delta\gamma}{\delta
(\phi_{0}^{*i})_x} \nonumber\\
&&\hspace*{-6mm}  -\frac{i}{2}
\Big(\frac{\delta^2\gamma}{\delta\phi_i \delta\phi^{*i}}\Big)^{-1}
- i M^{j\,i}
\Big(\frac{D^2}{16\partial^2}\Big)_x\Big(\frac{\delta^2\gamma}{\delta
(\phi_i)_x \delta(\phi_j)_y}\Big)^{-1}_{y=x} - i M^*_{j\,i}
\Big(\frac{\bar
D^2}{16\partial^2}\Big)_x\Big(\frac{\delta^2\gamma}{\delta
(\phi^{*i})_x \delta(\phi^{*j})_y}\Big)^{-1}_{y=x}
\Big\},\nonumber\\
\end{eqnarray}

\begin{figure}[h]
\begin{picture}(0,2.0)
\put(1.0,1.3){$\displaystyle{ \frac{\partial}{\partial \ln g}
\Big( \frac{1}{2} \int d^8x\, d^8y\,(\theta^4)_x (\theta^4)_y
\frac{d}{d\ln\Lambda} \frac{\delta^2 \Delta\Gamma}{\delta
\mbox{\boldmath$V$}_x \delta \mbox{\boldmath$V$}_y}\Big)\ =\
\frac{d}{d\ln\Lambda} \Bigg\{ }$} \put(10.85,1.3){$-2$}
\put(11.7,2.2){$(1)$}
\put(11.5,0.9){\includegraphics[scale=0.7]{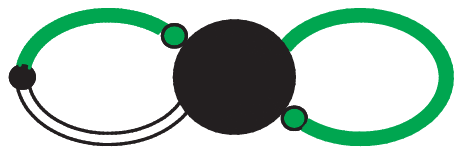}}
\put(1.2,0.3){$(2)$}
\put(1.3,-0.95){\includegraphics[scale=0.7]{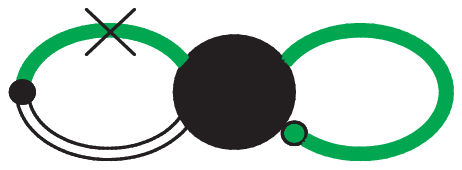}}
\put(5.8,0.3){$(3)$}
\put(11.1,-1.15){\includegraphics[scale=0.7]{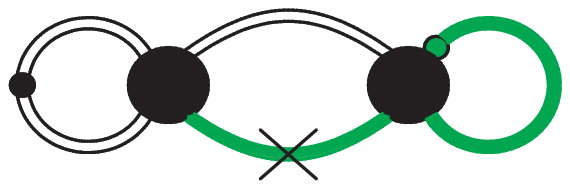}}
\put(11,0.3){$(4)$}
\put(5.8,-1){\includegraphics[scale=0.7]{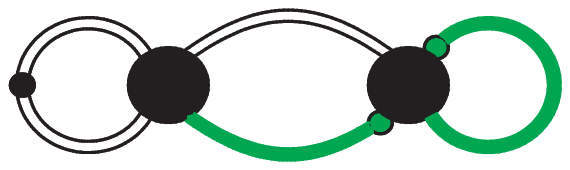}}
\put(1,-2.0){$(5)$}
\put(1.1,-4.1){\includegraphics[scale=0.7]{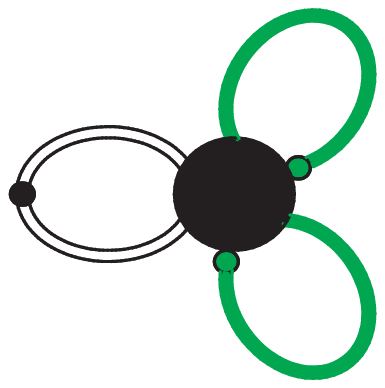}}
\put(4.3,-2.0){$(6)$}
\put(4.5,-4){\includegraphics[scale=0.7]{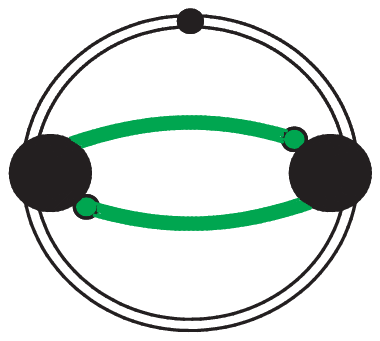}}
\put(7.8,-2.0){$(7)$}
\put(8.0,-4){\includegraphics[scale=0.7]{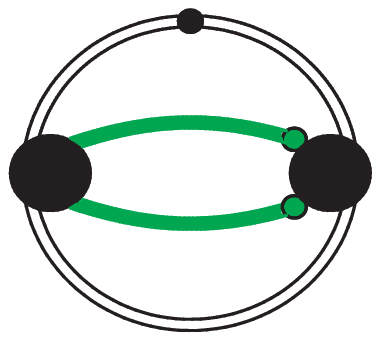}}
\put(11.2,-2.0){$(8)$}
\put(11.5,-3.4){\includegraphics[scale=0.7]{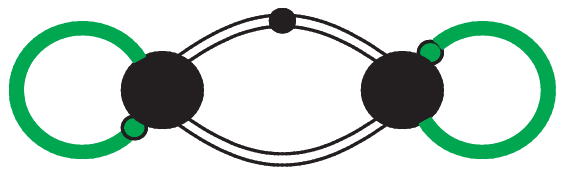}}
\put(0.5,-0.55){$-\ 2$} \put(5,-0.55){$-\ 2$} \put(10.3,-0.55){$-\
2$} \put(0.5,-2.95){$-$} \put(3.8,-2.95){$-\ 2$}
\put(7.3,-2.95){$-\ 2$} \put(10.8,-2.95){$-\ 2$}
\put(0.8,-4.8){$(9)$}\put(0.4,-5.4){$+\ 4$}
\put(1.2,-5.8){\includegraphics[scale=0.7]{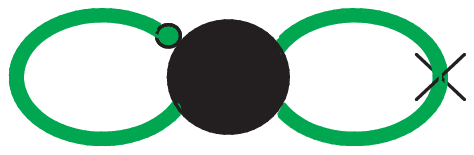}}
\put(4.7,-5.4){${\displaystyle\ \Bigg\}\ \  + \ \
\frac{\partial\Delta}{\partial\ln g}}$}
\end{picture}
\vspace*{6cm} \caption{These diagrams are obtained after
differentiating the effective diagram presented in Fig.
\ref{Figure_Double_Diagram} with respect to the parameter $g$.
Below we will see that this trick allows to write the result as an
integral of a double total derivative. The term
$\partial\Delta/\partial\ln g$ is given by Eq.
(\ref{Delta_Derivative}).} \label{Figure_Green_Diagrams}
\end{figure}

\noindent which is proved in Appendix
\ref{Appendix_G_Derivative_Routhian}. The technique constructed in
this appendix allows to calculate the derivative of Eq.
(\ref{SD_Equation_With_Two_Green_Lines}) with respect to the
parameter $g$. The result is presented in Fig.
\ref{Figure_Green_Diagrams}. The term $\partial\Delta/\partial\ln
g$ is calculated in Appendix \ref{Appendix_G_Derivative_One-Loop}
and is given by Eq. (\ref{Delta_Derivative}). The diagrams
presented in Fig. \ref{Figure_Green_Diagrams} are obtained by
differentiating the upper diagram in Fig.
\ref{Figure_Double_Diagram}. Technical details of this calculation
are presented in Appendixes \ref{Appendix_G_Derivative_Vertices}
and \ref{Appendix_G_Derivative_Lines}. Here we briefly discuss the
result. Diagrams (1)--(4) and (9) in this figure come from the
derivative of the green effective line. Actually, it is necessary
to differentiate the inverse Green functions inside
$\mbox{\boldmath$\delta$}/\mbox{\boldmath$\delta$} j$. All terms
corresponding to the two-loop effective diagrams are included into
diagrams (1), (2), and (9). Similarly, terms corresponding to
three-loop effective diagrams are included into diagrams (3) and
(4).

A green effective line with a cross (in diagrams (4) and (9))
corresponds to the operator

\begin{equation}\label{GreenWithCross}
\mbox{GreenWithCross}[1,2] = - i (M T)^{ij} \int d^8x\,
\bar\theta^{\dot a} (\gamma^\mu)_{\dot a}{}^b \theta_b
\Big(\frac{D^2\partial_\mu}{16\partial^4}
\frac{\mbox{\boldmath$\delta$}}{\mbox{\boldmath$\delta$}_2
j^i}\Big) \frac{\mbox{\boldmath$\delta$}}{
\mbox{\boldmath$\delta$}_1 j^j},\qquad
\end{equation}

\noindent which is obtained from the operator $\mbox{GreenLine}$
by keeping only terms proportional to the Pauli--Villars masses.
Integrating by parts and using Eq. (\ref{MT_Symmetry}) it is easy
to see that this operator is symmetric with respect to the
permutation of the points $1$ and $2$:

\begin{equation}
\mbox{GreenWithCross}[1,2] = \mbox{GreenWithCross}[2,1].
\end{equation}

\noindent Due to this symmetry of arguments we do not mark out one
of the ends by a circle. The lines with crosses appear, because
terms with the masses are quadratic in
$\mbox{\boldmath$\delta$}/\mbox{\boldmath$\delta$} j$, while the
other terms are linear in
$\mbox{\boldmath$\delta$}/\mbox{\boldmath$\delta$} j$.

A line with a dot corresponds to the operator

\begin{eqnarray}\label{LineWithDot}
&& \mbox{LineWithDot}[\alpha;1,2] \equiv \frac{i}{4} \int
d^8x\,(\alpha)_x\Bigg(
\frac{\mbox{\boldmath$\delta$}}{\mbox{\boldmath$\delta$}_2 j^*_i}
\frac{\mbox{\boldmath$\delta$}}{\mbox{\boldmath$\delta$}_1 j^i} +
\frac{\mbox{\boldmath$\delta$}}{\mbox{\boldmath$\delta$}_1 j^*_i}
\frac{\mbox{\boldmath$\delta$}}{\mbox{\boldmath$\delta$}_2 j^i}
+ M^{ij}
\Big(\frac{D^2}{8\partial^2}
\frac{\mbox{\boldmath$\delta$}}{\mbox{\boldmath$\delta$}_1
j^i}\Big)
\frac{\mbox{\boldmath$\delta$}}{\mbox{\boldmath$\delta$}_2 j^j}
\qquad\nonumber\\
&& +M^*_{ij} \Big(\frac{D^2}{8\partial^2}
\frac{\mbox{\boldmath$\delta$}}{\mbox{\boldmath$\delta$}_1
j^*_i}\Big)
\frac{\mbox{\boldmath$\delta$}}{\mbox{\boldmath$\delta$}_2 j^*_j}
+ M^{ij}
\Big(\frac{D^2}{8\partial^2}
\frac{\mbox{\boldmath$\delta$}}{\mbox{\boldmath$\delta$}_2
j^i}\Big)
\frac{\mbox{\boldmath$\delta$}}{\mbox{\boldmath$\delta$}_1 j^j} +
M^*_{ij} \Big(\frac{D^2}{8\partial^2}
\frac{\mbox{\boldmath$\delta$}}{\mbox{\boldmath$\delta$}_2
j^*_i}\Big)
\frac{\mbox{\boldmath$\delta$}}{\mbox{\boldmath$\delta$}_1 j^*_j}
\Bigg)\qquad
\end{eqnarray}

\noindent with $\alpha=1$. Diagram (1) contains the operator

\begin{eqnarray}\label{GreenWhiteLine}
&& \mbox{GreenWhiteLine}[1,2] \equiv -\frac{1}{2} \cdot
\mbox{GreenLine}[1,3]\cdot \mbox{UsualLine}[2,3]
\nonumber\\
&&\qquad\qquad\qquad\qquad\qquad\qquad \times \int d^8x\,
\Big(\phi_0^{*i} \phi_{0i} + M^{ij} \phi_{0i}
\frac{D^2}{8\partial^2} \phi_{0j} + M^*_{ij} \phi_0^{*i}
\frac{\bar D^2}{8\partial^2} \phi_0^{*j}\Big)_{[3]},\qquad
\end{eqnarray}

\noindent where

\begin{equation}
\mbox{UsualLine}[2,3]= \int d^8x\,
\Big(\frac{\delta}{\delta_3 \phi_{0i}} \cdot
\frac{\mbox{\boldmath$\delta$}}{\mbox{\boldmath$\delta$}_2
j^i} + \frac{\delta}{\delta_3 \phi^{*i}_{0}} \cdot
\frac{\mbox{\boldmath$\delta$}}{\mbox{\boldmath$\delta$}_2
j^*_i} \Big).
\end{equation}

\noindent In Eq. (\ref{GreenWhiteLine}) the subscript $[3]$ means
that only the derivatives $\delta/\delta_3 \phi_0$ nontrivially
act to the argument of the bracket. Similarly, diagram (2) in Fig.
\ref{Figure_Green_Diagrams} contains the operator

\begin{eqnarray}\label{GreenWithCrossWhite}
&& \mbox{GreenWithCrossWhite}[1,2] \equiv -\frac{1}{2} \cdot
\mbox{GreenWithCross}[1,3]\cdot \mbox{UsualLine}[2,3]
\nonumber\\
&&\qquad\qquad\qquad\qquad\qquad\qquad \times \int d^8x\,
\Big(\phi_0^{*i} \phi_{0i} + M^{ij} \phi_{0i}
\frac{D^2}{8\partial^2} \phi_{0j} + M^*_{ij} \phi_0^{*i}
\frac{\bar D^2}{8\partial^2} \phi_0^{*j}\Big)_{[3]}.\qquad
\end{eqnarray}

Diagrams (5)--(8) correspond to differentiation of the four-point
function (the large black circle in Fig.
\ref{Figure_Double_Diagram}). Technical details of the
corresponding calculation can be found in Appendix
\ref{Appendix_G_Derivative_Vertices}.

Below we try to avoid writing large analytical expressions
corresponding to effective Feynman diagrams. Instead of this, we
write numerical coefficients for all diagrams, so that the
analytical expression can be unambiguously constructed using the
definitions of the effective lines. As an example, here we present
an analytical expression for the sum of diagrams presented in Fig.
\ref{Figure_Green_Diagrams}:

\begin{eqnarray}
&& \frac{\partial}{\partial \ln g} \Big( \frac{1}{2} \int d^8x\,
d^8y\,(\theta^4)_x (\theta^4)_y \frac{d}{d\ln\Lambda}
\frac{\delta^2 \Delta\Gamma}{\delta \mbox{\boldmath$V$}_x
\delta \mbox{\boldmath$V$}_y}\Big)\nonumber\\
&& = \frac{d}{d\ln\Lambda}\Big(- 2\cdot \mbox{GreenWhiteLine}\cdot \mbox{GreenLine}\cdot
\gamma \vphantom{\Big(}\nonumber\\
&& - 2\cdot \mbox{GreenWithCrossWhite}\cdot
\mbox{GreenLine}\cdot \gamma \vphantom{\Big(}
\nonumber\\
&& - 2\cdot \mbox{GreenLine}[2,1]\, \mbox{UsualLine}[1,2]\,
\mbox{LineWithDot}[1; 1,1]\,
\mbox{GreenLine}[2,2] \Big(\gamma[1]\,\gamma[2]\Big) \nonumber\\
&& - 2\cdot \mbox{GreenWithCross}[2,1]\, \mbox{UsualLine}[1,2]\,
\mbox{LineWithDot}[1; 1,1]\,
\mbox{GreenLine}[2,2] \Big(\gamma[1]\,\gamma[2]\Big) \nonumber\\
&& - \mbox{LineWithDot}[1]\cdot (\mbox{GreenLine})^2\cdot
\gamma\vphantom{\Big(}
\nonumber\\
&& -2\cdot \mbox{LineWithDot}[1; 1, 2]\, \mbox{UsualLine}[1,2]\,
\mbox{GreenLine}[1,2]\,\mbox{GreenLine}[2,1]\,\Big(\gamma[1]\,\gamma[2]\Big)\nonumber\\
&& -2\cdot \mbox{LineWithDot}[1; 1, 2]\, \mbox{UsualLine}[1,2]\,
(\mbox{GreenLine}[2,1])^2\,\Big(\gamma[1]\,\gamma[2]\Big)\nonumber\\
&& -2\cdot \mbox{LineWithDot}[1; 1, 2]\,\mbox{UsualLine}[1,2]\,
\mbox{GreenLine}[1,1]\,\mbox{GreenLine}[2,2]\Big(\gamma[1]\,\gamma[2]\Big)
\nonumber\\
&& +4 \cdot \mbox{GreenLine}\cdot\mbox{GreenWithCross}\cdot
\gamma\Big) + \frac{\partial\Delta}{\partial\ln g}.
\end{eqnarray}

\noindent Expressions for all differential operators corresponding
to the various lines can be found in Fig.
\ref{Figure_Feynman_Rules}. Let us remind that if ends of an
effective line coincide, we sometimes omit numbers which numerate
them. In Sect. \ref{Section_Double_Commutators} we prove that the
sum of diagrams presented in Fig. \ref{Figure_Green_Diagrams} in
the momentum representation is given by integrals of double total
derivatives.

\begin{figure}[h]
\vspace*{2.2cm}
\begin{picture}(0,0.5)
\put(0,1.4){${\displaystyle \frac{d}{d\ln\Lambda}\Bigg\{ }$}
\put(2.0,1.9){$(1)$} \put(1.4,1.4){$-\ 2$}
\put(2.4,0.95){\includegraphics[scale=0.7]{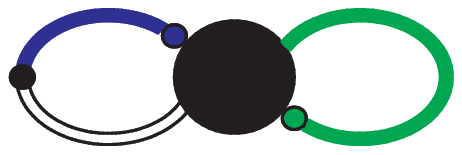}}
\put(6.8,1.9){$(2)$} \put(6.1,1.4){$-\ 2$}
\put(7.2,0.95){\includegraphics[scale=0.7]{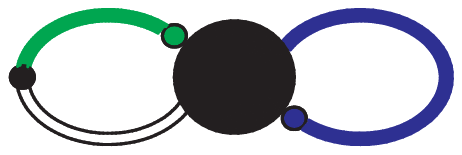}}
\put(11.7,1.9){$(3)$} \put(11.0,1.4){$-\ 2$}
\put(12.1,0.95){\includegraphics[scale=0.7]{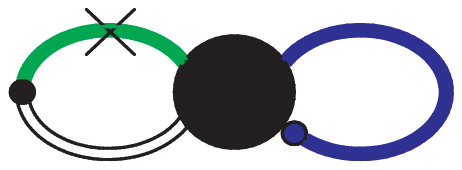}}
\put(0.4,0){$(4)$} \put(-0.1,-0.5){$-\ 2$}
\put(0.8,-1.15){\includegraphics[scale=0.7]{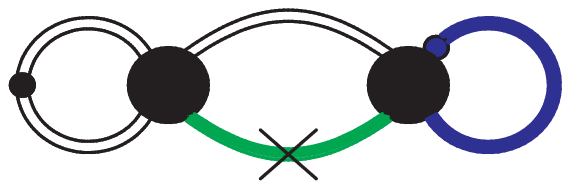}}
\put(5.7,0){$(5)$} \put(5.2,-0.5){$-\ 2$}
\put(6.1,-1){\includegraphics[scale=0.7]{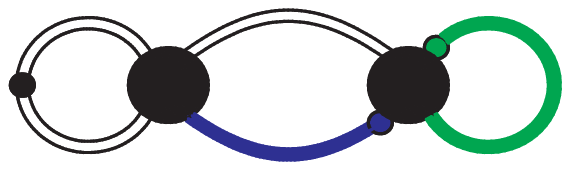}}
\put(10.9,0){$(6)$} \put(10.3,-0.5){$-\ 2$}
\put(11.2,-1){\includegraphics[scale=0.7]{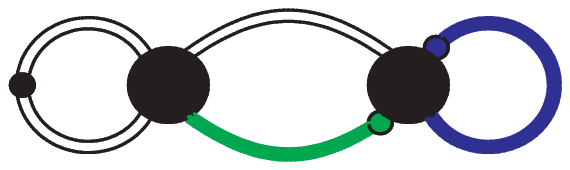}}
\put(0.4,-2){$(7)$} \put(-0.1,-3.0){$- 2$}
\put(0.5,-4.25){\includegraphics[scale=0.7]{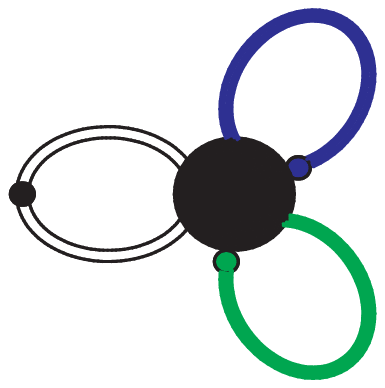}}
\put(4.1,-2){$(8)$} \put(3.3,-3.0){$-\ 4$}
\put(4.2,-4){\includegraphics[scale=0.7]{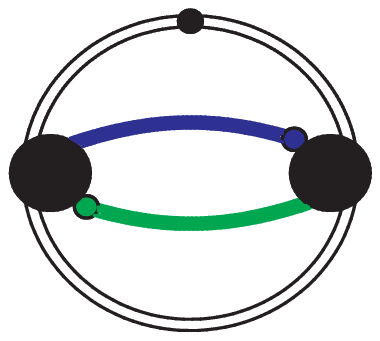}}
\put(7.8,-2){$(9)$} \put(7.1,-3.0){$-\ 4$}
\put(8.0,-4){\includegraphics[scale=0.7]{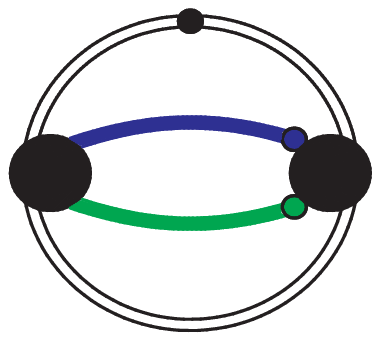}}
\put(11.3,-2.2){$(10)$} \put(10.9,-3.0){$-\ 4$}
\put(11.7,-3.4){\includegraphics[scale=0.7]{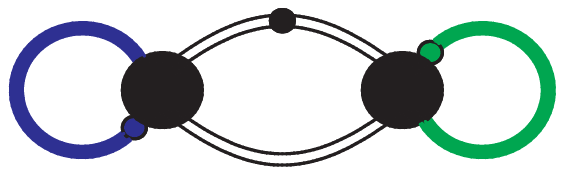}}
\put(0.6,-4.8){$(11)$}\put(0.4,-5.4){$+\ 4$}
\put(1.2,-5.8){\includegraphics[scale=0.7]{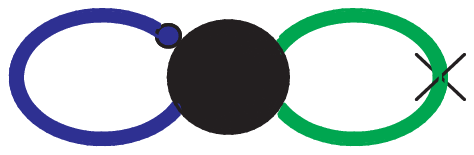}}
\put(4.9,-5.4){${\displaystyle \Bigg\}\ +\ \frac{\partial
\Delta_2}{\partial \ln g}}$}
\end{picture}
\vspace*{6cm} \caption{Diagrams obtained after differentiating the
effective diagram with a blue effective line presented in Fig.
\ref{Figure_Double_Diagram} with respect to the parameter $g$.
Here the blue effective line corresponds to the operator
$\mbox{Blueline}_{\dot b}[\theta^a \theta_a \bar\theta^{\dot b}]$
and the line with a dot corresponds to the operator
$\mbox{LineWithDot}[1]$.} \label{Figure_Blue_Green_Diagrams}
\end{figure}

The method considered in this paper also allows to prove the
identity presented in Fig. \ref{Figure_New_Identity} directly.
This is made in Sect. \ref{Subsection_New_Identity}. In
particular, we prove that the contribution of this effective
diagram to the $\beta$-function is given by a vanishing integral
of a total derivative. In order to do this, it is convenient to
differentiate the diagram with the blue effective line presented
in Fig. \ref{Figure_Double_Diagram} (or, equivalently, in Fig.
\ref{Figure_New_Identity}) with respect to the parameter $g$. In
the graphical form the result is given by a sum of diagrams
presented in Fig. \ref{Figure_Blue_Green_Diagrams}. The
corresponding analytical expression can be unambiguously
constructed using the expressions for the effective lines
presented in Fig. \ref{Figure_Feynman_Rules}.

\section{Total derivatives and the NSVZ $\beta$-function}
\label{Section_Commutators}

\subsection{The effective diagram with the yellow line}
\hspace{\parindent}\label{Subsection_Yellow_Diagram}

In order to prove that the $\beta$-function (defined in terms of
the bare coupling constant) is determined by integrals of total
derivatives, it is convenient to use the coordinate
representation. In the coordinate representation an integral of a
total derivative can be written as

\begin{equation}\label{Trace_Vanish}
\mbox{Tr}[x^\mu,\mbox{something}] = 0,
\end{equation}

\noindent where

\begin{equation}
\mbox{Tr}\, M \equiv \mbox{tr} \int d^8x\,M_{xx};\qquad\quad
[\alpha,A_{xy}]\equiv (\alpha)_x A_{xy} - A_{xy} (\alpha)_y,
\end{equation}

\noindent and $\mbox{tr}$ denotes the usual matrix trace. We will
try to present (the sum of) expressions for the effective diagrams
as such traces of commutators. First, as a simple example, we
consider a diagram with the yellow effective line presented in
Fig. \ref{Figure_Double_Diagram} and verify that the sum of the
this diagram and $\Delta_1$ is given by an integral of a total
derivative. This sum (for $N_f=1$) has been already calculated in
\cite{Stepanyantz:2004sg} by substituting solutions of the Ward
identities for the effective vertices. In this paper we reobtain
the result by a different method, which is also used for
calculation of the other diagram (which has not been calculated in
\cite{Stepanyantz:2004sg}.)

The expression for the considered diagram is written as

\begin{equation}\label{YellowBrown->YellowGreen}
-2\cdot \mbox{YellowLine}_\mu[\, \bar\theta^{\dot a}
(\gamma^\mu)_{\dot a}{}^b \theta_b]\cdot
\,\mbox{GreenLine}\cdot\gamma =
\mbox{YellowLine}_\mu[\theta^4]\cdot
\mbox{BrownLine}^\mu\cdot\gamma,
\end{equation}

\noindent where we used the identity proved in Appendix
\ref{Appendix_Identity1} and the notation

\begin{eqnarray}\label{BrownLine_Definition}
&& \mbox{BrownLine}_\mu[1,2] \equiv  \int d^8x\,\Bigg\{ (T)^j{}_i
\Big(-\frac{2i\partial_\mu}{\partial^2} - (\gamma^\mu)^{a{\dot b}}
\theta_a \frac{\bar D_{\dot b} D^2}{4\partial^2}\Big)
\frac{\mbox{\boldmath$\delta$}}{\mbox{\boldmath$\delta$}_2 j^j}
\frac{\delta}{\delta_1 \phi_{0i}}
\qquad\nonumber\\
&& + i (M T)^{ij}
\Big(\frac{D^2\partial_\mu}{16\partial^4}
\frac{\mbox{\boldmath$\delta$}}{ \mbox{\boldmath$\delta$}_2
j^i}\Big)
\frac{\mbox{\boldmath$\delta$}}{\mbox{\boldmath$\delta$}_1 j^j}
\Bigg\}.
\end{eqnarray}

\noindent This operator is very useful, because by the help of
this notation commutators of various Green functions with

\begin{equation}
(y_\mu)^* \equiv x_\mu - i\bar\theta^{\dot a} (\gamma_\mu)_{\dot
a}{}^b \theta_b
\end{equation}

\noindent can be written in a very compact form. The details of
the corresponding calculations are given in Appendix
\ref{Appendix_Commutators_Y}. Here we present only the results.
First, we introduce the following notation:

\begin{eqnarray}\label{T_Commutators}
&& \Big[(T)y_\mu^*, \frac{\delta^2\gamma}{\delta (\phi_{0i})_y
\delta (\phi_{0}^{*j})_z}\Big] \equiv - (T)_k{}^i (y_\mu^*)_y
\frac{\delta^2\gamma}{\delta (\phi_{0k})_y \delta
(\phi_{0}^{*j})_z} + (T)_j{}^k (y_\mu^*)_z
\frac{\delta^2\gamma}{\delta (\phi_{0i})_y
\delta (\phi_{0}^{*k})_z};\nonumber\\
&& \Big[(T)y_\mu^*, \frac{\delta^2\gamma}{\delta (\phi_{0i})_y
\delta (\phi_{0j})_z}\Big] \equiv - (T)_k{}^i (y_\mu^*)_y
\frac{\delta^2\gamma}{\delta (\phi_{0k})_y \delta (\phi_{0j})_z} -
(T)_k{}^j (y_\mu^*)_z \frac{\delta^2\gamma}{\delta (\phi_{0i})_y
\delta (\phi_{0k})_z},\qquad
\end{eqnarray}

\noindent e.t.c. Commutators with other Green functions (with an
arbitrary number of indexes) can be constructed similarly. (Each
index gives a term in the sum; for upper indexes the sign is
"$-$", and for lower indexes the sign is "$+$".) Then the result
of Appendix \ref{Appendix_Commutators_Y} can be written as

\begin{eqnarray}\label{Commutators}
&& \Big[(T) y_\mu^*, \frac{\delta^2\gamma}{\delta (\phi_{0k})_y
\delta (\phi^{*l}_{0})_z}\Big] = \mbox{BrownLine}_\mu \cdot
\frac{\delta^2\gamma}{\delta (\phi_{0k})_y \delta
(\phi^{*l}_{0})_z};\nonumber\\
&& \Big[(T)y_\mu^*, \frac{\delta^2\gamma}{\delta (\phi_{0k})_y
\delta (\phi_{0l})_z}\Big] = \mbox{BrownLine}_\mu \cdot
\frac{\delta^2\gamma}{\delta
(\phi_{0k})_y \delta (\phi_{0l})_z}\qquad
\end{eqnarray}

\begin{figure}[h]
\begin{picture}(0,2.2)
\put(0.2,0.45){${\displaystyle -2 \frac{d}{d\ln\Lambda}}$}
\put(2.0,0){\includegraphics[scale=0.8]{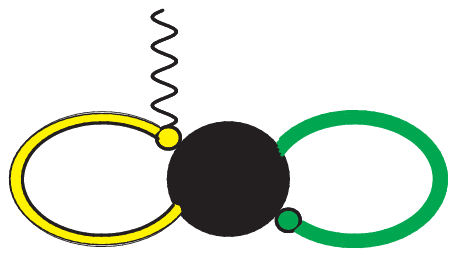}}
\put(5.9,0.45){${\displaystyle +\ \Delta_1 = \
\frac{d}{d\ln\Lambda}}$}
\put(8.8,0){\includegraphics[scale=0.8]{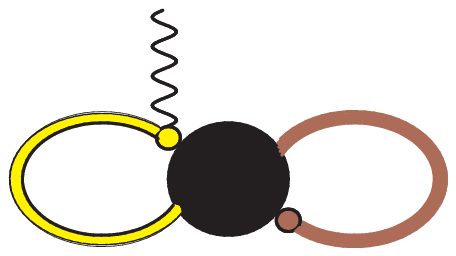}}
\put(10.3,2){$\theta^4$} \put(3.5,2){$\bar\theta^{\dot a}
(\gamma^\mu)_{\dot a}{}^b \theta_b$}  \put(12.8,0.5){$+\ \Delta_1
=$} \put(2.0,1.1){$\mu$} \put(8.8,1.1){$\mu$}
\put(12.3,1.1){$\mu$} \put(10.0,-0.9){$=\quad\mbox{integral of
total derivative}$}
\end{picture}
\vspace*{1cm} \caption{Using two-loop effective diagrams it is
possible to present the effective diagram with the yellow line
(including $\delta_1$) in Fig. \ref{Figure_SD_Result} (or a
corresponding diagram in Fig. \ref{Figure_Double_Diagram}) as an
integral of a total
derivative.}\label{Figure_YellowGreen_YellowBrown}
\end{figure}

\noindent e t.c., where all fields should be set to 0. These
identities allow to rewrite the considered contribution as an
integral of a total derivative in the momentum representation. A
simple graphical version of the result is presented in Fig.
\ref{Figure_YellowGreen_YellowBrown}. Below we prove the last
equality in this figure. For this purpose we substitute the
explicit expression for the operator
$\mbox{YellowLine}_\mu[\theta^4]$ into Eq.
(\ref{YellowBrown->YellowGreen}). Then the diagram with the brown
effective line is written in the form

\begin{equation}\label{Yellow_Diagram}
\frac{d}{d\ln\Lambda} \mbox{BrownLine}^\mu\cdot \int
d^8x\,\theta^4 \Bigg\{2i (T)^j{}_i \Big(
\frac{\partial_\mu}{\partial^2}
\frac{\mbox{\boldmath$\delta$}}{\mbox{\boldmath$\delta$} j^j}\Big)
\frac{\delta\gamma}{\delta \phi_{0i}} - i (MT)^{ij}
\Big(\frac{D^2\partial_\mu}{16\partial^4}
\frac{\mbox{\boldmath$\delta$}}{\mbox{\boldmath$\delta$} j^i}\Big)
\frac{\mbox{\boldmath$\delta$}\gamma}{\mbox{\boldmath$\delta$}
j^j}\Bigg\}.
\end{equation}

\noindent Using the identities (\ref{Commutators}) we can present
the expression (\ref{Yellow_Diagram}) in the form

\begin{eqnarray}\label{Yellow_Commutator}
&& \frac{d}{d\ln\Lambda} \int d^8x\,d^8y\,(\theta^4)_x \Bigg\{- 2i
(T)^k{}_i \Big(\frac{\partial_\mu}{\partial^2}\Big)_x
\Big(\frac{\delta^2\gamma}{\delta(\phi_k)_x \delta
(\phi^{*j})_y}\Big)^{-1} \Big[(T) y_\mu^*,
\frac{\delta^2\gamma}{\delta (\phi_0^{*j})_y \delta(\phi_{0i})_x}
\Big]\nonumber\\
&& - 2i (T)^k{}_i \Big(\frac{\partial_\mu}{\partial^2}\Big)_x
\Big(\frac{\delta^2\gamma}{\delta(\phi_k)_x \delta
(\phi_j)_y}\Big)^{-1} \Big[(T) y_\mu^*,
\frac{\delta^2\gamma}{\delta (\phi_{0j})_y \delta(\phi_{0i})_x}
\Big] - i \int d^8z\, (M T)^{ij}
\vphantom{\Bigg\{}\nonumber\\
&& \times\Bigg(\Big(\frac{\partial_\mu D^2}{16\partial^4}\Big)_x
\Big(\frac{\delta^2\gamma}{\delta(\phi_i)_x \delta
(\phi^{*m})_y}\Big)^{-1} \Big[(T) y_\mu^*,
\frac{\delta^2\gamma}{\delta(\phi^{*m}_{0})_y \delta
(\phi_{0n})_z}  \Big] \Big(\frac{\delta^2\gamma}{\delta (\phi_{j})_x
\delta(\phi_n)_z}\Big)^{-1}\vphantom{\Bigg\{}\qquad\nonumber\\
&& +\Big(\frac{\partial_\mu D^2}{16\partial^4}\Big)_x
\Big(\frac{\delta^2\gamma}{\delta(\phi_i)_x \delta
(\phi_{m})_y}\Big)^{-1} \Big[(T) y_\mu^*,
\frac{\delta^2\gamma}{\delta(\phi_{0m})_y \delta (\phi_{0n})_z}
\Big] \Big(\frac{\delta^2\gamma}{\delta (\phi_{j})_x
\delta(\phi_n)_z}\Big)^{-1}\vphantom{\Bigg\{}\nonumber\\
&& + \Big(\frac{\partial_\mu D^2}{16\partial^4}\Big)_x
\Big(\frac{\delta^2\gamma}{\delta(\phi_i)_x \delta
(\phi^{*m})_y}\Big)^{-1} \Big[(T) y_\mu^*,
\frac{\delta^2\gamma}{\delta(\phi^{*m}_{0})_y \delta
(\phi^{*n}_{0})_z}  \Big]
\Big(\frac{\delta^2\gamma}{\delta (\phi_{j})_x
\delta(\phi^{*n})_z}\Big)^{-1}\vphantom{\Bigg\{}\nonumber\\
&& +\Big(\frac{\partial_\mu D^2}{16\partial^4}\Big)_x
\Big(\frac{\delta^2\gamma}{\delta(\phi_i)_x \delta
(\phi_{m})_y}\Big)^{-1} \Big[(T) y_\mu^*,
\frac{\delta^2\gamma}{\delta(\phi_{0m})_y \delta
(\phi^{*n}_{0})_z} \Big]
\Big(\frac{\delta^2\gamma}{\delta (\phi_{j})_x
\delta(\phi^{*n})_z}\Big)^{-1} \Bigg) \Bigg\}.
\end{eqnarray}

\noindent In order to write Eq. (\ref{Yellow_Commutator}) as an
integral of a total derivative in the momentum space we
(anti)commute the generators with the Green functions. Using the
results of Appendix \ref{Appendix_Commutators_Inverse} we obtain

\begin{eqnarray}\label{Total_Derivative}
&& \frac{d}{d\ln\Lambda} \int d^8x\,(\theta^4)_x
C(R)_i{}^k \Bigg\{\int d^8y\,
\Bigg(-2i\Big(\frac{\partial_\mu}{\partial^2}\Big)_x
\Big(\frac{\delta^2\gamma}{\delta (\phi^{*j})_y
\delta(\phi_k)_x}\Big)^{-1} \Big[y_\mu^*,
\frac{\delta^2\gamma}{\delta (\phi_0^{*j})_y \delta(\phi_{0i})_x}
\Big]\qquad\nonumber\\
&& -2i \Big(\frac{\partial_\mu}{\partial^2}\Big)_x
\Big(\frac{\delta^2\gamma}{\delta (\phi_j)_y
\delta(\phi_k)_x}\Big)^{-1} \Big[y_\mu^*,
\frac{\delta^2\gamma}{\delta (\phi_{0j})_y \delta(\phi_{0i})_x}
\Big] + i M^{jn} M^{mi}
\Big(\frac{\delta^2\gamma}{\delta(\phi_k)_x
\delta(\phi_j)_y}\Big)^{-1}
\nonumber\\
&& \times \Big(\frac{\partial_\mu D^2}{16\partial^4}\Big)_x
\Big(\frac{\partial_\mu D^2}{8\partial^4}\Big)_y
\Big(\frac{\delta^2\gamma}{\delta(\phi_m)_x \delta
(\phi_n)_y}\Big)^{-1} \Bigg) + i M^{ij} \Big(\frac{\partial_\mu
D^2}{16\partial^4}\Big)_x \Big[y_\mu^*,
\Big(\frac{\delta^2\gamma}{\delta(\phi_j)_x \delta
(\phi_{k})_y}\Big)^{-1} \Big]_{y=x}
\nonumber\\
&& -i M^{ij} \Big(\frac{\partial^\mu}{\partial^2}\Big)_x
\Big(\frac{\partial_\mu D^2}{8\partial^4}\Big)_y
\Big(\frac{\delta^2\gamma}{\delta(\phi_j)_x \delta
(\phi_{k})_y}\Big)^{-1}_{y=x}\Bigg\}.
\end{eqnarray}

\noindent Then we add $\Delta_1$ to Eq. (\ref{Total_Derivative})
and write the result in the momentum representation substituting
explicit expressions for the (inverse) Green functions. Details of
this calculation are presented in Appendix
\ref{Appendix_Yellow_Diagram_Contribution}. Taking into account
Eqs. (\ref{G_And_J}) and (\ref{Propagator}) the result for the
considered contribution (written in the momentum representation
after the Wick rotation in the Euclidian space) can be written as

\begin{eqnarray}\label{NSVZ_Contribution}
&& {\cal V}_4 N_f \frac{d}{d\ln\Lambda} \int \frac{d^4q}{(2\pi)^4}
\frac{2 q^\mu}{q^4} \frac{\partial}{\partial
q^\mu}\Big\{2\ln G\nonumber\\
&& \qquad\qquad\qquad\qquad\qquad +\sum\limits_{I=1}^n
(-1)^{P_I}\Big(\ln(q^2 G^2 + M^2 J^2) + \frac{M^2 J}{q^2 G^2 + M^2
J^2}\Big)_I\Big\},\qquad
\end{eqnarray}

\noindent where we separate the main contribution of the fields
$\phi_\alpha$ and $\widetilde\phi_\alpha$ (corresponding to $I=0$)
and contributions of the Pauli--Villars fields (corresponding to
$I\ge 1$). This expression agrees with the result obtained in
\cite{Stepanyantz:2004sg} by a different method for $N_f=1$. (In
\cite{Stepanyantz:2004sg} the considered contribution was
calculated by substituting expressions for vertices obtained by
solving the Ward identities.)

Thus, the sum of the considered effective diagram and $\Delta_1$
is given by the integral of a total derivative, which can be
easily calculated using the identity

\begin{equation}\label{Momentum_Integration}
\int \frac{d^4q}{(2\pi)^4} \frac{q^\mu}{q^4}
\frac{\partial}{\partial q^\mu} f(q) = \frac{1}{(2\pi)^4}
\oint\limits_{S^3_\varepsilon} dS_\mu \frac{q^\mu}{q^4} f(q) = -
\frac{1}{8\pi^2} f(0) = - 2\pi^2 \int \frac{d^4q}{(2\pi)^4}
\delta^4(q) f(q),
\end{equation}

\noindent where $f(q^2)$ is a function which rapidly decreases at
the infinity, and $S^3_\varepsilon$ is a 3-sphere
in the momentum space surrounding the point $q=0$
with the radius $\varepsilon \to 0$.

Assuming that the other contributions vanish (we prove this
statement in the next section) we obtain the NSVZ relation for the
renormalization group functions defined in terms of the bare
coupling constant. Really, terms containing the Pauli--Villars
masses are convergent and finite beyond the one-loop
approximation, because these masses are proportional to the
parameter $\Lambda$. Therefore,

\begin{equation}
\frac{1}{2\pi}{\cal V}_4\cdot \frac{\beta(\alpha_0)}{\alpha_0^2} =
\frac{1}{2\pi^2} {\cal V}_4  N_f \Big(\sum\limits_{I=1}^n c_I -
\frac{d\ln G}{d\ln\Lambda}\Big|_{q=0}\Big) = \frac{1}{2\pi^2}
{\cal V}_4 N_f \Big(1-\gamma(\alpha_0)\Big).
\end{equation}

\noindent Thus, for ${\cal N}=1$ SQED with $N_f$ flavors we obtain
the NSVZ $\beta$-function

\begin{equation}
\beta(\alpha_0) = \frac{\alpha_0^2 N_f}{\pi}
\Big(1-\gamma(\alpha_0)\Big).
\end{equation}

\subsection{The effective diagram with the blue line}
\hspace{\parindent}\label{Subsection_New_Identity}

In order to prove that the $\beta$-function defined in terms of
the bare coupling constant is given by integrals of total
derivatives it is also necessary to present the expression for the
last diagram (with the blue effective line) in Fig.
\ref{Figure_SD_Result} plus $\delta_2$ as a trace of a commutator.
Calculations in the lowest orders allow to suggest that this
contribution is always given by integral of a total derivative and
vanishes \cite{Stepanyantz:2004sg}. An indirect proof of this fact
is actually given in \cite{Stepanyantz:2011jy} by a different
method. In this section we present a direct proof. In order to do
this, it is necessary to differentiate the generating functional
with respect to the auxiliary parameter $g$, introduced in Eq.
(\ref{Introducing_Of_G}). Next, we prove the identity presented in
Fig. \ref{Figure_New_Identity_Total_Derivatives} in a graphical
form. In this figure an arc with an arrow denotes a trace of a
commutator with $y_\mu^*$. Certainly, the corresponding analytical
expression can be easily constructed:

\begin{figure}[h]

\begin{picture}(0,0)
\put(0.7,-3){${\displaystyle 2\cdot
\frac{d}{d\ln\Lambda}\frac{\partial}{\partial \ln g}}$}
\put(4.0,-1.8){${\dot b}$}
\put(2.9,-3.9){\includegraphics[scale=0.4]{figure6_blue.eps}}
\put(2.7,-4.1){$\theta^a \theta_a \bar\theta^{\dot b}$}
\put(10.1,-3.8){${\dot b}$} \put(5.7,-4.1){$\theta^4$}

\put(5.7,-3){${\displaystyle +\ \delta_2\  = (\gamma^\mu)^{a{\dot
b}} \frac{d}{d\ln\Lambda}}$}
\put(9.3,-0.55){\includegraphics[angle=180,scale=0.8]{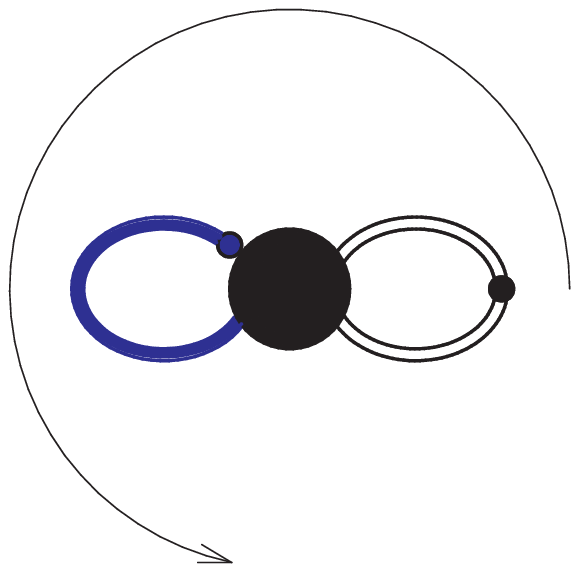}}
\put(11.4,-5.0){$y_\mu^*$} \put(13.5,-3.0){$\theta^4$}
\put(11.1,-2.2){$\theta_a$} \put(14.5,-3){$=0.$}
\end{picture}
\vspace*{5.0cm} \caption{This identity is needed for proving that
the $\beta$-function (defined in terms of the bare coupling
constant) is given by integrals of total derivatives and satisfies
the NSVZ relation.}\label{Figure_New_Identity_Total_Derivatives}
\end{figure}

\begin{eqnarray}\label{New_Identity}
&& 2 \frac{d}{d\ln\Lambda} \frac{\partial}{\partial \ln g} \int
d^8x\,(\theta^4)_x \mbox{BlueLine}_{\dot b}[\theta^a \theta_a
\bar\theta^{\dot b}]\cdot \frac{\delta\Gamma}{\delta
\mbox{\boldmath$V$}_x}
+\delta_2\nonumber\\
&&\qquad\qquad\qquad = (\gamma^\mu)^{a{\dot b}}
\frac{d}{d\ln\Lambda} \mbox{Tr} \Big[(T)
y_\mu^*,\,\mbox{LineWithDot}[\theta^4]\cdot \mbox{BlueLine}_{\dot
b}[\theta_a] \cdot\gamma \Big] = 0.\qquad
\end{eqnarray}

\noindent For the simplest cases the operation $[(T) \alpha,
\ldots ]$ was defined in the previous section. The natural
generalization of this definition can be formulated as follows: If
a tensor $B$ has a lower index $i$ corresponding to a superfield
in the point $x$, then $[(T)\alpha, B]$\footnote{For the case
$P_\alpha=P_B=1$ we use the notation $\{(T)\alpha, B\}$.} includes

\begin{equation}
(T)_i{}^j \alpha_x (B_j)_x.
\end{equation}

\noindent Similarly, for an upper index $i$ this expression
includes

\begin{equation}
- (T)_j{}^i \alpha_x (B^j)_x.
\end{equation}

\noindent For example, applying this rule we can easily obtain Eq.
(\ref{T_Commutators}). However, it is necessary to define the
operation $[(T) y_\mu^*,\ldots]$ (and other similar operations)
more accurately for diagrams containing closed loops of the matter
superfields. First, let us explain, how to construct the
expression in the right hand side of Eq. (\ref{New_Identity}). It
can be schematically written as

\begin{equation}
\mbox{Diagram}^\mu \equiv \int d\mu\,\mbox{Line}^\mu \cdot
\mbox{Vertex} \cdot \mbox{Line},
\end{equation}

\noindent where $d\mu$ denotes the integration measure, and we
omitted indexes for simplicity. The whole expression in the right
hand side (for simplicity, without the derivative $d/d\ln\Lambda$)
can be written as

\begin{eqnarray}\label{Trace_Notation}
&& \mbox{Tr} \Big[(T) y_\mu^*,\mbox{Diagram}^\mu \Big] \equiv \int
d\mu\, \Big[(T) y_\mu^*, \mbox{Line}^\mu\Big] \cdot \mbox{Vertex}
\cdot \mbox{Line} +  \int d\mu\, \mbox{Line}^\mu\qquad\nonumber\\
&& \cdot \Big[(T) y_\mu^*,\mbox{Vertex}\Big] \cdot \mbox{Line} +
\int d\mu\,\mbox{Line}^\mu \cdot \mbox{Vertex} \cdot
\Big[(T)y_\mu^*, \mbox{Line}\Big] = 0.\vphantom{\Big(}
\end{eqnarray}

\noindent (It is easy to see that all terms in this sum cancel
each other.) The operation $[(T)y_\mu^*,\ldots]$ in Eq.
(\ref{New_Identity}) is constructed formally according to Eq.
(\ref{Trace_Notation}). In the momentum representation the
expression (\ref{Trace_Notation}) is given by an integral of a
total derivative, because

\begin{eqnarray}
&& \int d^8x\, \Big[(T) y_\mu^*, X_{(i,x)}{}^{(i,x)}\Big] = \int
d^8x\,d^8y\,\delta^8_{xy}\Big((y_\mu^*)_x T_i{}^j
X_{(j,x)}{}^{(i,y)} - (y_\mu^*)_y T_j{}^i
X_{(i,x)}{}^{(j,y)}\Big)\nonumber\\
&& = \mbox{Tr}\Big[y_\mu^*,\,T_j{}^i X_{i}{}^{j}\Big] = -\int
\frac{d^4q}{(2\pi)^4} \frac{\partial}{\partial q^\mu} \int
d^4\theta\, T_j{}^i X_j{}^i(q,\theta).
\end{eqnarray}

\noindent (The last equation is written in the Euclidian space
after the Wick rotation.) Therefore, the equation presented in
Fig. \ref{Figure_New_Identity_Total_Derivatives} implies that the
derivative of the effective diagram in the left hand side with
respect to $\ln g$ is given by an integral of a total derivative.
Moreover, the result is 0, because the integrand does not contain
singularities. Then we integrate the considered equality over $\ln
g$ from $g=0$ to $g=1$. The theory corresponding to $g=0$ does not
contain quantum gauge field, and quantum corrections to the Green
function of the gauge superfield are given only by one-loop
diagrams. It is easy to see that in the one-loop approximation the
effective diagram in the left hand side of Fig.
\ref{Figure_New_Identity_Total_Derivatives} vanishes. Therefore,
(because the original theory corresponds to $g=1$) this effective
diagram is also given by an integral of a total derivative and is
equal to 0 for $g=1$.

Thus, taking into account the results of the previous section, the
identity presented in Fig.
\ref{Figure_New_Identity_Total_Derivatives} allows to prove that
the $\beta$-function defined in terms of the bare coupling
constant is given by integrals of total derivatives and satisfies
the NSVZ relation. Let us proceed to proving this identity. The
expressions

\begin{eqnarray}
\Big[(T) y_\mu^*,\mbox{Vertex}\Big];\qquad \Big[(T)
y_\mu^*,\mbox{Line}\Big];\qquad \Big[(T)
y_\mu^*,\mbox{Line}^\mu\Big]
\end{eqnarray}

\begin{figure}[h]
\begin{picture}(0,0)
\put(2.0,-2.9){${\displaystyle 0 = i \cdot \frac{d}{d\ln\Lambda}
\Bigg\{}$}
\put(4.8,-0.5){\includegraphics[angle=180,scale=0.8]{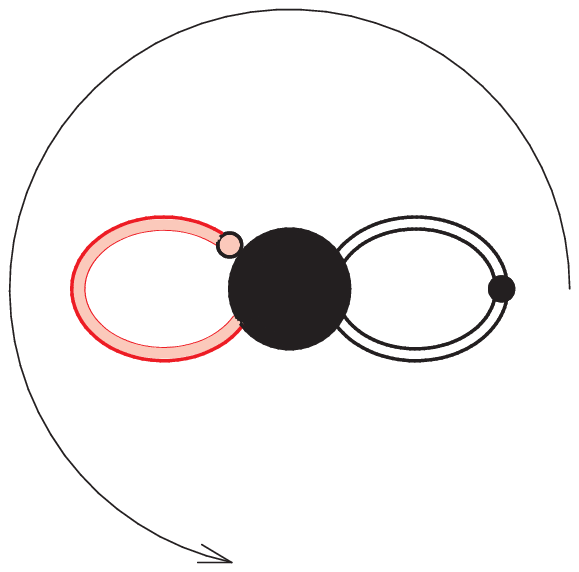}}
\put(6.9,-5.0){$\bar\theta^{\dot a}$} \put(9.1,-3.0){$\theta^4$}
\put(10.1,-2.9){$+$} \put(12.0,-4.4){$\bar\theta^{\dot a}$}
\put(5.8,-2.1){${\dot a}$} \put(11.6,-2.1){${\dot a}$}
\put(13.0,-2.9){$\theta^4$}
\put(10.7,-4.1){\includegraphics{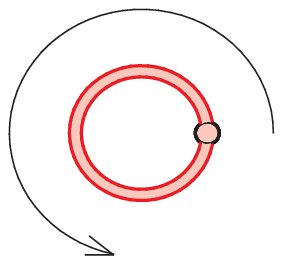}}
\put(13.8,-2.9){${\displaystyle \Bigg\}}$}
\end{picture}
\vspace*{5.0cm} \caption{We add these terms to the diagram in the
right hand side of Fig.
\ref{Figure_New_Identity_Total_Derivatives}.}
\label{Figure_Pink_Additional_Terms}
\end{figure}

\noindent (and other similar expressions) can be calculated using
the Schwinger--Dyson equations. The details of the corresponding
calculations are presented in Appendix \ref{Appendix_Commutators}.
It is convenient to add the (vanishing) diagrams presented in Fig.
\ref{Figure_Pink_Additional_Terms}, where

\begin{eqnarray}
&& \mbox{PinkLine}^{\dot b}[1,2] = \int d^8x\, \Bigg\{(T)^j{}_i
\Big(\theta^a \theta_a \frac{\bar D^{\dot b} D^2}{2\partial^2} - i
(\gamma^\mu)^{a{\dot b}} \theta_a \frac{\bar D^2
D^2\partial_\mu}{8\partial^4}\Big)
\frac{\mbox{\boldmath$\delta$}}{\mbox{\boldmath$\delta$}_2 j^j}
\cdot \frac{\delta}{\delta_1 \phi_{0i}}\quad\nonumber\\
&& - i (M T)^{ij} (\gamma^\mu)^{{\dot b}a} \theta_a
\Big(\frac{D^2\partial_\mu}{16\partial^4}
\frac{\mbox{\boldmath$\delta$}}{\mbox{\boldmath$\delta$}_2
j^i}\Big)\cdot \frac{\mbox{\boldmath$\delta$}}{
\mbox{\boldmath$\delta$}_1 j^j}\Bigg\},
\end{eqnarray}

\noindent to the diagram presented in the right hand side of Fig.
\ref{Figure_New_Identity_Total_Derivatives}. It is easy to see
that the diagrams presented in Fig.
\ref{Figure_Pink_Additional_Terms} encode the commutator

\begin{eqnarray}
&& - \frac{1}{2}\cdot\frac{d}{d\ln\Lambda} \mbox{Tr}\,(\theta^4)_x
\Big\{(T) \bar\theta^{\dot a},\,\mbox{PinkLine}_{\dot a} \cdot
\Big(\frac{\mbox{\boldmath$\delta$}^2\gamma}{\mbox{\boldmath$\delta$}
j^*_i \mbox{\boldmath$\delta$} j^i} +i \phi^{*i}_0 \phi_{0i} +
M^{ik} \Big(\frac{D^2}{8\partial^2}
\frac{\mbox{\boldmath$\delta$}}{\mbox{\boldmath$\delta$} j^i}\Big)
\frac{\mbox{\boldmath$\delta$}\gamma}{\mbox{\boldmath$\delta$}
j^k}\qquad\nonumber\\
&& + i M^{ik} \Big(\frac{D^2}{8\partial^2}\phi_{0i}\Big) \phi_{0k}
+ M^*_{ik} \Big(\frac{\bar D^2}{8\partial^2}
\frac{\mbox{\boldmath$\delta$}}{\mbox{\boldmath$\delta$}
j^*_i}\Big)
\frac{\mbox{\boldmath$\delta$}\gamma}{\mbox{\boldmath$\delta$}
j^*_k} + i M^*_{ik} \Big(\frac{\bar
D^2}{8\partial^2}\phi_{0}^{*i}\Big) \phi_{0}^{*k}
\Big)_{x}\Big\}.\qquad\quad
\end{eqnarray}

\begin{figure}[p]
\vspace*{-1.6cm}
\begin{picture}(0,0)
\put(3.5,-1.8){$(1)$} \put(0.3,-2.8){${\displaystyle
\frac{d}{d\ln\Lambda}\Bigg\{ (\gamma^\mu)^{a{\dot b}}\cdot
\theta^4 }$} \put(5.7,-1.2){$\mu$} \put(6.1,-4){${\dot b}$}
\put(4.5,-3.5){$\theta_a$}
\put(3.4,-4){\includegraphics[scale=0.7]{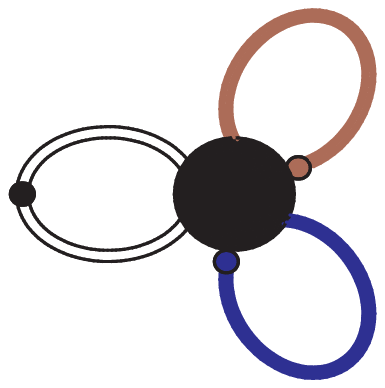}}
\put(7.5,-1.8){$(2)$} \put(6.6,-2.8){$+$}\put(7.3,-2.8){$2$}
\put(7.8,-3.85){\includegraphics[scale=0.7]{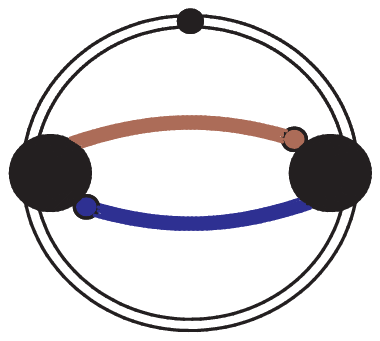}}
\put(11.0,-2.8){$+$}\put(11.7,-2.8){$2$} \put(12.0,-1.8){$(3)$}
\put(12.2,-3.85){\includegraphics[scale=0.7]{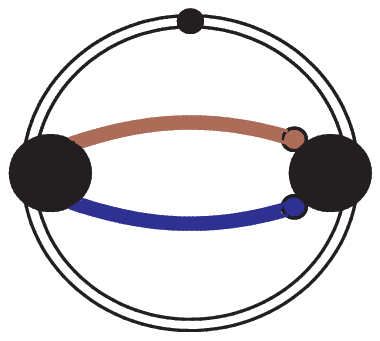}}
\put(1,-4.3){$(4)$} \put(0.8,-5.15){$+$}
\put(1.3,-5.6){\includegraphics[scale=0.7]{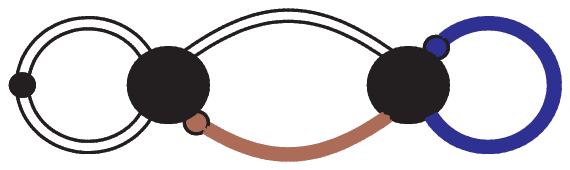}}
\put(6.5,-4.3){$(5)$} \put(5.6,-5.15){$+$}
\put(6.1,-5.6){\includegraphics[scale=0.7]{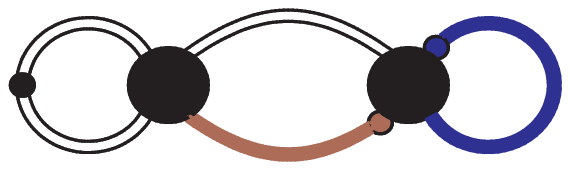}}
\put(11,-4.3){$(6)$} \put(10.35,-5.15){$+\ 2$}
\put(11.2,-5.6){\includegraphics[scale=0.7]{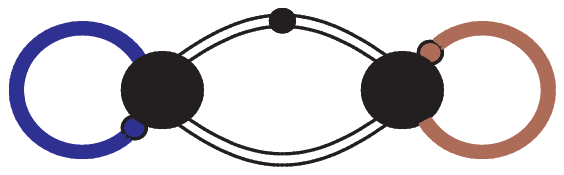}}
\put(1,-6.6){$(7)$} \put(0.8,-7.4){$+$}
\put(1.3,-7.85){\includegraphics[scale=0.7]{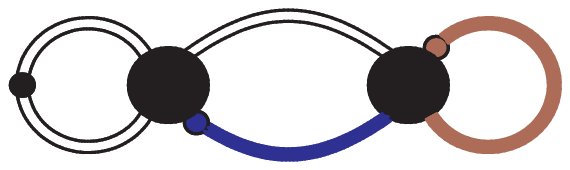}}
\put(6.6,-6.6){$(8)$} \put(5.9,-7.4){$+$}
\put(6.7,-7.85){\includegraphics[scale=0.7]{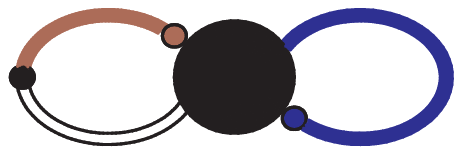}}
\put(11.2,-6.6){$(9)$} \put(10.55,-7.4){$+$}
\put(11.4,-7.8){\includegraphics[scale=0.7]{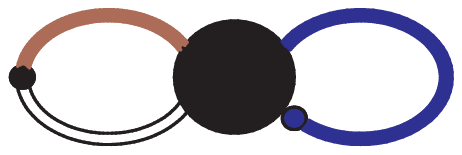}}
\put(2.5,-8.8){$(10)$} \put(1.2,-9.8){$- \ \ \theta^4$}
\put(5.0,-8.7){${\dot b}$} \put(4.9,-11){${\dot b}$}
\put(3.5,-10.5){$1$}
\put(2.2,-11){\includegraphics[scale=0.7]{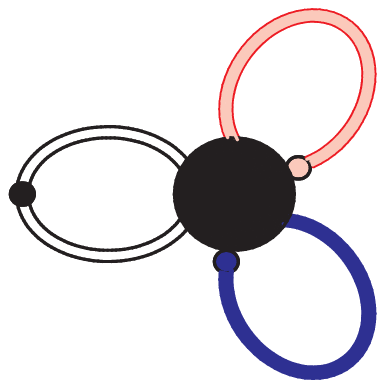}}
\put(6.6,-8.8){$(11)$} \put(5.8,-9.8){$-$}\put(6.6,-9.8){$2$}
\put(7.0,-10.85){\includegraphics[scale=0.7]{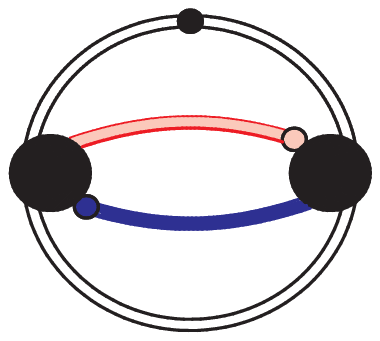}}
\put(11.2,-8.8){$(12)$} \put(10.4,-9.8){$-$}\put(11.2,-9.8){$2$}
\put(11.6,-10.85){\includegraphics[scale=0.7]{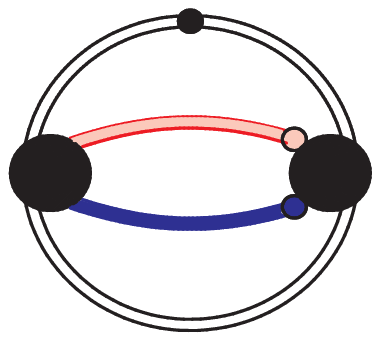}}
\put(1,-11.4){$(13)$} \put(0.8,-12.25){$-$}
\put(1.3,-12.7){\includegraphics[scale=0.7]{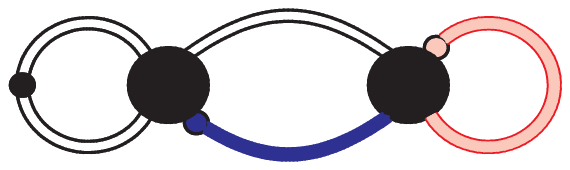}}
\put(5.9,-11.4){$(14)$} \put(5.6,-12.25){$-$}
\put(6.2,-12.7){\includegraphics[scale=0.7]{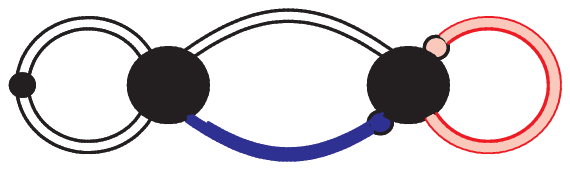}}
\put(10.9,-11.4){$(15)$} \put(10.5,-12.25){$-\ 2$}
\put(11.3,-12.7){\includegraphics[scale=0.7]{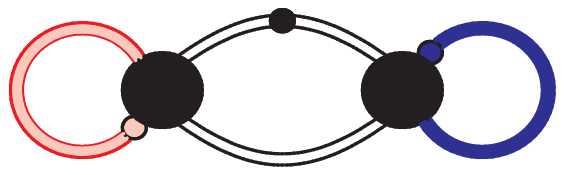}}
\put(0.9,-13.6){$(16)$} \put(0.8,-14.35){$-$}
\put(1.3,-14.8){\includegraphics[scale=0.7]{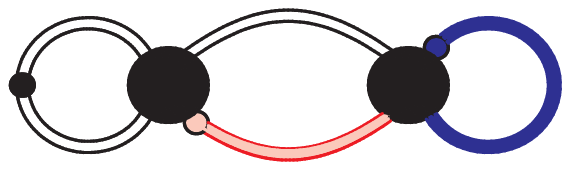}}
\put(5.9,-13.6){$(17)$} \put(5.6,-14.35){$-$}
\put(6.3,-14.95){\includegraphics[scale=0.7]{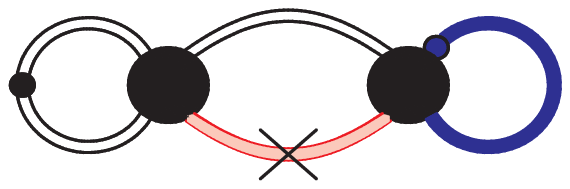}}
\put(11.5,-13.6){$(18)$} \put(10.9,-14.35){$-$}
\put(11.8,-14.8){\includegraphics[scale=0.7]{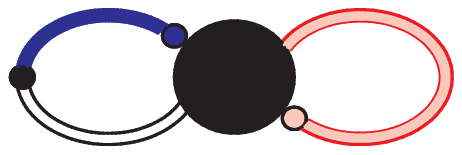}}
\put(1.3,-15.6){$(19)$} \put(1.2,-16.35){$-$}
\put(1.7,-16.8){\includegraphics[scale=0.7]{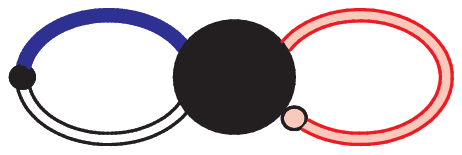}}
\put(6.5,-15.6){$(20)$} \put(6.1,-16.35){$-$}
\put(6.8,-16.8){\includegraphics[scale=0.7]{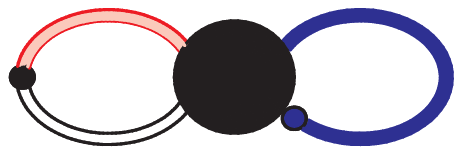}}
\put(11.5,-15.6){$(21)$} \put(10.9,-16.35){$-$}
\put(11.8,-16.8){\includegraphics[scale=0.7]{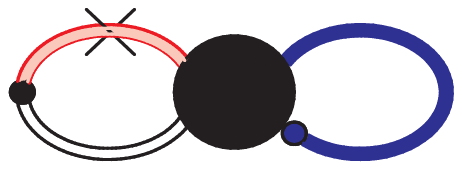}}
\put(1.2,-17.6){$(22)$} \put(1.2,-18.35){$-$}
\put(1.7,-18.75){\includegraphics[scale=0.7]{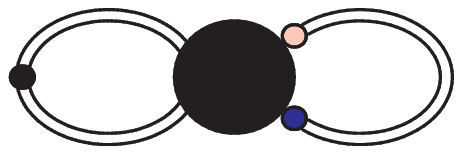}}
\put(6.2,-17.6){$(23)$} \put(6.2,-18.35){$+$}
\put(6.7,-18.75){\includegraphics[scale=0.7]{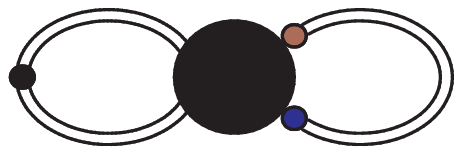}}

\put(10.7,-18.3){${\displaystyle +\  \int d^8x\, \Bigg\{
\frac{i}{2} M^{mk} C(R)_k{}^n  }$}

\put(0.2,-19.7){${\displaystyle \times \Bigg(
\mbox{LineWithDot}[\theta^4]\cdot
\Big(\frac{D^2}{8\partial^4}\Big)_x
\frac{\mbox{\boldmath$\delta$}^2 \gamma}{\mbox{\boldmath$\delta$}
(j^m)_x \mbox{\boldmath$\delta$} (j^n)_y}\Big|_{y=x} + \int d^8y\,
\Big(\frac{\partial_\mu}{\partial^2}\Big)_x
\Big(\frac{\delta^2\gamma}{\delta (\phi_i)_x \delta(\phi_m)_y}
\Big)^{-1}  }$}

\put(0.2,-21.1){${\displaystyle \times
\mbox{LineWithDot}[\theta^4]\cdot
\Big(\frac{D^2\partial_\mu}{4\partial^4}\Big)_y
\frac{\mbox{\boldmath$\delta$}}{\mbox{\boldmath$\delta$} (j^n)_y}
\frac{\delta\gamma}{\delta (\phi_{0i})_x} \Bigg) - \frac{i}{2}
(\theta^4)_x (\gamma^\mu)^{{\dot a}b} \mbox{BlueLine}_{\dot
a}[\theta_b] \Bigg( (MT)^{ik}   }$}

\put(0.2,-22.5){${\displaystyle \times
\Big(\frac{D^2\partial_\mu}{4\partial^4}\Big)_x
\frac{\mbox{\boldmath$\delta$}^2\gamma}{\mbox{\boldmath$\delta$}
(j^i)_x \mbox{\boldmath$\delta$} (j^k)_y} + (TM^*)_{ik}
\Big(\frac{\bar D^2\partial_\mu}{4\partial^4}\Big)_x
\frac{\mbox{\boldmath$\delta$}^2\gamma}{\mbox{\boldmath$\delta$}
(j^*_i)_x \mbox{\boldmath$\delta$} (j^*_k)_y} \Bigg)_{y=x} -
(\gamma^\mu)^{{\dot a}b} (\bar\theta^{\dot c} \bar\theta_{\dot c}
\theta_b)_x (T M^*)_{ik}  }$}

\put(0.2,-23.9){${\displaystyle \times \Big(\frac{i \bar
D^2\partial_\mu}{16\partial^4}\Big)_x \mbox{PinkLine}_{\dot a}
\frac{\mbox{\boldmath$\delta$}^2\gamma}{\mbox{\boldmath$\delta$}
(j^*_i)_x \mbox{\boldmath$\delta$} (j^*_k)_y}\Big|_{y=x} \Bigg\}
\Bigg\}. }$}

\end{picture}
\vspace*{24.2cm} \caption{These diagrams are obtained after
calculating the commutators in the right hand side of Figs.
\ref{Figure_New_Identity_Total_Derivatives} and
\ref{Figure_Pink_Additional_Terms}. In all these diagrams the line
with a dot corresponds to the operator
$\mbox{LineWithDot}[\theta^4]$.} \label{Figure_Blue_Diagrams}
\end{figure}

In the graphical form the result for the sum of all commutators is
presented in Fig. \ref{Figure_Blue_Diagrams}. The expressions for
the effective lines are collected in Fig.
\ref{Figure_Feynman_Rules}. Let us briefly explain, how the
diagrams presented in Fig. \ref{Figure_Blue_Diagrams} are
constructed.

1. A commutator $[(T)y_\mu^*,\mbox{Vertex}]$ (where
$\mbox{Vertex}$ denotes the four-point Green function in Fig.
\ref{Figure_New_Identity_Total_Derivatives}) is calculated
according to the rules derived in Appendix
\ref{Appendix_Commutators_Y}. From this commutator we obtain
diagrams (1) -- (5) in Fig. \ref{Figure_Blue_Diagrams}. A similar
commutator $[(T)\bar\theta^{\dot b},\mbox{Vertex}]$ (which appears
in the first diagram in Fig. \ref{Figure_Pink_Additional_Terms})
gives diagrams (10) -- (14).

2. It is also necessary to calculate commutators with the inverse
Green functions which are contained inside the derivatives
$\mbox{\boldmath$\delta$}/\mbox{\boldmath$\delta$} j$. The
calculation of these commutators is described in Appendix
\ref{Appendix_Commutators_Inverse}. Commuting $(T) y_\mu^*$ with
the operator $\mbox{LineWithDot}$ gives diagrams (6), (8), and
(9). In diagrams (8) and (9) the left effective line is given by
the operators

\begin{eqnarray}\label{BrownWhiteLine}
&& \mbox{BrownWhiteLine}_\mu[1,2] \equiv -\frac{1}{2} \cdot
\mbox{BrownLine}_\mu[1,3]\cdot \mbox{UsualLine}[2,3]
\nonumber\\
&&\qquad\qquad\qquad\qquad\quad \times \int d^8x\, \theta^4
\Big(\phi_0^{*i} \phi_{0i} + M^{ij} \phi_{0i}
\frac{D^2}{8\partial^2} \phi_{0j} + M^*_{ij}
\phi_0^{*i} \frac{\bar D^2}{8\partial^2} \phi_0^{*j}\Big)_{[3]}.\qquad\\
\nonumber\\
\label{WhiteBrownLine} && \mbox{WhiteBrownLine}_\mu[1,2] \equiv
-\frac{1}{2} \cdot \mbox{BrownLine}_\mu[3,1]\cdot \mbox{UsualLine}[2,3]
\nonumber\\
&&\qquad\qquad\qquad\qquad \times \int d^8x\,\theta^4
\Big(\phi_0^{*i} \phi_{0i} + M^{ij} \phi_{0i}
\frac{D^2}{8\partial^2} \phi_{0j} + M^*_{ij} \phi_0^{*i}
\frac{\bar D^2}{8\partial^2} \phi_0^{*j}\Big)_{[3]},\qquad
\end{eqnarray}

\noindent respectively. As earlier, the subscript $[3]$ means that
only the derivatives $\delta/\delta_3\phi_0$ nontrivially act on
the expression in the brackets.

Similarly, commuting $(T) \bar\theta^{\dot b}$ with the operator
$\mbox{LineWithDot}$ in the first diagram in Fig.
\ref{Figure_Pink_Additional_Terms} gives diagrams (15), (18), and
(19). In diagrams (18) and (19) we use the operators (with $\alpha
= \theta^4$ and $\beta=1$)

\begin{eqnarray}\label{BlueWhiteLine}
&& \mbox{BlueWhiteLine}^{\dot b}[\alpha, \beta; 1,2] \equiv
-\frac{1}{2} \cdot \mbox{BlueLine}^{\dot b}[\beta; 1,3]\cdot
\mbox{UsualLine}[2,3]
\nonumber\\
&&\qquad\qquad\qquad\qquad \times \int d^8x\,\alpha
\Big(\phi_0^{*i} \phi_{0i} + M^{ij} \phi_{0i}
\frac{D^2}{8\partial^2} \phi_{0j} + M^*_{ij}
\phi_0^{*i} \frac{\bar D^2}{8\partial^2} \phi_0^{*j}\Big)_{[3]};\qquad\\
\nonumber\\
\label{WhiteBlueLine} && \mbox{WhiteBlueLine}^{\dot
b}[\alpha,\beta; 1,2] \equiv -\frac{1}{2} \cdot
\mbox{BlueLine}^{\dot b}[\beta; 3,1]\cdot \mbox{UsualLine}[2,3]
\nonumber\\
&&\qquad\qquad\qquad\qquad \times \int d^8x\,\alpha
\Big(\phi_0^{*i} \phi_{0i} + M^{ij} \phi_{0i}
\frac{D^2}{8\partial^2} \phi_{0j} + M^*_{ij} \phi_0^{*i}
\frac{\bar D^2}{8\partial^2} \phi_0^{*j}\Big)_{[3]},\qquad
\end{eqnarray}

\noindent respectively. For example, the operator
$\mbox{WhiteBlueLine}$ can be explicitly written as

\begin{eqnarray}
&& \mbox{WhiteBlueLine}^{\dot b}[\alpha,\beta;1,2] = -\frac{1}{2}
(-1)^{P_\alpha (1+P_\beta)}\int d^8x\,\alpha \Big(\beta\,T^j{}_i
\Big(\frac{\bar D^{\dot b} D^2}{8\partial^2}
\frac{\mbox{\boldmath$\delta$}}{\mbox{\boldmath$\delta$}_1
j^j}\Big)
\frac{\mbox{\boldmath$\delta$}}{\mbox{\boldmath$\delta$}_2 j^*_i}
\qquad
\nonumber\\
&& -\beta\, (M T)^{ij} \frac{D^2}{8\partial^2}
\frac{\mbox{\boldmath$\delta$}}{\mbox{\boldmath$\delta$}_1 j^i}
\frac{\bar D^{\dot b} D^2}{8\partial^2}
\frac{\mbox{\boldmath$\delta$}}{\mbox{\boldmath$\delta$}_2 j^j} -
(M T)^{ij} \frac{D^2}{8\partial^2}\,\beta\,\frac{\bar D^{\dot b}
D^2}{8\partial^2}
\frac{\mbox{\boldmath$\delta$}}{\mbox{\boldmath$\delta$}_1 j^i}
\frac{\mbox{\boldmath$\delta$}}{\mbox{\boldmath$\delta$}_2 j^j}
\Big).
\end{eqnarray}

\noindent (The other similar operators have much more complicated
form.)

3. Commuting $(T) y_\mu^*$ with the inverse Green functions inside
the operator $\mbox{BlueLine}$ in a diagram presented in Fig.
\ref{Figure_New_Identity_Total_Derivatives} we obtain diagrams (7)
and (23). The effective line with two color disks in diagram (23)
encodes the operator

\begin{eqnarray}
&& - i (T)^j{}_i (T)^l{}_k \int d^8x\,d^8y\,\Big( \theta_a
\frac{\bar D_{\dot b} D^2}{8\partial^2}\Big)_x
\Big(-\frac{2i\partial_\mu}{\partial^2} - (\gamma_\mu)^{c{\dot
d}}\theta_c \frac{\bar D^{\dot d} D^2}{4\partial^2}\Big)_y
\nonumber\\
&& \qquad\qquad\qquad\qquad\qquad\qquad\qquad\qquad\qquad
\Big(\frac{\delta^2\gamma}{\delta (\phi_l)_y \delta (\phi_j)_x}
\Big)^{-1} \frac{\delta}{\delta (\phi_{0k})_y}
\frac{\delta}{\delta (\phi_{0i})_x}.\qquad
\end{eqnarray}

\noindent Moreover, we also obtain the commutator

\begin{eqnarray}\label{Pink_Commutator}
&& \Big[y_\mu^*, (\gamma^\mu)^{a{\dot b}} \theta_a \frac{\bar
D_{\dot b} D^2}{8\partial^2} \Big] (T)_j{}^k (T)^j{}_i
\frac{\mbox{\boldmath$\delta$}}{\mbox{\boldmath$\delta$}_2 j^k}
\cdot \frac{\delta}{\delta_1 \phi_{0i}} = (T)_j{}^k (T)^j{}_i
\Big\{\bar\theta_{\dot a}, i\theta^c \theta_c \frac{\bar D^{\dot
a} D^2}{2\partial^2} + (\gamma^\mu)^{{\dot a}b}\theta_b \qquad
\nonumber\\
&& \times \frac{\bar D^2 D^2\partial_\mu}{8\partial^4}\Big\}
\frac{\mbox{\boldmath$\delta$}}{\mbox{\boldmath$\delta$}_2 j^k}
\cdot \frac{\delta}{\delta_1 \phi_{0i}} + (T)_j{}^k (M T)^{ij}
\Big\{\bar\theta_{\dot a}, (\gamma^\mu)^{{\dot a}b} \theta_b
\frac{D^2\partial_\mu}{16\partial^4} \Big\}
\frac{\mbox{\boldmath$\delta$}}{\mbox{\boldmath$\delta$}_2
j^i}\cdot \frac{\mbox{\boldmath$\delta$}}{
\mbox{\boldmath$\delta$}_1 j^k}.
\end{eqnarray}

\noindent Note that the last term evidently vanishes. It is
included for the convenience, because due to its presence this
commutator cancels the corresponding contribution from
$\{(T)\bar\theta^{\dot a},\, \mbox{PinkLine}_{\dot a}\}$.
Moreover, commuting $(T)\bar\theta^{\dot a}$ with the inverse
Green functions inside the operator $\mbox{PinkLine}_{\dot a}$ in
the first diagram in Fig. \ref{Figure_Pink_Additional_Terms}, we
obtain diagrams (16), (17) and (22). The effective line with two
color disks in diagram (22) corresponds to

\begin{eqnarray}
&& - i (T)^j{}_i (T)^l{}_k \int d^8x\,d^8y\, \Big(\theta^c
\theta_c \frac{\bar D_{\dot b} D^2}{2\partial^2} - i
(\gamma^\mu)_{\dot b}{}^{c} \theta_c
\frac{\bar D^2 D^2\partial_\mu}{8\partial^4}\Big)_x \nonumber\\
&& \qquad\qquad\qquad\qquad\qquad\qquad\qquad \Big(\frac{\bar
D^{\dot b} D^2}{8\partial^2}\Big)_y
\Big(\frac{\delta^2\gamma}{\delta (\phi_l)_y \delta (\phi_j)_x}
\Big)^{-1} \frac{\delta}{\delta (\phi_{0k})_y}
\frac{\delta}{\delta (\phi_{0i})_x}.\qquad
\end{eqnarray}

\noindent Similar commutators in the second diagram in Fig.
\ref{Figure_Pink_Additional_Terms} give diagrams (20) and (21).

4. Some terms (for example, the effective one-loop diagrams) do
not have simple graphical interpretation. We write their sum
explicitly in Fig. \ref{Figure_Blue_Diagrams}.

In order to prove identity (\ref{New_Identity}), it is necessary
to verify that the sum of the diagrams presented in Fig.
\ref{Figure_Blue_Diagrams} coincides with the sum of the diagrams
presented in Fig. \ref{Figure_Blue_Green_Diagrams}. (Certainly,
the explicitly written terms should be also taken into account).
This is made in Appendix \ref{Appendix_Comparing_Of_Diagrams},
using the identity

\begin{eqnarray}\label{Pink_Identity}
&&\hspace*{-5mm} 2\cdot \mbox{BlueLine}_{\dot b}[\theta^a\theta_a
\bar\theta^{\dot b};1,2]\cdot \mbox{GreenLine}[3,4] + 2\cdot
\mbox{GreenLine}[1,2]\cdot \mbox{BlueLine}_{\dot
b}[\theta^a\theta_a \bar\theta^{\dot b};3,4] + O(\theta^3)
\vphantom{\Big(}\quad\nonumber\\
&&\hspace*{-5mm} = (\theta^4)_z \Big(\mbox{BlueLine}_{\dot
b}[1;1,2]\cdot \mbox{PinkLine}^{\dot b}[3,4] +
\mbox{PinkLine}_{\dot b}[1,2] \cdot
\mbox{BlueLine}^{\dot b}[1;3,4]\\
&&\hspace*{-5mm} - (\gamma^\mu)^{a{\dot b}} \mbox{BlueLine}_{\dot
b}[\theta_a;1,2]\cdot \mbox{BrownLine}_\mu[3,4] -
(\gamma^\mu)^{a{\dot b}} \mbox{BrownLine}_\mu[1,2]\cdot
\mbox{BlueLine}_{\dot b}[\theta_a;3,4] \Big). \nonumber
\end{eqnarray}

\begin{figure}[h]
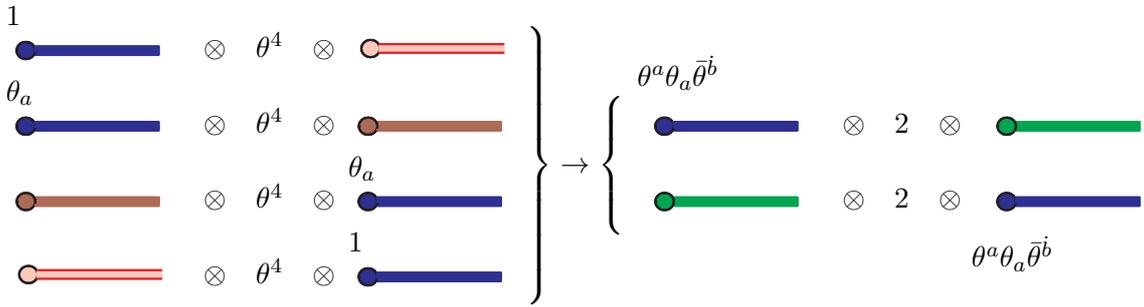

\begin{center}
\begin{picture}(0,0)
\put(-8,0){\includegraphics[angle=180]{line_blue.eps}}
\put(-4.9,-0.72){$\otimes\quad \theta^4\quad \otimes$}
\put(-2.9,-0.78){\includegraphics{line_pink.eps}}
\put(-8,-1){\includegraphics[angle=180]{line_blue.eps}}
\put(-4.9,-1.72){$\otimes\quad \theta^4\quad \otimes$}
\put(-3.5,-1){\includegraphics[angle=180]{line_brown.eps}}
\put(-8,-2){\includegraphics[angle=180]{line_brown.eps}}
\put(-4.9,-2.72){$\otimes\quad \theta^4\quad \otimes$}
\put(-3.5,-2){\includegraphics[angle=180]{line_blue.eps}}
\put(-7.4,-3.78){\includegraphics{line_pink.eps}}
\put(-4.9,-3.72){$\otimes\quad \theta^4\quad \otimes$}
\put(-3.5,-3){\includegraphics[angle=180]{line_blue.eps}}
\put(-0.7,-2.25){$\left. \vphantom{
\begin{array}{c}
1\\
1\\
1\\
1\\
1\\
1\\
1\\
1\\
\end{array}
} \right\} \to \left\{ \vphantom{
\begin{array}{c}
1\\
1\\
1\\
1\\
\end{array}
}\right. $} \put(-7.5,-0.3){$1$} \put(-3,-3.3){$1$}
\put(-7.5,-1.3){$\theta_a$} \put(-3,-2.3){$\theta_a$}
\put(0.8,-1.1){$\theta^a \theta_a \bar\theta^{\dot b}$}
\put(5.2,-3.5){$\theta^a \theta_a \bar\theta^{\dot b}$}
\put(0.4,-1){\includegraphics[angle=180]{line_blue.eps}}
\put(3.5,-1.72){$\otimes\quad 2\quad \otimes$}
\put(4.9,-1){\includegraphics[angle=180]{line_green.eps}}
\put(0.4,-2){\includegraphics[angle=180]{line_green.eps}}
\put(3.5,-2.72){$\otimes\quad 2\quad \otimes$}
\put(4.9,-2){\includegraphics[angle=180]{line_blue.eps}}
\end{picture}
\vspace*{3.5cm}
\end{center}
\caption{A graphical interpretation of the identity
(\ref{Pink_Identity}). (For simplicity we do not write indexes
corresponding to various lines.)}
\label{Figure_Auxiliary_Identity_Blue_Green}
\end{figure}

\begin{figure}[h]
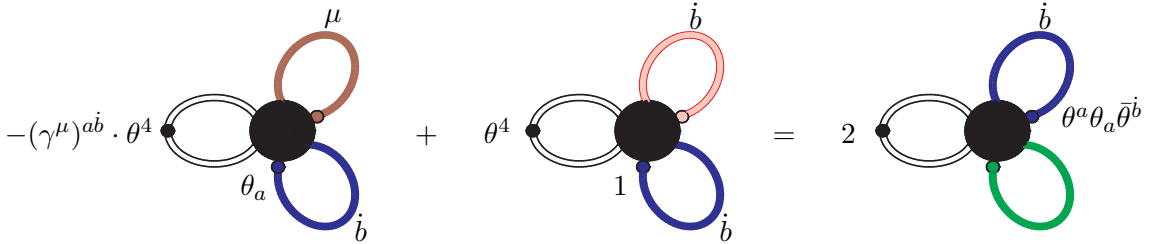

\begin{picture}(0,0)
\put(2.3,-3){\includegraphics[scale=0.7]{figure15diag1.eps}}
\put(7.1,-3){\includegraphics[scale=0.7]{figure15diag10.eps}}
\put(11.7,-3){\includegraphics[scale=0.7]{figure11diag7.eps}}
\put(5.7,-1.8){$+$} \put(10.4,-1.8){$=$}
\put(6.6,-1.8){$\theta^4$} \put(11.3,-1.8){$2$}
\put(8.3,-2.5){$1$} \put(3.4,-2.5){$\theta_a$}
\put(14.2,-1.6){$\theta^a \theta_a \bar\theta^{\dot b}$}
\put(0.3,-1.8){$-(\gamma^\mu)^{a{\dot b}}\cdot\theta^4$}
\put(4.5,-0.2){$\mu$} \put(4.9,-3.1){${\dot b}$}
\put(13.9,-0.3){${\dot b}$} \put(9.3,-0.3){${\dot b}$}
\put(9.7,-3.1){${\dot b}$}
\end{picture}
\vspace*{3cm} \caption{An example of applying the identity
(\ref{Pink_Identity}). This example illustrates how to prove an
identity presented in Fig.
\ref{Figure_New_Identity_Total_Derivatives}. (The diagrams in this
figure are symmetric with respect to permutations of the effective
lines.)} \label{Figure_Example}
\end{figure}

\noindent This identity is proved in Appendix
\ref{Appendix_Pink_Identity}. Its graphical version is presented
in Fig. \ref{Figure_Auxiliary_Identity_Blue_Green}. According to
this figure the sum of four diagrams with the same topology
containing the lines given in the left hand side of this figure is
equal to the sum of two diagrams containing the lines in the right
hand side of this figure. An example of applying this identity is
presented in Fig.\ref{Figure_Example}.

Thus, we have obtained that the $\beta$-function
(\ref{Beta_Definition}) is determined by the integrals of total
derivatives and coincides with the NSVZ $\beta$-function. Also we
have proved the identity (\ref{New_Identity_Analytical_Form})
directly.

\section{Double total derivatives}
\label{Section_Double_Commutators}

\subsection{Factorization of integrands into double total derivatives}
\hspace{\parindent}

\begin{figure}[h]

\vspace*{2.5cm}
\begin{picture}(0,0)
\put(1.2,0.2){${\displaystyle \frac{d}{d\ln\Lambda}\,
\frac{\partial}{\partial \ln g}}$}
\put(2.5,-0.8){\includegraphics[scale=0.4]{figure4.eps}}
\put(6.0,0.2){${\displaystyle = \frac{d}{d\ln\Lambda}}$}
\put(7,-2){\includegraphics{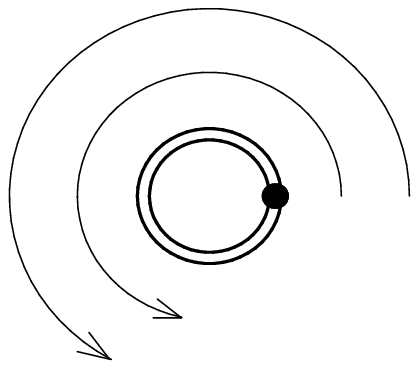}} \put(9.8,-1){$y_\mu^*$}
\put(9.0,-1.6){$y_\mu^*$} \put(10.7,0.17){$\theta^4$}
\put(12.4,0.2){$-\ \ \mbox{singularities}$}
\end{picture}
\vspace*{1.8cm}

\caption{A graphical presentation of the double total derivatives
in Eq.
(\ref{Double_Total_Derivative}).}\label{Figure_Double_Total_Derivatives}
\end{figure}

In the previous section we have proved that a $\beta$-function of
${\cal N}=1$ SQED, regularized by higher derivatives, is given by
integrals of total derivatives. This allows to calculate one of
the loop integrals and obtain the exact NSVZ relation for the
renormalization group functions defined in terms of the bare
coupling constant. However, according to
\cite{Smilga:2004zr,Stepanyantz:2011bz,Stepanyantz:2012zz,Stepanyantz:2012us,Stepanyantz:2011jy}
the $\beta$-function is given by integrals of double total
derivatives. In this section we prove this statement for ${\cal
N}=1$ SQED with $N_f$ flavors, regularized by higher derivatives,
in all orders. For this purpose we differentiate the two-point
Green function of the background gauge superfield with respect to
$\ln g$. The required statement follows from the identity

\begin{eqnarray}\label{Double_Total_Derivative}
&& \frac{d}{d\ln\Lambda} \frac{\partial}{\partial \ln g}
\Big(\frac{1}{2}\int d^8x\,d^8y\,(\theta^4)_x (\theta^4)_y
\frac{\delta^2\Delta\Gamma}{\delta \mbox{\boldmath$V$}_x \delta
\mbox{\boldmath$V$}_y}\Big)\nonumber\\
&& = \frac{i}{4} C(R)_i{}^j \frac{d}{d\ln\Lambda}
\mbox{Tr}\,(\theta^4)_x \Big[y_\mu^*,\Big[y_\mu^*,
\Big(\frac{\delta^2\gamma}{\delta(\phi_j)_x
\delta(\phi^{*i})_y}\Big)^{-1} + M^{ik}
\Big(\frac{D^2}{8\partial^2}\Big)_x\Big(\frac{\delta^2\gamma}{\delta
(\phi_k)_x \delta(\phi_j)_y}\Big)^{-1}\qquad\nonumber\\
&& + M^*_{jk} \Big(\frac{\bar
D^2}{8\partial^2}\Big)_x\Big(\frac{\delta^2\gamma}{\delta
(\phi^{*k})_x \delta(\phi^{*i})_y}\Big)^{-1}
\Big]\Big]_{y=x}-\mbox{singularities}\ =\ -\mbox{singularities},
\end{eqnarray}

\noindent where

\begin{equation}
[y_\mu^*,A_{xy}] \equiv (y_\mu^*)_x A_{xy} - A_{xy} (y_\mu^*)_y
\end{equation}

\noindent and "$-\,\mbox{singularites}$" means that singular
contributions containing $\delta$-functions (see below) should be
subtracted from this expression. (The expression in the right hand
side should be accurately defined for diagrams which include
closed loops of the matter superfields. We will discuss this
definition below.) Note that the last equality in Eq.
(\ref{Double_Total_Derivative}) evidently follows from Eq.
(\ref{Trace_Vanish}). In the graphical form the identity
(\ref{Double_Total_Derivative}) is presented in Fig.
\ref{Figure_Double_Total_Derivatives}. According to Eq.
(\ref{Double_Total_Derivative}) the $\beta$-function of ${\cal
N}=1$ SQED with $N_f$ flavors, regularized by higher derivatives,
in the momentum representation is given by not only by integrals
of total derivatives, but by integrals of {\it double} total
derivatives. (It is necessary to take into account that $\theta^4$
and $y_\mu^*$ commute.)

Using the operation $[(T)y_\mu^*,\ldots]$, defined by Eq.
(\ref{T_Commutators}), the equality
(\ref{Double_Total_Derivative}) can be rewritten in the form

\begin{eqnarray}\label{Double_Total_Derivative2}
&& \frac{d}{d\ln\Lambda} \frac{\partial}{\partial \ln g}
\Big(\frac{1}{2}\int d^8x\,d^8y\,(\theta^4)_x (\theta^4)_y
\frac{\delta^2\Delta\Gamma}{\delta \mbox{\boldmath$V$}_x \delta
\mbox{\boldmath$V$}_y}\Big)\nonumber\\
&& = \frac{i}{4}\cdot \frac{d}{d\ln\Lambda}
\mbox{Tr}\,(\theta^4)_x \Big[(T) y_\mu^*,\Big[(T) y_\mu^*,
\Big(\frac{\delta^2\gamma}{\delta(\phi_i)_x
\delta(\phi^{*i})_y}\Big)^{-1} + M^{ik}
\Big(\frac{D^2}{8\partial^2}\Big)_x\Big(\frac{\delta^2\gamma}{\delta
(\phi_k)_x \delta(\phi_i)_y}\Big)^{-1}\qquad\nonumber\\
&& + M^*_{ik} \Big(\frac{\bar
D^2}{8\partial^2}\Big)_x\Big(\frac{\delta^2\gamma}{\delta
(\phi^{*k})_x \delta(\phi^{*i})_y}\Big)^{-1} \Big]\Big]_{y=x}
-\mbox{singularities}.
\end{eqnarray}

\noindent Then the commutators in the right hand side of this
equation can be calculated using identities obtained in Appendix
\ref{Appendix_Commutators}. First, it is necessary to calculate
the inner commutator. The right hand side of Eq.
(\ref{Double_Total_Derivative2}) can be equivalently rewritten in
the form

\begin{eqnarray}\label{Single_Total_Derivative}
&& -\frac{i}{4}\cdot\frac{d}{d\ln\Lambda}\,
\mbox{Tr}\,(\theta^4)_x \Big[(T) y_\mu^*,\,\mbox{BrownLine}^\mu
\cdot
\Big(\frac{\mbox{\boldmath$\delta$}^2\gamma}{\mbox{\boldmath$\delta$}
j^*_i \mbox{\boldmath$\delta$} j^i} + i \phi^{*i}_0 \phi_{0i} +
M^{ik} \Big(\frac{D^2}{8\partial^2}
\frac{\mbox{\boldmath$\delta$}}{\mbox{\boldmath$\delta$} j^i}\Big)
\frac{\mbox{\boldmath$\delta$}\gamma}{\mbox{\boldmath$\delta$}
j^k}
\nonumber\\
&& + i M^{ik} \Big(\frac{D^2}{8\partial^2}\phi_{0i}\Big) \phi_{0k}
+ M^*_{ik} \Big(\frac{\bar D^2}{8\partial^2}
\frac{\mbox{\boldmath$\delta$}}{\mbox{\boldmath$\delta$}
j^*_i}\Big)
\frac{\mbox{\boldmath$\delta$}\gamma}{\mbox{\boldmath$\delta$}
j^*_k} + i M^*_{ik} \Big(\frac{\bar
D^2}{8\partial^2}\phi_{0}^{*i}\Big) \phi_{0}^{*k}  \Big)- (M
T)^{ik} \Big(\frac{D^2\partial^\mu}{4\partial^4}\Big)_x \qquad
\nonumber\\
&& \times\Big(\frac{\delta^2\gamma}{\delta (\phi_k)_x
\delta(\phi_i)_y}\Big)^{-1}_{y=x} - (T M^*)_{ik} \Big(\frac{\bar
D^2\partial^\mu}{4\partial^4}\Big)_x
\Big(\frac{\delta^2\gamma}{\delta (\phi^{*k})_x
\delta(\phi^{*i})_y}\Big)^{-1}_{y=x}\Big]
-\mbox{singularities}.\qquad
\end{eqnarray}

\begin{figure}[h]
\begin{picture}(0,0)
\put(0.0,-2.9){${\displaystyle \frac{d}{d\ln\Lambda}\,
\frac{\partial}{\partial \ln g}}$}
\put(1.3,-3.9){\includegraphics[scale=0.4]{figure4.eps}}
\put(4.0,-2.9){${\displaystyle + \delta = -\frac{1}{2}\cdot
\frac{d}{d\ln\Lambda}\Bigg\{ }$} \put(8.1,-2.1){$\mu$}
\put(13.5,-2.1){$\mu$}
\put(7.2,-0.5){\includegraphics[angle=180,scale=0.8]{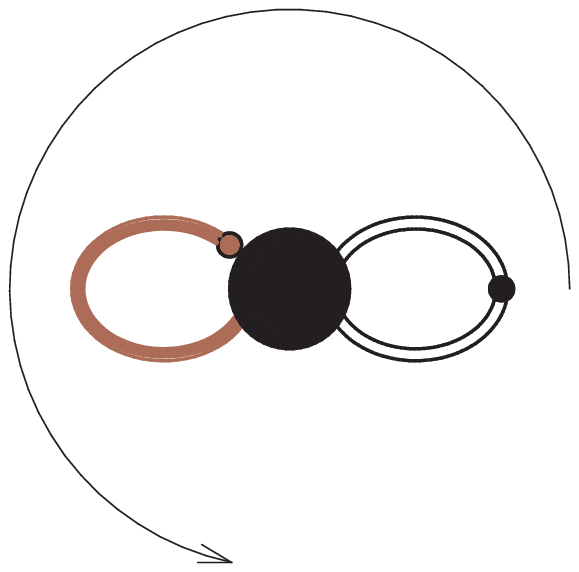}}
\put(9.3,-5.0){$y_\mu^*$} \put(11.5,-3.0){$\theta^4$}
\put(12.1,-2.9){$+$} \put(13.5,-4.4){$y_\mu^*$}
\put(15.5,-2.9){${\displaystyle \Bigg\}}$}
\put(14.9,-2.9){$\theta^4$}
\put(12.6,-4.1){\includegraphics{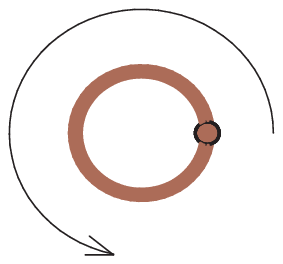}}
\put(0.0,-6.0){${\displaystyle + \frac{i}{4}\cdot
\frac{d}{d\ln\Lambda} \mbox{Tr}\,(\theta^4)_x \Big[(T) y_\mu^*,\,
(M T)^{ik} \Big(\frac{D^2\partial^\mu}{4\partial^4}\Big)_x
\Big(\frac{\delta^2\gamma}{\delta (\phi_k)_x
\delta(\phi_i)_y}\Big)^{-1}_{y=x}}$}
\put(6.1,-7.3){${\displaystyle  + (T M^*)_{ik} \Big(\frac{\bar
D^2\partial^\mu}{4\partial^4}\Big)_x
\Big(\frac{\delta^2\gamma}{\delta (\phi^{*k})_x
\delta(\phi^{*i})_y}\Big)^{-1}_{y=x}\Big]-
\mbox{singularities}.}$}
\end{picture}
\vspace*{7.5cm} \caption{A graphical presentation of total
derivatives which are obtained after calculating the inner
commutator in Eq. (\ref{Double_Total_Derivative}). This is a
graphical form of Eq. (\ref{Single_Total_Derivative}).
}\label{Figure_Single_Total_Derivatives}
\end{figure}

\noindent (This equation and Eq. (\ref{Trace_Notation}) accurately
define the expression in the right hand side for diagrams with
closed loops of the matter superfields.) The terms containing the
operator $\mbox{BrownLine}^\mu$ in this equality can be easily
presented in a graphical form. The result is shown in Fig.
\ref{Figure_Single_Total_Derivatives}. In order to avoid too large
number of effective lines some terms are written explicitly.

\begin{figure}[h]
\begin{picture}(0,0)
\put(2.1,-2.9){${\displaystyle 0 = \frac{i}{2}\cdot
\frac{d}{d\ln\Lambda} \Bigg\{}$}
\put(5.0,-0.5){\includegraphics[angle=180,scale=0.8]{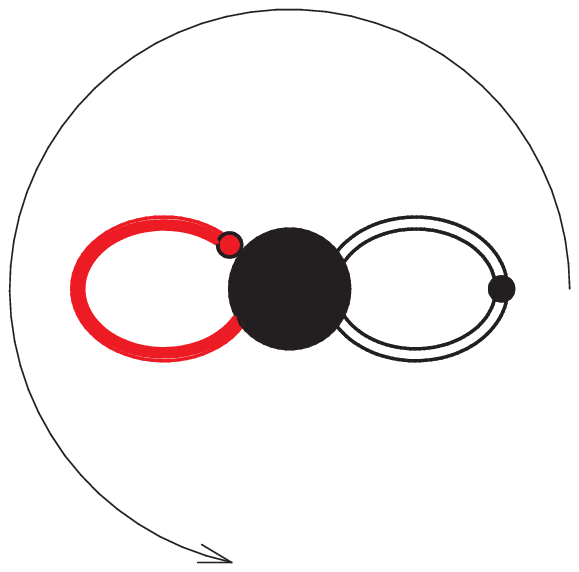}}
\put(7.1,-5.0){$\bar\theta^{\dot a}$} \put(9.3,-3.0){$\theta^4$}
\put(10.3,-2.9){$+$} \put(12.2,-4.4){$\bar\theta^{\dot a}$}
\put(6,-2.1){${\dot a}$} \put(11.8,-2.1){${\dot a}$}
\put(13.2,-2.9){$\theta^4$}
\put(10.9,-4.1){\includegraphics{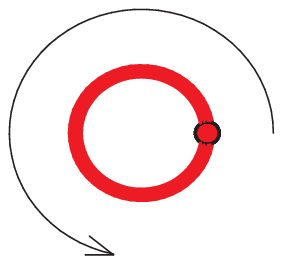}}
\put(14,-2.9){${\displaystyle \Bigg\}}$}
\end{picture}
\vspace*{5.0cm} \caption{It is convenient to add these terms
to the diagrams in the right hand side of
Fig. \ref{Figure_Single_Total_Derivatives}.}
\label{Figure_Additional_Terms}
\end{figure}

Then it is necessary to calculate the second commutator. This can
be done similarly to the calculation made in the previous section.
As earlier, it is convenient to add some terms to the expression
in the right hand side of Fig.
\ref{Figure_Single_Total_Derivatives}. They are presented in Fig.
\ref{Figure_Additional_Terms}. These effective diagrams correspond
to the analytical expression

\begin{eqnarray}
&& - \frac{1}{4}\cdot
\frac{d}{d\ln\Lambda}\,\mbox{Tr}\,(\theta^4)_x \Big[(T)
\theta^{\dot a},\,\mbox{RedLine}_{\dot a} \cdot
\Big(\frac{\mbox{\boldmath$\delta$}^2\gamma}{\mbox{\boldmath$\delta$}
j^*_i \mbox{\boldmath$\delta$} j^i} +i\, \phi^{*i}_0 \phi_{0i} +
M^{ik} \Big(\frac{D^2}{8\partial^2}
\frac{\mbox{\boldmath$\delta$}}{\mbox{\boldmath$\delta$} j^i}\Big)
\frac{\mbox{\boldmath$\delta$}\gamma}{\mbox{\boldmath$\delta$}
j^k}\qquad \nonumber\\
&& + i\, M^{ik} \Big(\frac{D^2}{8\partial^2}\phi_{0i}\Big)
\phi_{0k} + M^*_{ik} \Big(\frac{\bar D^2}{8\partial^2}
\frac{\mbox{\boldmath$\delta$}}{\mbox{\boldmath$\delta$}
j^*_i}\Big)
\frac{\mbox{\boldmath$\delta$}\gamma}{\mbox{\boldmath$\delta$}
j^*_k} + i\, M^*_{ik} \Big(\frac{\bar
D^2}{8\partial^2}\phi_{0}^{*i}\Big) \phi_{0}^{*k} \Big)\Big] =
0,\qquad
\end{eqnarray}

\noindent
where the operator

\begin{eqnarray}
&& \mbox{RedLine}^{\dot a}[1,2] = \int d^8x\, \Big((T)^j{}_i
\Big(\theta^b \theta_b \frac{\bar D^{\dot a} D^2}{\partial^2} - i
(\gamma^\mu)^{{\dot a}b} \theta_b \frac{\bar D^2
D^2\partial_\mu}{2\partial^4}\Big)
\frac{\mbox{\boldmath$\delta$}}{\mbox{\boldmath$\delta$}_2 j^j}
\cdot \frac{\delta}{\delta_1 \phi_{0i}}\qquad\nonumber\\
&& - i (M T)^{ij} (\gamma^\mu)^{{\dot a}b} \theta_b
\Big(\frac{D^2\partial_\mu}{4\partial^4}
\frac{\mbox{\boldmath$\delta$}}{\mbox{\boldmath$\delta$}_2
j^i}\Big)\cdot \frac{\mbox{\boldmath$\delta$}}{
\mbox{\boldmath$\delta$}_1 j^j}\Big)_{x}
\end{eqnarray}

\noindent is denoted by the red line.

\begin{figure}[p]
\vspace*{-1cm}
\begin{picture}(0,0)
\put(0,-0.9){${\displaystyle \frac{d}{d\ln\Lambda} \Bigg\{}$}
\put(1.4,-0.9){${\displaystyle -\frac{1}{2}\ \times}$}
\put(2.5,-0.9){$\theta^4$} \put(5.5,0.4){$\mu$}
\put(5.5,-2.0){$\mu$} \put(3.3,0.2){$(1)$}
\put(2.9,-2.1){\includegraphics[scale=0.7]{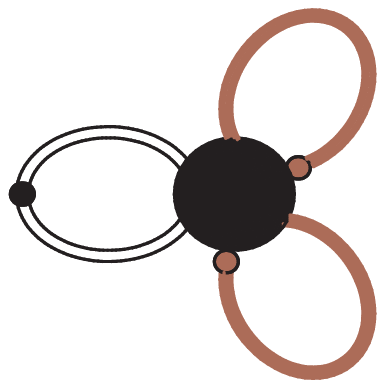}}
\put(6.2,-0.9){$-$} \put(7.0,0.2){$(2)$}
\put(7.3,-1.9){\includegraphics[scale=0.7]{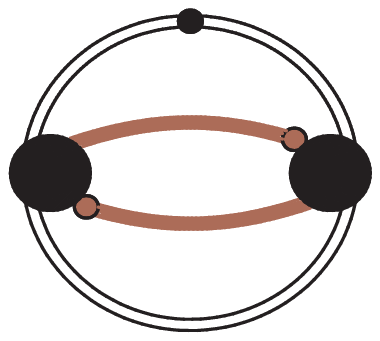}}
\put(10.5,-0.9){$-$} \put(11.3,0.2){$(3)$}
\put(11.6,-1.9){\includegraphics[scale=0.7]{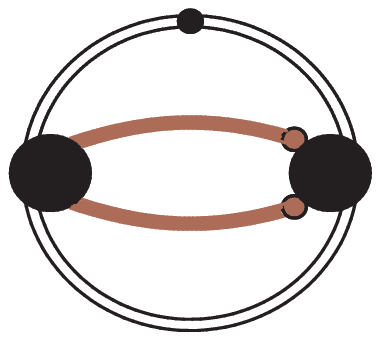}}
\put(0.5,-3.25){${\displaystyle -\ \frac{1}{2}}$}
\put(1.5,-2.4){$(4)$}
\put(1.4,-3.7){\includegraphics[scale=0.7]{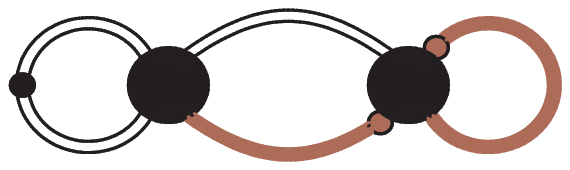}}
\put(5.7,-3.25){$-$} \put(6.5,-2.4){$(5)$}
\put(6.3,-3.7){\includegraphics[scale=0.7]{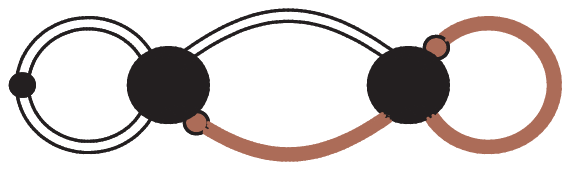}}
\put(10.6,-3.25){${\displaystyle -\ \frac{1}{2}}$}
\put(11.5,-2.4){$(6)$}
\put(11.4,-3.85){\includegraphics[scale=0.7]{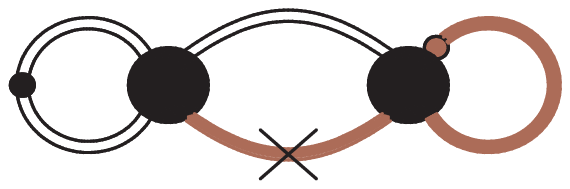}}
\put(0.5,-5.15){$-$} \put(1.5,-4.4){$(7)$}
\put(1.15,-5.6){\includegraphics[scale=0.7]{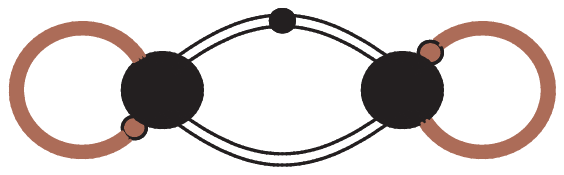}}
\put(5.8,-5.15){${\displaystyle -\ \ \frac{1}{2}}$}
\put(6.9,-4.4){$(8)$}
\put(6.8,-5.55){\includegraphics[scale=0.7]{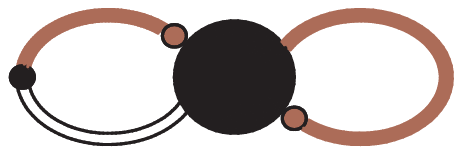}}
\put(10.6,-5.15){$-$} \put(11.5,-4.4){$(9)$}
\put(11.4,-5.55){\includegraphics[scale=0.7]{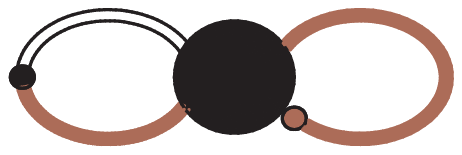}}
\put(0.7,-7.05){${\displaystyle -\ \frac{1}{2}}$}
\put(1.4,-6.3){$(10)$}
\put(1.6,-7.4){\includegraphics[scale=0.7]{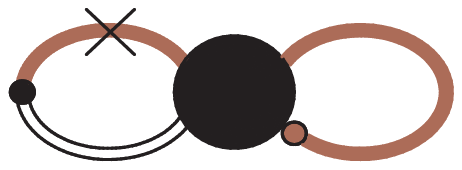}}
\put(5.9,-7.05){${\displaystyle -\ \frac{1}{2}}$}
\put(6.6,-6.3){$(11)$}
\put(6.8,-7.4){\includegraphics[scale=0.7]{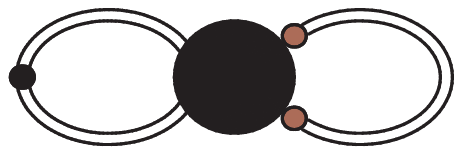}}
\put(10.6,-7.05){${\displaystyle -\ \frac{1}{2}}$}
\put(11.3,-6.3){$(12)$}
\put(11.5,-7.4){\includegraphics[scale=0.7]{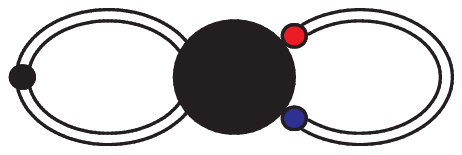}}
\put(0.6,-9.4){${\displaystyle -\ \frac{1}{2}\ \ \times}$}
\put(2.8,-8.4){$(13)$} \put(2.0,-9.4){$\theta^4$}
\put(5.0,-8.1){${\dot b}$} \put(5.1,-10.5){${\dot b}$}
\put(3.7,-10.1){$1$}
\put(2.4,-10.6){\includegraphics[scale=0.7]{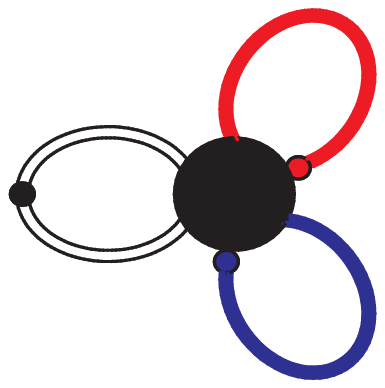}}
\put(6.0,-9.4){$-$} \put(6.5,-8.4){$(14)$}
\put(7.1,-10.5){\includegraphics[scale=0.7]{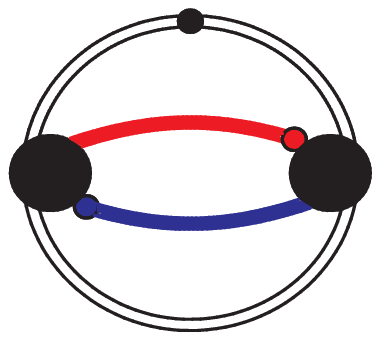}}
\put(10.5,-9.4){$-$} \put(11.3,-8.4){$(15)$}
\put(11.6,-10.5){\includegraphics[scale=0.7]{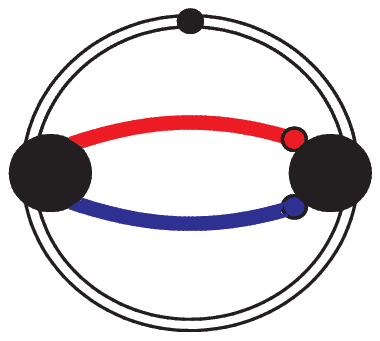}}
\put(0.4,-11.85){${\displaystyle -\ \frac{1}{2}}$}
\put(1.5,-11.0){$(16)$}
\put(1.3,-12.3){\includegraphics[scale=0.7]{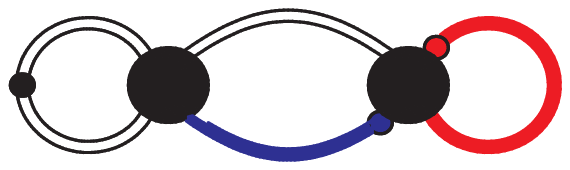}}
\put(5.5,-11.85){${\displaystyle -\ \frac{1}{2}}$}
\put(6.5,-11.0){$(17)$}
\put(6.4,-12.3){\includegraphics[scale=0.7]{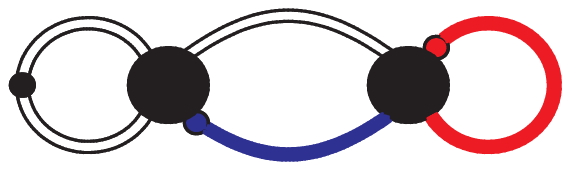}}
\put(10.6,-11.85){${\displaystyle -\ \frac{1}{2}}$}
\put(11.5,-11.0){$(18)$}
\put(11.4,-12.3){\includegraphics[scale=0.7]{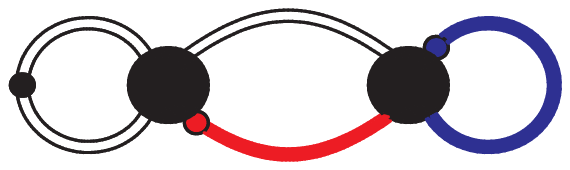}}
\put(0.4,-13.95){${\displaystyle -\ \frac{1}{2}}$}
\put(1.5,-13.1){$(19)$}
\put(1.3,-14.5){\includegraphics[scale=0.7]{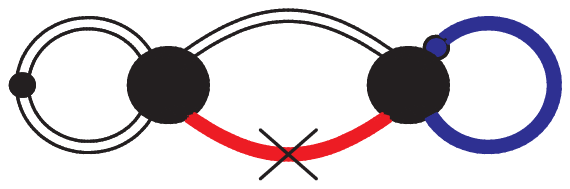}}
\put(5.9,-13.95){$-$} \put(6.5,-13.1){$(20)$}
\put(6.4,-14.4){\includegraphics[scale=0.7]{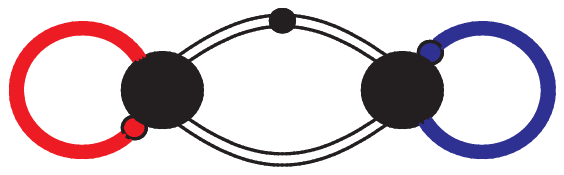}}
\put(10.9,-13.95){${\displaystyle -\ \frac{1}{2}}$}
\put(11.7,-13.1){$(21)$}
\put(11.8,-14.35){\includegraphics[scale=0.7]{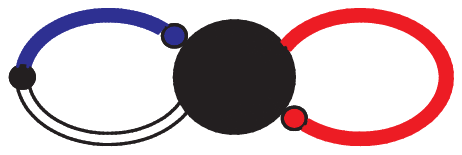}}
\put(0.9,-15.85){${\displaystyle -\ \frac{1}{2}}$}
\put(1.6,-15.1){$(22)$}
\put(1.8,-16.2){\includegraphics[scale=0.7]{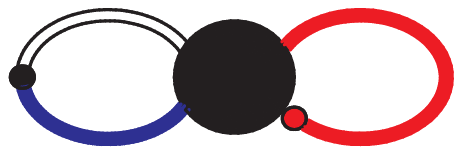}}
\put(5.9,-15.85){${\displaystyle -\ \frac{1}{2}}$}
\put(6.6,-15.1){$(23)$}
\put(6.8,-16.2){\includegraphics[scale=0.7]{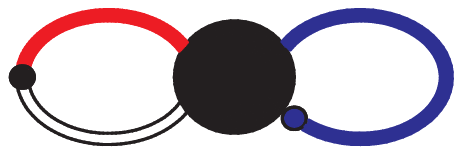}}
\put(10.9,-15.85){${\displaystyle -\ \frac{1}{2}}$}
\put(11.6,-15.1){$(24)$}
\put(11.8,-16.2){\includegraphics[scale=0.7]{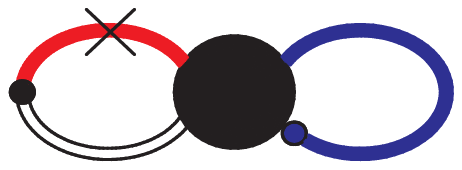}}

\put(0.2,-17.5){${\displaystyle +\mbox{One-Loop} + \frac{i}{2}
\int d^8x\,d^8y\, \Bigg( C(R)_k{}^n M^{jk} \delta^8_{xy}
\Big(\frac{D^2}{4\partial^4}\Big)_x + (MT)^{mn} (M T)^{ij}
\Big(\frac{D^2\partial^\mu }{8\partial^4}\Big)_y }$}

\put(0.2,-19.0){${\displaystyle \times
\Big(\frac{\delta^2\gamma}{\delta(\phi_i)_x
\delta(\phi_m)_y}\Big)^{-1} \Big(\frac{D^2
\partial_\mu}{8\partial^4}\Big)_x\Bigg)\cdot \mbox{LineWithDot}[\theta^4]
\cdot \frac{\mbox{\boldmath$\delta$}^2
\gamma}{\mbox{\boldmath$\delta$} (j^j)_x \mbox{\boldmath$\delta$}
(j^n)_y} + \frac{i}{2} \int d^8x\,(\theta^4)_x
\mbox{BrownLine}^\mu }$}

\put(0.2,-20.5){${\displaystyle \times \Big( (MT)^{ik}
\Big(\frac{D^2\partial_\mu}{4\partial^4}\Big)_x
\frac{\mbox{\boldmath$\delta$}^2\gamma}{\mbox{\boldmath$\delta$}
(j^i)_x \mbox{\boldmath$\delta$} (j^k)_y}\Big|_{y=x} + (TM^*)_{ik}
\Big(\frac{\bar D^2\partial_\mu}{4\partial^4}\Big)_x
\frac{\mbox{\boldmath$\delta$}^2\gamma}{\mbox{\boldmath$\delta$}
(j^*_i)_x \mbox{\boldmath$\delta$} (j^*_k)_y}\Big|_{y=x} \Big) -
\int d^8x\,(\gamma^\mu)^{{\dot a}b} }$}

\put(0.2,-22.0){${\displaystyle \times (\bar\theta^{\dot c}
\bar\theta_{\dot c} \theta_b)_x (T M^*)_{ik} \Big(\frac{i \bar
D^2\partial_\mu}{32\partial^4}\Big)_x \mbox{RedLine}_{\dot a}
\cdot
\frac{\mbox{\boldmath$\delta$}^2\gamma}{\mbox{\boldmath$\delta$}
(j^*_i)_x \mbox{\boldmath$\delta$} (j^*_k)_y}\Bigg|_{y=x}
\Bigg\}.}$}
\end{picture}
\vspace*{22.5cm} \caption{These diagrams are obtained by
calculating two commutators in Eq.
(\ref{Double_Total_Derivative}). Constructing expressions
corresponding to these diagrams we assume that the first spinor
index is lower and the second one is upper. The expression
$\mbox{One-Loop}$ is given by Eq. (\ref{One-Loop_Diagrams}).}
\label{Figure_Brown_Diagrams}
\end{figure}

The commutators are calculated using equations derived in Appendix
\ref{Appendix_Commutators}. In a graphical form the result is
presented in Fig. \ref{Figure_Brown_Diagrams}. (The singular
contributions will be calculated in the next section.) Let us
describe this calculation in details.

1. Commuting $(T) y_\mu^*$ with the four point function according
to the prescription presented in Appendix
\ref{Appendix_Commutators_Y} gives diagrams (1) --- (4) and $1/2$
of diagram (5) in Fig. \ref{Figure_Brown_Diagrams}.

2. Commuting $(T) \bar\theta^{\dot a}$ with the four-point
vertices in the first effective diagram in Fig.
\ref{Figure_Additional_Terms} we obtain diagrams (13) --- (17) in
Fig. \ref{Figure_Brown_Diagrams}. The details of this calculation
are presented in Appendix \ref{Appendix_Commutators_Theta}.

3. The other $1/2$ of diagram (5) and diagrams (6) and (11) in
Fig. \ref{Figure_Brown_Diagrams} are obtained if $(T) y_\mu^*$ is
commuted with the inverse Green functions coming from the
derivatives $\mbox{\boldmath$\delta$}/\mbox{\boldmath$\delta$} j$
which are contained in the brown effective line (in the first
effective diagram in Fig. \ref{Figure_Single_Total_Derivatives}).
In diagram (11) the effective line with two brown disks denotes
the operator

\begin{eqnarray}
&& - i (T)^j{}_i (T)^l{}_k \int
d^8x\,d^8y\,\Big(-\frac{2i\partial^\mu}{\partial^2} +
(\gamma^\mu)^{a{\dot b}}\theta_a \frac{\bar D^{\dot b}
D^2}{4\partial^2}\Big)_x \Big(-\frac{2i\partial_\mu}{\partial^2} +
(\gamma_\mu)^{c{\dot d}}\theta_c
\frac{\bar D^{\dot d} D^2}{4\partial^2}\Big)_y \nonumber\\
&& \times \Big(\frac{\delta^2\gamma}{\delta (\phi_l)_y \delta
(\phi_j)_x} \Big)^{-1} \frac{\delta}{\delta (\phi_{0k})_y}
\frac{\delta}{\delta (\phi_{0i})_x}.
\end{eqnarray}

Similarly, diagrams (12), (18), and (19) in Fig.
\ref{Figure_Brown_Diagrams} are obtained from the first effective
diagram in Fig. \ref{Figure_Additional_Terms}, if
$(T)\bar\theta^{\dot a}$ is commuted with the inverse Green
functions contained in the red effective line. In diagram (12) the
effective line with two color disks denotes the operator

\begin{eqnarray}
&& - i (T)^j{}_i (T)^l{}_k \int d^8x\,d^8y\,\Big(\frac{\bar
D_{\dot b} D^2}{8\partial^2}\Big)_x \Big(\theta^a \theta_a
\frac{\bar D^{\dot b} D^2}{2\partial^2}- i(\gamma^\mu)^{a{\dot
b}}\theta_a \frac{\bar D^2 D^2
\partial_\mu}{4\partial^2}\Big)_y\nonumber\\
&& \times \Big(\frac{\delta^2\gamma}{\delta (\phi_l)_y \delta
(\phi_j)_x} \Big)^{-1} \frac{\delta}{\delta (\phi_{0k})_y}
\frac{\delta}{\delta (\phi_{0i})_x}.
\end{eqnarray}

\noindent In addition to these diagrams the considered commutators
give terms containing the operator $\mbox{LineWithDot}[\theta^4]$
which are explicitly written in Fig. \ref{Figure_Brown_Diagrams}.

4. It is also necessary to commute $(T)y_\mu^*$ with the operators

\begin{equation}
i \frac{\bar D^2 D^2\partial^\mu }{8\partial^4}
-(\gamma^\mu)^{a{\dot b}} \theta_a \frac{\bar D_{\dot b}
D^2}{4\partial^2}\qquad \mbox{and} \qquad
\frac{D^2\partial_\mu}{16\partial^4},
\end{equation}

\noindent which are contained in the brown effective line. Taking
into account that

\begin{equation}\label{Delta_Commutator}
[x^\mu,\frac{\partial_\mu}{\partial^4}] =
[-i\frac{\partial}{\partial p_\mu}, -\frac{ip_\mu}{p^4}] = -2\pi^2
\delta^4(p_E) = -2\pi^2 i \delta^4(p) = -2\pi^2
i\delta^4(\partial),
\end{equation}

\noindent we obtain

\begin{eqnarray}\label{Red_Commutator}
&&\hspace*{-7mm} \Big[ y_\mu^*,\, i \frac{\bar D^2 D^2\partial^\mu
}{8\partial^4}-(\gamma^\mu)^{a{\dot b}} \theta_a \frac{\bar
D_{\dot b} D^2}{4\partial^2} \Big] = \Big\{\bar\theta_{\dot a},
-i\theta^c \theta_c \frac{\bar D^{\dot a} D^2}{\partial^2} -
(\gamma^\mu)^{{\dot a}b}\theta_b \frac{\bar D^2
D^2\partial_\mu}{2\partial^4}\Big\} +
\frac{\pi^2}{4}\delta^4(\partial) \bar D^2 D^2;\nonumber\\
&&\hspace*{-7mm} \Big[y_\mu^*,\frac{i D^2
\partial_\mu}{16\partial^4}\Big]
= - \Big\{\bar\theta_{\dot a}, (\gamma^\mu)^{{\dot a}b} \theta_b
\frac{D^2\partial_\mu}{4\partial^4} \Big\} +
\frac{\pi^2}{8}\delta^4(\partial) D^2.
\end{eqnarray}

\noindent As a consequence, the terms which do not contain
$\delta$-functions cancel the corresponding terms coming from the
diagrams presented in Fig. \ref{Figure_Additional_Terms}.

5. Commuting $(T) y_\mu^*$ with the operator
$\mbox{LineWithDot}[\theta^4]$ we obtain diagrams (7), (8), and
the other $1/2$ of diagram (9) in Fig.
\ref{Figure_Brown_Diagrams}. These diagrams are constructed using
the effective lines defined by Eqs. (\ref{BrownWhiteLine}) and
(\ref{WhiteBrownLine}). Also the considered commutators and the
commutators written in Fig. \ref{Figure_Single_Total_Derivatives}
explicitly give the terms containing the operator
$\mbox{BrownLine}^\mu$ in Fig. \ref{Figure_Brown_Diagrams}.

6. Diagrams (20), (21), and (22) are obtained if
$(T)\bar\theta^{\dot a}$ is commuted with the effective line
containing a dot. Expressions for diagrams (21) and (22) are
constructed using the notation (\ref{BlueWhiteLine}) and
(\ref{WhiteBlueLine}), respectively. Also we obtain a term
containing the operator $\mbox{RedLine}^{\dot a}$ explicitly
written in Fig. \ref{Figure_Brown_Diagrams}.

7. The second effective diagram in Fig.
\ref{Figure_Single_Total_Derivatives} gives $1/2$ of diagram (9)
and diagram (10). Similarly, diagrams (23) and (24) are obtained
from the second effective diagram in Fig.
\ref{Figure_Additional_Terms}. In these diagrams we use the
notation

\begin{eqnarray}\label{RedWhiteLine}
&& \mbox{RedWhiteLine}_{\dot a}[\theta^4; 1,2] \equiv -\frac{1}{2}
\cdot \mbox{RedLine}_{\dot a}[1,3]\cdot \mbox{UsualLine}[2,3]
\nonumber\\
&&\qquad\qquad\qquad\qquad \times \int d^8x\,(\theta^4)_x
\Big(\phi_0^{*i} \phi_{0i} + M^{ij} \phi_{0i}
\frac{D^2}{8\partial^2} \phi_{0j} + M^*_{ij}
\phi_0^{*i} \frac{\bar D^2}{8\partial^2} \phi_0^{*j}\Big)_{[3]};\qquad\\
\nonumber\\
\label{WhiteRedLine} && \mbox{WhiteRedLine}_{\dot a}[\theta^4;
1,2] \equiv -\frac{1}{2} \cdot \mbox{RedLine}_{\dot a}[3,1]\cdot
\mbox{UsualLine}[2,3]
\nonumber\\
&&\qquad\qquad\qquad\qquad \times \int d^8x\,(\theta^4)_x
\Big(\phi_0^{*i} \phi_{0i} + M^{ij} \phi_{0i}
\frac{D^2}{8\partial^2} \phi_{0j} + M^*_{ij} \phi_0^{*i}
\frac{\bar D^2}{8\partial^2} \phi_0^{*j}\Big)_{[3]}.\qquad
\end{eqnarray}

8. Also the considered diagrams and terms written explicitly in
Fig. \ref{Figure_Single_Total_Derivatives} give some contributions
which can be graphically presented as one-loop effective diagrams.
In Fig. \ref{Figure_Brown_Diagrams} they are denoted by
$\mbox{One-Loop}$. This expression has the following form:

\begin{eqnarray}\label{One-Loop_Diagrams}
&&\hspace*{-7mm} \mbox{One-Loop} = -i \frac{d}{d\ln\Lambda} \int
d^8x\,d^8y\, (\theta^4)_x \Bigg\{\Bigg( C(R)_j{}^i M^{mj}
\delta^8_{xy} \Big(\frac{D^2}{4\partial^4}\Big)_x - (MT)^{nm} (M
T)^{ij} \Big(\frac{D^2\partial^\mu }{8\partial^4}\Big)_y
\nonumber\\
&&\hspace*{-7mm} \times
\Big(\frac{\delta^2\gamma}{\delta(\phi_j)_x
\delta(\phi_n)_y}\Big)^{-1} \Big(\frac{D^2
\partial_\mu}{8\partial^4}\Big)_x\Bigg)\Bigg(- \Big(\frac{\delta^2\gamma}{\delta
(\phi_i)_x \delta (\phi_m)_y}\Big)^{-1} + \int d^8z\,
\Bigg[\,\frac{1}{4} \Big(\frac{\delta^2\gamma}{\delta (\phi_i)_y
\delta (\phi_k)_z}\Big)^{-1}\nonumber\\
&&\hspace*{-7mm} \times \Big(\frac{\delta^2\gamma}{\delta
(\phi^{*k})_z \delta (\phi_m)_x}\Big)^{-1} + \frac{1}{4}
\Big(\frac{\delta^2\gamma}{\delta (\phi_i)_y \delta
(\phi^{*k})_z}\Big)^{-1} \Big(\frac{\delta^2\gamma}{\delta
(\phi_m)_x \delta (\phi_{k})_z}\Big)^{-1} + M^{kl}
\Big(\frac{\delta^2\gamma}{\delta (\phi_m)_y \delta
(\phi_k)_z}\Big)^{-1}\nonumber\\
&&\hspace*{-7mm} \times \Big(\frac{D^2}{16\partial^2}\Big)_z
\Big(\frac{\delta^2\gamma}{\delta (\phi_l)_z \delta
(\phi_i)_x}\Big)^{-1} + M^*_{kl} \Big(\frac{\delta^2\gamma}{\delta
(\phi_m)_y \delta (\phi^{*k})_z}\Big)^{-1} \Big(\frac{\bar
D^2}{16\partial^2}\Big)_z \Big(\frac{\delta^2\gamma}{\delta
(\phi^{*l})_z \delta (\phi_i)_x}\Big)^{-1} \Bigg]  \Bigg)
\Bigg\}.\nonumber\\
\end{eqnarray}

The sum of diagrams presented in Fig. \ref{Figure_Brown_Diagrams}
should be compared with the sum of diagrams presented in Fig.
\ref{Figure_Green_Diagrams}. (Certainly, the explicitly written
terms should be also taken into account.) For this purpose it is
necessary to use the identity

\begin{eqnarray}\label{Red_Identity_Main}
&& (\theta^4)_z \Big( \mbox{BlueLine}_{\dot b}[1;1,2]\cdot
\mbox{RedLine}^{\dot b}[3,4] + \mbox{RedLine}_{\dot b}[1,2] \cdot
\mbox{BlueLine}^{\dot b}[1;3,4]\\
&& + 2\cdot \mbox{BrownLine}^\mu[1,2]\cdot
\mbox{BrownLine}_\mu[3,4] \Big)= 4\cdot\mbox{GreenLine}[1,2]\cdot
\mbox{GreenLine}[3,4] + O(\theta^3),\nonumber\vphantom{\Big(}\quad
\end{eqnarray}

\begin{figure}[h]
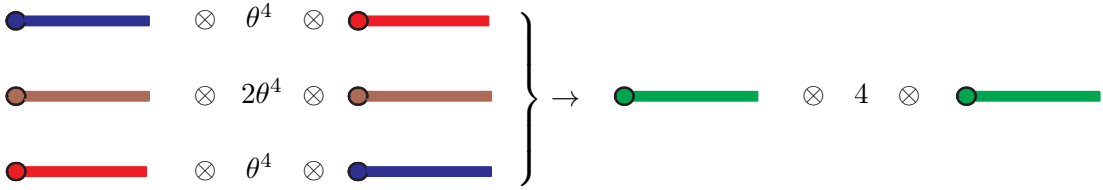

\begin{center}
\begin{picture}(0,0)

\put(-8,0){\includegraphics[angle=180]{line_blue.eps}}
\put(-4.9,-0.72){$\otimes\quad \theta^4\quad \otimes$}
\put(-3.5,0){\includegraphics[angle=180]{line_red.eps}}

\put(-8,-1){\includegraphics[angle=180]{line_brown.eps}}
\put(-4.9,-1.72){$\otimes\ \ \, 2\theta^4\ \  \otimes$}
\put(-3.5,-1){\includegraphics[angle=180]{line_brown.eps}}

\put(-8,-2){\includegraphics[angle=180]{line_red.eps}}
\put(-4.9,-2.72){$\otimes\quad \theta^4\quad \otimes$}
\put(-3.5,-2){\includegraphics[angle=180]{line_blue.eps}}

\put(-0.7,-1.77){$\left. \vphantom{
\begin{array}{c}
1\\
1\\
1\\
1\\
1\\
\end{array}
} \right\} \to$}

\put(0,-1){\includegraphics[angle=180]{line_green.eps}}
\put(3.1,-1.72){$\otimes\quad 4\quad \otimes$}
\put(4.5,-1){\includegraphics[angle=180]{line_green.eps}}

\end{picture}
\vspace*{2.5cm}
\end{center}
\caption{A graphical interpretation of the identity
(\ref{Red_Identity_Main}).} \label{Figure_Auxiliary_Identity}
\end{figure}

\noindent derived in Appendix \ref{Appendix_Red_Identity}. This
identity is graphically presented in Fig.
\ref{Figure_Auxiliary_Identity}. As earlier, this figure should be
understood as follows: we find a sum of three diagrams with the
same topology which contain effective lines presented in the left
hand side of Fig. \ref{Figure_Auxiliary_Identity} and $\theta^4$
in an auxiliary (but fixed) position. Then this sum can be
replaced by a single diagram with the same topology containing two
green effective lines. Using this identity we see that

1. The sum of diagrams (1) and (13) in Fig.
\ref{Figure_Brown_Diagrams} gives diagram (5) in
Fig.~\ref{Figure_Green_Diagrams}.

2. The sum of diagrams (2) and (14) in Fig.
\ref{Figure_Brown_Diagrams} gives diagram (6) in
Fig.~\ref{Figure_Green_Diagrams}.

3. The sum of diagrams (3) and (15) in Fig.
\ref{Figure_Brown_Diagrams} gives diagram (7) in
Fig.~\ref{Figure_Green_Diagrams}.

4. The sum of diagrams (7) and (20) in Fig.
\ref{Figure_Brown_Diagrams} gives diagram (8) in
Fig.~\ref{Figure_Green_Diagrams}.

5. The sum of the expression $\mbox{One-Loop}$ and the terms
containing the operator $\mbox{LineWithDot}$ explicitly written in
Fig. \ref{Figure_Brown_Diagrams} is equal to
$\partial\Delta/\partial\ln g$, which is calculated in Appendix
\ref{Appendix_G_Derivative_One-Loop}. (It is evident that
$\theta^4$ in the terms with the operator $\mbox{LineWithDot}$ can
be shifted to an arbitrary point of the diagram.)

\begin{figure}[h]
\begin{picture}(0,0)

\put(3.0,0.2){$(a)$}
\put(3.3,-0.9){\includegraphics[scale=0.7]{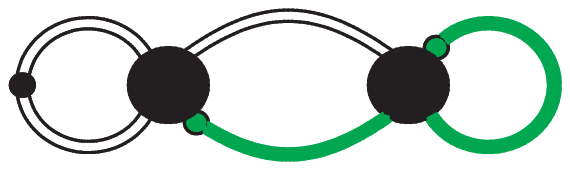}}
\put(8.7,-1.05){\includegraphics[scale=0.7]{figure10diag3.eps}}
\put(7.6,-0.5){$=$} \put(2.7,-0.5){$-2$} \put(8.1,-0.5){$-2$}

\put(0.8,-1.8){$(b)$}
\put(1.2,-2.9){\includegraphics[scale=0.7]{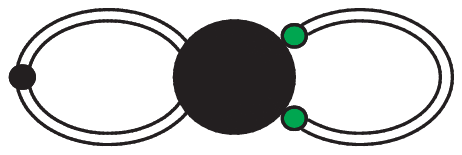}}
\put(4.7,-2.5){$ = \ 0;$} \put(0.6,-2.5){$-1$}

\put(6.7,-1.8){$(c)$}
\put(7.0,-2.9){\includegraphics[scale=0.7]{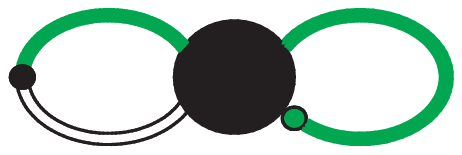}}
\put(11.4,-2.9){\includegraphics[scale=0.7]{figure10diag2.eps}}
\put(10.4,-2.5){$=$} \put(6.4,-2.5){$-2$} \put(10.8,-2.5){$-2$}

\put(3.0,-3.8){$(d)$}
\put(3.3,-4.9){\includegraphics[scale=0.7]{figure21diag4.eps}}
\put(8.7,-5.1){\includegraphics[scale=0.7]{figure21diag6.eps}}
\put(7.7,-4.5){$=$} \put(11.2,-5.1){$\mu$}
\put(8.3,-4.5){$\theta^4$} \put(12.7,-4.2){$\mu$}
\put(5.3,-5.2){$\mu$} \put(2.9,-4.5){$\theta^4$}
\put(7.35,-4.2){$\mu$}

\put(3.0,-5.7){$(e)$}
\put(8.7,-7.1){\includegraphics[scale=0.7]{figure21diag19.eps}}
\put(8.3,-6.5){$\theta^4$} \put(11.1,-7.25){${\dot b}$}
\put(12.7,-6.2){${\dot b}$} \put(11.5,-5.8){$1$}
\put(3.3,-6.9){\includegraphics[scale=0.7]{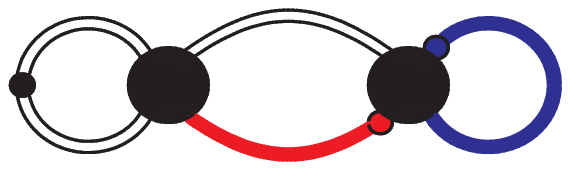}}
\put(7.7,-6.5){$=$} \put(2.9,-6.5){$\theta^4$}
\put(5.5,-7.25){${\dot b}$} \put(7.35,-6.2){${\dot b}$}
\put(6.1,-5.8){$1$}

\put(3.5,-7.8){$(f)$}
\put(4.0,-8.85){\includegraphics[scale=0.7]{figure21diag8.eps}}
\put(3.5,-8.5){$\theta^4$} \put(8.5,-8.5){$\theta^4$}
\put(7.8,-8.5){=}
\put(9.0,-8.85){\includegraphics[scale=0.7]{figure21diag10.eps}}

\put(3.5,-9.8){$(g)$}
\put(4.0,-10.85){\includegraphics[scale=0.7]{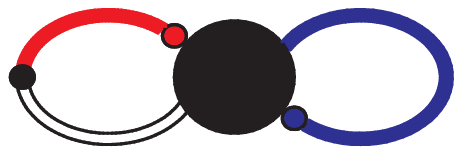}}
\put(3.5,-10.5){$\theta^4$} \put(8.5,-10.5){$\theta^4$}
\put(5.9,-11.1){$1$} \put(10.9,-11.1){$1$} \put(7.8,-10.5){=}
\put(9.0,-10.85){\includegraphics[scale=0.7]{figure21diag24.eps}}

\end{picture}
\vspace*{11cm} \caption{Some useful identities for the effective
diagrams.} \label{Figure_Zero_Diagrams2}
\end{figure}

6. The sum of diagrams (5), (17), and (18) in Fig.
\ref{Figure_Brown_Diagrams} is equal to diagram $(a)$ in Fig.
\ref{Figure_Zero_Diagrams2}. Although this diagram is absent in
Fig. \ref{Figure_Green_Diagrams}, it is equal to diagram (4) in
this figure (see the first string in Fig.
\ref{Figure_Zero_Diagrams2}). In order to see this, we note that
the left part of this diagram is proportional to

\begin{equation}
\Big(\frac{\partial}{\partial \ln g} - 1\Big)
\frac{\delta^2\gamma}{\delta (\phi_{0i})_x \delta (\phi^{*j})_y} -
\frac{\bar D_y^2}{8} \delta^8_{xy} \sim D_x^2 \delta^8_{xy} \qquad
\mbox{or}\qquad \Big(\frac{\partial}{\partial \ln g}-1\Big)
\frac{\delta^2\gamma}{\delta (\phi_{0i})_x \delta (\phi_j)_y} \sim
D_x^2 \bar D_x^2 \delta^8_{xy}.
\end{equation}

\vspace*{5mm}

\noindent In both cases there is the projector $D_x^2$ acting on
the remaining part of the green effective line:

\begin{equation}\label{Projector_To_Green}
D^2 \Big(\theta^a \theta_a \bar\theta^{\dot b} \frac{\bar D_{\dot
b} D^2}{4\partial^2} - i \bar\theta^{\dot a} (\gamma^\mu)_{\dot
a}{}^b \theta_b \frac{\bar D^2 D^2
\partial_\mu}{8\partial^4}\Big) = 0.
\end{equation}

\noindent Therefore, the part of the green line containing
$\delta/\delta \phi_0$ vanishes. The remaining part of the green
line is denoted by the green line with a cross. Thus, we prove the
identity presented in the first string of Fig.
\ref{Figure_Zero_Diagrams2}. Using this identity we see that the
sum of the considered diagrams gives diagram (4) in Fig.
\ref{Figure_Green_Diagrams}.

7. The sum of diagrams (11) and (12) is equal to diagram $(b)$ in
Fig. \ref{Figure_Zero_Diagrams2}, where the line with two green
disks corresponds to the operator

\begin{eqnarray}
&& - i (T)^j{}_i (T)^l{}_k \int
d^8x\,d^8y\,\Big(-2i\bar\theta^{\dot a} (\gamma^\mu)_{\dot a}{}^b
\theta_b \frac{\partial^\mu}{\partial^2} + \theta^a \theta_a
\bar\theta^{\dot b}
\frac{\bar D_{\dot b} D^2}{4\partial^2}\Big)_x\nonumber\\
&&\qquad\qquad \times \Big(-2i\bar\theta^{\dot c}
(\gamma^\nu)_{\dot c}{}^d \theta_d\frac{\partial_\nu}{\partial^2}
+ \theta^c \theta_c \bar\theta^{\dot d} \frac{\bar D_{\dot d}
D^2}{4\partial^2}\Big)_y \Big(\frac{\delta^2\gamma}{\delta
(\phi_l)_y \delta (\phi_j)_x} \Big)^{-1} \frac{\delta}{\delta
(\phi_{0k})_y} \frac{\delta}{\delta (\phi_{0i})_x}.\qquad\quad
\end{eqnarray}

\noindent Using the above arguments it is easy to prove that this
diagram vanishes. Really, its left side is proportional to

\begin{equation}
\Big( \frac{\partial}{\partial \ln g} - 1\Big)
\frac{\delta^2\gamma}{\delta (\phi_{0k})_y \delta (\phi_{0i})_x}.
\end{equation}

\noindent This Green function can contain parts proportional to
$\delta^8_{xy}$, $D^2 \delta^8_{xy}$, $\bar D^2 \delta^8_{xy}$,
$(D^2 \bar D^2)_x \delta^8_{xy}$, and $(\bar D^2 D^2)_x
\delta^8_{xy}$. However, it is easy to see that all these
structures give 0 if they act on the product

\begin{equation}
\Big(\theta^a \theta_a \bar\theta^{\dot b} \frac{\bar D_{\dot b}
D^2}{4\partial^2} - i \bar\theta^{\dot a} (\gamma^\mu)_{\dot
a}{}^b \theta_b \frac{\bar D^2 D^2
\partial_\mu}{8\partial^4}\Big)_x \Big(\theta^c \theta_c
\bar\theta^{\dot d} \frac{\bar D_{\dot d} D^2}{4\partial^2} - i
\bar\theta^{\dot c} (\gamma^\nu)_{\dot c}{}^d \theta_d \frac{\bar
D^2 D^2
\partial_\nu}{8\partial^4}\Big)_y.
\end{equation}

\noindent (It is necessary to take into account the integral over
$d^8x\,d^8y$ and note that terms which do not contain $\theta^4$
vanish.) Therefore, the sum of diagrams (11) and (12) vanishes.

8. The sum of diagrams (9), (22), and (23) gives diagram $(c)$ in
Fig. \ref{Figure_Zero_Diagrams2}. Using the identity
(\ref{Projector_To_Green}) we obtain that the considered sum of
diagrams is equal to diagram (2) in Fig.
\ref{Figure_Green_Diagrams}. This equality is presented in Fig.
\ref{Figure_Zero_Diagrams2}.

9. Let us consider a sum of diagrams (4), (6), (16), and (19) in
Fig. \ref{Figure_Brown_Diagrams}. First, it is necessary to note
that diagrams (4) and (6) are equal (see the third string in Fig.
\ref{Figure_Zero_Diagrams2}). In order to see this, we consider
terms which do not contain the masses in the left brown line. In
these terms the right vertex with the right brown line has the
following structure

\begin{equation}
(D^2)_y \Big[(T)y_\mu^*,\, \frac{\delta^2\gamma}{\delta
(\phi_{0i})_x \delta (\phi_0^{*j})_y}\Big] \sim D_x^2
\delta^8_{xy} \qquad \mbox{or}\qquad (\bar D^2)_y
\Big[(T)y_\mu^*,\, \frac{\delta^2\gamma}{\delta (\phi_{0i})_x
\delta (\phi_{0j})_y}\Big] \sim D_x^2 \bar D_x^2 \delta^8_{xy}.
\end{equation}

\noindent In order to verify the last equality we note that from
dimensional arguments and the Feynman rules

\begin{equation}\label{Aux_Two_Point}
\frac{\delta^2\gamma}{\delta (\phi_{0i})_x \delta (\phi_{0j})_y} =
f_1(\partial^2) D^2 \delta^8_{xy} + f_2(\partial^2) \bar D^2
\delta^8_{xy}.
\end{equation}

\noindent Therefore, this part of the diagram contains the
projector $D_x^2$ acting on the left brown line. Taking into
account that

\begin{equation}
D^2 \Big(i\frac{\bar D^2 D^2\partial_\mu}{8\partial^4} -
(\gamma_\mu)^{a{\dot b}} \theta_a \frac{\bar D_{\dot b}
D^2}{4\partial^2}\Big) = 0,
\end{equation}

\noindent we see that the part of the left brown line containing
$\delta/\delta\phi_0$ vanishes. The remaining part of the brown
line is equal to the brown line with a cross.

Moreover, diagram (19) is equal to diagram $(e)$ in Fig.
\ref{Figure_Zero_Diagrams2} multiplied by $-1/2$. In order to
prove this identity we again consider terms which do not contain
masses in the red effective line of diagram $(e)$. Then the right
vertex with the blue effective line is proportional to

\begin{equation}
(D^2)_y \Big[(T)\bar\theta^{\dot b},\,
\frac{\delta^2\gamma}{\delta (\phi_{0i})_x \delta
(\phi_0^{*j})_y}\Big] \sim \Big[\,\bar\theta^{\dot b},\,D_x^2
\delta^8_{xy}\Big] = 0 \qquad \mbox{or}\qquad (\bar D^2)_y
\Big[(T)\bar\theta^{\dot b},\, \frac{\delta^2\gamma}{\delta
(\phi_{0i})_x \delta (\phi_{0j})_y}\Big] = 0.
\end{equation}

\noindent (The last equality can be also verified using Eq.
(\ref{Aux_Two_Point}).) Therefore, the part of the red line
containing $\delta/\delta\phi_0$ vanishes. The remaining part of
the red line is equal to the red line with a cross.

Taking into account identities presented in the third and forth
lines of Fig. \ref{Figure_Zero_Diagrams2}, we obtain the sum of
effective diagrams to which we can apply the identity
(\ref{Red_Identity}). The result is given by diagram (3) in Fig.
\ref{Figure_Green_Diagrams}.

10. The sum of diagrams (8), (10), (21), and (24) is investigated
similarly to the previous group of diagrams. For this purpose it
is necessary to take into account identities presented in the last
two strings of Fig. \ref{Figure_Zero_Diagrams2}, which can be
proved exactly as in the previous case. As usually, the red
effective line in diagram $(g)$ corresponds to the operator
$\mbox{RedWhiteLine}_{\dot a}[\theta^4]$, and the red line with a
cross in diagram (24) corresponds to the operator

\begin{eqnarray}
&& \mbox{RedWhiteWithCross}_{\dot a}[\theta^4; 1,2] \equiv
-\frac{1}{2} \cdot \mbox{RedWithCross}_{\dot a}[1,3]\cdot
\mbox{UsualLine}[2,3]
\nonumber\\
&&\qquad\qquad\qquad\qquad\quad \times \int d^8x\,(\theta^4)_x
\Big(\phi_0^{*i} \phi_{0i} + M^{ij} \phi_{0i}
\frac{D^2}{8\partial^2} \phi_{0j} + M^*_{ij} \phi_0^{*i}
\frac{\bar D^2}{8\partial^2} \phi_0^{*j}\Big)_{[3]}.\qquad
\end{eqnarray}

\noindent Using these identities we see that the sum of the
considered diagrams is equal to diagram (1) in Fig.
\ref{Figure_Green_Diagrams}.

11. It is easy to see that terms proportional to $(T M^*)$
explicitly written in Fig. \ref{Figure_Brown_Diagrams} cancel each
other. For this purpose it is necessary to use the algebraic
identity

\begin{equation}\label{Simple_Algebraic_Identity}
\theta^c \theta_c A \theta_a = - (-1)^{P(A)} \theta_a A \theta^c
\theta_c + O(\theta)
\end{equation}

\noindent and its consequence, Eq. (\ref{RedLine_Equality}), which
is proved in Appendix \ref{Appendix_Identity1}.

12. The term proportional to $(M T)$ and containing the operator
$\mbox{BrownLine}_\mu$ explicitly written in Fig.
\ref{Figure_Brown_Diagrams} can be presented in the form

\begin{equation}
4\cdot \mbox{GreenLine}\cdot \mbox{GreenWithCross}\cdot \gamma
\end{equation}

\noindent using the identity (\ref{Simple_Algebraic_Identity}).
This expression coincides with diagram (9) in Fig.
\ref{Figure_Green_Diagrams}.

Collecting the results we see that the sum of diagrams presented
in Fig. \ref{Figure_Brown_Diagrams} is equal to the sum of
diagrams presented in Fig. \ref{Figure_Green_Diagrams}. This
completes the proof of the identity
(\ref{Double_Total_Derivative}).

\subsection{Derivation of the NSVZ $\beta$-function}
\hspace{\parindent}

Diagrams presented in Fig. \ref{Figure_Brown_Diagrams} (the sum of
which is equal to the sum of diagrams presented in Fig.
\ref{Figure_Green_Diagrams}) are obtained after calculating
commutators with $y_\mu^*$. However,

\begin{equation}
\mbox{Tr}[y_\mu^*, A] = 0.
\end{equation}

\noindent As a consequence, the sum of diagrams presented in Fig.
\ref{Figure_Green_Diagrams} is equal to the sum of terms
containing the $\delta$-singularities in Eq.
(\ref{Double_Total_Derivative}) with an opposite sign. Now, let us
calculate these singular contributions starting from Eq.
(\ref{Single_Total_Derivative}), which can be written in the form

\begin{eqnarray}
&&\hspace*{-7mm} -\frac{i}{4}\, \frac{d}{d\ln\Lambda}
\mbox{Tr}\,(\theta^4)_x \Big[(T) y_\mu^*,\,\mbox{BrownLine}^\mu
\cdot
\Big(\frac{\mbox{\boldmath$\delta$}^2\gamma}{\mbox{\boldmath$\delta$}
j^*_i \mbox{\boldmath$\delta$} j^i}  + M^{ik}
\Big(\frac{D^2}{8\partial^2}
\frac{\mbox{\boldmath$\delta$}}{\mbox{\boldmath$\delta$} j^i}\Big)
\frac{\mbox{\boldmath$\delta$}\gamma}{\mbox{\boldmath$\delta$}
j^k} + M^*_{ik} \Big(\frac{\bar D^2}{8\partial^2}
\frac{\mbox{\boldmath$\delta$}}{\mbox{\boldmath$\delta$}
j^*_i}\Big)
\frac{\mbox{\boldmath$\delta$}\gamma}{\mbox{\boldmath$\delta$}
j^*_k}\nonumber\\
&&\hspace*{-7mm} +i\, \phi^{*i}_0 \phi_{0i}\Big)  + (M
T)^{ki}\Bigg\{\Big(\frac{\bar D^2 D^2
\partial^\mu}{8\partial^4}\Big)_y
\Big(\frac{D^2}{8\partial^2}\Big)_x
\Big(\frac{\delta^2\gamma}{\delta(\phi_i)_y \delta
(\phi_k)_x}\Big)^{-1}_{y=x} + \int d^8y\,\Bigg\{M^{mn}
\Big(\frac{D^2\partial^\mu}{8\partial^4}\Big)_y\nonumber\\
&&\hspace*{-7mm} \times \Big(\frac{\delta^2\gamma}{\delta
(\phi_i)_y \delta (\phi_m)_x}\Big)^{-1}
\Big(\frac{D^2}{8\partial^2}\Big)_x
\Big(\frac{\delta^2\gamma}{\delta (\phi_k)_y \delta
(\phi_n)_x}\Big)^{-1} + M^*_{mn}
\Big(\frac{D^2\partial^\mu}{8\partial^4}\Big)_y
\Big(\frac{\delta^2\gamma}{\delta (\phi_i)_y \delta (\phi^{*m})_x
}\Big)^{-1}  \Big(\frac{\bar D^2}{8\partial^2}\Big)_x \nonumber\\
&&\hspace*{-7mm} \Big(\frac{\delta^2\gamma}{\delta (\phi_k)_y
\delta (\phi^{*n})_x}\Big)^{-1} \Bigg\}\Bigg\} -(M T)^{ik}
\Big(\frac{D^2\partial^\mu}{4\partial^4}\Big)_x
\Big(\frac{\delta^2\gamma}{\delta (\phi_k)_x
\delta(\phi_i)_y}\Big)^{-1}_{y=x} - (T M)^*_{ik}\Big(\frac{\bar
D^2\partial^\mu}{4\partial^4}\Big)_x \nonumber\\
&&\hspace*{-7mm} \times \Big(\frac{\delta^2\gamma}{\delta
(\phi^{*k})_x \delta(\phi^{*i})_y}\Big)^{-1}_{y=x}\Big]
-\mbox{singularities}.
\end{eqnarray}

\noindent The singular part of this expression is calculated using
Eq. (\ref{Delta_Commutator}). The result with the opposite sign
(which is equal to the sum of the considered diagrams) is given by

\begin{eqnarray}\label{Singularities}
&&\hspace*{-8mm} \frac{\pi^2}{8} C(R)_i{}^j \frac{d}{d\ln\Lambda}
\int d^8x\,\delta^4(\partial_\alpha)_x
\Bigg\{\Bigg(\frac{\delta}{\delta (\phi_{0i})_z} \Big( \bar D^2
D^2 \frac{\mbox{\boldmath$\delta$}}{\mbox{\boldmath$\delta$}
(j^j)_x}\Big) - \frac{1}{2} M^{ki} \Big(D^2
\frac{\mbox{\boldmath$\delta$}}{ \mbox{\boldmath$\delta$}
(j^k)_x}\Big)
\frac{\mbox{\boldmath$\delta$}}{\mbox{\boldmath$\delta$}
(j^j)_z} \Bigg)\nonumber\\
&&\hspace*{-8mm} \times \Bigg( \mbox{LineWithDot}[\theta^4]\cdot
\gamma - \frac{1}{2}\int d^8y\,(\theta^4 \phi_0^{*k} \phi_{0k})_y
\Bigg) - (\theta^4)_x M^*_{kj}(\bar D^2)_x
\Big(\frac{\delta^2\gamma}{\delta (\phi^{*i})_x
\delta(\phi^{*k})_z}\Big)^{-1}\nonumber\\
&&\hspace*{-8mm} - 2 (\theta^4)_x M^{ki}(D^2)_x
\Big(\frac{\delta^2\gamma}{\delta (\phi_j)_x
\delta(\phi_k)_z}\Big)^{-1} + M^{ki} (D^2)_x  \int d^8y\,
(\theta^4)_y \Bigg( M^{mn} \Big(\frac{\delta^2\gamma}{\delta
(\phi_j)_x
\delta (\phi_m)_y }\Big)^{-1} \nonumber\\
&&\hspace*{-8mm} \times \Big(\frac{D^2}{16\partial^2}\Big)_y
\Big(\frac{\delta^2\gamma}{\delta (\phi_k)_z \delta (\phi_n)_y
}\Big)^{-1} + M^*_{mn}\, \Big(\frac{\delta^2\gamma}{\delta
(\phi_j)_x \delta (\phi^{*m})_y}\Big)^{-1} \Big(\frac{\bar
D^2}{16\partial^2}\Big)_y  \Big(\frac{\delta^2\gamma}{\delta
(\phi_k)_z \delta (\phi^{*n})_y} \Big)^{-1}
\Bigg)\Bigg\}_{z=x}.\nonumber\\
\end{eqnarray}

\noindent This expression contains explicit dependence only on
$\theta^4$. Commutation of $\theta^4$ produces terms of the third
and lower degrees of $\theta$. All these terms vanish after
integrating over the anticommuting variables. Therefore, it is
possible to shift $\theta^4$ to an arbitrary point of the diagram.
Let us shift $\theta^4$ to the point $x$:

\begin{equation}
\mbox{LineWithDot}[\theta^4] \to (\theta^4)_x \cdot
\mbox{LineWithDot}[1].
\end{equation}

\noindent Next, we use the equalities

\begin{eqnarray}
&& \hspace*{-7mm} \frac{\delta^2}{\delta (\phi_{0i})_x \delta
(\phi_{0j})_y} \Big(\mbox{LineWithDot}[1]\cdot \gamma -
\frac{1}{2}\int d^8x\,\phi_0^{*k} \phi_{0k} \Big) = 2\Big(
\frac{\partial}{\partial \ln g} - 1\Big)
\frac{\delta^2\gamma}{\delta (\phi_{0i})_x
\delta (\phi_{0j})_y};\nonumber\\
&& \hspace*{-7mm} \frac{\delta^2}{\delta (\phi_{0i})_x \delta
(\phi_{0}^{*j})_y} \Big(\mbox{LineWithDot}[1]\cdot \gamma -
\frac{1}{2}\int d^8x\,\phi_0^{*k} \phi_{0k} \Big) = 2\Big(
\frac{\partial}{\partial \ln g} - 1\Big)
\frac{\delta^2\gamma}{\delta (\phi_{0i})_x \delta
(\phi_{0}^{*j})_y},\qquad
\end{eqnarray}

\noindent which are proved in Appendix
\ref{Appendix_G_Derivative}. It is easy to see that all terms
which do not contain the derivatives with respect to $\ln g$
vanish. Really, according to Eq. (\ref{Explicit_Green_Functions})

\begin{eqnarray}
(-2) \Big(\bar D^2 D^2
\frac{\mbox{\boldmath$\delta$}}{\mbox{\boldmath$\delta$}
(j^j)_x}\Big) \frac{\delta\gamma}{\delta (\phi_{0i})_y} = 2 (\bar
D^2 D^2)_x \delta^8_{xy} \delta_j^i +  2 M^{ik} (D^2)_y
\Big(\frac{\delta^2\gamma}{\delta (\phi_j)_x \delta
(\phi_k)_y}\Big)^{-1}.
\end{eqnarray}

\noindent (The first term in this expression vanishes after
differentiation with respect to $\ln\Lambda$.) Similarly,

\begin{eqnarray}
&& M^{ki} \Big(D^2 \frac{\mbox{\boldmath$\delta$}}{
\mbox{\boldmath$\delta$} (j^k)_x}\Big)
\frac{\mbox{\boldmath$\delta$}\gamma}{\mbox{\boldmath$\delta$}
(j^j)_x}  = M^{ki}\Bigg\{(D^2)_x \Big(\frac{\delta^2\gamma}{\delta
(\phi_j)_y \delta(\phi_k)_x}\Big)^{-1}_{y=x} - \int
d^8y\Bigg(M^{nm} (D^2)_x
\nonumber\\
&& \times  \Big(\frac{\delta^2\gamma}{\delta (\phi_k)_x \delta
(\phi_m)_y}\Big)^{-1} \Big(\frac{D^2}{16\partial^2}\Big)_y
\Big(\frac{\delta^2\gamma}{\delta (\phi_j)_x \delta
(\phi_n)_y}\Big)^{-1} + M^*_{nm} (D^2)_x
\Big(\frac{\delta^2\gamma}{\delta (\phi_k)_x \delta
(\phi^{*m})_y}\Big)^{-1} \nonumber\\
&& \times \Big(\frac{\bar D^2}{16\partial^2}\Big)_y
\Big(\frac{\delta^2\gamma}{\delta (\phi_j)_x \delta
(\phi^{*n})_y}\Big)^{-1} \Bigg)\Bigg\}.
\end{eqnarray}

\noindent Collecting all terms, the considered singular
contribution can be presented as

\begin{equation}
\frac{\pi^2}{4} C(R)_i{}^j \frac{d}{d\ln\Lambda} \int
d^8x\,\theta^4 \delta^4(\partial_\alpha)_x \Bigg(
\frac{\delta}{\delta (\phi_{0i})_y} \bar D^2 D^2
\frac{\mbox{\boldmath$\delta$}}{\mbox{\boldmath$\delta$} (j^j)_x}
- \frac{1}{2} M^{ki} (D^2)_x \frac{\mbox{\boldmath$\delta$}}{
\mbox{\boldmath$\delta$} (j^k)_x}
\frac{\mbox{\boldmath$\delta$}}{\mbox{\boldmath$\delta$} (j^j)_y}
\Bigg) \frac{\partial \gamma}{\partial \ln g}.
\end{equation}

\noindent Using the equation for the derivative of the inverse
matrix it is possible to rewrite this expression in a simpler form

\begin{eqnarray}
&& \frac{\pi^2}{4} C(R)_i{}^j \frac{d}{d\ln\Lambda} \int
d^8x\,\theta^4 \delta^4(\partial_\alpha)_x \Bigg(
\frac{\delta}{\delta (\phi_{0i})_y} \bar D^2 D^2
\frac{\mbox{\boldmath$\delta$}}{\mbox{\boldmath$\delta$} (j^j)_x}
\frac{\partial\gamma}{\partial\ln
g}\nonumber\\
&& \qquad\qquad\qquad\qquad\qquad\qquad\qquad\qquad\quad +
\frac{1}{2} M^{ki} (D^2)_x \frac{\partial}{\partial\ln g}
\Big(\frac{\delta^2\gamma}{\delta (\phi_j)_y
\delta(\phi_k)_x}\Big)^{-1}\Bigg)_{y=x}.\qquad\qquad
\end{eqnarray}

\noindent Then we express the derivatives with respect to sources
in terms of the derivatives with respect to fields and substitute
explicit expressions for the Green functions, for example,

\begin{equation}
\frac{\delta^2\gamma}{\delta (\phi_{i})_y \delta (\phi_0^{*j})_x}
= - \frac{1}{8} G_j{}^i \bar D_x^2 \delta^8_{xy};\qquad
\frac{\delta^2\gamma}{\delta (\phi_{i})_y \delta (\phi_{0j})_x} =
- \frac{1}{32 \partial^2} \Big((MJ)^{ji} - M^{ji}\Big) D_x^2 \bar
D_x^2 \delta^8_{xy}.
\end{equation}

\noindent After calculating the integrals over $d^4\theta$ in the
Euclidean space the result can be presented in the following form:

\begin{eqnarray}
&& \frac{\partial}{\partial \ln g} \frac{d}{d\ln\Lambda}
\Big(\frac{1}{2} \int d^8x\,d^8y\, (\theta^4)_x (\theta^4)_y
\frac{\delta^2\Delta\Gamma}{\delta \mbox{\boldmath$V$}_x \delta
\mbox{\boldmath$V$}_y}\Big) = - \frac{1}{4\pi^2} {\cal V}_4 N_f
\frac{\partial}{\partial\ln g}
\nonumber\\
&& \times \frac{d}{d\ln\Lambda} \Bigg(2\ln G + \sum\limits_{I=1}^n
(-1)^{P_I} \Big(\ln(q^2 G^2 + M^2 J^2) + \frac{M^2 J}{(q^2 G^2 +
M^2 J^2)} \Big)_I\Bigg)_{q=0}.\qquad
\end{eqnarray}

\noindent This result is very similar to Eq.
(\ref{NSVZ_Contribution}), which was found earlier by a different
method. The only difference is the presence of the derivative with
respect to $\ln g$. As earlier, we obtain

\begin{equation}
\frac{\partial}{\partial \ln g} \frac{d}{d\ln\Lambda}
\Big(\frac{1}{2} \int d^8x\,d^8y\, (\theta^4)_x (\theta^4)_y
\frac{\delta^2\Delta\Gamma}{\delta \mbox{\boldmath$V$}_x \delta
\mbox{\boldmath$V$}_y}\Big)= - \frac{1}{2\pi^2} {\cal V}_4  N_f
\frac{\partial}{\partial\ln g} \frac{d}{d\ln\Lambda} \ln G.
\end{equation}

\noindent Let us integrate this equation over $\ln g$ from $g=0$
to $g=1$. The considered theory coincides with ${\cal N}=1$ SQED
with $N_f$ flavors for $g=1$. Therefore, at the upper limit

\begin{eqnarray}
&& \frac{d}{d\ln\Lambda} \Big(\frac{1}{2} \int d^8x\,d^8y\,
(\theta^4)_x (\theta^4)_y \frac{\delta^2\Delta\Gamma}{\delta
\mbox{\boldmath$V$}_x \delta \mbox{\boldmath$V$}_y}\Big)_{g=1} =
\frac{1}{2\pi} {\cal V}_4 \cdot
\frac{\beta(\alpha_0)}{\alpha_0^2};\nonumber\\
&& \frac{d}{d\ln\Lambda} \ln G \Big|_{g=1} = \gamma(\alpha_0).
\end{eqnarray}

\noindent For $g=0$ the considered theory does not contain the
{\it quantum} gauge field. Therefore, for $g=0$ only one-loop
diagrams contribute to the two-point Green function of the gauge
superfield, and

\begin{eqnarray}
&& \frac{d}{d\ln\Lambda} \Big(\frac{1}{2} \int d^8x\,d^8y\,
(\theta^4)_x (\theta^4)_y \frac{\delta^2\Delta\Gamma}{\delta
\mbox{\boldmath$V$}_x \delta \mbox{\boldmath$V$}_y}\Big)_{g=0} =
\frac{1}{2\pi} {\cal V}_4 \cdot \frac{\beta_{\mbox{\scriptsize
1-loop}}(\alpha_0)}{\alpha_0^2};\nonumber\\
&& \frac{d}{d\ln\Lambda} \ln G \Big|_{g=0} = 0,
\end{eqnarray}

\noindent where

\begin{equation}
\beta_{\mbox{\scriptsize 1-loop}} = \frac{\alpha_0^2}{\pi} N_f.
\end{equation}

\noindent Thus, after the integration we obtain the NSVZ relation

\begin{equation}
\beta(\alpha_0) = \frac{\alpha_0^2 N_f}{\pi} (1-\gamma(\alpha_0))
\end{equation}

\noindent for the renormalization group functions defined in terms
of the bare coupling constant.

\section{Conclusion}
\hspace{\parindent}\label{Section_Conclusion}

In this paper we present a derivation of the NSVZ relation for
${\cal N}=1$ SQED with $N_f$ flavors, regularized by higher
derivatives, using a method based on the effective diagram
technique and the Schwinger--Dyson equations. We prove that with
this regularization the exact NSVZ $\beta$-function relates the
renormalization group functions defined in terms of the bare
coupling constant. (If the renormalization group functions are
defined in terms of the renormalized coupling constant, the NSVZ
scheme can be easily constructed by imposing the simple boundary
conditions (\ref{NSVZ_Scheme}) on the renormalization constants
\cite{Kataev:2013eta,Kataev:2013csa}.) The technique based on the
Schwinger--Dyson equations seems to be more convenient for
generalization of the results to the non-Abelian case than another
method discussed in \cite{Stepanyantz:2011jy}.

The method considered in this paper allows to easily calculate a
contribution to the $\beta$-function proportional to the anomalous
dimension of the matter superfields. For this purpose expressions
for the effective vertices are found by solving the Ward (or
Slavnov--Taylor) identities \cite{Stepanyantz:2004sg}. However, in
order to prove that the other contributions vanish for the
considered theory, it is necessary to essentially modify the
method. First, a $\beta$-function should be written in terms of
two-loop effective diagrams. Moreover, it is necessary to
introduce an auxiliary parameter $g$ and perform a differentiation
with respect to $\ln g$. The derivative of the two-point function
of the gauge superfield with respect to $\ln g$ can be presented
as a sum of three-loop effective diagrams. After these
modifications it is possible to find the remaining contribution to
the $\beta$-function defined in terms of the bare coupling
constant. In this paper we obtain that this contribution vanishes.
Moreover, we prove that the $\beta$-function is given by integrals
of double total derivatives in agreement with the results of
\cite{Smilga:2004zr,Stepanyantz:2011jy}. Such a structure allows
to calculate one of the loop integrals and obtain the NSVZ
$\beta$-function in all orders. The origin of the exact NSVZ
$\beta$-function can be easily explained, because taking one of
loop integrals we relate the $\beta$-function in a certain order
with the anomalous dimension in the previous order.

The results obtained in this paper can be verified by explicit
calculations in the lowest loops. The three-loop calculation will
be described in the forthcoming paper.

\bigskip
\bigskip

\noindent {\Large\bf Acknowledgements.}

\bigskip

\noindent The author is very grateful to A.L.Kataev for valuable
discussions. The work was supported by RFBR grant No 14-01-00695.

\appendix

\section{The Schwinger--Dyson equation for the two-point Green
function of the gauge superfield} \hspace{\parindent}
\label{Appendix_SD_Equation}

We are interested in the expression

\begin{equation}\label{We_Calculate_Appendix}
\frac{1}{2} \int d^8x\,d^8y\, \mbox{\boldmath$V$}_x
\mbox{\boldmath$V$}_y \frac{\delta^2\Delta\Gamma}{\delta
\mbox{\boldmath$V$}_x \delta \mbox{\boldmath$V$}_y},
\end{equation}

\noindent where the effective action $\Gamma$ is given by Eq.
(\ref{Gamma}) and all fields are set to 0. In order to calculate
this expression we differentiate the Schwinger--Dyson equation
(\ref{SD_Equation}) with respect to $\mbox{\boldmath$V$}_y$ and
set all fields to 0. Then the result is multiplied by
$\mbox{\boldmath$V$}_x \mbox{\boldmath$V$}_y/2$. After integrating
over $d^8x\,d^8y$ we rewrite Eq. (\ref{We_Calculate_Appendix}) as

\begin{equation}\label{SD_Middle}
\frac{1}{i} (T)^j{}_i \int d^8x\, d^8y\, \mbox{\boldmath$V$}_x
\mbox{\boldmath$V$}_y \frac{\delta}{\delta \mbox{\boldmath$V$}_y}
\Big(\frac{\delta}{\delta (j^j)_x} \frac{\delta\Gamma}{\delta
(\phi_{0i})_x}\Big).
\end{equation}

\noindent In order to present this expression as a sum of
effective diagrams, it is convenient to commute $\delta/\delta
\mbox{\boldmath$V$}$ and $\delta/\delta j$. Differentiating
inverse Green functions inside the derivatives $\delta/\delta j^j$
and simplifying the result using Eq. (\ref{Phi0_SD_Equation}), it
is easy to see that

\begin{eqnarray}
&& \Big[\frac{\delta}{\delta
\mbox{\boldmath$V$}_y},\frac{\delta}{\delta (j^j)_x}\Big] = \int
d^8z\,\Bigg( \frac{\delta}{\delta (j^j)_x}
\frac{\delta^2\Gamma}{\delta \mbox{\boldmath$V$}_y \delta
(\phi_{0k})_z} \cdot
\frac{\delta}{\delta (j^k)_z}\nonumber\\
&&\qquad\qquad\qquad + \frac{\delta}{\delta (j^j)_x}
\frac{\delta^2\Gamma}{\delta \mbox{\boldmath$V$}_y \delta
(\phi^{*k}_{0})_z} \cdot \frac{\delta}{\delta (j^*_{k})_z} +
\frac{\delta}{\delta (j^j)_x }\frac{\delta^2\Gamma}{\delta
\mbox{\boldmath$V$}_y \delta V_z}\cdot \frac{\delta}{\delta
J_z}\Bigg),\qquad
\end{eqnarray}

\noindent where the derivatives with respect to sources should be
expressed in terms of the derivatives with respect to fields.
Really, for example,

\begin{eqnarray}
&& \int d^8z\,\frac{\delta^2\Gamma}{\delta \mbox{\boldmath$V$}_y
\delta (\phi_{0k})_z} \cdot \frac{\delta}{\delta (j^k)_z} = \int
d^8z\,\frac{\delta}{\delta \mbox{\boldmath$V$}_y}
\Big(-\frac{1}{2}\bar D_z^2\Big) \frac{\delta\Gamma}{\delta
(\phi_{0k})_z}\cdot \frac{D^2}{8\partial^2} \frac{\delta}{\delta
(j^k)_z} \qquad
\nonumber\\
&& = \int d^8z\,\frac{\delta^2\Gamma}{\delta \mbox{\boldmath$V$}_y
\delta (\phi_{k})_z}\cdot \frac{D^2}{8\partial^2}
\frac{\delta}{\delta (j^k)_z}
\end{eqnarray}

\noindent due to the chirality of the derivative with respect to
the source $j$. All vertices containing odd degrees of the matter
superfields vanish after setting the fields to 0. Two-point Green
functions of the matter superfields constructed from the
functionals $\Gamma$ and $\gamma$ evidently coincide (again, after
setting the fields to 0). Using these facts Eq. (\ref{SD_Middle})
can be written in the form

\begin{eqnarray}\label{Original_Expression}
&& \frac{1}{2} \int d^8x\,d^8y\, \mbox{\boldmath$V$}_x
\mbox{\boldmath$V$}_y \frac{\delta^2\Delta\Gamma}{\delta
\mbox{\boldmath$V$}_y \delta \mbox{\boldmath$V$}_x} = -i (T)^j{}_i
\int d^8x\,d^8y\,d^8z\,\mbox{\boldmath$V$}_x \mbox{\boldmath$V$}_y
\Bigg( \frac{\mbox{\boldmath$\delta$}}{ \mbox{\boldmath$\delta$}
(j^j)_x} \frac{\delta^2\Gamma}{\delta \mbox{\boldmath$V$}_y
\delta(\phi_{0k})_z}\quad\nonumber\\
&& \times \Big[\delta^8_{xz} \delta^i_k +
\frac{\mbox{\boldmath$\delta$}}{\mbox{\boldmath$\delta$} (j^k)_z}
\frac{\delta\gamma}{\delta (\phi_{0i})_x}  \Big] +
\frac{\mbox{\boldmath$\delta$}}{\mbox{\boldmath$\delta$} (j^j)_x}
\frac{\delta^2\Gamma}{\delta \mbox{\boldmath$V$}_y
\delta(\phi^{*k}_{0})_z}\cdot
\frac{\mbox{\boldmath$\delta$}}{\mbox{\boldmath$\delta$}
(j_{k}^*)_z} \frac{\delta\gamma}{\delta (\phi_{0i})_x}
\Bigg),\qquad
\end{eqnarray}

\noindent where the derivative $\mbox{\boldmath$\delta$}/
\mbox{\boldmath$\delta$} j^i$ is defined by Eq.
(\ref{Bold_Derivative}). It differs from the derivative
$\delta/\delta j^i$, because all fields in the inverse two-point
Green functions are set to 0. The graphical interpretation of this
result is presented in Fig. \ref{Figure_Graphical_SD_Equation}.

\section{The identity for the effective lines}
\hspace{\parindent}\label{Appendix_Sum_Of_Lines}

In order to simplify the calculations we use the substitution
$\mbox{\boldmath$V$} \to \theta^4$. Then we find a sum of
effective lines presented in the left hand side of Fig.
\ref{Figure_Sum_Of_Lines}. In the analytical form this sum
corresponds to the expression

\begin{eqnarray}\label{GreenLine}
&& (T)^j{}_i \int d^8x\,(\theta^4)_x \Bigg\{
\frac{\mbox{\boldmath$\delta$}}{\mbox{\boldmath$\delta$} (j^j)_x}
\frac{\delta}{\delta (\phi_{0i})_x} + \int
d^8y\,\Bigg[\Big(\frac{\mbox{\boldmath$\delta$}}{\mbox{\boldmath$\delta$}
(j^k)_y} \frac{\delta\gamma}{\delta (\phi_{0i})_x}\Big) \cdot
\frac{\mbox{\boldmath$\delta$}}{\mbox{\boldmath$\delta$} (j^j)_x}
\frac{\delta}{\delta (\phi_{0k})_y}\nonumber\\
&& + \Big(\frac{\mbox{\boldmath$\delta$}}{
\mbox{\boldmath$\delta$} (j^*_k)_y}\frac{\delta\gamma}{\delta
(\phi_{0i})_x}\Big) \cdot
\frac{\mbox{\boldmath$\delta$}}{\mbox{\boldmath$\delta$} (j^j)_x}
\frac{\delta}{\delta (\phi_{0}^{*k})_y}\Bigg]\Bigg\},
\end{eqnarray}

\noindent where the fields in the two-point functions are set to
0. Note that in our notation effective lines include derivatives
which act on the vertices attached to the line. This allows to
considerably simplify expressions for the multiloop effective
diagrams.

Using Eq. (\ref{Phi0_SD_Equation}) and arguments based on
chirality it is easy to verify that

\begin{eqnarray}\label{Explicit_Green_Functions}
&& \frac{\mbox{\boldmath$\delta$}}{\mbox{\boldmath$\delta$}
(j^k)_y} \frac{\delta\gamma}{\delta (\phi_{0i})_x} =
\Big(\frac{\bar D^2 D^2}{16\partial^2}\Big)_y \delta^8_{xy}
\delta^i_k + M^{im} \Big(\frac{D^2}{16\partial^2}\Big)_x
\Big(\frac{\delta^2\gamma}{\delta(\phi_k)_y \delta(\phi_m)_x}
\Big)^{-1};\nonumber\\
&& \frac{\mbox{\boldmath$\delta$}}{\mbox{\boldmath$\delta$}
(j^*_k)_y} \frac{\delta\gamma}{ \delta (\phi_{0i})_x} =
M^{im}\Big(\frac{D^2}{16\partial^2}\Big)_x
\Big(\frac{\delta^2\gamma}{\delta(\phi^{*k})_y
\delta(\phi_m)_x}\Big)^{-1}.
\end{eqnarray}

\noindent (The fields are set to 0.) In order to prove these
equations, it is necessary to apply to them the operator $-\bar
D_x^2/2$. Substituting the Green functions
(\ref{Explicit_Green_Functions}) into Eq. (\ref{GreenLine}) after
some simple transformations we rewrite the considered expression
as

\begin{equation}\label{GreenLine2}
\int d^8x\,\Bigg\{(T)^j{}_i \Big[\Big(1+\frac{\bar D^2
D^2}{16\partial^2}\Big) \theta^4
\frac{\mbox{\boldmath$\delta$}}{\mbox{\boldmath$\delta$} j^j}\Big]
\frac{\delta}{\delta \phi_{0i}} + \theta^4 (M T)^{ij}
\frac{\mbox{\boldmath$\delta$}}{\mbox{\boldmath$\delta$} j^i}
\,\Big(\frac{D^2}{16\partial^2}
\frac{\mbox{\boldmath$\delta$}}{\mbox{\boldmath$\delta$} j^j}\Big)
\Bigg\}.
\end{equation}

\noindent The first term in this expression can be transformed
using the identity

\begin{equation}
\Big(1+\frac{\bar D^2 D^2}{16\partial^2}\Big) \theta^4
\frac{\mbox{\boldmath$\delta$}}{\mbox{\boldmath$\delta$} j^j} =
\Big(\theta^a\theta_a \bar\theta^{\dot b} \frac{\bar D_{\dot b}
D^2}{4\partial^2} + 2i\bar\theta^{\dot a} (\gamma^\mu)_{\dot
a}{}^{b} \theta_b \frac{\partial_\mu}{\partial^2} -
\frac{D^2}{4\partial^2}\, \theta^a \theta_a\Big)
\frac{\mbox{\boldmath$\delta$}}{\mbox{\boldmath$\delta$} j^j}.
\end{equation}

\noindent The degree of $\theta$ in the last term of Eq.
(\ref{GreenLine2}) can be also decreased by the help of Eq.
(\ref{MT_Symmetry}):

\begin{eqnarray}
&& \int d^8x\, \theta^4 (M T)^{ij}
\frac{\mbox{\boldmath$\delta$}}{\mbox{\boldmath$\delta$} j^i}
\Big(\frac{D^2}{16\partial^2}
\frac{\mbox{\boldmath$\delta$}}{\mbox{\boldmath$\delta$} j^j}\Big)
= - \int d^8x\, \theta^4 (M T)^{ij} \Big(\frac{\bar D^2
D^2}{16\partial^2}
\frac{\mbox{\boldmath$\delta$}}{\mbox{\boldmath$\delta$} j^i}\Big)
\Big(\frac{D^2}{16\partial^2}
\frac{\mbox{\boldmath$\delta$}}{\mbox{\boldmath$\delta$}
j^j}\Big)\nonumber\\
&& = \int d^8x\, \theta^a \theta_a \bar\theta^{\dot b} (M T)^{ij}
\Big(\frac{\bar D_{\dot b} D^2}{8\partial^2}
\frac{\mbox{\boldmath$\delta$}}{\mbox{\boldmath$\delta$} j^i}\Big)
\Big(\frac{D^2}{16\partial^2}
\frac{\mbox{\boldmath$\delta$}}{\mbox{\boldmath$\delta$}
j^j}\Big).
\end{eqnarray}

\noindent Then integrating by parts gives

\begin{eqnarray}
&& (M T)^{ij} \int d^8x\, \Big(\Big(\theta_a \bar\theta^{\dot b}
\frac{D^a \bar D_{\dot b} D^2}{32\partial^4} - \bar\theta^{\dot b}
\frac{\bar D_{\dot b} D^2}{32\partial^4}\Big)
\frac{\mbox{\boldmath$\delta$}}{\mbox{\boldmath$\delta$} j^i}
\Big)\frac{\mbox{\boldmath$\delta$}}{ \mbox{\boldmath$\delta$}
j^j} = (M T)^{ij} \int d^8x\,\Big(\Big(- i \bar\theta^{\dot a}
(\gamma^\mu)_{\dot a}{}^b \theta_b \qquad\nonumber\\
&& \times \frac{D^2\partial_\mu}{16\partial^4} -
\frac{D^2}{16\partial^4}\Big)
\frac{\mbox{\boldmath$\delta$}}{\mbox{\boldmath$\delta$} j^i}
\Big) \frac{\mbox{\boldmath$\delta$}}{\mbox{\boldmath$\delta$}
j^j} = - i (M T)^{ij} \int d^8x\,\bar\theta^{\dot a}
(\gamma^\mu)_{\dot a}{}^b \theta_b
\Big(\frac{D^2\partial_\mu}{16\partial^4}
\frac{\mbox{\boldmath$\delta$}}{\mbox{\boldmath$\delta$} j^i}\Big)
\frac{\mbox{\boldmath$\delta$}}{ \mbox{\boldmath$\delta$}
j^j}.\qquad
\end{eqnarray}

\noindent Therefore, the expression (\ref{GreenLine}) can be
rewritten in the following form:

\begin{eqnarray}\label{CompleteGreenLine}
&& \int d^8x\,\Bigg\{ (T)^j{}_i \Big(\theta^a\theta_a
\bar\theta^{\dot b} \frac{\bar D_{\dot b} D^2}{4\partial^2} +
2i\bar\theta^{\dot a} (\gamma^\mu)_{\dot a}{}^{b} \theta_b
\frac{\partial_\mu}{\partial^2} - \frac{D^2}{4\partial^2}\,
\theta^a \theta_a \Big)
\frac{\mbox{\boldmath$\delta$}}{\mbox{\boldmath$\delta$} j^j}
\frac{\delta}{\delta \phi_{0i}}
\nonumber\\
&&\qquad\qquad\qquad\qquad\qquad\qquad\qquad\qquad\qquad - i (M
T)^{ij} \bar\theta^{\dot a} (\gamma^\mu)_{\dot a}{}^b \theta_b
\Big(\frac{D^2\partial_\mu}{16\partial^4}
\frac{\mbox{\boldmath$\delta$}}{\mbox{\boldmath$\delta$} j^i}\Big)
\frac{\mbox{\boldmath$\delta$}}{ \mbox{\boldmath$\delta$} j^j}
\Bigg\}.\qquad\quad
\end{eqnarray}

\section{The Schwinger--Dyson equation in terms of two-loop effective diagrams}
\label{Appendix_Second_SD}

\subsection{$\beta$-function in terms of two-loop effective diagrams}
\hspace{\parindent}

In order to rewrite the Schwinger--Dyson equation as a sum of
two-loop effective diagrams, let us start with Eq.
(\ref{SD_Equation_With_Green_Line}) and substitute
$\delta\Gamma/\delta \mbox{\boldmath$V$}_y$ from Eq.
(\ref{SD_Equation}) \cite{Shevtsova}:

\begin{eqnarray}\label{One-Loop}
&&\hspace*{-7mm} \frac{1}{2\pi} {\cal V}_4\cdot
\frac{\beta(\alpha_0)}{\alpha_0^2} = -i \frac{d}{d\ln\Lambda} \int
d^8x\, d^8y\, \Big[ (T)^j{}_i \Big(\theta^a\theta_a
\bar\theta^{\dot b} \frac{\bar D_{\dot b} D^2}{4\partial^2}+
2i\bar\theta^{\dot a} (\gamma^\mu)_{\dot a}{}^{b} \theta_b
\frac{\partial_\mu}{\partial^2}- \frac{D^2}{4\partial^2}\,
\theta^a \theta_a \Big)
\frac{\mbox{\boldmath$\delta$}}{\mbox{\boldmath$\delta$} j^j} \nonumber\\
&&\hspace*{-7mm}\times \frac{\delta}{\delta \phi_{0i}} - i (M
T)^{ij} \bar\theta^{\dot a} (\gamma^\mu)_{\dot a}{}^b \theta_b
\Big(\frac{D^2\partial_\mu}{16\partial^4}
\frac{\mbox{\boldmath$\delta$}}{\mbox{\boldmath$\delta$} j^i}\Big)
\frac{\mbox{\boldmath$\delta$}}{ \mbox{\boldmath$\delta$} j^j}
\Big]_{x} (\theta^4)_y  (T)^l{}_k \Big[\frac{2}{i}
\frac{\delta}{\delta j^l} \frac{\delta\Gamma}{\delta \phi_{0k}} +
2 \frac{\delta\Gamma}{\delta
\phi_{0k}} (\phi_l+\phi_{0l}) \Big]_{y}.\nonumber\\
\end{eqnarray}

\noindent (Certainly, all fields in this expression should be set
to 0 after calculation of the derivatives.) First, let us find a
contribution of the second term in the last square brackets (or,
equivalently, terms containing $\phi_l+\phi_{0l}$). We will denote
this contribution by $\Delta$. After calculating the derivatives
nontrivial terms can be written as

\begin{eqnarray}\label{FirstTerm}
&& \Delta = 2i \frac{d}{d\ln\Lambda} \int d^8x\,d^8y\,(\theta^4)_y
(T)_k{}^l \Bigg\{ (T)^j{}_l \Big(\frac{D^2}{4\partial^2}\Big)_x
(\theta^a \theta_a)_x \Big(\frac{\mbox{\boldmath$\delta$}}{
\mbox{\boldmath$\delta$} (j^j)_x} \frac{\delta\gamma}{\delta
(\phi_{0k})_y}\Big)\delta^8_{xy}
\nonumber\\
&& + (T)^j{}_i \frac{\delta^2\gamma}{\delta (\phi_{0i})_x \delta
(\phi_{0k})_y} \Big(\theta^a\theta_a \bar\theta^{\dot b}
\frac{\bar D_{\dot b} D^2}{4\partial^2} + 2i\bar\theta^{\dot a}
(\gamma^\mu)_{\dot a}{}^{b} \theta_b
\frac{\partial_\mu}{\partial^2} - \frac{D^2}{4\partial^2}\,
\theta^a \theta_a\Big)_x \Big(\frac{\delta^2\gamma}{\delta
(\phi_l)_y \delta (\phi_j)_x}\Big)^{-1}\quad\nonumber\\
&& - i (M T)^{ij} \Big(\frac{\mbox{\boldmath$\delta$}}{
\mbox{\boldmath$\delta$} (j^j)_x} \frac{\delta\gamma}{\delta
(\phi_{0k})_y}\Big) \Big(\bar\theta^{\dot a} (\gamma^\mu)_{\dot
a}{}^b \theta_b \frac{D^2
\partial_\mu}{16\partial^4}\Big)_x
\Big(\frac{\delta^2\gamma}{\delta (\phi_l)_y \delta (\phi_i)_x}
\Big)^{-1} \Bigg\}.
\end{eqnarray}

\noindent Here instead of the effective action $\Gamma$ we use the
Routhian $\gamma$, because this considerably simplifies the
calculations. In order to rewrite all equations in terms of
$\gamma$, it is necessary to take into account that

\begin{equation}
\frac{\delta\Gamma}{\delta\phi_{0i}}[V,\phi,\phi_0] =
\frac{\delta\gamma}{\delta\phi_{0i}}[J,\phi,\phi_0].
\end{equation}

\noindent Note that some terms in Eq. (\ref{FirstTerm}) vanish,
because they are proportional to $\theta$ in a more than the forth
degree. For example, all terms in the second string of this
equation vanish due to this reason. Really, taking into account
that

\begin{equation}
\Big(\frac{\delta^2\gamma}{\delta (\phi_l)_y \delta
(\phi_j)_x}\Big)^{-1} \sim \bar D^2 \delta_{xy},
\end{equation}

\noindent we can commute $\theta_b$ with this Green function. Then
the result will be proportional to $(\theta^4)_y (\theta_b)_y =
0$. In order to simplify the remaining terms we use Eq.
(\ref{Explicit_Green_Functions}). Taking into account that some
terms vanish after differentiation with respect to $\ln\Lambda$ we
obtain

\begin{eqnarray}
&&\hspace*{-5mm} \Delta = \frac{d}{d\ln\Lambda}\int
d^8x\,d^8y\,(\theta^4)_y \Bigg\{ i C(R)_k{}^j M^{km}
\Big(\frac{D^2}{2\partial^2}\Big)_x (\theta^a \theta_a)_x
\Big(\frac{D^2}{16\partial^2}\Big)_y
\Big(\frac{\delta^2\gamma}{\delta (\phi_{m})_y
\delta(\phi_j)_x}\Big)^{-1}\delta^8_{xy} \nonumber\\
&&\hspace*{-5mm} + \Big[ \frac{\bar D^2 D^2}{8\partial^2}
\delta^8_{xy} (T)_j{}^l + (M T)^{ml}
\Big(\frac{D^2}{8\partial^2}\Big)_y \Big(\frac{\delta^2\gamma}{
\delta(\phi_m)_y \delta(\phi_j)_x}\Big)^{-1}\Big] (M T)^{ij}
(\bar\theta^{\dot a} (\gamma^\mu)_{\dot a}{}^b \theta_b)_x
\Big(\frac{D^2
\partial_\mu}{16\partial^4}\Big)_x\nonumber\\
&&\hspace*{-5mm} \times  \Big(\frac{\delta^2\gamma}{\delta
(\phi_l)_y \delta (\phi_i)_x}\Big)^{-1}\Bigg\}.\qquad
\end{eqnarray}

\noindent Shifting $(\theta^a \theta_a)_x$ and $(\bar\theta^{\dot
a} (\gamma^\mu)_{\dot a}{}^b \theta_b)_x$ to the point $y$ this
expression can be rewritten as

\begin{eqnarray}\label{One-Loop_Effective_Diagram}
&& \Delta \equiv -i \frac{d}{d\ln\Lambda}\int
d^8x\,d^8y\,(\theta^4)_y \Big(\frac{D^2}{4\partial^4}\Big)_x
\Big(\frac{\delta^2\gamma}{\delta (\phi_i)_x \delta
(\phi_j)_y}\Big)^{-1}\nonumber\\
&&\qquad\qquad\qquad \times \Big\{ C(R)_k{}^i M^{j k}
\delta^8_{xy} - (M T)^{im} (M T)^{lj}
\Big(\frac{D^2}{32\partial^2}\Big)_x
\Big(\frac{\delta^2\gamma}{\delta (\phi_m)_x \delta
(\phi_l)_y}\Big)^{-1}\Big\}.\qquad\quad
\end{eqnarray}

\noindent It is important that this expression does not contain
infrared singularities due to the differentiation with respect to
$\ln\Lambda$, which should be made before calculating the momentum
integral.

Let us now consider the remaining terms in Eq. (\ref{One-Loop})
(which are obtained from the first term in the last square
brackets). In order to rewrite them in a more convenient form, we
commute the derivatives $\delta/\delta j$ and
$\delta/\delta\phi_0$ (including $\delta/\delta\phi_0$ which is
contained in $\mbox{\boldmath$\delta$}/\mbox{\boldmath$\delta$}
j^i$) taking into account that

\begin{eqnarray}
&& \Big[\frac{\delta}{\delta (\phi_{0k})_z},\frac{\delta}{\delta
(j^n)_y}\Big\} \equiv \frac{\delta}{\delta
(\phi_{0k})_z}\frac{\delta}{\delta (j^n)_y} - (-1)^{P_k P_n}
\frac{\delta}{\delta
(j^n)_y} \frac{\delta}{\delta (\phi_{0k})_z}\nonumber\\
&& = (-1)^{P_k P_n}\int d^8w\,\Big(\frac{\delta}{\delta (j^n)_y}
\frac{\delta^2\gamma}{\delta (\phi_{0k})_z \delta (\phi_{0l})_w}
\cdot \frac{\delta}{\delta (j^l)_w} + \frac{\delta}{\delta
(j^n)_y} \frac{\delta^2 \gamma}{\delta (\phi_{0k})_z \delta
(\phi^{*l}_{0})_w} \cdot \frac{\delta}{\delta (j^*_l)_w}\Big).
\qquad\quad
\end{eqnarray}

\noindent Note that so far we did not set the fields to 0. Because
the Routhian $\gamma$ is used instead of the effective action
$\Gamma$, the expression in the right hand side does not contain
derivatives with respect to the gauge superfield $V$. (This is the
main reason, why in the subsequent equations we use the functional
$\gamma$.) As a consequence, setting all fields equal to 0 we
obtain

\begin{eqnarray}
&& \Big(\hat A \frac{\delta}{\delta j^j}\Big)_y
\frac{\delta}{\delta (\phi_{0i})_y}\,(T)^l{}_k \int
d^8x\,(\theta^4)_x \frac{\delta}{\delta (j^l)_x}
\frac{\delta\Gamma}{\delta
(\phi_{0k})_x}\nonumber\\
&& = (T)^l{}_k \int d^8x\,d^8z\,(\theta^4)_x \Big(
\Big[\delta^8_{xz} \delta_m^k +
\frac{\mbox{\boldmath$\delta$}}{\mbox{\boldmath$\delta$} (j^m)_z}
\frac{\delta\gamma}{\delta (\phi_{0k})_x}\Big] \Big(\hat A
\frac{\mbox{\boldmath$\delta$}}{\mbox{\boldmath$\delta$}
j^j}\Big)_y
\frac{\mbox{\boldmath$\delta$}}{\mbox{\boldmath$\delta$} (j^l)_x}
\frac{\delta^2\gamma}{\delta (\phi_{0m})_z \delta
(\phi_{0i})_y}\qquad\nonumber\\
&& + \frac{\mbox{\boldmath$\delta$}}{\mbox{\boldmath$\delta$}
(j_m^*)_z} \frac{\delta\gamma}{\delta (\phi_{0k})_x}\cdot
\Big(\hat A
\frac{\mbox{\boldmath$\delta$}}{\mbox{\boldmath$\delta$}
j^j}\Big)_y
\frac{\mbox{\boldmath$\delta$}}{\mbox{\boldmath$\delta$} (j^l)_x}
\frac{\delta^2\gamma}{\delta (\phi_{0}^{*m})_z \delta
(\phi_{0i})_y} \Big),
\end{eqnarray}

\noindent where $\hat A$ is an operator acting on the coordinates
$y$ which does not contain $\bar\theta^{\dot a} \bar\theta_{\dot
a}$. Its explicit form in the considered case can be found from
Eq. (\ref{One-Loop}). Then it is possible to substitute explicit
expressions for the two-point Green functions and repeat all
transformations made in Appendix \ref{Appendix_Sum_Of_Lines}. It
is easy to see that the result can be written as

\begin{equation}\label{J_Phi0_J}
\Big(\hat A \frac{\delta}{\delta j^j}\Big)_y \frac{\delta}{\delta
(\phi_{0i})_y}\, (T)^l{}_k \int d^8x\,(\theta^4)_x
\frac{\delta}{\delta (j^l)_x} \frac{\delta\Gamma}{\delta
(\phi_{0k})_x} = \mbox{GreenLine}\cdot \Big(\hat A
\frac{\mbox{\boldmath$\delta$}}{\mbox{\boldmath$\delta$}
j^j}\Big)_y \frac{\delta\gamma}{\delta (\phi_{0i})_y},
\end{equation}

\noindent where we take into account that only terms proportional
to $\bar\theta^2$ nontrivially contribute to the result. (In
particular, a term without $\bar\theta$ in Eq.
(\ref{CompleteGreenLine}) gives a vanishing contribution.)
Therefore, finally we obtain

\begin{eqnarray}
&&\hspace*{-7mm} \frac{1}{2\pi} {\cal V}_4 \cdot
\frac{\beta(\alpha_0)}{\alpha_0^2} = -2 \frac{d}{d\ln\Lambda}
(\mbox{GreenLine})^2 \gamma + \Delta,
\end{eqnarray}

\noindent where $\Delta$ is given by Eq.
(\ref{One-Loop_Effective_Diagram}). The first term in this
expression can be graphically presented as a two-loop effective
diagram, while the second one corresponds to a one-loop effective
diagram.

\subsection{The identity (\ref{New_Identity_Analytical_Form}) in terms
of two-loop effective diagrams}
\hspace{\parindent}\label{Appendix_Subsection_New_Identity}

In order to prove the identity
(\ref{New_Identity_Analytical_Form}), it is necessary to present
its left hand side as a sum of two-loop effective diagrams. For
this purpose we use the Schwinger--Dyson equation
(\ref{SD_Equation}). The first term in Eq.
(\ref{New_Identity_Analytical_Form}) can be presented as a
two-loop effective diagram

\begin{equation}
-4\cdot \mbox{BlueLine}_{\dot b}[\theta^a \theta_a
\bar\theta^{\dot b}] \cdot \mbox{GreenLine}\cdot \gamma
\end{equation}

\noindent similarly to the calculation made in the previous
section. In the graphical form this diagram (containing one blue
effective line and one green effective line) is shown in Fig.
\ref{Figure_Double_Diagram}. All terms which could be interpreted
as one-loop effective diagrams in this case vanish.

However, the second term in Eq.
(\ref{New_Identity_Analytical_Form}) can be written as a one-loop
effective diagram. Let us remind that a contribution of this term
is denoted by $\delta_2$. In order to write all two-loop
contributions in the same form we also use the notation $\Delta_2
= \delta_2$. Using the Schwinger--Dyson equation
(\ref{SD_Equation}) $\delta_2$ can be written as

\begin{equation}
i (T)^j{}_i (T)^l{}_k \frac{d}{d\ln\Lambda}  \int d^8x\,
d^8y\,(\theta^4)_y  \Big( \theta^a \theta_a
\frac{D^2}{2\partial^2}
\frac{\mbox{\boldmath$\delta$}}{\mbox{\boldmath$\delta$}
j^j}\Big)_x \frac{\delta}{\delta (\phi_{0i})_x} \Big[\frac{1}{i}
\frac{\delta}{\delta j^l} \frac{\delta\Gamma}{\delta \phi_{0k}} +
\frac{\delta\Gamma}{\delta \phi_{0k}} (\phi_l+\phi_{0l})
\Big]_{y}.
\end{equation}

\noindent The first term in the square brackets vanishes, because
it is proportional to the first degree of $\bar\theta$. This
follows from Eq. (\ref{J_Phi0_J}). Calculating the derivatives
with respect to $\phi_0$ it is easy to see that the only
nontrivial term is

\begin{equation}
- i (T)_i{}^j (T)_k{}^l \frac{d}{d\ln\Lambda}  \int d^8x\,
d^8y\,(\theta^4)_y  (\theta^a \theta_a)_x
\Big(\frac{D^2}{2\partial^2}\Big)_x
\Big(\frac{\delta^2\gamma}{\delta (\phi_{j})_x
(\phi_{l})_y}\Big)^{-1} \frac{\delta^2\gamma}{\delta (\phi_{0k})_y
\delta (\phi_{0i})_x}.
\end{equation}

\noindent Commuting $\theta$-s with the covariant derivatives and
using the identity

\begin{equation}
(T)_k{}^l \Big(\frac{\delta^2\gamma}{\delta (\phi_{j})_x
(\phi_{l})_y}\Big)^{-1} = - (T)_j{}^l
\Big(\frac{\delta^2\gamma}{\delta (\phi_{l})_x
(\phi_{k})_y}\Big)^{-1}
\end{equation}

\noindent we obtain

\begin{equation}
\Delta_2 = - 2i C(R)_i{}^j \frac{d}{d\ln\Lambda}  \int d^8x\,
d^8y\,(\theta^4)_y \frac{1}{\partial^2}
\Big(\frac{\delta^2\gamma}{\delta (\phi_{j})_x
(\phi_{k})_y}\Big)^{-1} \frac{\delta^2\gamma}{\delta (\phi_{0i})_x
\delta (\phi_{0k})_y}.
\end{equation}

\noindent This expression can be written in terms of the functions
$G$ and $J$. For this purpose we substitute the explicit
expression for the inverse Green function from Eq.
(\ref{A-1_Explicit}). Then, it is necessary to use Eq.
(\ref{Phi0_SD_Equation}), which allows to express the remaining
Green function in terms of $J$:

\begin{eqnarray}\label{Phi0->Phi}
&& (D^2)_y \frac{\delta^2\gamma}{\delta (\phi_0^{*j})_y \delta
(\phi_{0i})_x} = \frac{1}{4} G^i{}_j D^2
\delta^8_{xy};\nonumber\\
&& (\bar D^2)_x \frac{\delta^2\gamma}{\delta (\phi_{0j})_y \delta
(\phi_{0i})_x} = \Big((M J)^{ij} - M^{ij}\Big) \Big(\frac{\bar D^2
D^2}{16\partial^2}\Big)_x \delta^8_{xy}.
\end{eqnarray}

\noindent Calculating the integrals over $\theta$ after the Wick
rotation we obtain

\begin{equation}\label{Delta2_Explicit}
\Delta_2 = - {\cal V}_4 \cdot C(R)_i{}^j \frac{d}{d\ln\Lambda}
\int \frac{d^4q}{(2\pi)^4} \Big(\frac{2}{q^4(q^2 G^2 + |M J|^2)}
(M J)^*\Big)_{j\,k}\left((M J)^{k i} - M^{k i}\right).
\end{equation}

\section{Derivatives with respect to the parameter $g$}
\label{Appendix_G_Derivative}

\subsection{The derivative of the Routhian}
\hspace{\parindent}\label{Appendix_G_Derivative_Routhian}

Let us set the background gauge superfield to 0,
$\mbox{\boldmath$V$} = 0$, and differentiate the Routhian $\gamma$
with respect to the parameter $g$ using the identity

\begin{equation}
\frac{\partial\gamma}{\partial \ln g} = \frac{\partial W}{\partial
\ln g}.
\end{equation}

\noindent The result can be written as

\begin{equation}\label{Routhian_Derivative}
\frac{\partial\gamma}{\partial \ln g} = \Big\langle
\frac{g}{4}\sum\limits_{I=0}^n \sum\limits_{\alpha=1}^{N_f} \int
d^8x\,\Big\{(\phi^* +\phi_0^*) (e^{2V}-1) (\phi + \phi_0) +
(\widetilde\phi^* + \widetilde\phi_0^*) (e^{-2V}-1)(\widetilde\phi
+\widetilde\phi_0)\Big\}_{\alpha I} \Big\rangle.
\end{equation}

\noindent In terms of the Routhian $\gamma$ the right hand side of
this equation can be presented in the form

\begin{eqnarray}\label{Routhian_Derivative_Appendix}
&& \frac{\partial\gamma}{\partial\ln g} = \frac{1}{2} \int d^8x\,
\Big\{\frac{1}{i} (-1)_i{}^j \frac{\delta}{\delta (j^j)_x}
\frac{\delta\gamma}{\delta(\phi_{0i})_x} + (\phi_i + \phi_{0i})_x
\frac{\delta\gamma}{\delta (\phi_{0i})_x} + \frac{1}{i} (-1)_i{}^j
\frac{\delta}{\delta (j^*_i)_x}
\frac{\delta\gamma}{\delta(\phi_{0}^{*j})_x}
\quad\nonumber\\
&& + (\phi^{*i} + \phi_{0}^{*i})_x \frac{\delta\gamma}{\delta
(\phi_{0}^{*i})_x} - \frac{i}{2}
\Big(\frac{\delta^2\gamma}{\delta\phi^{*i} \delta\phi_i}\Big)^{-1}
-\frac{1}{2} (\phi^{*i} + \phi_0^{*i}) (\phi_i + \phi_{0i})
\Big\}, \qquad\quad
\end{eqnarray}

\noindent where

\begin{equation}
(-1)_i{}^j \equiv \delta_{ij}\cdot (-1)^{P_i}.
\end{equation}

\noindent Let us consider the first term in this expression.
Integrating by parts and using Eq. (\ref{Phi0_SD_Equation}) we
obtain

\begin{eqnarray}
&&\hspace*{-8mm} \int d^8x\,(-1)_i{}^j \frac{\delta}{\delta j^j}
\frac{\delta\gamma}{\delta\phi_{0i}} =  - \int d^8x\, (-1)_i{}^j
\Big(\frac{D^2}{16\partial^2}\frac{\delta}{\delta j^j}\Big) \bar
D^2 \frac{\delta\gamma}{\delta \phi_{0i}} = \int d^8x\,(-1)_i{}^j
\Big(\frac{D^2}{8\partial^2}\frac{\delta}{\delta j^i}\Big)
\Big(\frac{\delta\gamma}{\delta \phi_{i}}
\nonumber\\
&&\hspace*{-8mm} - \frac{1}{2} M^{ij} \phi_j\Big) = \int
d^8x\,\Big\{(-1)_i{}^i \Big(\frac{D^2 \bar
D^2}{16\partial^2}\Big)_x \delta^8_{xy} + M^{j\,i}
\Big(\frac{D^2}{16\partial^2}\Big)_x
\Big(\frac{\delta^2\gamma}{\delta (\phi_i)_x
\delta(\phi_j)_y}\Big)^{-1}\Big\}_{x=y},
\end{eqnarray}

\noindent where we take into account that $\delta\gamma/\delta
\phi_i = - j^i$. Note that the first term in this expression
vanishes, because

\begin{equation}
(-1)_i{}^i = \sum\limits_{\alpha=1}^{N_f} \sum\limits_{I=0}^n
(-1)^{P_I} = N_f \Big(1 - \sum\limits_{I=1}^n c_I\Big) = 0
\end{equation}

\noindent due to the first equality in Eq. (\ref{C_Conditions}).
The third term in Eq. (\ref{Routhian_Derivative_Appendix}) can be
considered similarly. Thus, we obtain

\begin{eqnarray}\label{Gamma_Derivative_Appendix}
&&\hspace*{-6mm} \frac{\partial\gamma}{\partial\ln g} =
\frac{1}{2} \int d^8x\, \Big\{ -\frac{1}{2} (\phi^{*i} +
\phi_0^{*i}) (\phi_i + \phi_{0i}) + (\phi_i + \phi_{0i})_x
\frac{\delta\gamma}{\delta (\phi_{0i})_x} + (\phi^{*i} +
\phi_{0}^{*i})_x \frac{\delta\gamma}{\delta
(\phi_{0}^{*i})_x} \nonumber\\
&&\hspace*{-6mm}  -\frac{i}{2}
\Big(\frac{\delta^2\gamma}{\delta\phi_i \delta\phi^{*i}}\Big)^{-1}
- i M^{j\,i}
\Big(\frac{D^2}{16\partial^2}\Big)_x\Big(\frac{\delta^2\gamma}{\delta
(\phi_i)_x \delta(\phi_j)_y}\Big)^{-1}_{y=x} - i M^*_{j\,i}
\Big(\frac{\bar
D^2}{16\partial^2}\Big)_x\Big(\frac{\delta^2\gamma}{\delta
(\phi^{*i})_x \delta(\phi^{*j})_y}\Big)^{-1}_{y=x}
\Big\}.\nonumber\\
\end{eqnarray}

\subsection{Derivatives of effective vertices}
\hspace{\parindent}\label{Appendix_G_Derivative_Vertices}

Using the derivative of the Routhian $\gamma$ with respect to $\ln
g$ given by Eq. (\ref{Gamma_Derivative_Appendix}) we can easily
calculate the derivatives of various Green functions. For example,

\begin{equation}
\frac{\partial}{\partial\ln g} \frac{\delta^2\gamma}{\delta
(\phi_{0i})_y \delta (\phi_{0j})_z} = \frac{\delta^2}{\delta
(\phi_{0i})_y \delta (\phi_{0j})_z}
\frac{\partial\gamma}{\partial\ln g}.
\end{equation}

\noindent Substituting $\delta\gamma/\delta\ln g$ from Eq.
(\ref{Gamma_Derivative_Appendix}), differentiating, and then
setting all fields to 0, we obtain

\begin{eqnarray}\label{Two-Point_Function_Derivative}
&& \frac{\partial}{\partial\ln g} \frac{\delta^2\gamma}{\delta
(\phi_{0i})_y \delta (\phi_{0j})_z} = \frac{\delta^2\gamma}{\delta
(\phi_{0i})_y \delta (\phi_{0j})_z} + \frac{i}{4} \int
d^8x\,\Big(\frac{\mbox{\boldmath$\delta$}}{\mbox{\boldmath$\delta$}
j^k} \frac{\mbox{\boldmath$\delta$}}{\mbox{\boldmath$\delta$}
j^*_k} \nonumber\\
&&\qquad\qquad\qquad\qquad\quad + M^{lk}
\Big(\frac{D^2}{8\partial^2}
\frac{\mbox{\boldmath$\delta$}}{\mbox{\boldmath$\delta$} j^k}\Big)
\frac{\mbox{\boldmath$\delta$}}{\mbox{\boldmath$\delta$} j^l} +
M^*_{lk} \Big(\frac{\bar D^2}{8\partial^2}
\frac{\mbox{\boldmath$\delta$}}{\mbox{\boldmath$\delta$}
j^*_k}\Big)
\frac{\mbox{\boldmath$\delta$}}{\mbox{\boldmath$\delta$} j^*_l}
\Big)_{x} \cdot \frac{\delta^2\gamma}{\delta (\phi_{0i})_y \delta
(\phi_{0j})_z}. \qquad\quad
\end{eqnarray}

\noindent Similar expressions can be written for the derivatives
of the other two-point functions. In order to rewrite the result
in a more compact form, we use the notation (\ref{LineWithDot}).
Then the derivatives of the two-point functions can be written as

\begin{eqnarray}\label{Green_Functions_Derivatives}
&& \frac{\partial}{\partial \ln g}\Big(g^{-1}
\frac{\delta^2\gamma}{\delta (\phi_{0i})_y \delta
(\phi_{0j})_z}\Big) = \frac{1}{2g}\, \mbox{LineWithDot}[1] \cdot
\frac{\delta^2\gamma}{\delta (\phi_{0i})_y \delta
(\phi_{0j})_z};\nonumber\\
&& \frac{\partial}{\partial \ln g}\Big(g^{-1}
\frac{\delta^2\gamma}{\delta (\phi_{0i})_y \delta
(\phi^{*j}_{0})_z}\Big) = -\frac{1}{4 g} \delta^8_{yz} \delta^i_j
+\frac{1}{2g}\, \mbox{LineWithDot}[1]\cdot
\frac{\delta^2\gamma}{\delta (\phi_{0i})_y \delta
(\phi^{*j}_{0})_z}.\qquad
\end{eqnarray}

\noindent (The second equality is derived using the same method.)
Derivatives of the other two-point functions can be written in a
similar form.

In order to calculate the derivatives of the four-point functions
we again use this method. It is convenient to introduce the
notation

\begin{equation}
\mbox{UsualLine}[1,2] \equiv \int d^8x\,
\Big(\frac{\delta}{\delta_2 \phi_{0i}} \cdot
\frac{\mbox{\boldmath$\delta$}}{\mbox{\boldmath$\delta$}_1
j^i} + \frac{\delta}{\delta_2 \phi^{*i}_{0}} \cdot
\frac{\mbox{\boldmath$\delta$}}{\mbox{\boldmath$\delta$}_1
j^*_i} \Big)_{x} =
\mbox{UsualLine}[2,1].
\end{equation}

\noindent Then it is possible to write the derivatives of the
four-point functions in a rather simple form. For example, taking
into account that any Green function with an odd number of legs
corresponding to the matter superfields vanishes, we obtain

\begin{eqnarray}
&&\hspace*{-4mm} \frac{\partial}{\partial \ln g}\Big(g^{-2}
\frac{\delta^4\gamma}{\delta (\phi_{0i})_x \delta (\phi_{0j})_y
\delta (\phi_{0k})_z \delta (\phi_{0l})_w}\Big) = \frac{1}{2g^{2}}
\Bigg(\mbox{LineWithDot}[1]\cdot\\
&&\hspace*{-4mm} \times \frac{\delta^4\gamma}{\delta (\phi_{0i})_x
\delta (\phi_{0j})_y \delta (\phi_{0k})_z \delta (\phi_{0l})_w} +
2\cdot\mbox{LineWithDot}[1;1,2]\cdot \mbox{UsualLine}[1,2]
\nonumber\\
&&\hspace*{-4mm} \times \Big(\frac{\delta^2\gamma[1]}{\delta
(\phi_{0i})_x \delta (\phi_{0j})_y}
\frac{\delta^2\gamma[2]}{\delta (\phi_{0k})_z \delta
(\phi_{0l})_w} + \frac{\delta^2\gamma[1]}{\delta (\phi_{0i})_x
\delta (\phi_{0k})_z}\cdot \frac{\delta^2\gamma[2]}{\delta
(\phi_{0j})_y \delta (\phi_{0l})_w}\cdot (-1)^{P_j P_k}\nonumber\\
&& + \frac{\delta^2\gamma[1]}{\delta (\phi_{0j})_y \delta
(\phi_{0k})_z}\cdot \frac{\delta^2\gamma[2]}{\delta (\phi_{0i})_x
\delta (\phi_{0l})_w}\cdot (-1)^{P_i P_j + P_i
P_k}\Big)\Bigg).\qquad\nonumber
\end{eqnarray}

\noindent In this expression the derivatives
$\mbox{\boldmath$\delta$}/\mbox{\boldmath$\delta$}_1 j$ act on
$\gamma[1]$, and the derivatives
$\mbox{\boldmath$\delta$}/\mbox{\boldmath$\delta$}_2 j$ act on
$\gamma[2]$. The derivatives of the other four-point functions
with respect to $\ln g$ can be written in a similar form.

\subsection{Derivatives of effective lines}
\hspace{\parindent}\label{Appendix_G_Derivative_Lines}

Expressions for the derivatives of effective vertices obtained in
the previous section allow to find derivatives of effective
diagrams. The effective diagrams include effective lines.
Therefore, it is desirable to find derivatives of these effective
lines, which, in particular, contain inverse Green functions
inside the derivatives
$\mbox{\boldmath$\delta$}/\mbox{\boldmath$\delta$} j$. The
derivatives of the inverse Green functions can be easily
calculated using Eq. (\ref{Green_Functions_Derivatives}):

\begin{eqnarray}\label{Inverse_Derivative}
&&\hspace*{-5mm} \frac{\partial}{\partial\ln g} \Bigg(g
\Big(\frac{\delta^2\gamma}{\delta (\phi_i)_x \delta
(\phi_j)_y}\Big)^{-1}\Bigg) = g \int d^8z\,\Bigg[ M^{lk}
\Big(\frac{\delta^2\gamma}{\delta (\phi_i)_x \delta
(\phi_k)_z}\Big)^{-1} \Big(\frac{D^2}{16\partial^2}\Big)_z
\Big(\frac{\delta^2\gamma}{\delta (\phi_l)_z \delta
(\phi_j)_y}\Big)^{-1}\nonumber\\
&&\hspace*{-5mm} + M^*_{lk} \Big(\frac{\delta^2\gamma}{\delta
(\phi_i)_x \delta (\phi^{*k})_z}\Big)^{-1} \Big(\frac{\bar
D^2}{16\partial^2}\Big)_z \Big(\frac{\delta^2\gamma}{\delta
(\phi^{*l})_z \delta (\phi_j)_y}\Big)^{-1} +\frac{1}{4}
\Big(\frac{\delta^2\gamma}{\delta (\phi_i)_x \delta
(\phi_k)_z}\Big)^{-1} \Big(\frac{\delta^2\gamma}{\delta
(\phi^{*k})_z \delta (\phi_j)_y}\Big)^{-1}\nonumber\\
&&\hspace*{-5mm} +\frac{1}{4} \Big(\frac{\delta^2\gamma}{\delta
(\phi^{*k})_z \delta (\phi_i)_x} \Big)^{-1}
\Big(\frac{\delta^2\gamma}{\delta (\phi_{k})_z \delta
(\phi_j)_y}\Big)^{-1}\Bigg] - \frac{g}{2}
\mbox{LineWithDot}[1]\cdot \frac{\mbox{\boldmath$\delta$}^2
\gamma}{\mbox{\boldmath$\delta$} (j^j)_y \mbox{\boldmath$\delta$}
(j^i)_x}.
\end{eqnarray}

\noindent Similarly,

\begin{eqnarray}\label{Inverse_Derivative2}
&&\hspace*{-5mm} \frac{\partial}{\partial\ln g} \Bigg(g
\Big(\frac{\delta^2\gamma}{\delta (\phi_i)_x \delta
(\phi^{*j})_y}\Big)^{-1}\Bigg) = g \int d^8z\,\Bigg[ M^{lk}
\Big(\frac{\delta^2\gamma}{\delta (\phi_i)_x \delta
(\phi_k)_z}\Big)^{-1} \Big(\frac{D^2}{16\partial^2}\Big)_z
\Big(\frac{\delta^2\gamma}{\delta (\phi_l)_z \delta
(\phi^{*j})_y}\Big)^{-1}\nonumber\\
&&\hspace*{-5mm} + M^*_{lk} \Big(\frac{\delta^2\gamma}{\delta
(\phi_i)_x \delta (\phi^{*k})_z}\Big)^{-1} \Big(\frac{\bar
D^2}{16\partial^2}\Big)_z \Big(\frac{\delta^2\gamma}{\delta
(\phi^{*l})_z \delta (\phi^{*j})_y}\Big)^{-1} +\frac{1}{4}
\Big(\frac{\delta^2\gamma}{\delta (\phi_i)_x \delta
(\phi_k)_z}\Big)^{-1} \Big(\frac{\delta^2\gamma}{\delta
(\phi^{*k})_z \delta (\phi^{*j})_y}\Big)^{-1}\nonumber\\
&&\hspace*{-5mm} +\frac{1}{4} \Big(\frac{\delta^2\gamma}{\delta
(\phi^{*k})_z \delta (\phi_i)_x}\Big)^{-1}
\Big(\frac{\delta^2\gamma}{\delta (\phi_{k})_z \delta
(\phi^{*j})_y}\Big)^{-1}\Bigg] - \frac{g}{2}
\mbox{LineWithDot}[1]\cdot \frac{\mbox{\boldmath$\delta$}^2
\gamma}{\mbox{\boldmath$\delta$} (j^*_j)_y
\mbox{\boldmath$\delta$} (j^i)_x}.
\end{eqnarray}

\noindent From these equations we obtain

\begin{eqnarray}\label{Delta_Delta_J_Derivative}
&&\hspace*{-5mm} \Big[\frac{\partial}{\partial\ln g},\, g
\frac{\mbox{\boldmath$\delta$}}{\mbox{\boldmath$\delta$}
(j^i)_x}\Big] = - \int d^8y\,\Bigg(\frac{\partial}{\partial \ln
g}\Big[g \Big(\frac{\delta^2\gamma}{\delta (\phi_i)_x \delta
(\phi_j)_y}\Big)^{-1}\Big] \frac{\delta}{\delta
(\phi_{0j})_y}\nonumber\\
&&\hspace*{-5mm} + \frac{\partial}{\partial \ln g}\Big[g
\Big(\frac{\delta^2\gamma}{\delta (\phi_i)_x \delta
(\phi^{*j})_y}\Big)^{-1}\Big] \frac{\delta}{\delta
(\phi_{0}^{*j})_y}\Bigg) = \frac{g}{4} \int d^8z\,\Bigg\{\Big(
2\cdot \mbox{LineWithDot}[1]\cdot \frac{\mbox{\boldmath$\delta$}
}{\mbox{\boldmath$\delta$} (j^i)_x} \frac{\delta \gamma}{\delta
(\phi_{0k})_z}
\nonumber\\
&&\hspace*{-5mm} + \Big(\frac{\delta^2\gamma}{\delta (\phi^{*k})_z
\delta(\phi_i)_x}\Big)^{-1} + M^{kl}
\Big(\frac{\delta^2\gamma}{\delta(\phi_i)_x \delta
(\phi_{l})_z}\Big)^{-1} \frac{D^2}{4\partial^2} \Big)\cdot
\frac{\mbox{\boldmath$\delta$}}{\mbox{\boldmath$\delta$} (j^k)_z}
+ \Big( 2\cdot \mbox{LineWithDot}[1]\cdot
\frac{\mbox{\boldmath$\delta$}}{\mbox{\boldmath$\delta$} (j^i)_x}
\nonumber\\
&&\hspace*{-5mm} \times \frac{\delta \gamma}{\delta (\phi_{0}^{*k})_z} +
\Big(\frac{\delta^2\gamma}{\delta(\phi_i)_x \delta
(\phi_k)_z}\Big)^{-1} + M^*_{kl}
\Big(\frac{\delta^2\gamma}{\delta(\phi_i)_x \delta
(\phi^{*l})_z}\Big)^{-1} \frac{\bar D^2}{4\partial^2}\Big) \cdot
\frac{\mbox{\boldmath$\delta$}}{\mbox{\boldmath$\delta$}
(j_k^*)_z} \Bigg\}.
\end{eqnarray}

\noindent We can use this equation for differentiating the
operator $\mbox{GreenLine}$ (or other similar operators) with
respect to $\ln g$. Let us consider, for the definiteness, the
derivative of the green effective line. More exactly, let us
differentiate

\begin{eqnarray}
&& g\cdot \mbox{GreenLine}[1,2] = \int d^8x\,\Big\{g (T)^j{}_i
\Big(\theta^a \theta_a \bar\theta^{\dot b} \frac{\bar D_{\dot b}
D^2}{4\partial^2} + 2i \bar\theta^{\dot a} (\gamma^\mu)_{\dot
a}{}^b \theta_b \frac{\partial_\mu}{\partial^2}\Big)
\frac{\mbox{\boldmath$\delta$}}{\mbox{\boldmath$\delta$}_2
j^j}\cdot \frac{\delta}{\delta_1 \phi_{0i}}
\qquad\nonumber\\
&& - i\, g^{-1} g (M T)^{ij}\, \bar\theta^{\dot a}
(\gamma^\mu)_{\dot a}{}^b \theta_b \Big(\frac{D^2
\partial_\mu}{16\partial^4}
\frac{\mbox{\boldmath$\delta$}}{\mbox{\boldmath$\delta$}_2 j^i}
\Big)\cdot g
\frac{\mbox{\boldmath$\delta$}}{\mbox{\boldmath$\delta$}_1 j^j}
\Big\}_{x}.
\end{eqnarray}

\noindent It is convenient to write the result in the following
form:

\begin{eqnarray}
&& \Big[\frac{\partial}{\partial\ln g},\,g\cdot
\mbox{GreenLine}[1,2]\Big] = \frac{g}{2} \Big\{
\mbox{GreenWhiteLine}[1,2]
+ \mbox{GreenWithCrossWhite}[2,1] \qquad\nonumber\\
&& - 2\cdot \mbox{GreenWithCross}[1,2]
+ \mbox{LineWithDot}[1;3,3]\cdot \Big(\mbox{GreenLine}[1,3] \cdot
\mbox{UsualLine}[2,3] \nonumber\\
&& + \mbox{GreenWithCross}[2,3]\cdot
\mbox{UsualLine}[1,3]\Big) \cdot \gamma[3]
\Big\}.\qquad
\end{eqnarray}

\noindent Let us briefly explain the derivation of this identity.
In this expression all terms without the operator
$\mbox{LineWithDot[1]}$ are included into the effective lines

\begin{eqnarray}
&&\hspace*{-3mm} \mbox{GreenWhiteLine}[1,2] \equiv  \frac{1}{2}
\int d^8x\,d^8z\, \Bigg\{(T)^j{}_i\Big(\theta^a \theta_a
\bar\theta^{\dot b} \frac{\bar D_{\dot b} D^2}{4\partial^2} + 2i
\bar\theta^{\dot a} (\gamma^\mu)_{\dot a}{}^b \theta_b
\frac{\partial_\mu}{\partial^2}\Big)_x
\nonumber\\
&&\hspace*{-3mm} \times \Bigg(\Big(\frac{\delta^2\gamma}{\delta
(\phi^{*k})_z \delta(\phi_j)_x}\Big)^{-1}
\frac{\mbox{\boldmath$\delta$}}{\mbox{\boldmath$\delta$}_2
(j^k)_z} + \Big(\frac{\delta^2\gamma}{\delta(\phi_j)_x \delta
(\phi_k)_z}\Big)^{-1}
\frac{\mbox{\boldmath$\delta$}}{\mbox{\boldmath$\delta$}_2
(j_k^*)_z} + M^{lk} \Big(\frac{\delta^2\gamma}{\delta(\phi_j)_x
\delta (\phi_{k})_z}\Big)^{-1}\nonumber\\
&&\hspace*{-3mm} \times \frac{D^2}{4\partial^2}
\frac{\mbox{\boldmath$\delta$}}{\mbox{\boldmath$\delta$}_2
(j^l)_z} + M^*_{lk} \Big(\frac{\delta^2\gamma}{\delta(\phi_j)_x
\delta (\phi^{*k})_z}\Big)^{-1} \frac{\bar D^2}{4\partial^2}
\frac{\mbox{\boldmath$\delta$}}{\mbox{\boldmath$\delta$}_2
(j_l^*)_z} \Bigg) \frac{\delta}{\delta_1 (\phi_{0i})_x} - i
(MT)^{ij}
\bar\theta^{\dot a} (\gamma^\mu)_{\dot a}{}^b \theta_b \nonumber\\
&&\hspace*{-3mm} \times
\Big(\frac{D^2\partial_\mu}{16\partial^4}\Big)_x
\Bigg(\Big(\frac{\delta^2\gamma}{\delta (\phi^{*k})_z
\delta(\phi_i)_x}\Big)^{-1}
\frac{\mbox{\boldmath$\delta$}}{\mbox{\boldmath$\delta$}_2
(j^k)_z} + \Big(\frac{\delta^2\gamma}{\delta(\phi_i)_x \delta
(\phi_k)_z}\Big)^{-1}
\frac{\mbox{\boldmath$\delta$}}{\mbox{\boldmath$\delta$}_2
(j_k^*)_z} + M^{lk} \Big(\frac{\delta^2\gamma}{\delta(\phi_i)_x
\delta (\phi_{k})_z}\Big)^{-1}\nonumber\\
&&\hspace*{-3mm} \times  \frac{D^2}{4\partial^2}
\frac{\mbox{\boldmath$\delta$}}{\mbox{\boldmath$\delta$}_2
(j^l)_z} + M^*_{lk} \Big(\frac{\delta^2\gamma}{\delta(\phi_i)_x
\delta (\phi^{*k})_z}\Big)^{-1}\frac{\bar D^2}{4\partial^2}
\frac{\mbox{\boldmath$\delta$}}{\mbox{\boldmath$\delta$}_2
(j_l^*)_z} \Bigg)
\frac{\mbox{\boldmath$\delta$}}{\mbox{\boldmath$\delta$}_1
(j^j)_x} \Bigg\}\quad
\end{eqnarray}

\noindent and

\begin{eqnarray}
&&\hspace*{-3mm} \mbox{GreenWithCrossWhite}[1,2] \equiv -
\frac{i}{2} \int d^8x\,d^8z\, \Bigg\{ (MT)^{ij} \bar\theta^{\dot
a} (\gamma^\mu)_{\dot a}{}^b \theta_b
\Big(\frac{D^2\partial_\mu}{16\partial^4}\Big)_x
\nonumber\\
&&\hspace*{-3mm} \times
\Bigg(\Big(\frac{\delta^2\gamma}{\delta(\phi_i)_x \delta
(\phi^{*k})_z}\Big)^{-1}
\frac{\mbox{\boldmath$\delta$}}{\mbox{\boldmath$\delta$}_2
(j^k)_z} + \Big(\frac{\delta^2\gamma}{\delta
(\phi_k)_z \delta(\phi_i)_x}\Big)^{-1}
\frac{\mbox{\boldmath$\delta$}}{\mbox{\boldmath$\delta$}_2
(j_k^*)_z} + M^{kl} \Big(\frac{\delta^2\gamma}{\delta(\phi_i)_x
\delta (\phi_{k})_z}\Big)^{-1}\qquad\nonumber\\
&&\hspace*{-3mm} \times  \frac{D^2}{4\partial^2}
\frac{\mbox{\boldmath$\delta$}}{\mbox{\boldmath$\delta$}_2
(j^l)_z} + M^*_{kl} \Big(\frac{\delta^2\gamma}{\delta(\phi_i)_x
\delta (\phi^{*k})_z}\Big)^{-1}\frac{\bar D^2}{4\partial^2}
\frac{\mbox{\boldmath$\delta$}}{\mbox{\boldmath$\delta$}_2
(j_l^*)_z} \Bigg)
\frac{\mbox{\boldmath$\delta$}}{\mbox{\boldmath$\delta$}_1
(j^j)_x} \Bigg\}.\quad
\end{eqnarray}

\noindent These expressions can be written in a more compact form,
Eqs. (\ref{GreenWhiteLine}) and (\ref{GreenWithCrossWhite}),
respectively. The term with the operator $\mbox{GreenWithCross}$
appears when we differentiate the factor $g^{-1}$ in the mass
term. The other terms containing the operator $\mbox{LineWithDot}$
are obtained from the derivatives of $g\,
\mbox{\boldmath$\delta$}/\mbox{\boldmath$\delta$} j$. However, it
is necessary to take into account that the terms without masses
are linear in this derivative, while the terms containing the
masses are quadratic. That is why the result contains the sum of
the operators $\mbox{GreenLine}$ and $\mbox{GreenWithCross}$.

Derivatives of the other effective lines can be constructed
similarly.

\subsection{Derivative of $\Delta$}
\hspace{\parindent}\label{Appendix_G_Derivative_One-Loop}

The derivative of the additional contribution $\Delta$, given by
Eq. (\ref{Delta_Definition}), with respect to $\ln g$ can be
calculated using Eq. (\ref{Inverse_Derivative}). The result is

\begin{eqnarray}\label{Delta_Derivative}
&&\hspace*{-4mm} \frac{\partial \Delta}{\partial\ln g} = -i
\frac{d}{d\ln\Lambda} \int d^8x\,d^8y\, (\theta^4)_x \Bigg\{\Bigg(
C(R)_k{}^i M^{jk} \delta^8_{xy}
\Big(\frac{D^2}{4\partial^4}\Big)_x - (MT)^{im} (M T)^{nj}
\Big(\frac{D^2\partial^\mu }{4\partial^4}\Big)_y
\nonumber\\
&&\hspace*{-4mm} \times
\Big(\frac{\delta^2\gamma}{\delta(\phi_m)_x
\delta(\phi_n)_y}\Big)^{-1} \Big(\frac{D^2
\partial_\mu}{16\partial^4}\Big)_x\Bigg)  \Bigg( - \frac{1}{2}
\mbox{LineWithDot}[1]\cdot  \frac{\mbox{\boldmath$\delta$}^2
\gamma}{\mbox{\boldmath$\delta$} (j^j)_y
\mbox{\boldmath$\delta$} (j^i)_x} + \int d^8z\,\nonumber\\
&&\hspace*{-4mm} \times \Bigg[\,\frac{1}{4}
\Big(\frac{\delta^2\gamma}{\delta (\phi_i)_x \delta
(\phi_k)_z}\Big)^{-1} \Big(\frac{\delta^2\gamma}{\delta
(\phi^{*k})_z \delta (\phi_j)_y}\Big)^{-1} + \frac{1}{4}
\Big(\frac{\delta^2\gamma}{\delta (\phi_i)_x \delta
(\phi^{*k})_z}\Big)^{-1} \Big(\frac{\delta^2\gamma}{\delta (\phi_j)_y
\delta (\phi_{k})_z}\Big)^{-1}\nonumber\\
&&\hspace*{-4mm} + M^{lk} \Big(\frac{\delta^2\gamma}{\delta
(\phi_i)_x \delta (\phi_k)_z}\Big)^{-1}
\Big(\frac{D^2}{16\partial^2}\Big)_z
\Big(\frac{\delta^2\gamma}{\delta (\phi_l)_z \delta
(\phi_j)_y}\Big)^{-1} + M^*_{lk} \Big(\frac{\delta^2\gamma}{\delta
(\phi_i)_x \delta (\phi^{*k})_z}\Big)^{-1} \Big(\frac{\bar
D^2}{16\partial^2}\Big)_z\nonumber\\
&&\hspace*{-4mm} \times \Big(\frac{\delta^2\gamma}{\delta
(\phi^{*l})_z \delta (\phi_j)_y}\Big)^{-1} \Bigg] -
\Big(\frac{\delta^2\gamma}{\delta (\phi_i)_x \delta
(\phi_j)_y}\Big)^{-1} \Bigg) \Bigg\}.
\end{eqnarray}

\noindent The terms containing the operator
$\mbox{LineWithDot[1]}$ can be graphically presented as two-loop
effective diagrams. The other terms correspond to one-loop
effective diagrams.

\section{Calculation of commutators}
\label{Appendix_Commutators}

\subsection{Commutators with $(T) y_\mu^*$}
\hspace{\parindent} \label{Appendix_Commutators_Y}

In this section we calculate commutators of $(T) y_\mu^*$ with
various Green functions. For this purpose we can use the
Schwinger--Dyson equation (\ref{SD_Equation}), which is valid for
any values of $g$. Let us multiply this equation by $y^*_\mu$ and
integrate the result over $d^8x$:

\begin{equation}\label{SD_Integration_LHS}
\int d^8x\,(y_\mu^*)_x \frac{\delta(\Delta\Gamma)}{\delta
\mbox{\boldmath$V$}_x} = \int d^8x\,(y_\mu^*)_x (T)^j{}_i\Big(
\frac{2}{i} \frac{\delta}{\delta
(j^*_i)_x}\frac{\delta\gamma}{\delta (\phi^{*j}_{0})_x} +
2 \frac{\delta\gamma}{\delta
(\phi_{0}^{*j})_x} (\phi^{*i}+\phi_{0}^{i*})_x\Big).
\end{equation}

\noindent The first term in this expression vanishes. Really,
using antichirality of the derivative with respect to $j^*$ and
integrating by parts this term can be rewritten as

\begin{equation}
\frac{2}{i} \int d^8x\,(y_\mu^*)_x (T)^j{}_i \Big(-\frac{D^2 \bar
D^2}{16\partial^2}\frac{\delta}{\delta
j^*_i}\Big)_x\frac{\delta\gamma}{\delta (\phi^{*j}_{0})_x} =
\frac{2}{i} \int d^8x\,(y_\mu^*)_x (T)^j{}_i \Big(-\frac{\bar
D^2}{16\partial^2}\frac{\delta}{\delta j^*_i}\Big)_x
D_x^2\frac{\delta\gamma}{\delta (\phi^{*j}_{0})_x}.
\end{equation}

\noindent Taking into account Eq. (\ref{Phi0_SD_Equation}) we
obtain

\begin{eqnarray}
&& \frac{2}{i} \int d^8x\,(y_\mu^*)_x (T)^j{}_i \Big(\frac{\bar
D^2}{8\partial^2}\frac{\delta}{\delta j^*_i}\Big)_x
\Big(- j_j^* -\frac{1}{2} M^*_{jk} \phi^{*k}\Big)_x\nonumber\\
&&\qquad\qquad = i (T M^*)_{ik} \int d^8x\,(y_\mu^*)_x
\Big(\frac{\bar D^2}{8\partial^2} \Big)_x \Big(\frac{D^2 \bar
D^2}{16\partial^2}\Big)_y \Big(\frac{\delta^2\gamma}{\delta
(\phi^{*k})_y \delta (\phi^{*i})_x} \Big)_{x=y}^{-1} = 0.\qquad
\end{eqnarray}

\noindent (The term with the derivative of $j^*$ vanishes due to
the evident identity $T^i{}_i = 0$. The last equality can be
obtained by integrating $D^2$ by parts and taking into account Eq.
(\ref{MT_Symmetry}).) Note that the fields here are not yet set to
0. From the other side,

\begin{equation}\label{SD_Integration_RHS}
\int d^8x\,(y_\mu^*)_x \frac{\delta(\Delta\Gamma)}{\delta
\mbox{\boldmath$V$}_x} = \int d^8x\,(y_\mu^*)_x
(T)^j{}_i \Big(\frac{2}{i} \frac{\delta}{\delta (j^j)_x}
\frac{\delta\gamma}{\delta (\phi_{0i})_x} + 2
\frac{\delta\gamma}{\delta (\phi_{0i})_x}
(\phi_j+\phi_{0j})_x\Big).
\end{equation}

\noindent Comparing Eq. (\ref{SD_Integration_LHS}) and Eq.
(\ref{SD_Integration_RHS}) we obtain

\begin{eqnarray}\label{Ward_Identity_With_Y}
&& \int d^8x\,(y_\mu^*)_x (T)_i{}^j\Big(
(\phi^{i*}+\phi_{0}^{i*})_x \frac{\delta\gamma}{\delta
(\phi_{0}^{*j})_x}-  (\phi_j+\phi_{0j})_x
\frac{\delta\gamma}{\delta (\phi_{0i})_x}  \Big)\nonumber\\
&& \qquad\qquad\qquad\qquad\qquad\qquad\quad = -i \int
d^8x\,(y_\mu^*)_x (T)^j{}_i \Big(\frac{\delta}{\delta
(j^j)_x}\frac{\delta\gamma}{\delta (\phi_{0i})_x}\Big).\qquad
\end{eqnarray}

\noindent Differentiating Eq. (\ref{Ward_Identity_With_Y}) with
respect to various fields, it is possible to find

\begin{eqnarray}
&& \Big[(T) y_\mu^*, \frac{\delta^2\gamma}{\delta (\phi_{0k})_y
\delta (\phi_{0}^{*l})_z} \Big] \equiv - (y_\mu^*)_y (T)_i{}^k
\frac{\delta^2\gamma}{\delta (\phi_{0i})_y \delta
(\phi_{0}^{*l})_z} + (y_\mu^*)_z (T)_l{}^i
\frac{\delta^2\gamma}{\delta (\phi_{0k})_y \delta
(\phi_{0}^{*i})_z};\nonumber\\
&& \Big[(T) y_\mu^*, \frac{\delta^2\gamma}{\delta (\phi_{0k})_y
\delta (\phi_{0l})_z} \Big] \equiv - (y_\mu^*)_y (T)_i{}^k
\frac{\delta^2\gamma}{\delta (\phi_{0i})_y \delta (\phi_{0l})_z} -
(y_\mu^*)_z (T)_i{}^l \frac{\delta^2\gamma}{\delta (\phi_{0k})_y
\delta (\phi_{0i})_z}\qquad\quad
\end{eqnarray}

\noindent and other similar expressions. (All fields in this
commutators are set to 0.) For example, differentiating Eq.
(\ref{Ward_Identity_With_Y}) with respect to $(\phi_{0k})_y$ and
$(\phi_{0l})_z$ and setting all fields to 0 we obtain

\begin{equation}
\Big[(T) y_\mu^*, \frac{\delta^2\gamma}{\delta (\phi_{0k})_y
\delta (\phi_{0l})_z} \Big] = -i \int d^8x\,(y_\mu^*)_x (T)^j{}_i
\frac{\delta^2}{\delta (\phi_{0k})_y \delta (\phi_{0l})_z}\cdot
\frac{\delta}{\delta (j^j)_x}\frac{\delta\gamma}{\delta
(\phi_{0i})_x}.
\end{equation}

\noindent Commuting the derivative with respect to the source
$j^j$ with the derivatives with respect to the fields $\phi_0$ and
taking into account that all Green functions with an odd number of
$\phi$-lines vanish, this expression can be presented in the form

\begin{eqnarray}
&& -i \int d^8x\,d^8w\,(y_\mu^*)_x (T)^j{}_i
\Bigg[\frac{\mbox{\boldmath$\delta$}}{\mbox{\boldmath$\delta$}
(j^j)_{x}} \frac{\delta^3\gamma}{\delta (\phi_{0k})_{y} \delta
(\phi_{0l})_{z}  \delta (\phi_{0m})_{w}}\cdot \Big(\delta^8_{xw}
\delta_m^i +
\frac{\mbox{\boldmath$\delta$}}{\mbox{\boldmath$\delta$}
(j^m)_{w}} \frac{\delta\gamma}{\delta (\phi_{0i})_x} \Big)\qquad
\nonumber\\
&&  + \frac{\mbox{\boldmath$\delta$}}{\mbox{\boldmath$\delta$}
(j^j)_{x}} \frac{\delta^3\gamma}{\delta (\phi_{0k})_{y} \delta
(\phi_{0l})_{z} \delta (\phi^{*m}_{0})_{w}}\cdot
\Big(\frac{\mbox{\boldmath$\delta$}}{\mbox{\boldmath$\delta$}
(j^*_m)_w} \frac{\delta\gamma}{(\phi_{0i})_x}\Big) \Bigg].\qquad
\end{eqnarray}

\noindent After substituting the two-point Green functions from
Eq. (\ref{Explicit_Green_Functions}) the result is written as

\begin{eqnarray}
&& -i \int d^8x\,\Bigg[ (T)^j{}_i \Big(\Big(1+ \frac{\bar D^2
D^2}{16\partial^2}\Big) y_\mu^*
\frac{\mbox{\boldmath$\delta$}}{\mbox{\boldmath$\delta$}
j^j}\Big)_{x} \frac{\delta^3\gamma}{\delta (\phi_{0l})_{z} \delta
(\phi_{0k})_{y} \delta (\phi_{0i})_{x}}
\nonumber\\
&&\qquad\qquad\qquad\qquad\qquad\quad - (M T)^{ij} (y_\mu^*)_x
\Big(\frac{D^2}{16\partial^2}\frac{\mbox{\boldmath$\delta$}}{\mbox{\boldmath$\delta$}
j^i} \Big)_{x} \frac{\mbox{\boldmath$\delta$}}{
\mbox{\boldmath$\delta$} (j^j)_{x}} \frac{\delta^2\gamma}{\delta
(\phi_{0k})_{y} \delta (\phi_{0l})_{z}}\Bigg].\qquad\quad
\end{eqnarray}

\noindent In order to simplify this expression we note that

\begin{equation}
\Big(1+ \frac{\bar D^2 D^2}{16\partial^2}\Big) y_\mu^*
\frac{\mbox{\boldmath$\delta$}}{\mbox{\boldmath$\delta$} j^j} =
\Big(\frac{2\partial_\mu}{\partial^2} - i(\gamma^\mu)^{{\dot a}b}
\theta_b \frac{\bar D_{\dot a} D^2}{4\partial^2}\Big)
\frac{\mbox{\boldmath$\delta$}}{\mbox{\boldmath$\delta$} j^j}
\end{equation}

\noindent and

\begin{eqnarray}
&&\hspace*{-7mm} \int d^8x\, (M T)^{ij} y_\mu^*\,
\frac{D^2}{16\partial^2}\frac{\mbox{\boldmath$\delta$}}{\mbox{\boldmath$\delta$}
j^i} \cdot \frac{\mbox{\boldmath$\delta$}}{
\mbox{\boldmath$\delta$} j^j}  = - \int d^8x\, (M T)^{ij}
y_\mu^*\, \frac{D^2}{16\partial^2}
\frac{\mbox{\boldmath$\delta$}}{\mbox{\boldmath$\delta$} j^i}\cdot
\frac{\bar D^2 D^2}{16\partial^2}\frac{\mbox{\boldmath$\delta$}}{
\mbox{\boldmath$\delta$} j^j} = - \int d^8x\, \nonumber\\
&&\hspace*{-7mm} \times (M T)^{ij} \bar D^{\dot a} y_\mu^*\,
\frac{\bar D_{\dot a} D^2}{16\partial^2}
\frac{\mbox{\boldmath$\delta$}}{\mbox{\boldmath$\delta$} j^i}\cdot
\frac{D^2}{16\partial^2}\frac{\mbox{\boldmath$\delta$}}{
\mbox{\boldmath$\delta$} j^j} = 8 \int d^8x\, (M T)^{ij}
\theta^a\,(\gamma_\mu \gamma^\nu)_a{}^b \frac{D_b
\partial_\nu}{16\partial^2}
\frac{\mbox{\boldmath$\delta$}}{\mbox{\boldmath$\delta$} j^i}\cdot
\frac{D^2}{16\partial^2}\frac{\mbox{\boldmath$\delta$}}{
\mbox{\boldmath$\delta$} j^j} \nonumber\\
&&\hspace*{-7mm} = \int d^8x\, (M T)^{ij}
\frac{D^2\partial_\mu}{16\partial^4}\frac{\mbox{\boldmath$\delta$}}{
\mbox{\boldmath$\delta$} j^i}\cdot
\frac{\mbox{\boldmath$\delta$}}{\mbox{\boldmath$\delta$} j^j}.
\end{eqnarray}

\noindent Using these identities we finally obtain

\begin{equation}\label{Y_Commutator1}
\Big[(T) y_\mu^*, \frac{\delta^2\gamma}{\delta (\phi_{0k})_y
\delta (\phi_{0l})_z}\Big] = \mbox{BrownLine}_\mu \cdot
\frac{\delta^2\gamma}{\delta (\phi_{0k})_y \delta (\phi_{0l})_z},
\end{equation}

\noindent where we have defined the operator

\begin{eqnarray}
&& \mbox{BrownLine}_\mu[1,2] \equiv \int d^8x\,\Big[ (T)_i{}^j
\Big(\Big(-\frac{2i\partial_\mu}{\partial^2} - (\gamma^\mu)^{{\dot
a}b} \theta_b \frac{\bar D_{\dot a} D^2}{4\partial^2}\Big)
\frac{\mbox{\boldmath$\delta$}}{\mbox{\boldmath$\delta$}_2
j^j}\Big)_x \frac{\delta}{\delta_1 (\phi_{0i})_x}\qquad\nonumber\\
&& + i(M T)^{ij}
\Big(\frac{D^2\partial_\mu}{16\partial^4}\frac{\mbox{\boldmath$\delta$}}{
\mbox{\boldmath$\delta$}_2 j^i}\Big)_x
\frac{\mbox{\boldmath$\delta$}}{\mbox{\boldmath$\delta$}_1
(j^j)_x} \Big].
\end{eqnarray}

Similarly, we can differentiate Eq. (\ref{Ward_Identity_With_Y})
with respect to $(\phi_{0k})_y$ and $(\phi_0^{*l})_z$ and set all
fields equal to 0. Then, repeating the same operations as above,
we find

\begin{equation}\label{Y_Commutator2}
\Big[(T) y_\mu^*, \frac{\delta^2\gamma}{\delta (\phi_{0k})_y
\delta (\phi^{*l}_{0})_z}\Big] = \mbox{BrownLine}_\mu \cdot
\frac{\delta^2\gamma}{\delta (\phi_{0k})_y \delta
(\phi^{*l}_{0})_z}.
\end{equation}

The commutators of $(T) y_\mu^*$ with four-point Green functions
can be calculated by the same method. Differentiating  Eq.
(\ref{Ward_Identity_With_Y}) with respect to $(\phi_{0i})_x$,
$(\phi_{0j})_y$, $(\phi_{0k})_z$, and $(\phi_{0l})_w$ and setting
all fields equal to 0, we obtain

\begin{eqnarray}
&&\hspace*{-3mm} \Big[(T) y_\mu^*, \frac{\delta^4\gamma}{\delta
(\phi_{0i})_x \delta (\phi_{0j})_y \delta (\phi_{0k})_z \delta
(\phi_{0l})_w} \Big] = \mbox{BrownLine}_\mu \cdot
\frac{\delta^4\gamma}{\delta (\phi_{0i})_x \delta (\phi_{j0})_y
\delta (\phi_{k0})_z \delta
(\phi_{l0})_w}\vphantom{\Bigg(}\nonumber\\
&&\hspace*{-3mm} + \mbox{BrownLine}_\mu[1,2]\cdot
\mbox{UsualLine}[1,2]\cdot \Bigg( \frac{\delta^2\gamma[1]}{\delta
(\phi_{0i})_x \delta (\phi_{0j})_y}\cdot
\frac{\delta^2\gamma[2]}{\delta (\phi_{0k})_z \delta
(\phi_{0l})_w}  + \frac{\delta^2\gamma[1]}{\delta (\phi_{0i})_x
\delta
(\phi_{0k})_z}\nonumber\\
&&\hspace*{-3mm} \times \frac{\delta^2\gamma[2]}{\delta
(\phi_{0j})_y \delta (\phi_{0l})_w} (-1)^{P_j P_k} +
\frac{\delta^2\gamma[1]}{\delta (\phi_{0i})_x \delta
(\phi_{0l})_w}\cdot \frac{\delta^2\gamma[2]}{\delta (\phi_{0j})_y
\delta (\phi_{0k})_z} (-1)^{P_j P_l + P_k P_l} +
\Big([1]\leftrightarrow [2]\Big)\Bigg).\nonumber\\
\end{eqnarray}

\noindent Similar identities can be written for the derivatives
with respect to $\phi_0^*$. In this case in the left hand side for
each $\phi_0^*$ it is necessary to replace $T_i{}^j$ by
$-(T)_j{}^i$.

\subsection{Commutators with $(T)\bar\theta$}
\hspace{\parindent} \label{Appendix_Commutators_Theta}

Commutators with $(T)\bar\theta$ can be calculated by the same
method as  the commutators with $(T) y_\mu^*$. Exactly as in the
previous section we obtain

\begin{eqnarray}\label{Ward_Identity_With_Theta}
&& \int d^8x\,(\bar\theta^{\dot a})_x (T)_i{}^j\Big(
(\phi^{i*}+\phi_{0}^{i*})_x \frac{\delta\gamma}{\delta
(\phi_{0}^{*j})_x}- (\phi_j+\phi_{0j})_x
\frac{\delta\gamma}{\delta (\phi_{0i})_x}  \Big)\nonumber\\
&& \qquad\qquad\qquad\qquad\qquad\qquad\quad = -i \int
d^8x\,(\bar\theta^{\dot a})_x (T)^j{}_i \Big(\frac{\delta}{\delta
(j^j)_x}\frac{\delta\gamma}{\delta (\phi_{0i})_x}\Big).\qquad
\end{eqnarray}

\noindent Differentiating this equation we obtain commutators with
various Green functions. For example, the commutator with the
two-point Green function can be written as

\begin{eqnarray}
&& \Big[(T) \bar\theta^{\dot a}, \frac{\delta^2\gamma}{\delta
(\phi_{0k})_y \delta (\phi_{0l})_z} \Big] = -i \int d^8x\,\Bigg[
(T)_i{}^j \Big(\Big(1+ \frac{\bar D^2 D^2}{16\partial^2}\Big)
\bar\theta^{\dot a}
\frac{\mbox{\boldmath$\delta$}}{\mbox{\boldmath$\delta$}
j^j}\Big)_{x} \frac{\delta^3\gamma}{\delta (\phi_{0k})_{y} \delta
(\phi_{0l})_{z}
\delta (\phi_{0i})_{x}}\qquad\nonumber\\
&& - (M T)^{ij} (\bar\theta^{\dot a})_x
\Big(\frac{D^2}{16\partial^2}\frac{\mbox{\boldmath$\delta$}}{\mbox{\boldmath$\delta$}
j^i} \Big)_{x} \frac{\mbox{\boldmath$\delta$}}{
\mbox{\boldmath$\delta$} (j^j)_{x}} \frac{\delta^2\gamma}{\delta
(\phi_{0k})_{y} \delta (\phi_{0l})_{z}}\Bigg].\qquad
\end{eqnarray}

\noindent This equation can be simplified using the relations

\begin{eqnarray}
&& \Big(1+ \frac{\bar D^2 D^2}{16\partial^2}\Big) \bar\theta^{\dot
a} \frac{\mbox{\boldmath$\delta$}}{\mbox{\boldmath$\delta$} j^j} =
\frac{\bar D^{\dot a} D^2}{8\partial^2}
\frac{\mbox{\boldmath$\delta$}}{\mbox{\boldmath$\delta$} j^j};\nonumber\\
&& \int d^8x\, (M T)^{ij}\, \bar\theta^{\dot a}\,
\frac{D^2}{16\partial^2}\frac{\mbox{\boldmath$\delta$}}{\mbox{\boldmath$\delta$}
j^i} \cdot \frac{\mbox{\boldmath$\delta$}}{
\mbox{\boldmath$\delta$} j^j} = \int d^8x\, (M T)^{ij}\,
\bar\theta^{\dot a}\,
\frac{\mbox{\boldmath$\delta$}}{\mbox{\boldmath$\delta$} j^i}
\cdot \frac{D^2}{16\partial^2} \frac{\mbox{\boldmath$\delta$}}{
\mbox{\boldmath$\delta$} j^j}  = 0.\qquad
\end{eqnarray}

\noindent (The last equality follows from Eq.
(\ref{MT_Symmetry}).) Then we obtain

\begin{equation}
\Big[(T) \bar\theta^{\dot a}, \frac{\delta^2\gamma}{\delta
(\phi_{0k})_y \delta (\phi_{0l})_z} \Big] = -i\cdot
\mbox{BlueLine}^{\dot a}[1] \cdot \frac{\delta^2\gamma}{\delta
(\phi_{0k})_y \delta (\phi_{0l})_z},
\end{equation}

\noindent where we use the notation

\begin{equation}
\mbox{BlueLine}^{\dot a}[\alpha;1,2] = \int d^8x\,(\alpha)_x
(T)^j{}_i \Big(\frac{\bar D^{\dot a} D^2}{8\partial^2}
\frac{\mbox{\boldmath$\delta$}}{\mbox{\boldmath$\delta$}_2
j^j}\cdot \frac{\delta}{\delta_1 \phi_{0i}}\Big)_{x}.
\end{equation}

\noindent As always, the subscripts $1$ and $2$ denote end points
of the line. If these points coincide, we sometimes for simplicity
omit these indexes. Other commutators can be found similarly. For
example,

\begin{equation}
\Big[(T) \bar\theta^{\dot a}, \frac{\delta^2\gamma}{\delta
(\phi_{0k})_y \delta (\phi_{0}^{*l})_z} \Big] = -i\cdot
\mbox{BlueLine}^{\dot a}[1] \cdot \frac{\delta^2\gamma}{\delta
(\phi_{0k})_y \delta (\phi_{0}^{*l})_z}
\end{equation}

\noindent for the two-point function, or

\begin{eqnarray}
&&\hspace*{-5mm} \Big[(T) \bar\theta^{\dot a},
\frac{\delta^4\gamma}{\delta (\phi_{0i})_x \delta (\phi_{0j})_y
\delta \phi_{0k})_z \delta (\phi_{0l})_w} \Big] = -i\cdot
\mbox{BlueLine}^{\dot a}[1] \cdot \frac{\delta^4\gamma}{\delta
(\phi_{0i})_x \delta (\phi_{j0})_y \delta (\phi_{k0})_z \delta
(\phi_{l0})_w}\vphantom{\Bigg(}\nonumber\\
&&\hspace*{-5mm} - i\cdot \mbox{BlueLine}^{\dot a}[1; 1,2]\cdot
\mbox{UsualLine}[1,2]\cdot \Bigg( \frac{\delta^2\gamma[1]}{\delta
(\phi_{0i})_x \delta (\phi_{0j})_y}\cdot
\frac{\delta^2\gamma[2]}{\delta (\phi_{0k})_z \delta
(\phi_{0l})_w}+ \frac{\delta^2\gamma[1]}{\delta (\phi_{0i})_x
\delta (\phi_{0k})_z}\nonumber\\
&&\hspace*{-5mm} \times \frac{\delta^2\gamma[2]}{\delta
(\phi_{0j})_y \delta (\phi_{0l})_w} (-1)^{P_j P_k} +
\frac{\delta^2\gamma[1]}{\delta (\phi_{0i})_x \delta
(\phi_{0l})_w}\cdot \frac{\delta^2\gamma[2]}{\delta (\phi_{0j})_y
\delta (\phi_{0k})_z}(-1)^{P_j P_l + P_k P_l} +
\Big([1]\leftrightarrow [2]\Big) \Bigg)\nonumber\\
\end{eqnarray}

\noindent for the four-point function.

\subsection{Commutators with propagators}
\hspace{\parindent} \label{Appendix_Commutators_Inverse}

In order to calculate commutators of $(T) y_\mu^*$ or $(T)
\bar\theta$ with various Feynman diagrams according to the
prescription (\ref{Trace_Notation}), it is necessary to commute
$(T) y_\mu^*$ or $(T) \bar\theta$ with inverse Green functions.
Here we demonstrate, how this can be made. As a starting point we
consider the identities (\ref{Explicit_Green_Functions}), which
can be presented in the following matrix form:

\begin{eqnarray}\label{Matrix_Equation}
&&\hspace*{-7mm} \int d^8z\,\left(\begin{array}{cc} {\displaystyle
\frac{\delta^2\gamma}{\delta(\phi_{0}^{*i})_x \delta(\phi_{0k})_z}
\vphantom{\Bigg(} } & {\displaystyle
\frac{\delta^2\gamma}{\delta(\phi_{0}^{*i})_x
\delta(\phi_{0}^{*k})_z} + M^*_{ki} \frac{\bar D^2}{16\partial^2}
\delta^8_{xz}}
\\
{\displaystyle \frac{\delta^2\gamma}{\delta(\phi_{0i})_x
\delta(\phi_{0k})_z} + M^{ki} \frac{D^2}{16\partial^2}
\delta^8_{xz} } & {\displaystyle
\frac{\delta^2\gamma}{\delta(\phi_{0i})_x \delta(\phi_{0}^{*k})_z}
\vphantom{\Bigg(} }
\end{array}\right)\nonumber\\
&&\vphantom{1}\\
&&\hspace*{-7mm} \times \left(
\begin{array}{cc}
{\displaystyle \Big(\frac{\delta^2\gamma}{\delta(\phi_{k})_z
\delta(\phi^{*j})_y}\Big)^{-1}} &
{\displaystyle \Big(\frac{\delta^2\gamma}{\delta(\phi_{k})_z
\delta(\phi_{j})_y}\Big)^{-1} \vphantom{\Bigg(} }\\
{\displaystyle \Big(\frac{\delta^2\gamma}{\delta(\phi^{*k})_z
\delta(\phi^{*j})_y}\Big)^{-1}} & {\displaystyle
\Big(\frac{\delta^2\gamma}{\delta(\phi^{*k})_z
\delta(\phi_{j})_y}\Big)^{-1} \vphantom{\Bigg(} }
\end{array}
\right) = \left(\begin{array}{cc} {\displaystyle - \delta_i^j \Big(\frac{\bar
D^2 D^2}{16\partial^2}\Big)_x \vphantom{\Bigg(}} &
{\displaystyle 0}\\
{\displaystyle 0} & {\displaystyle - \delta_j^i\Big(\frac{D^2 \bar
D^2}{16\partial^2}\Big)_x \vphantom{\Bigg(} }
\end{array}\right) \delta^8_{xy}.\nonumber
\end{eqnarray}

\noindent Let us commute this equation with $(T) y_\mu^*$ or
$(T)\bar\theta^a$, taking into account that the commutators in the
right hand side do not vanish. We use the identities

\begin{eqnarray}
&& \Big[y_\mu^*, \frac{\bar D^2 D^2}{16\partial^2}\Big] =
\frac{\bar D^2 D^2\partial_\mu}{8\partial^4} +
i(\gamma_\mu)^{a{\dot b}} \theta_a \frac{\bar D_{\dot b}
D^2}{4\partial^2}; \qquad\qquad \Big[\bar\theta^{\dot a},
\frac{\bar D^2 D^2}{16\partial^2}\Big] = - \frac{\bar D^{\dot a}
D^2}{8\partial^2};\qquad
\nonumber\\
&& D^2 \Big[y_\mu^*,\frac{\bar D^2}{16\partial^2}\Big] D^2 = -
[y_\mu^*, D^2]  = 0; \qquad\qquad\qquad\qquad D^2
\Big[\bar\theta^{\dot a},\frac{\bar D^2}{16\partial^2}\Big] D^2 =
0.
\end{eqnarray}

\noindent For example, let us consider the commutator

\begin{eqnarray}
&& \Big[(T) y_\mu^*, \Big(\frac{\delta^2\gamma}{\delta(\phi_{i})_x
\delta(\phi^{*j})_y}\Big)^{-1} \Big]\nonumber\\
&&\qquad\qquad \equiv (y_\mu^*)_x (T)_i{}^m
\Big(\frac{\delta^2\gamma}{\delta(\phi_{m})_x
\delta(\phi^{*j})_y}\Big)^{-1} -
\Big(\frac{\delta^2\gamma}{\delta(\phi_{i})_x
\delta(\phi^{*m})_y}\Big)^{-1} (T)_m{}^j (y_\mu^*)_y.\qquad
\end{eqnarray}

\noindent Commuting $(T) y_\mu^*$ with Eq. (\ref{Matrix_Equation})
after some simple transformations we obtain

\begin{eqnarray}
&& \Big[(T) y_\mu^*, \Big(\frac{\delta^2\gamma}{\delta(\phi_{i})_x
\delta(\phi^{*j})_y}\Big)^{-1} \Big] = - \int
d^8z\,d^8w\,\nonumber\\
&& \times \Bigg\{\Big(\frac{\delta^2\gamma}{\delta(\phi_{i})_x
\delta(\phi_k)_z}\Big)^{-1}\Big[ (T) y_\mu^*,
\frac{\delta^2\gamma}{\delta(\phi_{0k})_z
\delta(\phi_0^{*l})_w}\Big]
\Big(\frac{\delta^2\gamma}{\delta(\phi^{*l})_w
\delta(\phi^{*j})_y}\Big)^{-1}\nonumber\\
&& + \Big(\frac{\delta^2\gamma}{\delta(\phi_{i})_x
\delta(\phi_k)_z}\Big)^{-1} \Big[(T) y_\mu^*,
\frac{\delta^2\gamma}{\delta(\phi_{0k})_z
\delta(\phi_{0l})_w}\Big]
\Big(\frac{\delta^2\gamma}{\delta(\phi_{l})_w
\delta(\phi^{*j})_y}\Big)^{-1}\nonumber\\
&& + \Big(\frac{\delta^2\gamma}{\delta(\phi_{i})_x
\delta(\phi^{*k})_z}\Big)^{-1} \Big[ (T) y_\mu^*,
\frac{\delta^2\gamma}{\delta(\phi^{*k}_{0})_z
\delta(\phi_0^{*l})_w} \Big]
\Big(\frac{\delta^2\gamma}{\delta(\phi^{*l})_w
\delta(\phi^{*j})_y}\Big)^{-1}\nonumber\\
&& + \Big(\frac{\delta^2\gamma}{\delta(\phi_{i})_x
\delta(\phi^{*k})_z}\Big)^{-1} \Big[(T) y_\mu^*,
\frac{\delta^2\gamma}{\delta(\phi^{*k}_{0})_z
\delta(\phi_{0l})_w}\Big]
\Big(\frac{\delta^2\gamma}{\delta(\phi_{l})_w
\delta(\phi^{*j})_y}\Big)^{-1} \Bigg\}\nonumber\\
&& + (T)_i{}^m \Big(\frac{2 \partial_\mu}{\partial^2} -
i(\gamma_\mu)^{a{\dot b}}\theta_a \frac{\bar D_{\dot b}
D^2}{4\partial^2}\Big)_x
\Big(\frac{\delta^2\gamma}{\delta(\phi_{m})_x
\delta(\phi^{*j})_y}\Big)^{-1}\nonumber\\
&& + \int d^8z\,\Big(\frac{\delta^2\gamma}{\delta (\phi_i)_x
\delta (\phi_k)_z}\Big)^{-1} (M T)^{lk}
\Big(\frac{D^2\partial_\mu}{8\partial^4}\Big)_z
\Big(\frac{\delta^2\gamma}{\delta (\phi_l)_z \delta
(\phi^{*j})_y}\Big)^{-1}.
\end{eqnarray}

\noindent All commutators here can be calculated according to the
prescription obtained in the previous section:

\begin{eqnarray}
&& \Big[(T) y_\mu^*, \Big(\frac{\delta^2\gamma}{\delta(\phi_{i})_x
\delta(\phi^{*j})_y}\Big)^{-1} \Big] = - \int
d^8z\,d^8w\,\nonumber\\
&& \times \Bigg\{\Big(\frac{\delta^2\gamma}{\delta(\phi_{i})_x
\delta(\phi_k)_z}\Big)^{-1} \mbox{BrownLine}_\mu \cdot
\frac{\delta^2\gamma}{\delta(\phi_{0k})_z \delta(\phi_0^{*l})_w}
\Big(\frac{\delta^2\gamma}{\delta(\phi^{*l})_w
\delta(\phi^{*j})_y}\Big)^{-1}\nonumber\\
&& + \Big(\frac{\delta^2\gamma}{\delta(\phi_{i})_x
\delta(\phi_k)_z}\Big)^{-1}  \mbox{BrownLine}_\mu \cdot
\frac{\delta^2\gamma}{\delta(\phi_{0k})_z \delta(\phi_{0l})_w}
\Big(\frac{\delta^2\gamma}{\delta(\phi_{l})_w
\delta(\phi^{*j})_y}\Big)^{-1}\nonumber\\
&& + \Big(\frac{\delta^2\gamma}{\delta(\phi_{i})_x
\delta(\phi^{*k})_z}\Big)^{-1} \mbox{BrownLine}_\mu \cdot
\frac{\delta^2\gamma}{\delta(\phi^{*k}_{0})_z
\delta(\phi_0^{*l})_w}
\Big(\frac{\delta^2\gamma}{\delta(\phi^{*l})_w
\delta(\phi^{*j})_y}\Big)^{-1}\nonumber\\
&& + \Big(\frac{\delta^2\gamma}{\delta(\phi_{i})_x
\delta(\phi^{*k})_z}\Big)^{-1} \mbox{BrownLine}_\mu \cdot
\frac{\delta^2\gamma}{\delta(\phi^{*k}_{0})_z \delta(\phi_{0l})_w}
\Big(\frac{\delta^2\gamma}{\delta(\phi_{l})_w
\delta(\phi^{*j})_y}\Big)^{-1} \Bigg\}\nonumber\\
&& + (T)_i{}^m \Big(\frac{2 \partial_\mu}{\partial^2} -
i(\gamma_\mu)^{a{\dot b}}\theta_a \frac{\bar D_{\dot b}
D^2}{4\partial^2}\Big)_x
\Big(\frac{\delta^2\gamma}{\delta(\phi_{m})_x
\delta(\phi^{*j})_y}\Big)^{-1}\nonumber\\
&& + \int d^8z\,\Big(\frac{\delta^2\gamma}{\delta (\phi_i)_x
\delta (\phi_k)_z}\Big)^{-1} (M T)^{lk}
\Big(\frac{D^2\partial_\mu}{8\partial^4}\Big)_z
\Big(\frac{\delta^2\gamma}{\delta (\phi_l)_z \delta
(\phi^{*j})_y}\Big)^{-1}.
\end{eqnarray}

\noindent This expression can be written in a very compact form

\begin{eqnarray}\label{Commutator_With_Inverse_Function}
&& \Big[(T) y_\mu^*, \Big(\frac{\delta^2\gamma}{\delta(\phi_{i})_x
\delta(\phi^{*j})_y}\Big)^{-1} \Big] = - \mbox{BrownLine}_\mu
\cdot \Big(\frac{\mbox{\boldmath$\delta$}^2\gamma}{
\mbox{\boldmath$\delta$} (j^*_{j})_y
\mbox{\boldmath$\delta$} (j^i)_x} + i (\phi^{*j}_0)_y (\phi_{0i})_x \Big).\qquad\quad
\end{eqnarray}

\noindent (The fields here should be set to 0.) The other
commutators can be considered similarly. Commutators with $(T)
\bar\theta_{\dot a}$ are calculated by the exactly the same
method. The result can be obtained by the substitution

\begin{equation}
\mbox{BrownLine}_\mu \to -i\cdot\mbox{BlueLine}_{\dot a}[1].
\end{equation}

\noindent In particular, there are no terms containing the masses
in the commutators with $(T) \bar\theta^{\dot a}$.

\section{Identities for effective lines}
\label{Appendix_Auxiliary_Identities}

\subsection{Proof of the identity (\ref{YellowBrown->YellowGreen})}
\hspace{\parindent}\label{Appendix_Identity1}

In order to present the two-loop effective diagram with the yellow
line presented in Fig. \ref{Figure_Double_Diagram} (plus
$\Delta_1$) as an integral of a total derivative, we use the
identity

\begin{equation}\label{Green-Brown_Identity}
2\cdot \mbox{YellowLine}_\mu[\, \bar\theta^{\dot a}
(\gamma^\mu)_{\dot a}{}^b \theta_b]\cdot
\,\mbox{GreenLine}\cdot\gamma = -
\mbox{YellowLine}_\mu[\theta^4]\cdot
\mbox{BrownLine}^\mu\cdot\gamma.
\end{equation}

\noindent In this section we prove this equality. Using the same
method in the subsequent sections we prove more complicated
identities relating the effective lines. First, we note that both
sides of the considered equation are quadratic in $\bar\theta$. As
a consequence, it is possible to shift $\bar\theta$ to an
arbitrary point of a diagram, because the integral over $d^8x$
does not vanish only if the integrand contains $\theta^4$.
Comparing definitions of the operators $\mbox{GreenLine}$ and
$\mbox{RedLine}_{\dot a}$ and taking into account the possibility
of shifting $\bar\theta^{\dot a}$ we easily obtain (omitting
unessential terms which do not contain $\bar\theta$)

\begin{equation}\label{GreenLine->RedLine}
4\cdot \mbox{GreenLine}[1,2] = (\bar\theta^{\dot c})_z \cdot
\mbox{RedLine}_{\dot c}[1,2],
\end{equation}

\noindent where $z$ is an arbitrary point of the considered
supergraph. Therefore, taking into account that $\bar\theta^{\dot
c} \bar\theta_{\dot a} = \delta^{\dot c}_{\dot a}\,
\bar\theta^{\dot b} \bar\theta_{\dot b}/2$, we see that the
equality (\ref{Green-Brown_Identity}) is equivalent to the
identity

\begin{equation}\label{RedLine_Equality}
(\gamma^\mu)^{{\dot a}b} (\theta_b)_z \cdot \mbox{RedLine}_{\dot
a}[1,2] = 4 (\theta^a \theta_a)_z\cdot \mbox{BrownLine}^\mu[1,2] +
O(\theta).
\end{equation}

\noindent Let us compare terms quadratic in $\theta$. In these
terms we can make arbitrary shifts of $\theta$, because terms
proportional to the first degree of $\theta$ (denoted by
$O(\theta)$) vanish after integrating over $d^4\theta$. Then the
considered terms in the left hand side of Eq.
(\ref{RedLine_Equality}) can be written as

\begin{eqnarray}
&& (\gamma_\mu)^{{\dot a}b} (\theta_b)_z \cdot \int d^8x
(\gamma^\nu)_{\dot a}{}^{c} (\theta_c)_x \Big(8i (T)^j{}_i
\frac{\partial_\nu}{\partial^2}
\frac{\mbox{\boldmath$\delta$}}{\mbox{\boldmath$\delta$} j^j}
\frac{\delta}{\delta\phi_{0i}} - i(MT)^{ij}
\frac{D^2\partial_\nu}{4\partial^4}
\frac{\mbox{\boldmath$\delta$}}{\mbox{\boldmath$\delta$} j^i}
\frac{\mbox{\boldmath$\delta$}}{\mbox{\boldmath$\delta$} j^j}
\Big)\qquad\nonumber\\
&& = - 4 (\theta^b\theta_b)_z \cdot \int d^8x \Big(2i (T)^j{}_i
\frac{\partial_\mu}{\partial^2}
\frac{\mbox{\boldmath$\delta$}}{\mbox{\boldmath$\delta$} j^j}
\frac{\delta}{\delta\phi_{0i}} - i(MT)^{ij}
\frac{D^2\partial_\mu}{16\partial^4}
\frac{\mbox{\boldmath$\delta$}}{\mbox{\boldmath$\delta$} j^i}
\frac{\mbox{\boldmath$\delta$}}{\mbox{\boldmath$\delta$} j^j}
\Big) + O(\theta)
\end{eqnarray}

\noindent and coincide with the terms quadratic in $\theta$ in the
right hand side of Eq. (\ref{RedLine_Equality}). Terms
proportional to the third degree of $\theta$ can be investigated
similarly. In the left hand side such terms are given by

\begin{equation}
(\gamma^\mu)^{{\dot a}b} (\theta_b)_z \cdot \int d^8x\,(T)^j{}_i\,
(\theta^c \theta_c)_x \frac{\bar D_{\dot a} D^2}{\partial^2 }
\frac{\mbox{\boldmath$\delta$}}{\mbox{\boldmath$\delta$} j^j}
\frac{\delta}{\delta\phi_{0i}}.
\end{equation}

\noindent In a Feynman graph the points $z$ and $x$ are connected
by a sequence of vertices and propagators. This allows to write
the left part of the above expression in the form ($P_A=0$)

\begin{equation}\label{Aux_Identity}
(\gamma^\mu)^{{\dot a}b} \theta_b\, A\, \theta^c \theta_c =
(\gamma^\mu)^{{\dot a}b} [\theta_b,\, A]\, \theta^c \theta_c =
(\gamma^\mu)^{{\dot a}b} \theta^c \theta_c\, [\theta_b,\, A] +
O(\theta) = - (\gamma^\mu)^{{\dot a}b} \theta^c \theta_c\, A
\theta_b + O(\theta),
\end{equation}

\noindent where $A$ is a differential operator, which does not
explicitly depend on $\theta$. (This operator encodes the sequence
of vertices and propagators which connect the points $z$ and $x$.)
Similarly, for an arbitrary $P_A$

\begin{equation}\label{Qubic_Identity}
\theta_b\, A\, \theta^c \theta_c = - (-1)^{P_A} \theta^c
\theta_c\, A\, \theta_b + O(\theta).
\end{equation}

\noindent Terms $O(\theta)$ vanish after integration over
$d^4\theta$. Omitting these terms we see that the last expression
in Eq. (\ref{Aux_Identity}) corresponds to

\begin{equation}
- (\theta^c\theta_c)_z \cdot \int d^8x\,(T)^j{}_i\,
(\gamma^\mu)^{{\dot a}b} (\theta_b)_x \frac{\bar D_{\dot a}
D^2}{\partial^2}
\frac{\mbox{\boldmath$\delta$}}{\mbox{\boldmath$\delta$} j^j}
\frac{\delta}{\delta\phi_{0i}}
\end{equation}

\noindent and coincides with the terms cubic in $\theta$ in the
right hand side of Eq. (\ref{RedLine_Equality}). Thus, we have
proved Eq. (\ref{RedLine_Equality}) and Eq.
(\ref{Green-Brown_Identity}).

\subsection{Auxiliary identities} \hspace{\parindent}
\label{Appendix_Auxiliary_Algebra}

In order to compare different groups of effective diagrams it is
necessary to use some identities which relate various effective
lines. All these identities follow from some simple commutators of
$\theta^a$ with differential operators containing supersymmetric
covariant derivatives and usual derivatives. In this subsection we
prove some simple algebraic equalities which allow to relate
various effective lines:

\begin{eqnarray}\label{Auxiliary_Identity1}
&& \theta_a AB \theta^b \theta_b + (-1)^{P_A+P_B} A \theta^b
\theta_b B \theta_a - \theta_a A \theta^b \theta_b B =O(\theta);
\vphantom{\Big(}\\
\label{Auxiliary_Identity2} && \theta^b \theta_b A B \theta_a +
(-1)^{P_A+P_B}\theta_a A\theta^b \theta_b B - A \theta^b \theta_b
B \theta_a = O(\theta);
\vphantom{\Big(}\\
\label{Auxiliary_Identity3} && \theta^a\theta_a AB
\theta^b\theta_b + 2(-1)^{P_A+P_B}\theta^a A \theta^b\theta_b B
\theta_a - \theta^a\theta_a A \theta^b\theta_b B -
A\theta^a\theta_a B \theta^b\theta_b
=O(\theta),\quad\vphantom{\Big(}
\end{eqnarray}

\noindent where $A$ and $B$ are differential operators which do
not explicitly depend on $\theta$. Actually, these operators
correspond to sequences of vertices and propagators in a Feynman
graph connecting two fixed points.

In order to prove the first identity we rewrite its left hand side
as

\begin{equation}
[\theta_a,A\} B \theta^b\theta_b + (-1)^{P_A} A [\theta_a, B\}
\theta^b \theta_b + (-1)^{P_A+P_B} A\, \theta^b \theta_b
[B,\theta_a\} -[\theta_a, A\} \theta^b \theta_b B.
\end{equation}

\noindent Evidently, $[A,\theta_a\}$ and $[B,\theta_a\}$ do not
explicitly depend on $\theta$. Therefore, the whole expression is
quadratic in $\theta$, and shifts of $\theta^a\theta_a$ can change
only the terms $O(\theta)$. As a consequence, the considered
expression can be rewritten as

\begin{equation}
\theta^b \theta_b \Big([\theta_a,A\} B + (-1)^{P_A} A[\theta_a,
B\} + (-1)^{P_A+P_B} A [B,\theta_a\} -[\theta_a, A\} B\Big)
+O(\theta) = O(\theta).
\end{equation}

\noindent The second and third identities can be proved similarly.
For example, the left hand side of the third identity can be
presented in the form

\begin{equation}
[\theta^a,[\theta_a, AB\}\} \theta^b\theta_b + 2 (-1)^{P_A+P_B}
[\theta^a,A\} \theta^b \theta_b [B,\theta_a\} -
[\theta^a,[\theta_a, A\}\} \theta^b \theta_b B - A
[\theta^a,[\theta_a,B\}\} \theta^b \theta_b.
\end{equation}

\noindent Again, shifts of $\theta^b\theta_b$ change only the
terms $O(\theta)$ and this expression can be presented as

\begin{eqnarray}
&& \theta^b \theta_b \Big( [\theta^a,[\theta_a, A\}\} B + 2
(-1)^{P_A} [\theta^a, A\} [\theta_a,B\} + A [\theta^a, [\theta_a,
B\}\} -  2(-1)^{P_A} [\theta^a,A\} [\theta_a,
B\}\quad\nonumber\\
&& - [\theta^a,[\theta_a, A\}\} B - A
[\theta^a,[\theta_a,B\}\}\Big) +O(\theta)= O(\theta).
\end{eqnarray}

\subsection{Proof of identity presented in Fig.
\ref{Figure_Auxiliary_Identity_Blue_Green}}\label{Appendix_Pink_Identity}
\hspace{\parindent}

Using the identities (\ref{Auxiliary_Identity1}) ---
(\ref{Auxiliary_Identity3}) it is possible to prove the identity
presented in Fig. \ref{Figure_Auxiliary_Identity_Blue_Green},
which can be written as

\begin{eqnarray}\label{Pink_Identity_Appendix}
&&\hspace*{-7mm} (\theta^4)_z \Big(\mbox{BlueLine}_{\dot
b}[1;1,2]\cdot \mbox{PinkLine}^{\dot b}[3,4] +
\mbox{PinkLine}_{\dot b}[1,2] \cdot
\mbox{BlueLine}^{\dot b}[1;3,4]\nonumber\\
&&\hspace*{-7mm} - (\gamma^\mu)^{a{\dot b}} \mbox{BlueLine}_{\dot
b}[\theta_a;1,2]\cdot \mbox{BrownLine}_\mu[3,4] -
(\gamma^\mu)^{a{\dot b}} \mbox{BrownLine}_\mu[1,2]\cdot
\mbox{BlueLine}_{\dot b}[\theta_a;3,4] \Big)
\nonumber\\
&&\hspace*{-7mm} = 2\cdot \mbox{BlueLine}_{\dot
b}[\theta^a\theta_a \bar\theta^{\dot b};1,2]\cdot
\mbox{GreenLine}[3,4] + 2\cdot \mbox{GreenLine}[1,2]\cdot
\mbox{BlueLine}_{\dot b}[\theta^a\theta_a
\bar\theta^{\dot b};3,4]\vphantom{\Big(}\nonumber\\
&& \hspace*{-7mm} + O(\theta^3),\vphantom{\Big(}
\end{eqnarray}

\noindent where $z$ is an arbitrary point of the considered
supergraph. (Certainly, we assume that all effective lines are
included into a connected Feynman graph.)

Both sides of the considered identity are quadratic in
$\bar\theta$. Therefore, it is possible to shift $\bar\theta$ to
an arbitrary point of the supergraph, because the terms
$O(\bar\theta)$ vanish after the integration over $d^4\theta$.
Using Eq. (\ref{GreenLine->RedLine}) one can equivalently rewrite
the identity (\ref{Pink_Identity_Appendix}) in the form

\begin{eqnarray}\label{Pink_Identity_Without_Bar_Theta_Appendix}
&&\hspace*{-5mm} (\theta^a \theta_a)_z \Big(\mbox{BlueLine}_{\dot
b}[1;1,2]\cdot \mbox{PinkLine}^{\dot b}[3,4] +
\mbox{PinkLine}_{\dot b}[1,2] \cdot
\mbox{BlueLine}^{\dot b}[1;3,4]\nonumber\\
&&\hspace*{-5mm} - (\gamma^\mu)^{a{\dot b}} \mbox{BlueLine}_{\dot
b}[\theta_a;1,2]\cdot \mbox{BrownLine}_\mu[3,4] -
(\gamma^\mu)^{a{\dot b}} \mbox{BrownLine}_\mu[1,2]\cdot
\mbox{BlueLine}_{\dot b}[\theta_a;3,4] \Big)
\nonumber\\
&&\hspace*{-5mm} =\frac{1}{4}\cdot \mbox{BlueLine}_{\dot
b}[\theta^a\theta_a;1,2]\cdot \mbox{RedLine}^{\dot b}[3,4] +
\frac{1}{4}\cdot\mbox{RedLine}_{\dot b}[1,2]\cdot
\mbox{BlueLine}^{\dot b}[\theta^a\theta_a;3,4] +
O(\theta).\vphantom{\Big(}\quad\nonumber\\
\end{eqnarray}

This identity contains terms cubic in $\theta$ and terms quartic
in $\theta$, which will be considered separately. We start with
the cubic terms. Let $A_{\dot b}$ denotes a sequence of lines and
vertices connecting the points $z$ and $x$ which also includes
terms coming from the operator

\begin{equation}
\int d^8x\,\Big((T)^j{}_i \Big(\frac{\bar D_{\dot b}
D^2}{8\partial^2}
\frac{\mbox{\boldmath$\delta$}}{\mbox{\boldmath$\delta$}_2
j^j}\Big) \frac{\delta}{\delta_1 \phi_{0i}}\Big)_x.
\end{equation}

\noindent Similarly, let $B$ denotes a sequence of line and
vertices connecting the points $z$ and $y$. Note that in this
operator we do not include terms coming from the operator

\begin{equation}
\int d^8y\,\Big( (T)^j{}_i\, \frac{2i\partial_\mu}{\partial^2}
\frac{\mbox{\boldmath$\delta$}}{\mbox{\boldmath$\delta$}_4 j^j}
\frac{\delta}{\delta_3 \phi_{0i}} - i(MT)^{ij}
\frac{D^2\partial_\mu}{16\partial^4}
\frac{\mbox{\boldmath$\delta$}}{\mbox{\boldmath$\delta$}_4 j^i}
\frac{\mbox{\boldmath$\delta$}}{\mbox{\boldmath$\delta$}_3
j^j}\Big)_{y}
\end{equation}

\noindent (Evidently, $P_A=1$ and $P_B=0$.) Then the contributions
of the first and third terms in the left hand side of Eq.
(\ref{Pink_Identity_Without_Bar_Theta_Appendix}) can be formally
written in the form

\begin{equation}
A_{\dot b} \theta^a \theta_a B (\gamma^\mu)^{{\dot b}c} \theta_c +
(\gamma^\mu)^{{\dot b}c} \theta_c A_{\dot b} \theta^a \theta_a B.
\end{equation}

\noindent Using the identity (\ref{Auxiliary_Identity2}) this
expression can be rewritten as

\begin{equation}
(\gamma^\mu)^{{\dot b}c} \theta^a \theta_a A_{\dot b} B \theta_c +
O(\theta)
\end{equation}

\noindent and coincides with the contribution of the first term in
the right hand side of Eq.
(\ref{Pink_Identity_Without_Bar_Theta_Appendix}). (So far we
discuss only terms cubic in $\theta$.) Using the same method we
prove that the second and fourth terms in the left hand side of
Eq. (\ref{Pink_Identity_Without_Bar_Theta_Appendix}) give the
second term in the right hand side. Therefore, the terms cubic in
$\theta$ coincide. Let us now verify that terms quartic in
$\theta$ are also the same in both sides of Eq.
(\ref{Pink_Identity_Without_Bar_Theta_Appendix}). In this case the
operators $A_{\dot b}$ and $B$ are defined exactly as earlier. In
particular, $B$ denotes a sequence of line and vertices connecting
the points $z$ and $y$ and does not include terms coming from the
operator

\begin{equation}
\int d^8y\,\Big( (T)^j{}_i \Big(\frac{\bar D_{\dot b}
D^2}{8\partial^2}
\frac{\mbox{\boldmath$\delta$}}{\mbox{\boldmath$\delta$}_4 j^j}
\Big) \frac{\delta}{\delta_3 \phi_{0i}}\Big)_{y}.
\end{equation}

\noindent (As earlier, in this case $P_A=1$, $P_B=0$.) Then terms
quartic in $\theta$ can be formally presented in the form

\begin{equation}\label{Forth_Order_Terms}
4 A_{\dot b}\, \theta^c \theta_c\, B\, \theta^d\theta_d + 4
\theta^c \theta_c\, A_{\dot b}\, \theta^d \theta_d B + 4
(\gamma^\mu)_{\dot d}{}^a \theta_a A^{\dot d}\, \theta^c \theta_c
B (\gamma_\mu)_{\dot b}{}^{e} \theta_e.
\end{equation}

\noindent It is easy to see that the last term in this equation
can be equivalently rewritten as

\begin{equation}
8\, \theta^a A_{\dot b} \theta^c \theta_c B \theta_a.
\end{equation}

\noindent This allows to apply Eq. (\ref{Auxiliary_Identity3}). As
a result, we obtain that terms of the forth order in $\theta$ are
given by the expression

\begin{equation}
4\, \theta^c \theta_c A_{\dot b} B \theta^d\theta_d + O(\theta),
\end{equation}

\noindent which coincides with the corresponding terms in the
right hand side of Eq.
(\ref{Pink_Identity_Without_Bar_Theta_Appendix}). Taking into
account that the cubic terms also coincide, we conclude that the
identity (\ref{Pink_Identity}) is proved.

\subsection{Proof of identity presented in Fig. \ref{Figure_Auxiliary_Identity}}
\hspace{\parindent}\label{Appendix_Red_Identity}

In order to verify that a $\beta$-function is given by integrals
of double total derivative we use the identity

\begin{eqnarray}\label{Red_Identity}
&& (\theta^4)_z \Big( \mbox{BlueLine}_{\dot b}[1;1,2]\cdot
\mbox{RedLine}^{\dot b}[3,4] + \mbox{RedLine}_{\dot b}[1,2] \cdot
\mbox{BlueLine}^{\dot b}[1;3,4]\\
&& + 2\cdot \mbox{BrownLine}^\mu[1,2]\cdot
\mbox{BrownLine}_\mu[3,4] \Big)= 4\cdot\mbox{GreenLine}[1,2]\cdot
\mbox{GreenLine}[3,4] + O(\theta^3),\nonumber\vphantom{\Big(}\quad
\end{eqnarray}

\noindent which is proved in this section. Exactly as in the
previous section we note that both sides of this identity are
quadratic in $\bar\theta$ and, therefore, it is possible to shift
$\bar\theta$ to an arbitrary point of the supergraph. Using Eq.
(\ref{GreenLine->RedLine}) it is easy to see that Eq.
(\ref{Red_Identity}) can be equivalently written in the form

\begin{eqnarray}
&& (\theta^a \theta_a)_z \Big( \mbox{BlueLine}_{\dot
b}[1;1,2]\cdot \mbox{RedLine}^{\dot b}[3,4] + \mbox{RedLine}_{\dot
b}[1,2] \cdot \mbox{BlueLine}^{\dot b}[1;3,4]\\
&& + 2\cdot \mbox{BrownLine}^\mu[1,2]\cdot
\mbox{BrownLine}_\mu[3,4] \Big)= -
\frac{1}{8}\cdot\mbox{RedLine}^b[1,2]\cdot \mbox{RedLine}_b[3,4] +
O(\theta).\nonumber\vphantom{\Big(}\quad
\end{eqnarray}

\noindent This equation contains terms quadratic, cubic, and
quartic in $\theta$. The quartic terms in both sides are equal to
the corresponding quartic terms in Eq.
(\ref{Pink_Identity_Without_Bar_Theta_Appendix}), which are
considered in the previous section, multiplied by 2. Similarly,
terms cubic in $\theta$ are obtained by multiplying cubic terms in
Eq. (\ref{Pink_Identity_Without_Bar_Theta_Appendix}) by 4.
Therefore, it is necessary to consider only terms quadratic in
$\theta$. It is evident that in such terms $\theta$-s can be
shifted to an arbitrary point of the supergraph. Then the required
equality of the quadratic terms follows from the algebraic
identity

\begin{eqnarray}
2\cdot \theta^a \theta_a \eta^{\mu\nu} = - \frac{1}{8}\cdot
4(\gamma^\mu)^{{\dot b}c} \theta_c \cdot 4(\gamma^\nu)_{\dot
b}{}^d \theta_d.
\end{eqnarray}

\noindent (In the right hand side we shift both $\theta$-s to the
same point $z$.)

\section{Derivation of the expression (\ref{NSVZ_Contribution})}
\hspace{\parindent}\label{Appendix_Yellow_Diagram_Contribution}

The sum of $\Delta_1$, which is given by Eq. (\ref{Delta1}), and
the expression (\ref{Total_Derivative}) is

\begin{eqnarray}\label{Total_Derivative_Appendix}
&&\hspace*{-8mm} i \frac{d}{d\ln\Lambda} \int d^8x\,(\theta^4)_x
C(R)_i{}^k \Bigg\{\int d^8y\,
\Bigg(-\Big(\frac{2\partial_\mu}{\partial^2}\Big)_x
\Big(\frac{\delta^2\gamma}{\delta (\phi^{*j})_y
\delta(\phi_k)_x}\Big)^{-1} \Big[y_\mu^*,
\frac{\delta^2\gamma}{\delta (\phi_0^{*j})_y \delta(\phi_{0i})_x}
\Big]\qquad\nonumber\\
&&\hspace*{-8mm} - \Big(\frac{2\partial_\mu}{\partial^2}\Big)_x
\Big(\frac{\delta^2\gamma}{\delta (\phi_j)_y
\delta(\phi_k)_x}\Big)^{-1} \Big[y_\mu^*,
\frac{\delta^2\gamma}{\delta (\phi_{0j})_y \delta(\phi_{0i})_x}
\Big] + \Big(\frac{D^2}{4\partial^4}\Big)_x
\Big(\frac{\delta^2\gamma}{\delta(\phi_k)_x
\delta(\phi_j)_y}\Big)^{-1}
\frac{\delta^2\gamma}{\delta(\phi_i)_x \delta
(\phi_{0j})_y}\Bigg)
\nonumber\\
&&\hspace*{-8mm} + M^{ij} \Big[y_\mu^*,
\Big(\frac{\partial_\mu D^2}{16\partial^4}\Big)_y
\Big(\frac{\delta^2\gamma}{\delta(\phi_j)_x \delta
(\phi_{k})_y}\Big)^{-1} \Big]_{y=x}
- M^{ij} \Big(\frac{D^2}{8\partial^4}\Big)_x
\Big(\frac{\delta^2\gamma}{\delta(\phi_j)_x \delta
(\phi_{k})_y}\Big)^{-1}_{y=x}\Bigg\}.
\end{eqnarray}

\noindent Substituting the inverse Green functions from Eq.
(\ref{A-1_Explicit}) we obtain

\begin{eqnarray}
&&\hspace*{-7mm} i C(R)_i{}^k \frac{d}{d\ln\Lambda}\int
d^8x\,(\theta^4)_x \Bigg\{\int d^8y\, \Bigg(
\Big(\frac{\partial_\mu \bar D^2 D^2}{2\partial^2(\partial^2 G^2
+ |M J|^2)} G\Big)_k{}^j\, \delta^8_{xy} \Big[y_\mu^*,
\frac{\delta^2\gamma}{\delta (\phi_0^{*j})_y \delta(\phi_{0i})_x}
\Big]\nonumber\\
&&\hspace*{-7mm} + \Big(\frac{2\partial_\mu \bar D^2}{\partial^2
(\partial^2 G^2 + |M J|^2)} (M J)^* \Big)_{jk} \delta^8_{xy}
\Big[y_\mu^*, \frac{\delta^2\gamma}{\delta (\phi_{0j})_y
\delta(\phi_{0i})_x} \Big] - \Big(\frac{D^2 \bar
D^2}{4 \partial^4 (\partial^2 G^2 + |M J|^2)}(M J)^* \Big)_{kj}
\nonumber\\
&&\hspace*{-7mm} \times \delta^8_{xy}\,
\frac{\delta^2\gamma}{\delta(\phi_{i})_x \delta (\phi_{0j})_y}\Bigg)
+ M^{ij} \Big[y_\mu^*,
\Big(\frac{\partial_\mu D^2 \bar
D^2}{16\partial^4(\partial^2 G^2 + |M J|^2)} (M J)^* \Big)_{jk} \Big]_x
\delta^8_{xy}\Big|_{y=x}
\nonumber\\
&& \hspace*{-7mm} + M^{ij}
\Big(\frac{1}{\partial^2 G^2 + |M J|^2} (M J)^* \Big)_{jk}
\Big(\frac{D^2 \bar D^2}{8\partial^4}\Big)_x
\delta^8_{xy}\Big|_{y=x} \Bigg\},
\end{eqnarray}

\noindent where all derivatives act on the point $x$. Taking into
account that $[y_\mu^*,D^2]=0$ and $(\theta^4)_x [y_\mu^*,\bar
D^2\delta^8_{xy}] = 0$, it is possible to use Eq.
(\ref{Phi0->Phi}), which allows to express the remaining Green
functions in terms of $G$ and $J$:

\begin{eqnarray}
&& i C(R)_i{}^k \frac{d}{d\ln\Lambda} \int d^8x\,(\theta^4)_x
\Bigg\{2\int d^8y\,\Bigg( \Big(\frac{\bar D^2 D^2
\partial_\mu}{8\partial^2(\partial^2 G^2 + |M
J|^2)}G\Big)_k{}^j\, \delta^8_{xy} \Big[y_\mu^*, G^i{}_j\Big]_y
\delta^8_{xy}
\nonumber\\
&& + \Big(\frac{\bar D^2 D^2
\partial_\mu}{8\partial^2(\partial^2 G^2 + |M
J|^2)}(M J)^*\Big)_{jk}\delta^8_{xy}\, \Big[y_\mu^*, \frac{(M
J)^{ij}}{\partial^2} - \frac{M^{ij}}{\partial^2} \Big]_y
\delta^8_{xy}\Bigg)\nonumber\\
&& - \Big(\frac{D^2 \bar D^2}{8\partial^4(\partial^2 G^2 + |M
J|^2)}(M J)^*\Big)_{jk} \Big((M J)^{ij} - 2 M^{ij}\Big)
\delta^8_{xy}\Big|_{y=x}\nonumber\\
&& +  M^{ij} \Big[y_\mu^*, \Big(\frac{\partial_\mu D^2 \bar
D^2}{16\partial^4(\partial^2 G^2 + |M J|^2)} (M
J)^*\Big)_{kj}\Big]_x \delta^8_{xy} \Big|_{y=x} \Bigg\}.
\end{eqnarray}

\noindent Calculating the integrals over the anticommuting
variables and using the identities

\begin{equation}
\bar D^2 D^2 \delta^4(\theta_x-\theta_y)\Big|_{\theta_x=\theta_y}
= 4;\qquad \int d^4\theta\,\theta^4 = 4,
\end{equation}

\noindent we obtain (in the Euclidian space after the Wick rotation)

\begin{eqnarray}
&&\hspace*{-5mm} {\cal V}_4\cdot C(R)_i{}^k \frac{d}{d\ln\Lambda}
\int \frac{d^4q}{(2\pi)^4} \Bigg\{\frac{\partial G^i{}_j}{\partial
q^\mu} \Big(\frac{2 q_\mu}{q^2(q^2 G^2 + |M J|^2)} G\Big)_k{}^j  +
\Bigg(\frac{\partial}{\partial q^\mu} \Big(\frac{(M
J)^{ij}}{q^2}\Big)
\nonumber\\
&&\hspace*{-5mm} + \frac{q^\mu (M J)^{ij}}{q^4}\Bigg) \Big(\frac{2
q_\mu}{q^2(q^2 G^2 + |M J|^2)}(M J)^* \Big)_{jk} +
\frac{q_\mu}{q^4} \frac{\partial}{\partial q^\mu}\Big(
\frac{1}{(q^2 G^2 + |M J|^2)} (M J)^*\Big)_{jk} M^{ij}
\Bigg\}.\qquad\quad
\end{eqnarray}

\noindent Taking into account Eqs. (\ref{G_And_J}),
(\ref{Propagator}), and (\ref{C(R)_Definition}) this expression
can be presented as an integral of a total derivative:

\begin{equation}
{\cal V}_4\cdot N_f \sum\limits_{I=0}^n (-1)^{P_I}
\frac{d}{d\ln\Lambda} \int \frac{d^4q}{(2\pi)^4} \frac{2
q^\mu}{q^4} \frac{\partial}{\partial q^\mu} \Big( \ln(q^2 G^2 +
M^2 J^2) + \frac{M^2 J}{q^2 G^2 + M^2 J^2} \Big)_I,
\end{equation}

\noindent where $P_I$ is a Grassmanian parity of the superfields
$\phi_I$ and $\widetilde\phi_I$. This expression coincides with
Eq. (\ref{NSVZ_Contribution}).

\section{Relation between diagrams presented in Figs. \ref{Figure_Blue_Green_Diagrams}
and \ref{Figure_Blue_Diagrams}.}
\label{Appendix_Comparing_Of_Diagrams} \hspace{\parindent}

Now, let us compare sums of the diagrams presented in Figs.
\ref{Figure_Blue_Green_Diagrams} and \ref{Figure_Blue_Diagrams}
(including the terms written explicitly) using the identity
(\ref{Pink_Identity}).

1. The sum of diagrams (1) and (10) in Fig.
\ref{Figure_Blue_Diagrams} is equal to diagram (7) in
Fig.~\ref{Figure_Blue_Green_Diagrams}. This equality is
illustrated in Fig. \ref{Figure_Example}.

2. The sum of diagrams (2) and (11) in Fig.
\ref{Figure_Blue_Diagrams} gives diagram (8) in
Fig.~\ref{Figure_Blue_Green_Diagrams}.

3. The sum of diagrams (3) and (12) in Fig.
\ref{Figure_Blue_Diagrams} gives diagram (9) in
Fig.~\ref{Figure_Blue_Green_Diagrams}.

4. The sum of diagrams (6) and (15) in Fig.
\ref{Figure_Blue_Diagrams} gives diagram (10) in
Fig.~\ref{Figure_Blue_Green_Diagrams}.

\begin{figure}[h]
\begin{picture}(0,0.8)

\put(3,0.2){$(a)$}
\put(3.3,-0.9){\includegraphics[scale=0.7]{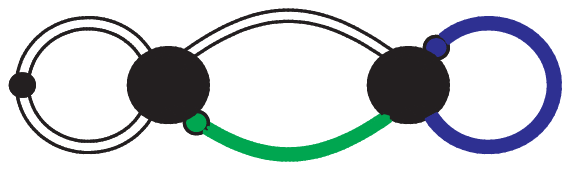}}
\put(8.7,-1.05){\includegraphics[scale=0.7]{figure11diag4.eps}}
\put(7.7,-0.5){$= -2$} \put(2.7,-0.5){$-2$}
\put(11.4,0.25){$\theta^a \theta_a \bar\theta^{\dot b}$}
\put(12.75,-0.7){${\dot b}$} \put(6.0,0.25){$\theta^a \theta_a
\bar\theta^{\dot b}$} \put(7.35,-0.7){${\dot b}$}

\put(2.7,-2.4){$-2$} \put(3.1,-1.8){$(b)$}
\put(5.0,-3.2){$\theta^a \theta_a \bar\theta^{\dot b}$}
\put(6.45,-1.9){${\dot b}$}
\put(3.4,-2.8){\includegraphics[scale=0.7]{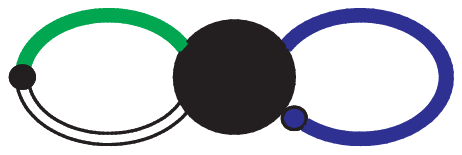}}
\put(7.0,-2.4){$= -2$} \put(9.7,-3.2){$\theta^a \theta_a
\bar\theta^{\dot b}$} \put(11.1,-1.9){${\dot b}$}
\put(8.1,-2.8){\includegraphics[scale=0.7]{figure11diag3.eps}}

\put(3,-3.8){$(c)$}
\put(8.9,-5.05){\includegraphics[scale=0.7]{figure15diag17.eps}}
\put(11.2,-5.25){${\dot b}$} \put(12.9,-4.2){${\dot b}$}
\put(11.7,-3.9){$1$}
\put(3.3,-4.9){\includegraphics[scale=0.7]{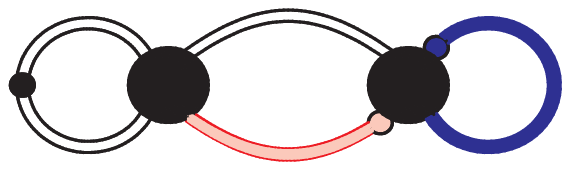}}
\put(7.7,-4.5){$= - \theta^4$} \put(2.5,-4.5){$- \theta^4$}
\put(5.4,-5.25){${\dot b}$} \put(7.3,-4.2){${\dot b}$}
\put(6.1,-3.9){$1$}

\put(3,-5.8){$(d)$}
\put(8.2,-6.87){\includegraphics[scale=0.7]{figure15diag21.eps}}
\put(7.0,-6.5){$= - \theta^4$} \put(10.2,-7.2){$1$}
\put(11.1,-5.9){${\dot b}$} \put(9.3,-5.85){${\dot b}$}
\put(3.3,-6.9){\includegraphics[scale=0.7]{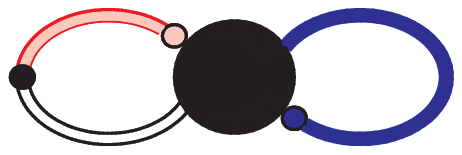}}
\put(2.5,-6.5){$- \theta^4$} \put(5.25,-7.2){$1$}
\put(6.1,-5.9){${\dot b}$} \put(4.2,-5.85){${\dot b}$}

\put(1.6,-7.8){$(e)$}
\put(1.9,-8.9){\includegraphics[scale=0.7]{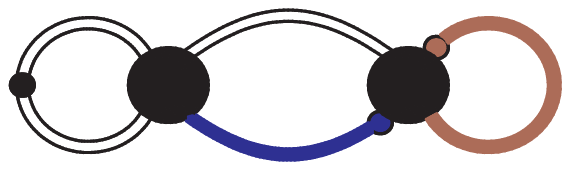}}
\put(3.4,-9.2){${\dot b}$} \put(4.6,-9.1){$\theta_a$}
\put(-0.1,-8.5){$-(\gamma^\mu)^{a{\dot b}} \cdot\theta^4$}
\put(6.0,-8.2){$\mu$} \put(8.9,-7.8){${\dot b}$}

\put(8.1,-7.9){$(f)$}\put(6.3,-8.5){$- (\gamma^\mu)^{a{\dot b}}
\cdot \theta^4$} \put(11.6,-8.2){$\mu$} \put(9.4,-7.8){$\theta_a$}
\put(12.0,-8.5){$= \mbox{Eq.\ }(\ref{Diagrams_Group1})$}
\put(8.3,-8.9){\includegraphics[scale=0.7]{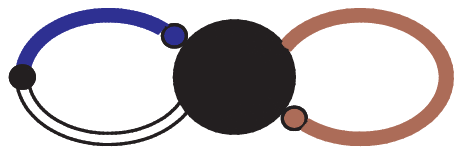}}

\put(0.9,-9.8){$(g)$}
\put(1.2,-10.9){\includegraphics[scale=0.7]{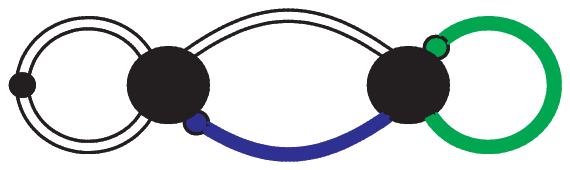}}
\put(0.4,-10.5){$- \ 2$} \put(2.1,-11.3){$\theta^a \theta_a
\bar\theta^{\dot b}$} \put(3.7,-11.15){${\dot b}$}

\put(6.0,-9.9){$(h)$} \put(5.4,-10.5){$-\ 2$}
\put(6.3,-10.9){\includegraphics[scale=0.7]{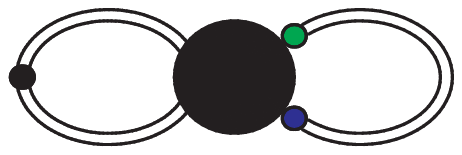}}
\put(9.7,-10.5){${\displaystyle +\ \mbox{Eq.}\
(\ref{Second_Explicit_Term})\, =\, \mbox{Eq.}\
(\ref{Diagrams_Group3}) }$}

\put(2.6,-11.9){$(i)$}  \put(0.0,-12.5){$- (\gamma^\mu)^{a{\dot
b}} \cdot (\theta^4,\theta_a)$} \put(6.2,-12.2){$\mu$}
\put(3.9,-11.8){${\dot b}$}
\put(2.9,-12.9){\includegraphics[scale=0.7]{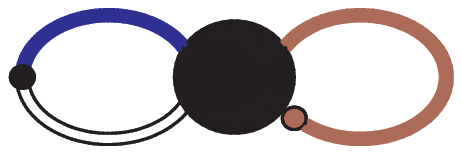}}

\put(9.1,-11.9){$(k)$} \put(6.6,-12.5){$-\ 2\cdot (1;\,\theta^a
\theta_a \bar\theta^b)$} \put(10.4,-11.8){${\dot b}$}
\put(9.4,-12.9){\includegraphics[scale=0.7]{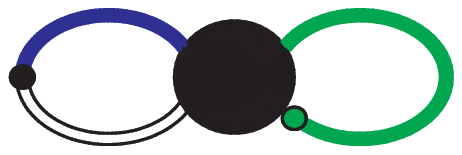}}

\put(7.3,-13.9){${\displaystyle +\ \mbox{Eq.}\ (\ref{M^*_Terms})\
+\ \mbox{Eq.}\ (\ref{First_Explicit_Term})\ =\ \mbox{Eq.}\
(\ref{Diagrams_Group4})}$}

\put(14.1,-11.2){${\displaystyle \left. \vphantom{\begin{array}{c} 1\\
1\\
1\\
1\\
1\\
1\\
1\\
1\\
1\\
1\\
1\\
1
\end{array}} \right\} = \frac{\partial\Delta_2}{\partial\ln g} }$}

\end{picture}
\vspace*{14.3cm} \caption{Some relations between effective
diagrams needed for proving identity presented in Fig.
\ref{Figure_New_Identity_Total_Derivatives}. (For simplicity we
omit the derivatives $d/d\ln\Lambda$ acting on these diagrams.)}
\label{Figure_Zero_Diagrams1}
\end{figure}

5. Let us consider the sum of diagrams (4), (7), (13), and (16) in
Fig. \ref{Figure_Blue_Diagrams}. Using the identity
(\ref{Pink_Identity}) we see that this sum is equal to the sum of
diagrams $(a)$ and $(g)$ in Fig. \ref{Figure_Zero_Diagrams1}.

Let us consider terms containing the derivative
$\delta/\delta\phi_0$ inside the operator $\mbox{GreenLine}$ in
diagram $(a)$. Then according to Eq.
(\ref{Green_Functions_Derivatives}) the left part of the diagram
is proportional to

\begin{equation}
\Big(\frac{\partial}{\partial \ln g} - 1\Big)
\frac{\delta^2\gamma}{\delta (\phi_{0i})_x \delta (\phi^{*j})_y} -
\frac{\bar D_y^2}{8} \delta^8_{xy} \sim D_x^2 \delta^8_{xy} \qquad
\mbox{or}\qquad \Big(\frac{\partial}{\partial \ln g}-1\Big)
\frac{\delta^2\gamma}{\delta (\phi_{0i})_x \delta (\phi_j)_y} \sim
D_x^2 \bar D_x^2 \delta^8_{xy}.
\end{equation}

\noindent In both cases the projector $D_x^2$ acts on the green
effective line. Taking into account the identity
(\ref{Projector_To_Green}) we see that the part of the green line
containing $\delta/\delta \phi_0$ vanishes. The remaining part of
the green line is denoted by the green line with a cross. Thus, we
prove the identity presented in the first string of Fig.
\ref{Figure_Zero_Diagrams1} and obtain diagram (4) in Fig.
\ref{Figure_Blue_Green_Diagrams}.

6. Let us consider a sum of diagrams (5), (14), and (17) in Fig.
\ref{Figure_Blue_Diagrams}. First, it is necessary to prove that
diagram (17) is equal to diagram $(c)$ in Fig.
\ref{Figure_Zero_Diagrams1}. For this purpose we consider terms
which do not contain the masses in the operator $\mbox{PinkLine}$.
Then it is easy to see that the right vertex with effective blue
line is proportional to

\begin{equation}
\Big[(T)\bar\theta^{\dot b},\, \frac{\delta^2\gamma}{\delta
(\phi_{0i})_x \delta (\phi^{*j})_y}\Big] = 0 \qquad
\mbox{or}\qquad (\bar D^2)_y \Big[(T)\bar\theta^{\dot b},\,
\frac{\delta^2\gamma}{\delta (\phi_{0i})_x \delta
(\phi_{0j})_y}\Big] = 0.
\end{equation}

\noindent Therefore, all such terms vanish. The remaining part of
the operator $\mbox{PinkLine}$ is denoted by the pink line with a
cross. Thus, we verify the identity presented in the third string
of Fig. \ref{Figure_Zero_Diagrams1}. Using this result we can
apply the identity (\ref{Pink_Identity}) to the considered sum of
diagrams. Then we obtain diagrams (5) and (6) in Fig.
\ref{Figure_Blue_Green_Diagrams} and diagram $(e)$ in Fig.
\ref{Figure_Zero_Diagrams1}.

7. Diagrams (8), (18), and (21) are considered similarly. Exactly
as in the previous item we prove that diagram (21) is equal to
diagram $(d)$ in Fig. \ref{Figure_Zero_Diagrams1}. Then applying
the identity (\ref{Pink_Identity}) we obtain diagrams (1) and (2)
in Fig. \ref{Figure_Blue_Green_Diagrams} and diagram $(f)$ in Fig.
\ref{Figure_Zero_Diagrams1}.

8. Applying the identity (\ref{Pink_Identity}) to the sum of
diagrams (9), (19), and (20) we obtain diagrams $(b)$, $(i)$, and
$(k)$ in Fig. \ref{Figure_Zero_Diagrams1}. Using the equality
(\ref{Projector_To_Green}) we obtain that diagram $(b)$ is equal
to diagram (3) in Fig. \ref{Figure_Blue_Green_Diagrams}.

9. The sum of diagrams (22) and (23) gives diagram $(h)$, which
corresponds to

\begin{eqnarray}
&&\hspace*{-5mm} 2 i (T)^j{}_i (T)^l{}_k \frac{d}{d\ln\Lambda}
\int d^8x\,d^8y\,\Big(\theta^c \theta_c \bar\theta^{\dot d}
\frac{\bar D_{\dot d} D^2}{4\partial^2} + 2i \bar\theta^{\dot c}
(\gamma^\mu)_{\dot c}{}^d \theta_d
\frac{\partial_\mu}{\partial^2}\Big)_y \Big(\theta^a\theta_a
\bar\theta^{\dot b} \frac{\bar D_{\dot b}
D^2}{8\partial^2}\Big)_x  \nonumber\\
&&\hspace*{-5mm} \times \Big(\frac{\delta^2\gamma}{\delta
(\phi_l)_y \delta (\phi_j)_x} \Big)^{-1}\cdot
\mbox{LineWithDot}[1]\cdot \frac{\delta^2\gamma}{\delta
(\phi_{0k})_y \delta (\phi_{0i})_x}.
\end{eqnarray}

\noindent Using Eqs. (\ref{Green_Functions_Derivatives}) and
commuting $(\theta^a\theta_a)_x$ with $(D^2)_x$ this expression
can be presented in the form

\begin{eqnarray}
&&\hspace*{-5mm} -4 g\, C(R)_i{}^j \frac{d}{d\ln\Lambda} \int
d^8x\,d^8y\, \Big(\theta^c \theta_c \bar\theta^{\dot d} \frac{\bar
D_{\dot d} D^2}{4\partial^2} + 2i \bar\theta^{\dot c}
(\gamma^\mu)_{\dot c}{}^d \theta_d
\frac{\partial_\mu}{\partial^2}\Big)_y \Big(\bar\theta^{\dot a}
(\gamma^\mu)_{\dot a}{}^b \theta^b
\frac{\partial_\nu}{\partial^2}\Big)_x
\nonumber\\
&&\hspace*{-5mm} \times \Big(\frac{\delta^2\gamma}{\delta
(\phi_k)_y \delta (\phi_j)_x} \Big)^{-1} \frac{\partial}{\partial
\ln g}\Big(g^{-1} \frac{\delta^2\gamma}{\delta (\phi_{0k})_y
\delta (\phi_{0i})_x}\Big).
\end{eqnarray}

\noindent Taking into account the identity (\ref{Qubic_Identity})
after some simple transformations we write the result in the form

\begin{eqnarray}\label{Diagram_With_Dots}
&& - 2i g\, C(R)_i{}^j \frac{d}{d\ln\Lambda} \int
d^8x\,d^8y\,(\theta^4)_x \Big( \frac{2\partial_\mu}{\partial^2}-i
(\gamma_\mu)^{a{\dot b}} \theta_a \frac{\bar D_{\dot b}
D^2}{4\partial^2} \Big)_y \Big(
\frac{\partial_\mu}{\partial^2}\Big)_x
\Big(\frac{\delta^2\gamma}{\delta (\phi_k)_y \delta (\phi_j)_x}
\Big)^{-1} \nonumber\\
&& \times \frac{\partial}{\partial \ln g}\Big(g^{-1}
\frac{\delta^2\gamma}{\delta (\phi_{0k})_y \delta (\phi_{0i})_x}
\Big).
\end{eqnarray}

10. Let us now consider the term

\begin{eqnarray}
- i \frac{d}{d\ln\Lambda} \int d^8x\,(\theta^4)_x
(\gamma^\mu)^{{\dot a}b} (MT)^{ik}
\Big(\frac{D^2\partial_\mu}{8\partial^4}
\frac{\mbox{\boldmath$\delta$}}{\mbox{\boldmath$\delta$} j^i}\Big)
\frac{\mbox{\boldmath$\delta$}}{\mbox{\boldmath$\delta$} j^k}
\cdot \mbox{BlueLine}_{\dot a}[\theta_b]\cdot \gamma,
\end{eqnarray}

\noindent which is written explicitly in Fig.
\ref{Figure_Blue_Diagrams}. This expression is quadratic in
$\bar\theta$. Therefore, it is possible to shift any $\bar\theta$
to an arbitrary point of the graph. Then using the identity
(\ref{Qubic_Identity}) it is easy to see that the considered
expression can be rewritten in the form

\begin{eqnarray}
&& - i \frac{d}{d\ln\Lambda} \int d^8x\,(\bar\theta^{\dot c}
(\gamma^\mu)_{\dot c}{}^d \theta_d)_x (MT)^{ik}
\Big(\frac{D^2\partial_\mu}{4\partial^4}
\frac{\mbox{\boldmath$\delta$}}{\mbox{\boldmath$\delta$} j^i}\Big)
\frac{\mbox{\boldmath$\delta$}}{\mbox{\boldmath$\delta$} j^k}\cdot
\mbox{BlueLine}_{\dot b}[\theta^a
\theta_a \bar\theta^{\dot b}] \cdot \gamma \nonumber\\
&& = \frac{d}{d\ln\Lambda}\Big( 4\cdot \mbox{GreenWithCross} \cdot
\mbox{BlueLine}_b[\theta^a \theta_a \bar\theta^b]\cdot
\gamma\Big).
\end{eqnarray}

\noindent This result coincides with diagram $(11)$ in Fig.
\ref{Figure_Blue_Green_Diagrams}.

11. The terms containing $M^*$, which are presented in Fig.
\ref{Figure_Blue_Diagrams} in the explicit form, are given by

\begin{eqnarray}\label{M^*_Terms}
&&\hspace*{-8mm} -i (\gamma^\mu)^{{\dot a}b} (TM^*)_{ik}
\frac{d}{d\ln\Lambda} \int d^8x\,\Big( 2 (\theta^4)_x
\mbox{BlueLine}_{\dot a}[\theta_b] + (\bar\theta^{\dot c}
\bar\theta_{\dot c} \theta_b)_x \mbox{PinkLine}_{\dot a}\Big)
\Big(\frac{D^2\partial_\mu}{16\partial^4}
\frac{\mbox{\boldmath$\delta$}}{\mbox{\boldmath$\delta$}
j^*_i}\Big)
\frac{\mbox{\boldmath$\delta$}}{\mbox{\boldmath$\delta$} j^*_k}
\cdot \gamma \nonumber\\
&&\hspace*{-8mm} = - i (TM^*)_{ik} \frac{d}{d\ln\Lambda} \int
d^8x\, (\theta^4)_x \mbox{BrownLine}^\mu \cdot
\Big(\frac{D^2\partial_\mu}{16\partial^4}
\frac{\mbox{\boldmath$\delta$}}{\mbox{\boldmath$\delta$}
j^*_i}\Big)
\frac{\mbox{\boldmath$\delta$}}{\mbox{\boldmath$\delta$} j^*_k}
\cdot \gamma \nonumber\\
&&\hspace*{-8mm} = i \frac{d}{d\ln\Lambda} \int d^8x\,
(\theta^4)_x (T M^*)_{ik} \Big(\frac{\bar D^2
\partial^\mu}{16\partial^4}\Big)_x \Bigg(\Big[(T) y_\mu^*,\,
\Big(\frac{\delta^2\gamma}{\delta (\phi^{*k})_y \delta
(\phi^{*i})_x}\Big)^{-1} \Big]
\nonumber\\
&&\hspace*{-8mm}\qquad\qquad\qquad - \int d^8z\, (MT)^{mn}
\Big(\frac{D^2\partial^\mu}{8\partial^4}\Big)_z
\Big(\frac{\delta^2\gamma}{\delta (\phi_m)_z \delta
(\phi^{*k})_y}\Big)^{-1} \Big(\frac{\delta^2\gamma}{\delta
(\phi_n)_z \delta (\phi^{*i})_x}\Big)^{-1} \Bigg)_{y=x},
\end{eqnarray}

\noindent where we use the results for $(T)y_\mu^*$ commutators
with inverse Green functions obtained in Appendix
\ref{Appendix_Commutators_Inverse} for deriving the last equality.

12. The first two terms explicitly written in Fig.
\ref{Figure_Blue_Diagrams} can be presented in the following form:

\begin{eqnarray}\label{First_Explicit_Term}
&&\hspace*{-11mm} \frac{i}{2}\cdot \frac{d}{d\ln\Lambda} \int
d^8x\, M^{mk} C(R)_k{}^n \Big(\frac{D^2}{8\partial^4}\Big)_x
\mbox{LineWithDot}[\theta^4]\cdot
\frac{\mbox{\boldmath$\delta$}^2\gamma}{\mbox{\boldmath$\delta$}
(j^m)_x \mbox{\boldmath$\delta$} (j^n)_y}\Bigg|_{y=x} = -i
C(R)_k{}^i M^{jk}
\nonumber\\
&&\hspace*{-11mm} \times \frac{d}{d\ln\Lambda} \int
d^8x\,(\theta^4)_x \Bigg\{g^{-1}
\Big(\frac{D^2}{8\partial^4}\Big)_x \frac{\partial}{\partial\ln
g}\Bigg( g\, \Big(\frac{\delta^2\gamma}{\delta (\phi_{i})_x \delta
(\phi_{j})_y}\Big)^{-1}\Bigg)_{y=x} - \int
d^8z\, \Bigg(\Big(\frac{D^2}{16\partial^4}\Big)_x \nonumber\\
&&\hspace*{-11mm}\times \Big(\frac{\delta^2\gamma}{\delta
(\phi_{m})_z \delta (\phi_{i})_x}\Big)^{-1}
\Big(\frac{\delta^2\gamma}{\delta (\phi^{*m})_z \delta
(\phi_{j})_x}\Big)^{-1} + M^{mn}
\Big(\frac{D^2}{16\partial^4}\Big)_x
\Big(\frac{\delta^2\gamma}{\delta (\phi_{i})_x \delta
(\phi_{m})_z}\Big)^{-1}
\Big(\frac{D^2}{8\partial^2}\Big)_z \nonumber\\
&&\hspace*{-11mm} \times \Big(\frac{\delta^2\gamma}{\delta
(\phi_{n})_z \delta (\phi_{j})_x}\Big)^{-1} + M^*_{mn}
\Big(\frac{D^2}{16\partial^4}\Big)_x
\Big(\frac{\delta^2\gamma}{\delta (\phi_{i})_x \delta
(\phi^{*m})_z}\Big)^{-1} \Big(\frac{\bar D^2}{8\partial^2}\Big)_z
\Big(\frac{\delta^2\gamma}{\delta (\phi^{*n})_z \delta
(\phi_{j})_x}\Big)^{-1} \Bigg) \Bigg\};\quad
\end{eqnarray}

\begin{eqnarray}\label{Second_Explicit_Term}
&& \hspace*{-5mm} \frac{i}{2}\cdot \frac{d}{d\ln\Lambda} \int
d^8x\,d^8y\,\Bigg\{ M^{mk} C(R)_k{}^n
\Big(\frac{\partial_\mu}{\partial^2}\Big)_x
\Big(\frac{\delta^2\gamma}{\delta (\phi_i)_x \delta(\phi_m)_y}
\Big)^{-1} \mbox{LineWithDot}[\theta^4]\cdot
\Big(\frac{D^2\partial_\mu}{4\partial^4}\Big)_y\nonumber\\
&& \hspace*{-5mm} \times
\frac{\mbox{\boldmath$\delta$}}{\mbox{\boldmath$\delta$} (j^n)_y}
\frac{\delta\gamma}{\delta (\phi_{0i})_x}\Bigg\}   = - ig
\frac{d}{d\ln\Lambda} \int d^8x\,d^8y\,d^8z\,(\theta^4)_x
\Bigg\{M^{mk} C(R)_k{}^n
\Big(\frac{\partial_\mu}{\partial^2}\Big)_x
\Big(\frac{\delta^2\gamma}{\delta (\phi_i)_x \delta(\phi_m)_y}
\Big)^{-1} \nonumber\\
&& \hspace*{-5mm} \times
\Big(\frac{D^2\partial_\mu}{4\partial^4}\Big)_y
\Bigg[\Big(\frac{\delta^2\gamma}{\delta(\phi^{*p})_z \delta
(\phi_n)_y} \Big)^{-1} \Bigg(\frac{\partial}{\partial\ln g} \Big(
g^{-1} \frac{\delta^2\gamma}{\delta (\phi_{0i})_x \delta
(\phi_{0}^{*p})_z}\Big) + \frac{1}{4g} \delta^8_{xz}
\delta_p^i\Bigg) + \Big(\frac{\delta^2\gamma}{\delta(\phi_p)_z
\delta (\phi_n)_y} \Big)^{-1}\nonumber\\
&& \hspace*{-5mm}  \times \frac{\partial}{\partial\ln g} \Big(
g^{-1} \frac{\delta^2\gamma}{\delta (\phi_{0i})_x \delta
(\phi_{0p})_z}\Big) \Bigg]\Bigg\}.
\end{eqnarray}

13. We have already obtained all diagrams presented in Fig.
\ref{Figure_Blue_Green_Diagrams}. However, there are also some
additional contributions. Let us verify that the sum of them gives
$\partial \Delta_2/\partial\ln g$.

We will start with calculating the sum of diagrams $(e)$ and $(f)$
in Fig. \ref{Figure_Zero_Diagrams1}. It is easy to see that the
right part of these diagrams contains the operator $D^2$ acting on
the right part of the blue effective line. Using the identity

\begin{equation}
D^2 \Big((\gamma^\mu)^{a{\dot b}}\theta_a \frac{\bar D_{\dot b}
D^2}{8\partial^2} +\frac{i\partial^\mu}{\partial^2}\Big)=0
\end{equation}

\noindent similar to item 6 it is possible to present the
considered sum in the form:

\begin{eqnarray}
&& \frac{1}{2}\cdot \frac{d}{d\ln\Lambda}
\Big(\mbox{LineWithDot}[\theta^4;1,1]\cdot
\mbox{YellowLine}^\mu[1,M=0;2,1] \cdot \mbox{UsualLine}[1,2] \cdot
\vphantom{\frac{1}{2}\cdot}\nonumber\\
&&  \times \mbox{BrownLine}_\mu[2,2]\cdot \gamma[1]\cdot \gamma[2]
+ \mbox{YellowWhiteLine}^\mu[\theta^4,M=0]\cdot
\mbox{BrownLine}_\mu \cdot
\gamma\Big),\qquad\vphantom{\frac{1}{2}\cdot}
\end{eqnarray}

\noindent where $M=0$ means that in the expression for the
effective line it is necessary to set masses to 0. The operator
$\mbox{BrownLine}_\mu$ acting on derivatives of the Routhian
$\gamma$ gives commutators with $(T)y_\mu^*$, see Eqs.
(\ref{Y_Commutator1}) and (\ref{Y_Commutator2}). Taking into
account Eqs. (\ref{Inverse_Derivative}) and
(\ref{Inverse_Derivative2}) we see that the other operators give
the derivative with respect to $\ln g$ and the considered sum of
diagrams $(e)$ and $(f)$ can be rewritten as

\begin{eqnarray}
&&\hspace*{-7mm} 2i\, (T)^l{}_i \frac{d}{d\ln\Lambda} \int
d^8x\,d^8y\,(\theta^4)_x\,\frac{1}{g} \Bigg\{
\frac{\partial}{\partial\ln
g}\Bigg(g\Big(\frac{\delta^2\gamma}{\delta(\phi_l)_x
\delta(\phi_j)_y}\Big)^{-1}\Bigg)
\Big(\frac{\partial_\mu}{\partial^2}\Big)_x \Big[(T) y_\mu^*,\,
\frac{\delta^2\gamma}{\delta
(\phi_{0j})_y \delta(\phi_{0i})_x}\Big]\nonumber\\
&&\hspace*{-7mm} + \frac{\partial}{\partial\ln
g}\Bigg(g\Big(\frac{\delta^2\gamma}{\delta(\phi_l)_x
\delta(\phi^{*j})_y}\Big)^{-1}\Bigg)
\Big(\frac{\partial_\mu}{\partial^2}\Big)_x \Big[y_\mu^*,\,
\frac{\delta^2\gamma}{\delta (\phi_0^{*j})_y
\delta(\phi_{0i})_x}\Big] \Bigg\}.
\end{eqnarray}

\noindent In order to calculate this expression we substitute the
inverse Green functions from Eq. (\ref{A-1_Explicit}) and use Eqs.
(\ref{Phi0_SD_Equation}) and (\ref{A_Definition}). The result in
the momentum representation is

\begin{eqnarray}\label{Diagrams_Group1}
&& {\cal V}_4 \cdot 2 C(R)_k{}^i \frac{d}{d\ln\Lambda} \int
\frac{d^4q}{(2\pi)^4} \frac{q^\mu}{g q^2} \Bigg\{
\frac{\partial}{\partial\ln g}\Big( \frac{g}{q^2 G^2 + |M J|^2} G
\Big)_i{}^j\, \frac{\partial G^k{}_j}{\partial q^\mu}
\nonumber\\
&&\qquad\quad\qquad\qquad\qquad\qquad +
\frac{\partial}{\partial\ln g} \Big(\frac{g}{q^2 G^2 + |M J|^2}(M
J)^*\Big)_{ij}\, \frac{\partial}{\partial q^\mu} \Bigg(\frac{(M
J)^{jk} - M^{jk}}{q^2}\Bigg) \Bigg\}.\qquad\quad
\end{eqnarray}

14. In order to simplify diagram $(g)$ in Fig.
\ref{Figure_Zero_Diagrams1} we use Eq. (\ref{Projector_To_Green}).
Then repeating the same arguments as for diagram (a) the
derivative of this diagram with respect to $\ln\Lambda$ can be
written as

\begin{eqnarray}
&& \frac{d}{d\ln\Lambda}\, \mbox{LineWithDot}[1;1,1]\cdot
\mbox{UsualLine}[1,2]\vphantom{\frac{1}{2}}\nonumber\\
&&\qquad\qquad\qquad\qquad \times
\mbox{YellowLine}_\mu[\bar\theta^{\dot a} (\gamma^\mu)_{\dot
a}{}^b \theta_b, M=0;1,2]\cdot \mbox{GreenLine}[2,2]\cdot
\gamma[1] \cdot \gamma[2] \nonumber\\
&&  = -\frac{1}{8}\cdot \frac{d}{d\ln\Lambda}\,
\mbox{LineWithDot}[1;1,1]\cdot
\mbox{UsualLine}[1,2]\nonumber\\
&&\qquad\qquad\qquad\qquad  \times
\mbox{YellowLine}_\mu[\bar\theta^{\dot c} \bar\theta_{\dot c}
(\gamma^\mu)^{{\dot a}b} \theta_b, M=0;1,2]\cdot
\mbox{RedLine}_{\dot a}[2,2] \cdot \gamma[1] \cdot \gamma[2]\nonumber\\
&&  =  -\frac{1}{2}\cdot \frac{d}{d\ln\Lambda}\,
\mbox{LineWithDot}[1;1,1] \cdot \mbox{UsualLine}[1,2]
\vphantom{\frac{1}{2}}\nonumber\\
&&\qquad\qquad\qquad\qquad \times \mbox{YellowLine}_\mu[\theta^4,
M=0;1,2]\cdot \mbox{BrownLine}^\mu[2,2] \cdot \gamma[1] \cdot
\gamma[2].
\end{eqnarray}

\noindent This expression contains the operator
$\mbox{BrownLine}_\mu$ acting on the two-point Green functions,
which is related with $(T)y_\mu^*$ commutators according to the
results obtained in Appendix \ref{Appendix_Commutators_Y}.
Similarly, the operator $\mbox{LineWithDot}[1]$ acting on
two-point functions is related with the derivatives of these
functions with respect to $\ln g$ according to the results
obtained in Appendix \ref{Appendix_G_Derivative_Vertices}. Using
these relations it is possible to rewrite diagram $(g)$ as

\begin{eqnarray}
&&\hspace*{-6mm} 2i g (T)^n{}_m \frac{d}{d\ln\Lambda} \int
d^8x\,d^8y\,(\theta^4)_y\,\Bigg\{ \frac{\partial}{\partial\ln
g}\Bigg[\Big(g^{-1} \frac{\delta^2\gamma}{\delta(\phi_{0m})_y
\delta (\phi_0^{*i})_x}\Big) + \frac{1}{4g} \delta^8_{xy}
\delta^m_i \Bigg]
\Big(\frac{\partial^\mu}{\partial^2}\Big)_y \Bigg(\Big[(T)y_\mu^*,\, \nonumber\\
&&\hspace*{-6mm} \Big(\frac{\delta^2\gamma}{\delta(\phi^{*i})_x
\delta(\phi_n)_y}\Big)^{-1} \Big] - (T)_n{}^j
\Big(\frac{2\partial_\mu}{\partial^2}\Big)_y
\Big(\frac{\delta^2\gamma}{\delta(\phi^{*i})_x
\delta(\phi_j)_y}\Big)^{-1} - (M T)^{kl} \int d^8z\,
\Big(\frac{D^2\partial_\mu}{8\partial^4}\Big)_z
\nonumber\\
&&\hspace*{-6mm} \times
\Big(\frac{\delta^2\gamma}{\delta(\phi_k)_z
\delta(\phi^{*i})_x}\Big)^{-1}
\Big(\frac{\delta^2\gamma}{\delta(\phi_l)_z
\delta(\phi_n)_y}\Big)^{-1}\Bigg) + \frac{\partial}{\partial\ln
g}\Big(g^{-1} \frac{\delta^2\gamma}{\delta(\phi_{0m})_y \delta
(\phi_{0i})_x} \Big) \Big(\frac{\partial_\mu}{\partial^2}\Big)_y
\Bigg(\Big[(T)y_\mu^*,\,\nonumber\\
&&\hspace*{-6mm} \Big(\frac{\delta^2\gamma}{\delta(\phi_i)_x
\delta(\phi_n)_y}\Big)^{-1} \Big] - (T)_n{}^j
\Big(\frac{2\partial_\mu}{\partial^2}\Big)_y
\Big(\frac{\delta^2\gamma}{\delta(\phi_i)_x
\delta(\phi_j)_y}\Big)^{-1} - (T)_i{}^j
\Big(\frac{2\partial_\mu}{\partial^2} -i(\gamma_\mu)^{a{\dot b}}
\theta_a
\frac{\bar D_{\dot b} D^2}{4\partial^2} \Big)_x \nonumber\\
&&\hspace*{-6mm} \times
\Big(\frac{\delta^2\gamma}{\delta(\phi_j)_x
\delta(\phi_n)_y}\Big)^{-1} - (M T)^{kl} \int d^8z\,
\Big(\frac{D^2\partial_\mu}{8\partial^4}\Big)_z
\Big(\frac{\delta^2\gamma}{\delta(\phi_k)_z
\delta(\phi_i)_x}\Big)^{-1}
\Big(\frac{\delta^2\gamma}{\delta(\phi_l)_z
\delta(\phi_n)_y}\Big)^{-1}  \Bigg) \Bigg\}.
\end{eqnarray}

\noindent Adding to this expression the contributions
(\ref{Diagram_With_Dots}) and (\ref{Second_Explicit_Term}) we
obtain a simpler expression

\begin{eqnarray}
&&\hspace*{-8mm} 2i g (T)^n{}_m \frac{d}{d\ln\Lambda} \int
d^8x\,d^8y\,(\theta^4)_y\, \Bigg\{\frac{\partial}{\partial\ln
g}\Bigg[\Big(g^{-1} \frac{\delta^2\gamma}{\delta(\phi_{0m})_y
\delta (\phi_0^{*i})_x}\Big) + \frac{1}{4g} \delta^8_{xy}
\delta^m_i \Bigg] \Big(\frac{\partial^\mu}{\partial^4}\Big)_y
\Big[(T)y_\mu^*,\,(\partial^2)_x \nonumber\\
&&\hspace*{-8mm} \times
\Big(\frac{\delta^2\gamma}{\delta(\phi^{*i})_x
\delta(\phi_n)_y}\Big)^{-1} \Big] + \frac{\partial}{\partial\ln
g}\Big(g^{-1} \frac{\delta^2\gamma}{\delta(\phi_{0m})_y \delta
(\phi_{0i})_x} \Big) \Big(\frac{\partial_\mu}{\partial^4}\Big)_y
\Big[(T)y_\mu^*,\,(\partial^2)_x
\Big(\frac{\delta^2\gamma}{\delta(\phi_i)_x \delta(\phi_n)_y}
\Big)^{-1} \Big]  \Bigg\}.\nonumber\\
\end{eqnarray}

\noindent Substituting the inverse Green functions and using Eq.
(\ref{Phi0_SD_Equation}) after the Wick rotation in the Euclidian
space it can be rewritten in the form

\begin{eqnarray}\label{Diagrams_Group3}
&& {\cal V}_4 \cdot 2 C(R)_k{}^i \frac{d}{d\ln\Lambda} \int
\frac{d^4q}{(2\pi)^4} \frac{q^\mu}{q^2} \Bigg\{- \frac{1}{q^2}
\frac{\partial}{\partial q^\mu}\Big( \frac{g q^2}{q^2 G^2 + |M
J|^2} G\Big)^j{}_i \frac{\partial }{\partial \ln g} \Big(g^{-1}
G_j{}^k - g^{-1} \delta_j^k \Big)
\nonumber\\
&& - \frac{\partial}{\partial q^\mu} \Big(\frac{g q^2}{q^2 G^2 +
|M J|^2}(M J)^*\Big)_{ij}\, \frac{\partial}{\partial \ln
g}\Big(\frac{(M J)^{jk} - M^{jk}}{g q^4}\Big) \Bigg\}.
\end{eqnarray}

15. The sum of the expressions (\ref{Diagrams_Group1}) and
(\ref{Diagrams_Group3}) is

\begin{eqnarray}\label{Main_Sum}
&& {\cal V}_4 \cdot 2 N_f \sum\limits_{I=0}^n (-1)^{P_I}
\frac{d}{d\ln\Lambda} \int \frac{d^4q}{(2\pi)^4}
\Bigg\{\frac{\partial}{\partial\ln g}\Big(\frac{M^2 J (-2 J +
4)}{q^4(q^2 G^2 + M^2 J^2)}\Big)
\nonumber\\
&&\qquad\qquad\qquad\qquad\qquad\qquad\qquad\qquad\qquad\qquad\quad
- \frac{q^\mu}{q^4} \frac{\partial}{\partial q^\mu}\Big(\frac{2q^2
G + 2 M^2 J}{q^2 G^2 + M^2 J^2}\Big) \Bigg\}_I.\qquad\quad
\end{eqnarray}

16. Diagram $(k)$ contains

\begin{eqnarray}
&& -2\cdot \mbox{WhiteBlueLine}_{\dot b}[1,\theta^a\theta_a
\bar\theta^{\dot b}]\vphantom{\frac{1}{2}}\nonumber\\
&& = \int d^8x\,\theta^a\theta_a \bar\theta^{\dot b}
\Big\{(T)^k{}_i \frac{\bar D_{\dot b} D^2}{8\partial^2}
\frac{\mbox{\boldmath$\delta$}}{\mbox{\boldmath$\delta$} j^k}
\cdot \frac{\mbox{\boldmath$\delta$}}{\mbox{\boldmath$\delta$}
j^*_i} - (M T)^{ij} \frac{\bar D_{\dot b} D^2}{8\partial^2}
\frac{\mbox{\boldmath$\delta$}}{\mbox{\boldmath$\delta$} j^i}
\frac{D^2}{4\partial^2}
\frac{\mbox{\boldmath$\delta$}}{\mbox{\boldmath$\delta$} j^j}
\Big\}\nonumber\\
&& = -i \int d^8x\,\bar\theta^{\dot a} (\gamma^\mu)_{\dot a}{}^b
\theta_b \Big\{(T)^k{}_i \frac{\partial_\mu}{\partial^2}
\frac{\mbox{\boldmath$\delta$}}{\mbox{\boldmath$\delta$} j^k}
\cdot \frac{\mbox{\boldmath$\delta$}}{\mbox{\boldmath$\delta$}
j^*_i} - (M T)^{ij} \frac{D^2 \partial_\mu}{4\partial^4}
\frac{\mbox{\boldmath$\delta$}}{\mbox{\boldmath$\delta$} j^i}
\cdot \frac{\mbox{\boldmath$\delta$}}{\mbox{\boldmath$\delta$}
j^j} \Big\}.\qquad
\end{eqnarray}

\noindent Taking into account that the considered graph is
quadratic in $\bar\theta$ and using Eq. (\ref{Qubic_Identity}) we
obtain

\begin{eqnarray}
&& (\bar\theta^{\dot a} (\gamma^\mu)_{\dot a}{}^b \theta_b)_x
\times \mbox{GreenLine} = - \frac{1}{8} (\gamma^\mu)^{a{\dot b}}
(\bar\theta^{\dot c} \bar\theta_{\dot c} \theta_a)_x \times
\mbox{RedLine}_{\dot
b} +\ldots\nonumber\\
&& = - \frac{1}{2} (\theta^4)_x \times \mbox{BrownLine}^\mu +
\ldots,
\end{eqnarray}

\noindent where dots denote terms vanishing after integration over
$d^4\theta$. Using this equation the derivative of the considered
diagram with respect to $\ln\Lambda$ can be written in the form

\begin{eqnarray}
\frac{i}{2}\cdot \frac{d}{d\ln\Lambda} \int d^8x\, (\theta^4)_x
\Big((T)^k{}_i \frac{\partial_\mu}{\partial^2}
\frac{\mbox{\boldmath$\delta$}}{\mbox{\boldmath$\delta$} j^k}
\cdot \frac{\mbox{\boldmath$\delta$}}{\mbox{\boldmath$\delta$}
j^*_i} - (M T)^{ij} \frac{D^2 \partial_\mu}{4\partial^4}
\frac{\mbox{\boldmath$\delta$}}{\mbox{\boldmath$\delta$} j^i}
\cdot \frac{\mbox{\boldmath$\delta$}}{\mbox{\boldmath$\delta$}
j^j} \Big) \cdot \mbox{BrownLine}^\mu \cdot\gamma.
\end{eqnarray}

\noindent In this expression the operator $\mbox{BrownLine}^\mu$
acts on two-point Green functions. This allows to present the
result in the form

\begin{eqnarray}\label{Diagram_I}
&&\hspace*{-5mm} \frac{i}{2}\cdot \frac{d}{d\ln\Lambda} \int
d^8x\, (\theta^4)_x \Bigg\{- (T)^n{}_m
\Big(\frac{\partial^\mu}{\partial^2}\Big)_x \Bigg(\Big[(T)
y_\mu^*,\, \Big(\frac{\delta^2\gamma}{\delta (\phi^{*m})_y
\delta(\phi_n)_x}\Big)^{-1} \Big] - (T)_n{}^p
\Big(\frac{2\partial_\mu}{\partial^2}\Big)_x
\nonumber\\
&&\hspace*{-5mm} \times \Big(\frac{\delta^2\gamma}{\delta
(\phi^{*m})_y \delta(\phi_p)_x}\Big)^{-1} - (M T)^{pq} \int d^8z\,
\Big(\frac{D^2 \partial_\mu}{8\partial^4}\Big)_z
\Big(\frac{\delta^2\gamma}{\delta(\phi_p)_z \delta
(\phi^{*m})_y}\Big)^{-1}
\Big(\frac{\delta^2\gamma}{\delta(\phi_q)_z \delta
(\phi_n)_x}\Big)^{-1}  \Bigg)
\nonumber\\
&&\hspace*{-5mm} + (M T)^{nm} \Big(\frac{\partial_\mu}{\partial^2}
\Big)_x \Big(\frac{D^2}{4\partial^2}\Big)_y \Bigg(\Big[(T)
y_\mu^*,\, \Big(\frac{\delta^2\gamma}{\delta(\phi_m)_y \delta
(\phi_n)_x}\Big)^{-1} \Big] - (T)_n{}^p
\Big(\frac{2\partial_\mu}{\partial^2}\Big)_x
\Big(\frac{\delta^2\gamma}{\delta(\phi_m)_y \delta
(\phi_p)_x}\Big)^{-1}
\nonumber\\
&&\hspace*{-5mm} - (M T)^{pq} \int d^8z\, \Big(\frac{D^2
\partial_\mu}{8\partial^4}\Big)_z \Big(\frac{\delta^2\gamma}{\delta(\phi_p)_z \delta
(\phi_m)_y}\Big)^{-1} \Big(\frac{\delta^2\gamma}{\delta(\phi_q)_z
\delta (\phi_n)_x}\Big)^{-1} \Bigg)\Bigg\}_{y=x}.
\end{eqnarray}

17. Diagram $(i)$ contains

\begin{eqnarray}
&& - (\gamma^\mu)^{a{\dot b}} \mbox{WhiteBlueLine}_{\dot
b}[\theta^4,\theta_a] = - \frac{1}{2} (\gamma^\mu)^{a{\dot b}}\int
d^8x\,\theta^4 (M T)^{ij} \Big(\frac{D^2}{8\partial^2}\,\theta_a
\frac{\bar D_{\dot b} D^2}{8\partial^2}
\frac{\mbox{\boldmath$\delta$}}{\mbox{\boldmath$\delta$} j^i}\Big)
\frac{\mbox{\boldmath$\delta$}}{\mbox{\boldmath$\delta$}
j^j}\qquad\nonumber\\
&& = i  \int d^8x\,\theta^4 (M T)^{ij} \Big(\frac{
D^2\partial_\mu}{16\partial^4}
\frac{\mbox{\boldmath$\delta$}}{\mbox{\boldmath$\delta$} j^i}
\Big) \frac{\mbox{\boldmath$\delta$}}{\mbox{\boldmath$\delta$}
j^j}.
\end{eqnarray}

\noindent Therefore, the derivative of this diagram with respect
to $\ln\Lambda$ can be written as

\begin{eqnarray}\label{Diagram_G}
&& \frac{d}{d\ln \Lambda} \int d^8x\, (\theta^4)_x (M T)^{nm}
\frac{i D^2
\partial_\mu}{16\partial^4}
\frac{\mbox{\boldmath$\delta$}}{\mbox{\boldmath$\delta$} j^n}
\cdot \frac{\mbox{\boldmath$\delta$}}{\mbox{\boldmath$\delta$}
j^m} \cdot \mbox{BrownLine}^\mu \cdot\gamma  =
\frac{d}{d\ln\Lambda} \int d^8x\,
(\theta^4)_x \qquad\nonumber\\
&& \times (M T)^{nm} \Big(\frac{i D^2
\partial_\mu}{16\partial^4}\Big)_x \Bigg(- \Big[(T) y_\mu^*,\,
\Big(\frac{\delta^2\gamma}{\delta (\phi_m)_y \delta
(\phi_n)_x}\Big)^{-1} \Big] + (T)_n{}^p
\Big(\frac{2\partial_\mu}{\partial^2}\Big)_x
\Big(\frac{\delta^2\gamma}{\delta (\phi_m)_y \delta
(\phi_p)_x}\Big)^{-1} \nonumber\\
&& + (MT)^{pq} \int d^8z\,
\Big(\frac{D^2\partial^\mu}{8\partial^4}\Big)_z
\Big(\frac{\delta^2\gamma}{\delta (\phi_p)_z \delta
(\phi_m)_y}\Big)^{-1} \Big(\frac{\delta^2\gamma}{\delta (\phi_q)_z
\delta (\phi_n)_x}\Big)^{-1}\Bigg)_{y=x}.
\end{eqnarray}

18. It is easy to see that the sum of Eq. (\ref{M^*_Terms}),
(\ref{First_Explicit_Term}), (\ref{Diagram_I}), and
(\ref{Diagram_G}) is given by

\begin{eqnarray}\label{Diagrams_Group4}
&& {\cal V}_4 \cdot 2 N_f \frac{d}{d\ln\Lambda}
\sum\limits_{I=0}^n (-1)^{P_I} \int \frac{d^4q}{(2\pi)^4}\Bigg\{ -
\frac{\partial}{\partial\ln g}\Bigg(\frac{2 M^2 J}{q^4 (q^2 G^2 +
M^2 J^2)}\Bigg)\nonumber\\
&& + \frac{q^\mu}{q^4} \frac{\partial}{\partial q^\mu} \Bigg(
\frac{2q^2 G + 2 M^2 J}{q^2 G^2 + |M J|^2} \Bigg)
\Bigg\}_I.\qquad\quad
\end{eqnarray}

19. The sum of Eqs. (\ref{Main_Sum}) and (\ref{Diagrams_Group4})
is equal to $\partial\Delta_2/\partial\ln g$ ($\Delta_2$ is given
by Eq. (\ref{Delta2}) or Eq. (\ref{Delta2_Explicit})):

\begin{eqnarray}
- {\cal V}_4 \cdot 2 N_f \frac{d}{d\ln\Lambda} \sum\limits_{I=0}^n
(-1)^{P_I} \frac{\partial}{\partial\ln g} \int
\frac{d^4q}{(2\pi)^4} \Big(\frac{2 M^2 J (J - 1)}{q^4 (q^2 G^2 +
M^2 J^2)}\Big)_I.
\end{eqnarray}

\noindent This completes the prove that the sum of diagrams
presented in Fig. \ref{Figure_Blue_Green_Diagrams} is equal to the
sum of diagrams presented in Fig. \ref{Figure_Blue_Diagrams}.


\end{document}